\documentclass[12pt]{article}
\pdfoutput=1
\usepackage[utf8]{inputenc}
\usepackage[english]{babel}
\usepackage{amsmath, amssymb}
\usepackage{hyperref}
\usepackage{epsfig}

\usepackage[usenames, dvipsnames, svgnames]{xcolor}
\usepackage{graphicx}
\graphicspath{{figures/}}
\usepackage{lineno}
\usepackage{cite}

\begingroup\expandafter\expandafter\expandafter\endgroup
\expandafter\ifx\csname pdfsuppresswarningpagegroup\endcsname\relax
\else
\fi

\newcommand{\lsim}{\raisebox{-0.13cm}{~\shortstack{$<$ \\[-0.07cm] $\sim$}}~} 
\newcommand{\gsim}{\raisebox{-0.13cm}{~\shortstack{$>$ \\[-0.07cm] $\sim$}}~} 
\newcommand{\tb}{\ensuremath{\tan\beta}}

\newcommand{\pt}{\ensuremath{p_\text{T}}}
\newcommand{\ttbar}{\ensuremath{{t\bar t}}}
\newcommand{\mtt}{\ensuremath{m_{\ttbar}}}

\newcommand{\intlumi}{\ensuremath{\mathcal L}}

\definecolor{orange}{rgb}{1,0.5,0}

\begin{document} 

\rightline{CERN-TH/2019-001, KCL-PH-TH/2018-05, LAPTH/001/19}
\vskip 1in

\begin{center}

{\large\bf Interference Effects in $t{\bar t}$ Production at the LHC \\ 
\vspace{0.2cm}
as a Window on New Physics}  

\vspace{1cm}

{\sc Abdelhak Djouadi$^{1,2}$, John Ellis$^{2,3,4}$, Andrey Popov$^{5,6}$ and J\'er\'emie Quevillon$^{7}$}

\vspace{1cm}

{\small {\it
$^1$ Universit\'e Grenoble Alpes, USMB, CNRS, LAPTh, F-74000 Annecy, France.\\
\vspace{0.15cm}

$^2$ NICPB, R{\"a}vala pst. 10, 10143 Tallinn, Estonia.\\
\vspace{0.15cm}

$^3$ Theoretical Particle Physics and Cosmology Group, Physics Department, \\
King's College London, London WC2R 2LS, UK. \\
\vspace{0.15cm}

$^4$ Theoretical Physics Department, CERN, CH-1211 Geneva 23, Switzerland. \\
\vspace{0.15cm}

$^5$ Institut de Physique Nucl\'eaire de Lyon, CNRS, 4 rue Enrico Fermi, 69622 Villeurbanne, France. \\
\vspace{0.15cm}

$^6$ Lomonosov Moscow State University, SINP MSU, 1(2) Leninskie gory, GSP-1 119991 Moscow, Russia. \\
\vspace{0.15cm}

$^7$ Laboratoire de Physique Subatomique et de Cosmologie, Universit\'e
Grenoble-Alpes, CNRS/IN2P3, 53 Avenue des Martyrs, 38026 Grenoble, France. } }

\end{center}

\vspace{1.cm}

\begin{abstract} 
\noindent 

Many extensions of the Standard Model (SM) contain (pseudo)scalar bosons with
masses in the TeV range. At hadron colliders, such particles would predominantly
be produced in gluon fusion and would decay into top quark pair final sates, a
signal that interferes with the large QCD background $gg \to t\bar t$. This
phenomenon is of interest for searches for by the LHC experiments. Here, we consider  the
signal and background interference in this process and study it in various
benchmark scenarios, including models with extra singlet (pseudo)scalar resonances,
two--Higgs doublet models (2HDM), and the minimal supersymmetric extension of the
SM with parameters chosen to obtain the measured light Higgs mass (the hMSSM).  
We allow for the possible exchanges of beyond the SM vector--like particles
as well as scalar quarks. We calculate the possible interference effects
including realistic estimates of the attainable detection efficiency and mass
resolution. Studies of our benchmark scenarios indicate that searches with an
LHC detector could permit the observation of the $t\bar t$ final states
or constrain significantly large regions of the parameter spaces of the
benchmark scenarios.

\end{abstract}

\newpage

\section{Introduction}

Searches for new physics at the CERN LHC exploit a variety of different
signatures, including searches for resonant enhancements  in various final
states, searches for non-resonant excesses in the distributions of one, two or
more particles, and events with missing transverse momentum. All of these
signatures have contributed to the discovery and characterization of the Higgs
boson \cite{Aad:2012tfa,Chatrchyan:2012xdj}, and are being pursued in ongoing
searches for new physics such as new $Z'$ or $W'$ gauge bosons
\cite{Aaboud:2017sjh,Sirunyan:2018exx,Khachatryan:2017wny,Aaboud:2018jux,Sirunyan:2018lbg}
or supersymmetry \cite{Aaboud:2018mna,Sirunyan:2018lul}. Many authors have
pointed out that interference effects in the spectra of specific particle pairs
are of particular interest for the discovery and exploration of spin-0 states.
One example is the constraint that off-shell interference in $ZZ$ final states
imposes on the total width of the observed Higgs boson
\cite{Caola:2013yja,Kauer:2012hd} and many authors have also emphasized the
potential utility of interference effects in $\gamma \gamma$
\cite{Dixon:2003yb,Martin:2013ula,Jung:2015sna} and $t{\bar t}$
\cite{Gaemers:1984sj,Dicus:1994bm,Moretti:2012mq,Frederix:2007gi,Hespel:2016qaf}
final states for constraining the properties of heavier spin-0 bosons. Some of
these studies were stimulated by the ill-fated hint of a 750-GeV feature in the
$\gamma \gamma$ spectrum at the LHC \cite{Jung:2015etr,Djouadi:2016ack}, but
their principles have broader applicability.

There are many scenarios for physics beyond the Standard Model (BSM) that
feature additional scalar or pseudoscalar bosons with masses in the few hundred
to the TeV range that could be accessible to experiments at the LHC. The
simplest of these models contain an additional SM-singlet (pseudo)scalar boson,
and many others postulate a second SU(2)-doublet of Higgs bosons (two--Higgs
doublet models or 2HDMs) \cite{Branco:2011iw}. Prominent among the latter is the
minimal supersymmetric extension of the SM (the MSSM) \cite{Djouadi:2005gj},
whose parameters may be constrained so that it predicts correctly the mass
$\simeq 125$~GeV of the observed Higgs boson (the hMSSM)
\cite{Djouadi:2013uqa,Djouadi:2015jea}. The interferences with SM backgrounds
that such bosons  would produce  in various two--body channels provide an
interesting physics opportunity for the LHC experiments and a novel window on
possible BSM physics.

Of particular interest are $t{\bar t}$ final states, for several reasons. If the
sought-for spin-0 boson has a mass in the range of several hundred GeV, the
decays into $t{\bar t}$ final states would be kinematically allowed with a
branching ratio that may be large, yielding a signal-to-background ratio that is
more favourable than for decays into lighter fermions, for instance. Moreover,
unlike the case of a spin-1 boson, there is no good reason why the couplings of
spin-0 bosons to fermions should be generation-independent. Indeed, this is
known not to be the case for the SM Higgs boson, which couples to particles
proportionally to their mass and, hence, has a very large $t{\bar t}$ coupling. 
This is also the case in 2HDMs such as the MSSM if the ratio, $\tb$, of the two
Higgs vacuum expectation values (vevs) is not very large, and the additional
scalar and pseudoscalar states $\Phi=H/A$ states are relatively heavy, in which
case the decay modes $\Phi \rightarrow t\bar t$ tend to dominate. In such a case
the $gg\rightarrow H/A$ production cross sections are still substantial, thanks
to the large Higgs coupling to the top quarks that mediate the production
process through triangle diagrams, and may be further enhanced if there are
additional vector--like quarks (VLQs) or scalar quarks (such as the stop squarks
in the MSSM). Therefore, searches at the LHC for heavy scalar bosons decaying
into $t{\bar t}$  final states are of special interest, even mandatory as the
only way to probe regions of the 2HDM or MSSM parameter space with large
$M_{H/A}$ and small $\tb$ values.

As we show in this paper using various benchmark scenarios, including a model
with an extra scalar singlet and models with two Higgs doublets such as  the 
2HDM and the MSSM, interference signatures in the cross section for producing
$t{\bar t}$ final states at the LHC provide distinctive and sensitive signatures
for searches of new physics beyond the SM.

Unlike the case of a spin--one electroweak resonance, a peak in the $t{\bar t}$
invariant mass distribution,  as expected in the narrow--width approximation, is
not the only possible signature of a resonance. Indeed, in the case  of
resonances such as heavy neutral gauge bosons $Z'$ or bosonic Kaluza--Klein
excitations, this situation does not occur: these states are mainly produced in
the $q\bar q$ annihilation channel and the electroweak process $q \bar q
\rightarrow V \rightarrow t \bar t$ does not interfere with the  background from
the colored $q\bar q \rightarrow t \bar t$ channel due to $s$-channel gluon
exchange, leading one to expect only a peak  in the invariant mass distribution
in these cases~\footnote{However, in the general electroweak processes $q\bar q
\to f\bar f$, with $f=e,\mu$ for instance, there is also an interference between
the  contribution of the spin-1 $V$ boson and the $Z,\gamma$ exchange
contributions; but as the  new resonance $V$ is expected to be rather heavy and
narrow, the effect of the interference is  entirely negligible if one restricts
to invariant fermion masses close to the resonance mass
\cite{Accomando:2011eu,Accomando:2013sfa}. This interference can be  important
in $e^ + e^-$ collisions though, in particular to detect the resonance effects
below the kinematical threshold where it can be produced on--shell, see for
instance Ref.~\cite{Djouadi:1991sx}. Note that interference effects in the
production of Kaluza-Klein gluons in $gg\!\to\! g_{KK}\! \to \!t\bar t$  have
been discussed in e.g. Ref.~\cite{Djouadi:2007eg} but they are small.}. In turn,
$gg\rightarrow H/A$ production  followed by the $H/A \to t \bar t$  decay in a
2HDM for instance will interfere with the QCD  background for top quark pair
production that, at LHC energies, is mainly generated by the gluon--fusion
process $gg \rightarrow t{\bar t}$. The signal-to-background  interference
pattern will depend on the CP nature(s) of the $\Phi=H/A$ boson(s) as well as
mass(es) and total decay width(s) \cite{Gaemers:1984sj,Dicus:1994bm}. The
interference may be either constructive or destructive, varying across the
resonance mass, leading to a rather complex signature exhibiting a peak-and-dip
structure in the $t{\bar t}$ invariant mass distribution.

These issues were discussed at the theoretical level in
Refs.~\cite{Djouadi:2013vqa,Djouadi:2015jea} in the context of the MSSM, which
contains, in addition to a light CP--even $h$ boson that is identified with the
125 GeV Higgs state observed at the LHC, another CP--even or scalar particle
$H$, a CP--odd or pseudoscalar particle $A$ and two charged Higgs states
$H^\pm$.   In this case, the situation is relatively simple to discuss as the
Higgs sector can be described using only two input parameters,  namely $\tb$ and
$M_A$, if the value $M_h=125$ GeV is accounted for by the large radiative
corrections that occur in the Higgs sector, the so-called hMSSM approach
\cite{Djouadi:2013uqa,Djouadi:2015jea}.  When $M_A \gsim 2m_t$, the MSSM
approaches the decoupling limit where $h$ is SM--like while $M_H \approx
M_{H^\pm} \approx M_A$ and $A/H$ have similar couplings. One can thus describe
the $gg \to H/A \to  t\bar t$ process  simply in terms of the two inputs $M_A$
and $\tb$, and hence  in the  $[M_A, \tb ]$ parameter space, like other
Higgs-sector processes in the MSSM. 

However, in a general 2HDM there are in principle seven parameters: the four
Higgs masses, two mixing angles  $\alpha$ and $\beta$ and the mixing mass
$m_{12}$, which render a complete analysis a daunting task.  We therefore need a
benchmark scenario suitable for experimental studies of the process and which
illustrates the main important points. Here, we show that one can describe the
$t{\bar t}$ search in some generality by:\vspace*{-3mm}

\begin{itemize}

\item assuming the alignment limit $\alpha \simeq \beta-\frac{\pi}{2}$, which
renders the lighter $h$ boson SM--like as favoured by LHC data
\cite{Aad:2015pla,CMS:2016qbe}, which suppresses many  decay channels that might
otherwise compete with the $H/A \rightarrow t\bar{t}$ mode, such as
$H\rightarrow WW,ZZ,hh$ or $A\rightarrow hZ$, and makes gluon fusion (together
with associated production with $t,b$ quarks) the only $H/A$ production
mode;\vspace*{-3mm}  

\item assuming a small difference between the $H$ and $A$ masses, $|M_H - M_A|
\lsim M_Z$ in most cases,  so as to be consistent with the electroweak precision
data (particularly the constraint from the $\rho$ parameter)
\cite{deBlas:2016ojx,Haller:2018nnx}, thereby suppressing many other decay
channels that might compete with $H/A \rightarrow t{\bar t}$, such as
$H\rightarrow AZ$ or $A\rightarrow HZ$; we also assume that $H^\pm$ is heavier
than max$(M_A,M_H)$ in order to suppress decays like $H/A \rightarrow
WH^+$;\vspace*{-3mm}

\item finally, ignoring the mass mixing parameter $m_{12}$, as it enters only in
the Higgs self-couplings that do not affect our discussion here.\vspace*{-3mm}

\end{itemize}

This benchmark scenario is almost as simple as the hMSSM, and allows us to study
the process $pp \rightarrow H/A\rightarrow t{\bar t}$ also in the $[M_A, \tb]$
plane. In this initial exploration, in addition to the assumptions above, we
restrict our attention to the range $M_{A,H} \gsim 350$ GeV in which the
two--body $A/H \to t\bar t$ is kinematically allowed~\footnote{The three-body 
decay $H/A \to t t^* \to tbW$  below the kinematic threshold is in general
suppressed, but can be significant for values of $\tb$ close and below unity;
see for instance Ref.~\cite{Djouadi:1995gv}.} and to values $\frac12 \leq \tb
\leq 5$--7 for which the $\Phi t\bar t$ couplings are large and, hence, the
$t\bar t$ branching fractions are significant. 

As well as the case of the SM but with the addition of an isosinglet scalar or
pseudoscalar boson that interacts with the top quark  with a coupling simply
$\propto m_{t}/v$, we also consider the possibility of an enhancement of the $gg
\to H/A$ triangular loop contribution by additional matter particles. We will
take two examples for illustration: $i)$ a vector--like  quark (VLQ) with a mass
in the TeV range and general couplings to the (pseudo)scalar resonance, and
$ii)$ the scalar top squark in the MSSM, which   can be relatively light and has
large couplings to the CP--even $H$ boson. The  shape and magnitude of the
interference between signal and backgrounds can be  significantly modified  in
these cases. 

The rest of the paper is organised as follow. In the next Section, we describe
and give the analytical forms at leading order of the pure signal, pure
background and interference contribution in the $gg\to t\bar t$ process. In
Section~\ref{Sec:Models}, we introduce the three benchmark scenarios that we
will consider, the isosinglet resonance, the 2HDM and MSSM cases, and the models
in which extra contributions from VLQs or scalar top squarks are present. In
Section~\ref{sec:Simulation}, we present the method of computation of the
expected sensitivity and exclusion potential for the $t\bar t$ signal. This
analysis targets the $\ell + \text{jets}$ final state and exploits the
\mtt~distribution smeared to emulate the detector resolution. In
Section~\ref{Sec:Results}, we present the potential sensitivities in all the
benchmark scenarios that we consider. A brief summary is given in
Section~\ref{Sec:Summary}.  

\section{The \texorpdfstring{$\mathbf{gg \rightarrow H/A \rightarrow t\bar{t}}$}{gg -> H/A -> tt} process at the LHC}
\label{Sec:Process}

The discussion in the Introduction suggests that the gluon fusion production
processes  $gg \rightarrow \Phi=H/A$, followed by decays of the Higgs bosons
into top quark pairs, $H/A \to t \bar t$, are promising channels for exploring
the parameter spaces of 2HDMs such as the MSSM. This is the case particularly
for low $\tb$ values: the $\Phi t\bar t$ couplings are only slightly suppressed
for $\tb \lsim 3$--5 and enhanced for $\tb \lsim 1$,  compared to those in the
SM, and the $\Phi \rightarrow t\bar t$ decay modes are by far the dominant
ones.   The search for a peak in the invariant mass distribution of the top
quark pairs produced at the LHC,  which would signal the presence of a resonance
that couples to $t\bar t$ states,  is thus a priority. It would complement the
search for resonances decaying into $\tau\tau$ pairs that,  as is well known,
are more sensitive to the high- and moderate-$\tb$ regions of the MSSM. 
However, a peak in the invariant mass of the $t\bar t$ system, which  one
generally expects in the  narrow--width approximation that is valid in the case
of  the models we study, as the $H/A$ total decay widths are in general small, 
is not the only possible signature of a Higgs resonance. 

\begin{figure}[!ht]
\vspace*{-4mm}
\begin{center}
\mbox{ \epsfig{file=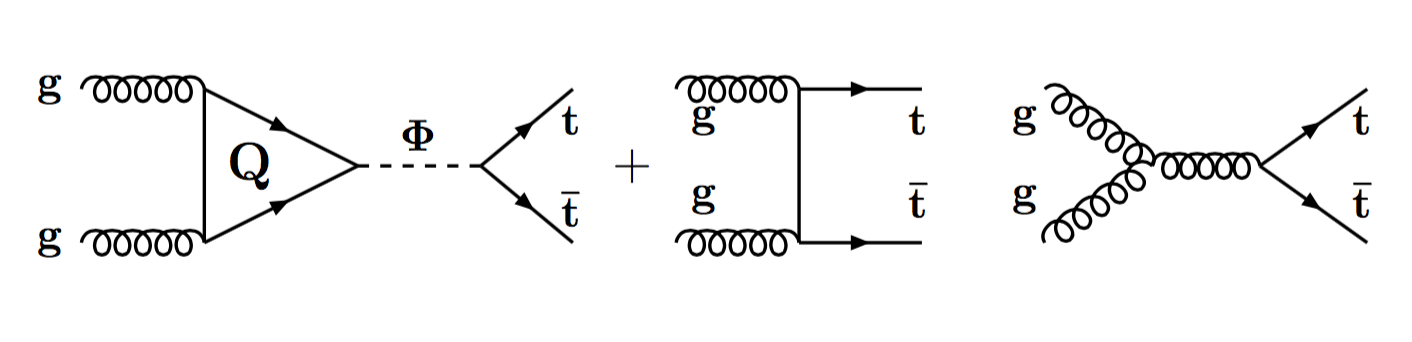,width=16cm} }
\end{center}
\vspace*{-1.4cm}
\caption{\it Feynman diagrams for the signal process $gg \rightarrow \Phi
\rightarrow t\bar t$ and the QCD process $gg \rightarrow t\bar t$ that is the
dominant background at the LHC. The state $\Phi$ may represent either a CP--even
state $H$ or a CP--odd $A$ state.}
\label{fig:ggtt}
\end{figure}

This is because the amplitude for the Higgs signal due to the gluon fusion
process $gg\rightarrow \Phi$ interferes with the QCD amplitude for the
pair-production of top quarks, which is mediated by the $q\bar q \rightarrow
t\bar t$ annihilation and $gg\rightarrow t\bar t$ fusion channels,  the latter
being  by far the  dominant component at the LHC. The two Feynman diagrams for
the QCD $gg\rightarrow t\bar t$ process at leading order are shown in 
Fig.~\ref{fig:ggtt}, together with the Feynman diagram for the resonant signal.
The interference depends on the CP nature of the $\Phi$ boson as well as on its
mass and total width, and may be either constructive or destructive, leading to
a rather complex signature with a  peak--dip structure in the $t\bar t$
invariant mass distribution, as already mentioned.

This feature has been known for some time, and has in particular been discussed
in the context of a heavy  SM Higgs
boson~\cite{Gaemers:1984sj,Dicus:1994bm,Moretti:2012mq} and, by extension,  also
for the case of a CP--even Higgs boson in a 2HDM~\cite{Hespel:2016qaf}. However,
in this case the situation is more complicated, as there are two possible
resonances with different CP quantum numbers and hence two possible peaks if the
interference is positive and/or dips if the interference is negative,  which
would be close together if the $H$ and $A$ states are nearly degenerate.
Compared to the SM Higgs case, this situation has been relatively rarely
discussed  in the literature, though see for instance 
Refs.~\cite{Bernreuther:1997gs,Barger:2006hm,Barcelo:2010bm,Figy:2011yu,Craig:2015jba} 
and the detailed analyses of Refs.~\cite{Frederix:2007gi,Carena:2016npr}, in
which many resonances  including CP--even and CP--odd scalar particles have been
studied. In the following we summarize briefly the main aspects.

\begin{figure}[!ht]
\vspace*{-2.3cm}
\begin{center}
\mbox{ \hspace*{-2cm}
\epsfig{file=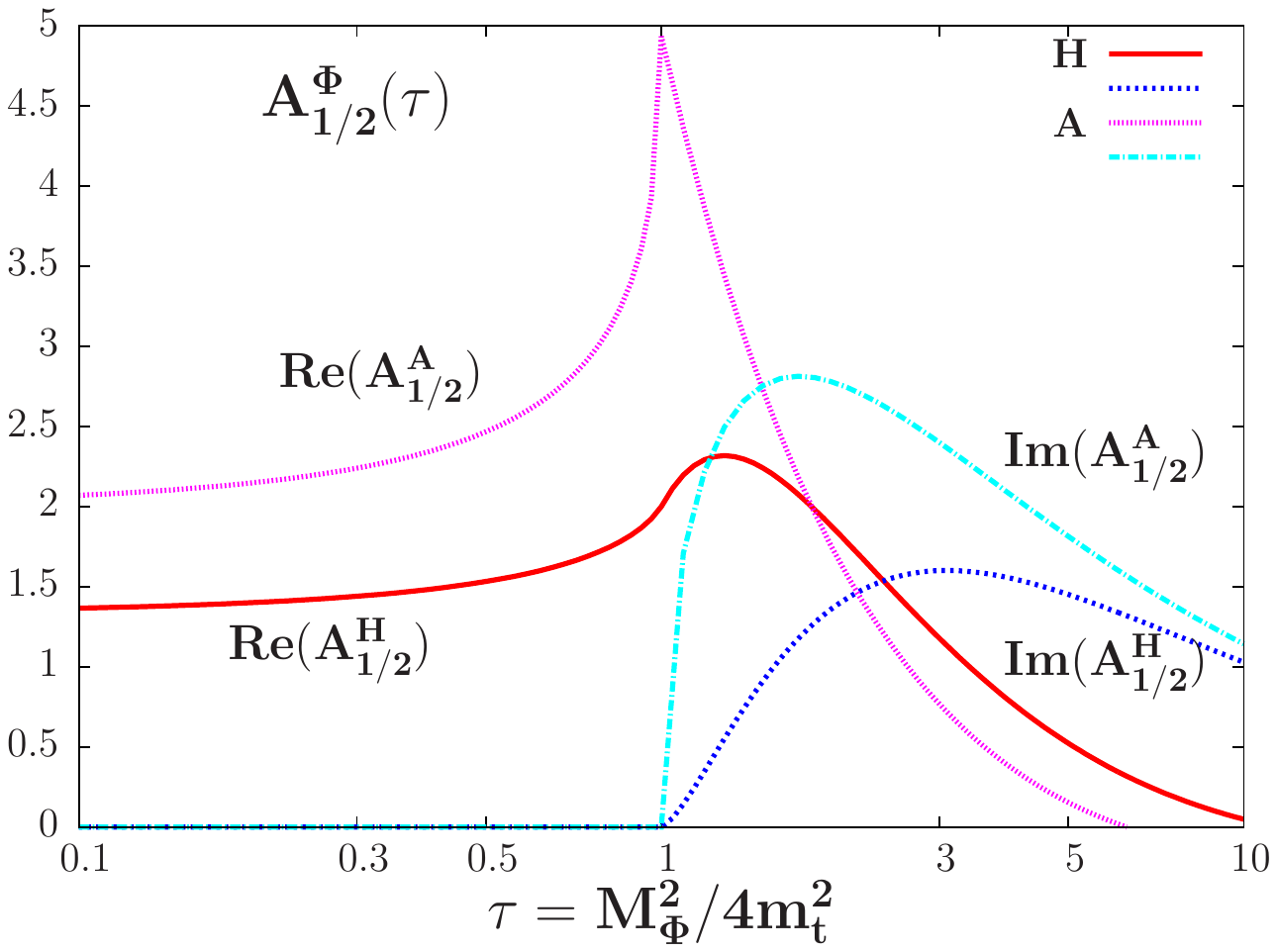,width=16.3cm} }
\end{center}
\vspace*{-13.2cm}
\caption{\it The  real  and  imaginary  parts  of  the  form  factors
$A_{1/2}^{\Phi}$ for the contributions of top quarks as functions of the
variable $\tau \equiv M_\Phi^2 / 4 m_t^2$ in the CP--even $H$ and CP--odd $A$
cases.} 
\label{fig:ffA}
\end{figure}

The amplitude for the $gg (\rightarrow \Phi ) \rightarrow t\bar{t}$ process,
including the contributions of both the resonant signal production of the state
$\Phi$ with mass $M_\Phi$ and total decay width $\Gamma_\Phi$, and the continuum
background, may be written as
\begin{equation}
{\cal A}^{\Phi }_{gg \rightarrow t\bar t} \; =    \;
- \sum_\Phi   \frac{ {\cal A}_{gg\Phi}   \, \hat s {\cal A}_{\Phi tt} } { \hat s -M_\Phi^2 +i  M_\Phi \Gamma_\Phi }  +  {\cal A}_{ggtt} \, ,
\end{equation}
where $\hat s$ is the partonic centre-of-mass energy-squared and ${\cal
 A}_{ggtt}$ is the tree-level QCD background amplitude.  The amplitude ${\cal
 A}_{gg\Phi}$ for resonance production $gg \rightarrow \Phi$ is induced at LO by
 loops of  heavy quarks $Q$ and is given by \cite{Djouadi:2005gi}
\begin{equation}
\label{A12}
{\cal A}_{gg \Phi} \; = \; \frac{\alpha_s}{8 \pi v} \hat s \sum_{Q} \hat g_{\Phi QQ} A_{1/2}^\Phi (\tau_Q) \, ,
\end{equation}
where the  form factors $A_{1/2}^{\Phi}$ for the contributions of
spin--$\frac12$ quarks as functions of the variable $\tau_Q \equiv M_\Phi^2 / 4
m_Q^2$ are given in the CP--even $H$ and CP--odd $A$ cases by 
\begin{eqnarray}
& A_{1/2}^{H}(\tau) \;  = \; 2 \left[  \tau +( \tau -1) f(\tau)\right]  \tau^{-2} \, , \ \  
A_{1/2}^{A} (\tau) \; = \; 2 \tau^{-1} f(\tau) \, :  \label{eq:Af}  
\end{eqnarray}
with
\begin{eqnarray}
& f(\tau) \ \equiv \; \left\{ \begin{array}{ll}  \displaystyle \arcsin^2\sqrt{\tau } & 
{\rm for} \;  \tau \leq 1 \, , \\
\displaystyle -\frac{1}{4}\left[ \log\frac{1+\sqrt{1-\tau^{-1} }} 
{1-\sqrt{1-\tau^{-1}}}-i\pi \right]^2\hspace{0.5cm} & {\rm for} \; \tau > 1 \, .
\end{array} \right.
\label{eq:formfactors}
\end{eqnarray}
The form factors are displayed in Fig.~\ref{fig:ffA} as a function of the
variable $\tau$. They vanish in the zero--mass limit for the fermions, while in
the infinite-mass limit they reach constant values $A_{1/2}^H \rightarrow
\frac43$ and $A_{1/2}^A \rightarrow 2$. They are real below the kinematical
threshold $M_\Phi = 2m_Q$ and develop imaginary parts above, reaching their
maximum values just above the threshold.  

The partonic differential cross section can be written in a convenient way as 
\begin{eqnarray} 
\frac{ {\rm d}\hat \sigma }{{\rm d} z} \; = \; \frac{ {\rm d}\hat \sigma_B }{{\rm d} z}
+ \frac{ {\rm d}\hat \sigma_S }{{\rm d} z} + \frac{ {\rm d}\hat \sigma_I }{{\rm d} z} \, ,
\label{ggparton}
\end{eqnarray}
where $z \equiv \cos\theta$, with $\theta$ the scattering angle in the
parton-parton centre-of-mass between an incoming gluon and the top quark. The
various terms in eq.~(\ref{ggparton}) can be written as~\cite{Dicus:1994bm}
\begin{equation}
\begin{aligned}
 \frac{ {\rm d}\hat \sigma_B }{{\rm d} z} &=  \frac{\pi \alpha_s^2}{6 \hat s }  \hat \beta_t \left( \frac {1}{1- \hat \beta_t^2 z^2} -\frac{9}{16}  \right) \bigg[ 3+ \hat \beta_t^2 z^2 - 2 \hat \beta_t^2  - \frac{ 2 (1- \hat \beta_t^2)^2}{1- \hat \beta_t^2 z^2} \bigg] \, , \\
 \frac{ {\rm d}\hat \sigma_S }{{\rm d} z} &=  \frac{3 \alpha_s^2 G_F^2 m_t^2}{8192 \pi^3} 
\hat s^2  \sum_\Phi  \frac{ \hat \beta_t^{p_\Phi} | \hat g_{\Phi t\bar t}^{2}  
A_{1/2}^\Phi (\tau_t) |^2} {(s- M_\Phi^2) ^2+ \Gamma_\Phi^2 M_\Phi^2}  \, , \\
 \frac{ {\rm d}\hat \sigma_I }{{\rm d} z} &=  - \frac{\alpha_s^2 G_F m_t^2}{64\sqrt 2  \pi} 
\frac {1}{1- \hat \beta_t^2 z^2} {\rm Re} \bigg[ \sum_\Phi \frac{ \hat \beta_t^{p_\Phi} 
\hat g_{\Phi t\bar t}^{2}  A_{1/2}^\Phi (\tau_t) }
{s- M_\Phi^2+i \Gamma_\Phi M_\Phi} \bigg] \, ,
\label{Dicus}
\end{aligned}
\end{equation}
where the velocity of the final top quark in the parton-parton centre-of-mass
frame is $\hat \beta_t \equiv \sqrt{1-4m_t^2 /\hat{s}}$, and  $p_\Phi =3$~(1)
for the CP--even (CP--odd) scalar. With our convention for the $\Phi t \bar t$
coupling, the SM Higgs coupling corresponds to $\hat g_{h t \bar t}=1$.  The
expressions eq.~(\ref{Dicus}) involve the energy-dependent heavy (pseudo)scalar
partial width~\cite{Djouadi:2005gi}:
\begin{eqnarray}
\Gamma (\Phi \rightarrow t \bar{t} ) \; \equiv \; \Gamma_{\Phi} (\hat{s}) \;  = \; 3 \frac{ G_F m_t^2}{4\sqrt{2} \pi}
\, \hat g_{\Phi t\bar t}^2 \, \beta^{p_\Phi}_t \frac{\hat{s}}{M_{\Phi}} \, .
\end{eqnarray}
The total cross sections for the signal, the background and interference are
then obtained by integrating the partonic cross sections over the scattering
angle $\theta$  and folding them with the $gg$ luminosity function; see
eq.~(\ref{Eq:PDFConvF}).

\begin{figure}[!ht]
 \centerline{
 \includegraphics[width=0.5\textwidth]{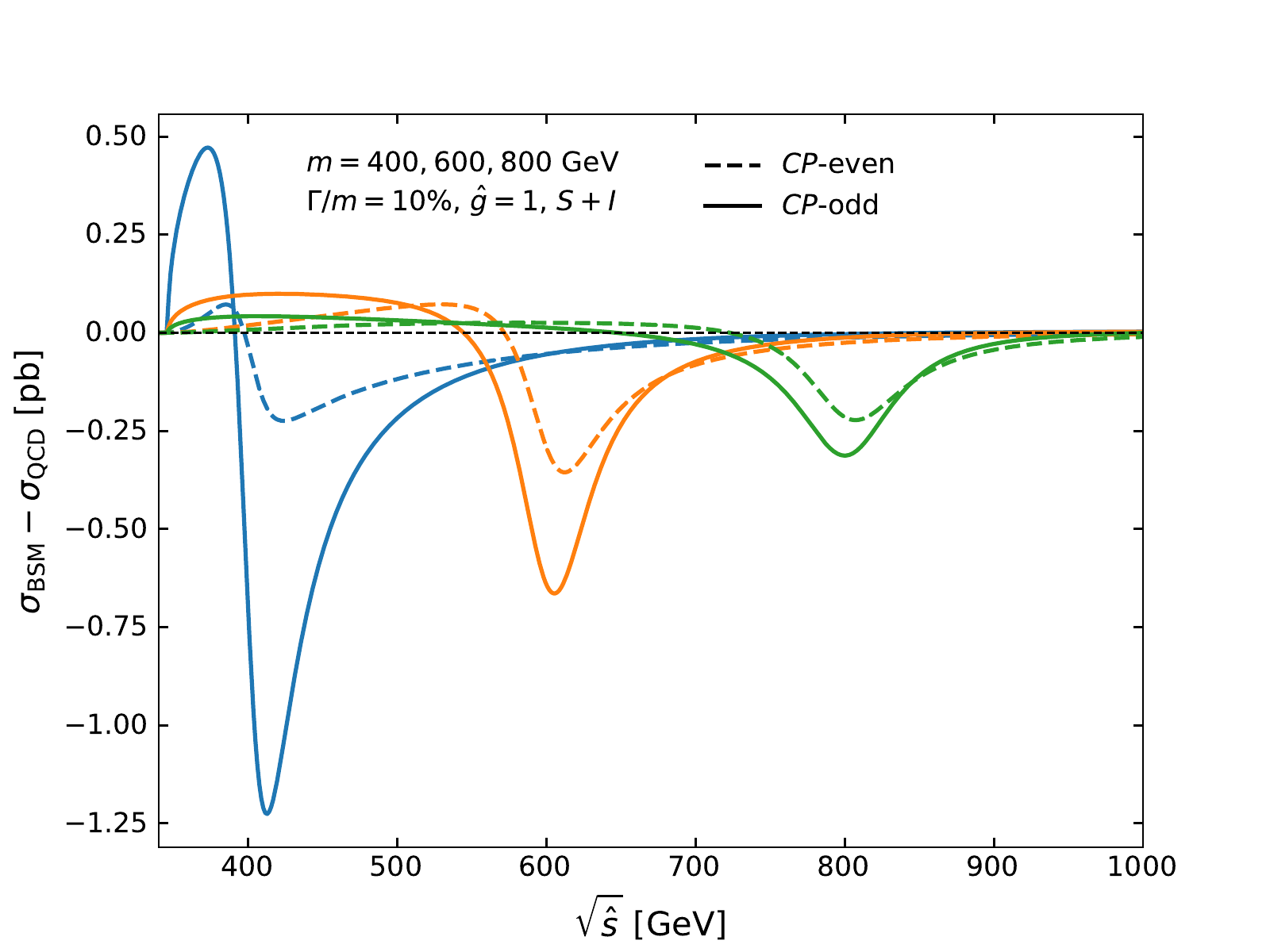}
 \includegraphics[width=0.5\textwidth]{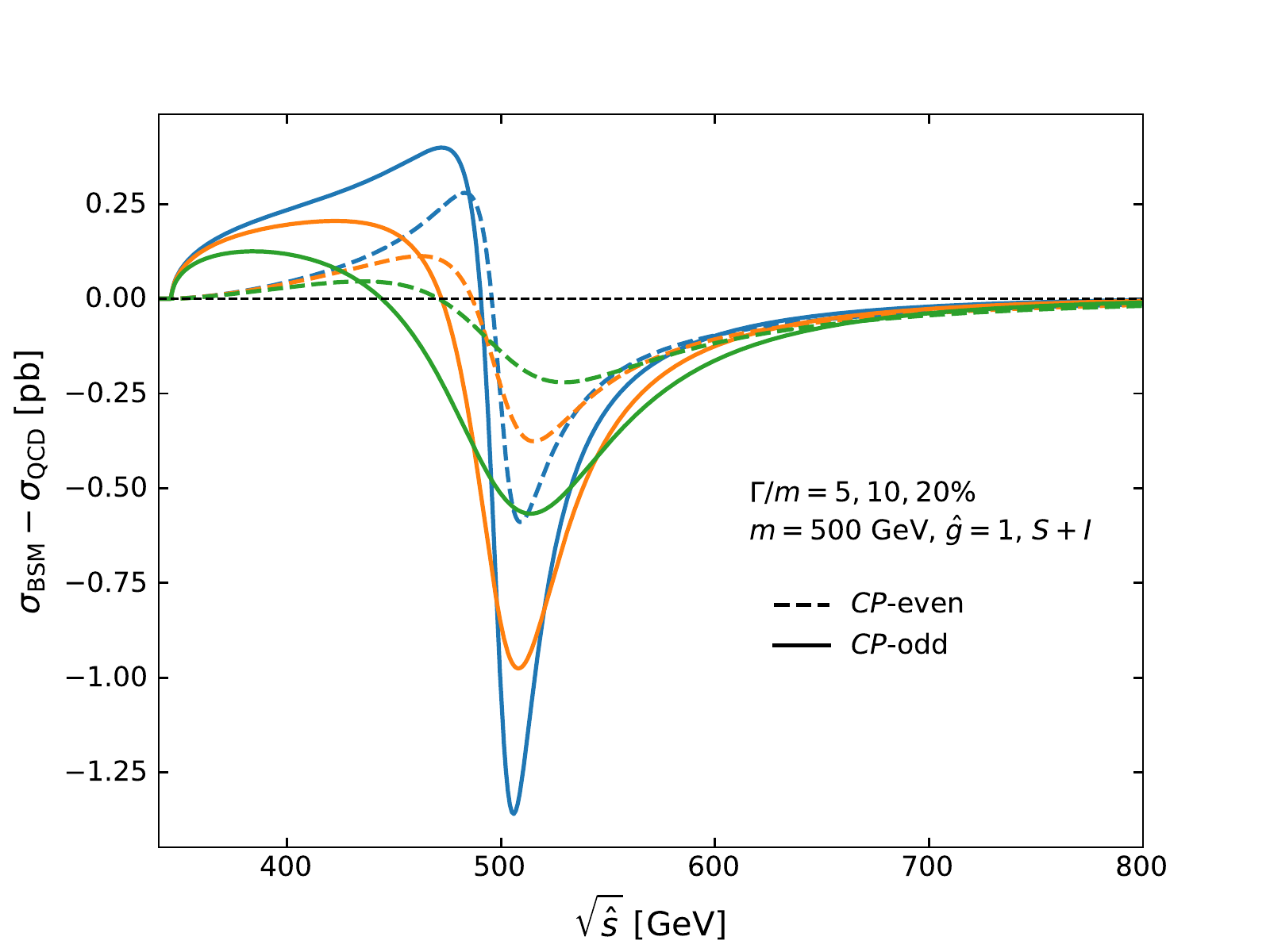} }
 \centerline{
 \includegraphics[width=0.5\textwidth]{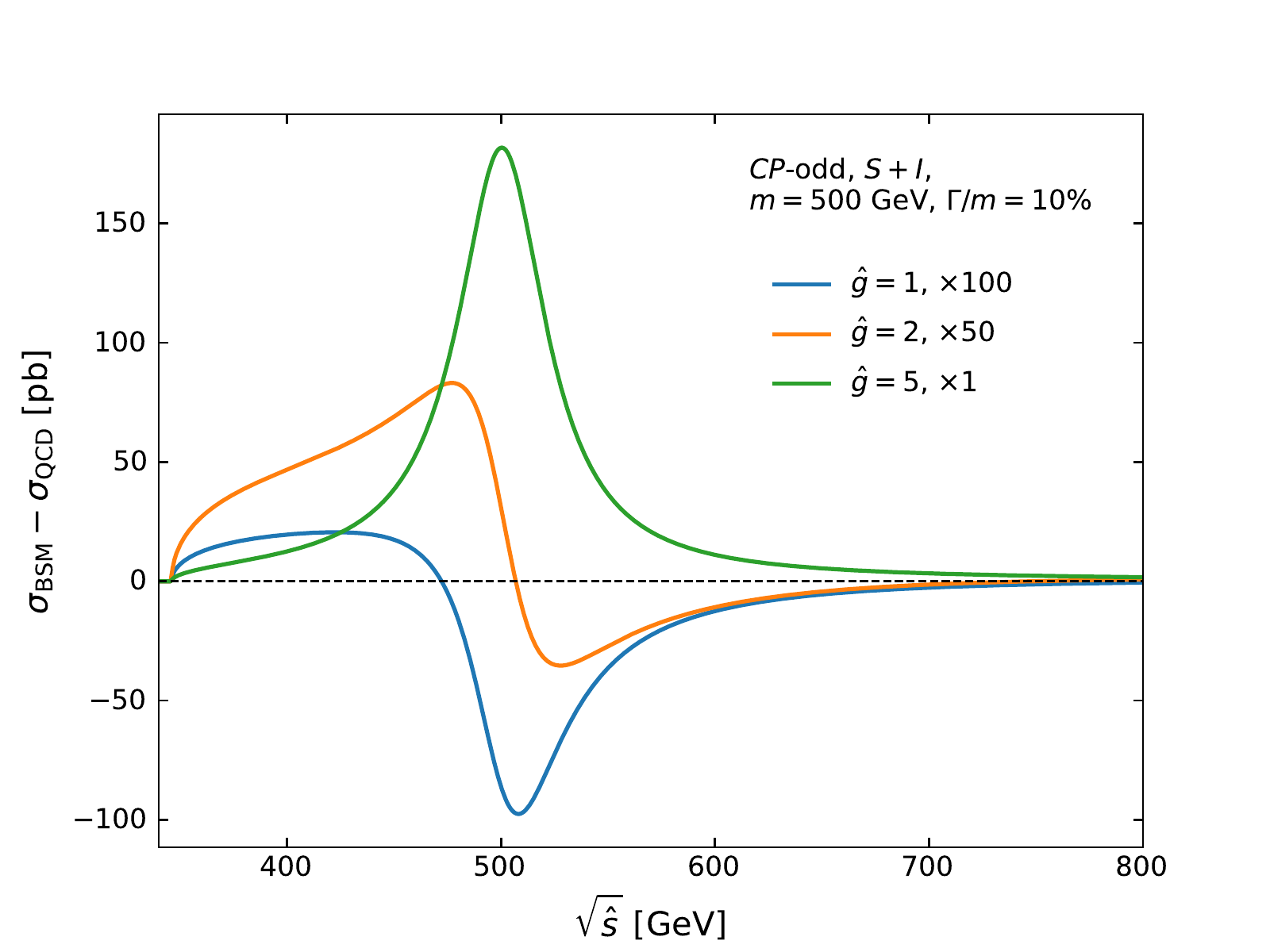}
 \includegraphics[width=0.5\textwidth]{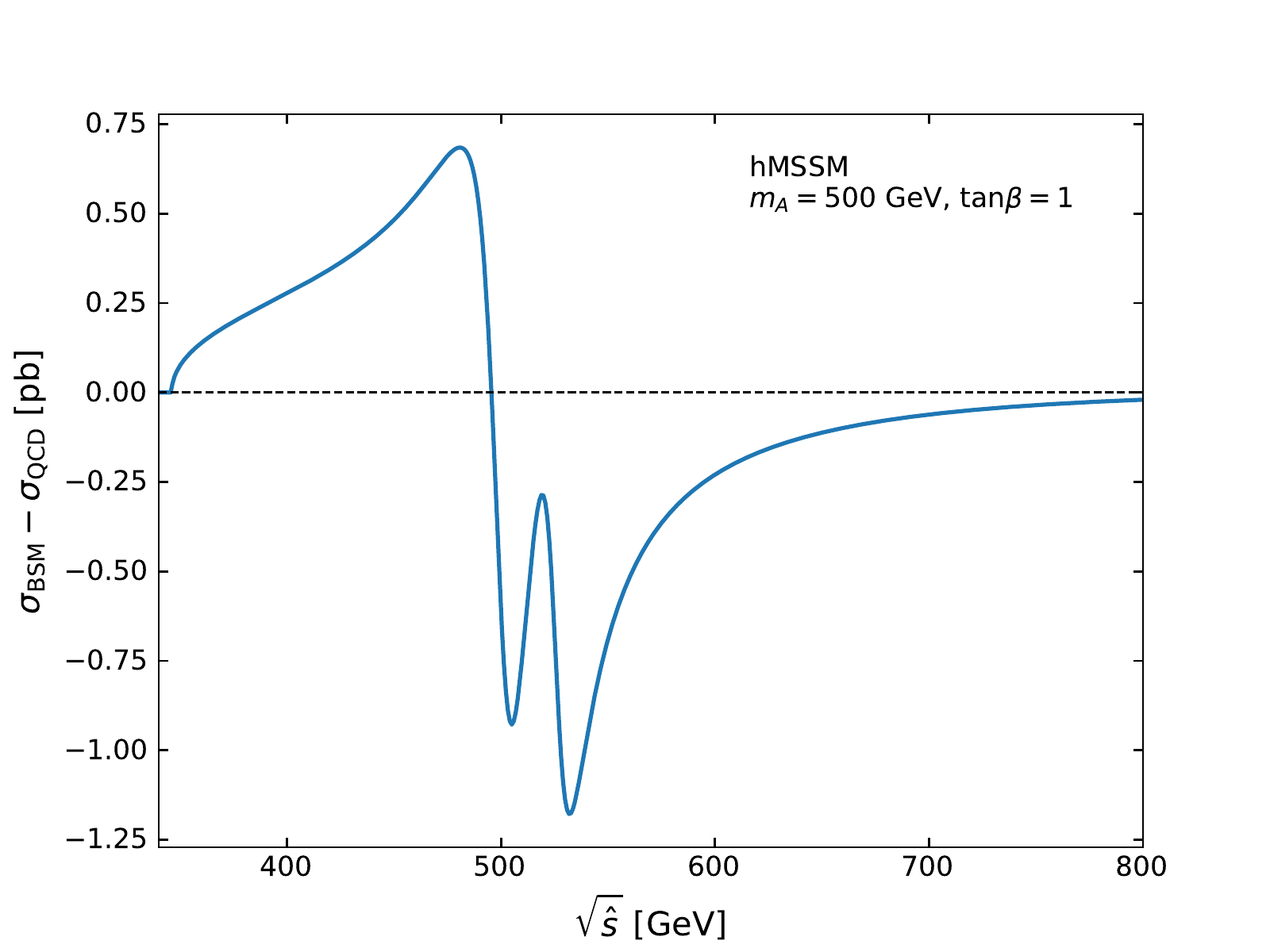} }
 \caption{\it Examples of parton-level cross sections for the sum of the signal
process and the interference.  The top row shows cross sections for different
masses (left) and total widths (right). The lower left plot shows the dependence
on the value of the $\hat g_{\Phi t\bar t}$ coupling, with the cross sections
scaled by additional factors specified in the legend. The lower right plot shows
the cross section for a sample point in the hMSSM.} 
\label{Fig:PartonXSec}
 \end{figure}

Figure~\ref{Fig:PartonXSec} shows examples of parton-level cross sections
computed according to eqs.~(\ref{Dicus}) but with fixed widths. These plots
illustrate the richness of possible effects due to interference. The top left
plot shows the parton-level cross section for the sum of the signal process and
the interference  for different masses of the extra singlet scalar (dashed line)
or pseudo-scalar (full line). The top right plot shows the parton-level cross
section for the sum of the signal process and the interference  for a given mass
of $500$~GeV for the extra singlet scalar (dashed line) or pseudo-scalar (full
line) but for different values of its width. The lower left plot shows the
dependence on the value of the $\hat g_{\Phi tt}$ coupling, with the cross
sections  scaled by additional factors for convenience, as specified in the
legend. Examples of the effects of the smearing  of peak and dip effects due to
the mass resolution will be shown later. Other models to be discussed later
exhibit analogous features.

In the case where the quark mass is such that $2m_Q>\sqrt{\hat s}$, the
amplitude $A_{1/2}^\Phi (\tau)$ (or the function $f(\tau)$ defined earlier) is
real, and the interference given by the last term in eq.~(\ref{Dicus}) vanishes
at $\hat s= M_\Phi^2$. On the other hand,  for $2m_Q <\sqrt{\hat s}$, which is
always the case of the bottom--quark loop and the  top--quark loop above the
$M_\Phi = 2m_t$ threshold, the form factor $A_{1/2}^\Phi (\tau)$ develops an
imaginary part that  leads to a non--zero interference term when $\hat s
=M_\Phi^2$. The magnitude of this  interference is controlled  essentially by
the  total decay width of the $\Phi$ boson,  $\Gamma_\Phi$, and is larger for
smaller value of $\Gamma_\Phi$. In the cases that we consider here, i.e., for
$\tb$ not much smaller than unity, $\Gamma_\Phi/M_\Phi$ is indeed small, namely
$\lsim 5$\%~\footnote{This is  in  contrast to the case of a heavy SM--like
Higgs boson, which would have had a large total decay width, ${\cal O}$(500 GeV)
for a mass $\sim 1$~TeV~\cite{Djouadi:1997yw,Baglio:2010ae}. This would have
made the interference between signal and background irrelevant in that case.}.  

Since the interference term may be large, the narrow--width approximation with a
signal characterized by an excess in  $\sigma(gg\rightarrow \Phi) \times {\rm
BR}(\Phi \rightarrow t\bar t)$ on top of the QCD background is not adequate in
general, and one would observe a more complicated structure in the $t\bar t$
invariant-mass distribution. Depending on the Higgs masses and couplings,  a
peak and/or dip structure should be present, as seen in
Fig.~\ref{Fig:PartonXSec}. In the following discussion we quantify the 
sensitivity to the Higgs signal in the process $pp \rightarrow \Phi \rightarrow
t\bar{t}$ with $\Phi=H/A$ including the interference terms. 

All the amplitudes quoted above were given at leading order (LO) in perturbation
theory and important higher-order effects need to be included.  The QCD
corrections to the $gg \rightarrow \Phi$ production cross section, beyond the 
$M_\Phi =2m_t$ threshold that is relevant here, have been calculated only at NLO
\cite{Spira:1995rr} and lead to a $K$--factor of about 1.6. It turns out,
however, that the result in the heavy top quark limit is a good approximation 
at this order \cite{Djouadi:1991tka,Dawson:1990zj,Spira:1993bb}  (at least for
the real part). In this approximation, the corrections have been calculated at
NNLO in the scalar and pseudoscalar cases
\cite{Harlander:2002wh,Anastasiou:2002yz,Ravindran:2003um,Harlander:2002vv} (and
in the former case even at N3LO \cite{Anastasiou:2015ema}). They have been
implemented in recent years in the code {\tt
SusHi}~\cite{Harlander:2012pb,Harlander:2016hcx} and lead to a $K$-factor of
about 2 in total. The QCD corrections to the decay $\Phi \rightarrow t\bar t$
are known  to NNLO in QCD \cite{Bernreuther:2018ynm} and increase the LO partial
width by approximately a factor $\sim 1.5$.  In the case of the $pp\rightarrow
t\bar t$ QCD background process, the QCD corrections at NLO have been known for
a long time, and the resulting $K$--factor is $K_{\rm NLO}^{\rm QCD} \approx
1.3$~\cite{Nason:1987xz,Beenakker:1988bq}. The NNLO QCD corrections to the
process have also been completed recently \cite{Czakon:2013goa}  and they 
increase the total cross section slightly beyond the NLO value.  The electroweak
corrections are rather small in both the  signal and background processes, and
can be ignored to first approximation. 

The NLO QCD corrections to the interference between signal and background in
resonant the process $gg\to H/A \to t\bar t$ have been calculated recently using
an effective field theory approach \cite{BuarqueFranzosi:2017jrj}. In our case,
we take into account the QCD corrections  simply by rescaling the Higgs signal
and the QCD background cross sections by the respective NNLO $K$-factors $K^{\rm
S}_{\rm NNLO}$ (computed using the program {\tt SusHi}) and  $K^{\rm QCD}_{\rm
NNLO}$.  We then rescale the interference term by the geometrical average of the
signal and QCD background NNLO $K$--factors~\cite{Hespel:2016qaf},  namely
$K^{I}_{\rm NNLO}=\sqrt{K^{\rm QCD}_{\rm NNLO} \times K^{S}_{\rm NNLO}}$ (see
Section~\ref{sec:Simulation} for more details about the analysis). 

\section{Benchmark models}
\label{Sec:Models}

\subsection{Extended Higgs sectors}

\subsubsection{The SM with an extra singlet (pseudo)scalar}
We consider first a minimal benchmark model, namely the SM supplemented by just
one of  the following terms for the interaction of a heavy scalar $H$ or
pseudoscalar $A$ with the top quark:
\begin{eqnarray}
\mathcal{L}^{\rm new Yukawa}\supset -g_{Ht\bar t} \bar{t}t H \quad {\rm or} \quad i g_{At\bar t} \bar{t} \gamma_5 t A \, .
\label{SP}
\end{eqnarray}
In a complete model, the field $H$ or $A$ should be integrated into an
electroweak doublet, or the interactions should be generated via a dimension-5
(or greater) interaction including an SM Higgs doublet, but we ignore such
complications here, for the purposes of illustration.

We use the SM--like Higgs coupling to fermions as a reference, expressing these
new scalar  and pseudoscalar couplings to the top quark in the form:
\begin{eqnarray} 
g_{\Phi t \bar t}= \frac{m_t}{v} \times \hat g_{\Phi t \bar t}    
\label{eq:yukawa}
\end{eqnarray} 
where $\Phi = H$ or $A$, and with the vev $v=1/\sqrt{\sqrt{2}G_F}=246$ GeV. With
this definition, the SM Higgs coupling $g_{h t \bar t}= {m_t}/{v}$
corresponds to $\hat g_{h t \bar t}=1$. As usual, the Feynman rules associated
with these couplings are obtained by multiplying $g_{\Phi t \bar t}$ by $(-i)$.
These new interactions induce $gg \Phi$ couplings via quantum corrections.  

\subsubsection{Two Higgs doublet models}

An excellent benchmark for studying extended Higgs sectors with a richer
phenomenology is a model  in which two Higgs doublets $\Phi_1$ and $\Phi_2$
break the electroweak symmetry (for a review on 2HDMs, see
Ref.~\cite{Branco:2011iw}, for example). It leads to five physical states:  two
CP--even neutral bosons, $h$ and $H$,  a  CP--odd boson, $A$, and two charged
$H^\pm$ bosons. In the general case,  the masses $M_h, M_H, M_A$ and $M_{H^\pm}$
are free parameters, and one assumes that $h$ is the observed Higgs boson  with
mass $M_h=125$ GeV. At least two additional mixing parameters $\beta$ and
$\alpha$ are needed to characterize fully the model:  $\tb = v_2/v_1$ is the
ratio of the vacuum expectation values of the two fields with $v_1^2\!+\!v_2^2
\! = \! v^2 \! = \! {\rm (246~GeV)^2}$, while  $\alpha$ diagonalises the
CP--even $h$ and $H$ mass matrix. In a general 2HDM,  there is also an
additional mass parameter linking the $\Phi_1, \Phi_2$ fields that enters only
in the quartic couplings among Higgs bosons, and can safely be ignored in our
present  discussion.  

\begin{table}[!ht]
\begin{center}
\renewcommand{\arraystretch}{1.4}
\begin{tabular}{|c|c|c|c|c|c|c|} \hline
\ \ $\Phi$ \ \  &\multicolumn{2}{c|}{$\hat g_{\Phi u\bar{u}}$}&  
                  \multicolumn{2}{c|}{$\hat g_{\Phi d \bar{d}}$}&  
$\hat g_{ \Phi VV} $ \\ \hline
& Type I & Type II & Type I & Type II & Type I/II \\   \hline
$h$  &  $\; \cos\alpha/\sin\beta       \; $ 
     &  $\; \cos\alpha/\sin\beta       \; $  
     &  $ \; \cos\alpha/\sin\beta \; $  
     &  $ \; -\sin\alpha/\cos\beta \; $  
     &  $ \; \sin(\beta-\alpha) \; $  \\
$H$  &  $\; \sin\alpha/\sin\beta \; $  
     &  $\; \sin\alpha/\sin\beta \; $  
     &  $ \; \sin\alpha/ \sin\beta \; $  
     &  $ \; \cos\alpha/ \cos\beta \; $  
     &  $ \; \cos(\beta-\alpha) \; $  \\
$A$  &  $\; \cot \beta \; $ 
     &  $\; \cot \beta \; $ 
     &  $ \; \cot \beta \; $    
     &  $\; \tan \beta \; $ 
     &  $ \; 0 \; $ \ \\ \hline
\end{tabular}
\end{center}
\vspace{-3mm}
\caption[]{\it The couplings of the  $h,H,A$ bosons to fermions and gauge bosons in Type-I and -II 2HDMs relative to SM Higgs couplings; the $H^\pm$ couplings to fermions have factors similar to those of the CP--odd boson $A$.} 
\label{Tab:cpg-2HDM}
\vspace{-3mm}
\end{table}

The couplings of the neutral $h,H,A$ bosons to massive gauge bosons and to 
fermions,  normalised to those of the SM Higgs boson, are given in Table
\ref{Tab:cpg-2HDM}.  There is no coupling of the CP--odd boson $A$ to the vector
bosons $V=W,Z$ when CP is conserved, as we consider here,  but the CP--even $h$
and $H$ states share the couplings of the SM Higgs particle to vector boson
pairs $VV$:
\begin{eqnarray}
\hat g_{hVV} \; = \; \frac{g_{hVV}}{g_{H_{\rm SM}VV}} \; = \; \sin(\beta-\alpha) \ , ~~~
\hat g_{HVV} \; = \; \frac{g_{HVV}}{g_{H_{\rm SM}VV}} \; = \; \cos(\beta-\alpha) \, .
\end{eqnarray}

Taking into account the fact that the couplings of the $h$ boson have been
measured at the LHC, and found to be SM--like within 10\% accuracy
\cite{Khachatryan:2016vau,Sirunyan:2018koj}, we infer that $\hat g^2_{hVV} \gsim
0.9$ and hence $\cos^2 (\beta-\alpha) \lsim 0.1$. This constraint can be
accommodated naturally  in a 2HDM by invoking the alignment
limit~\cite{Bernon:2015qea} in which one has $\cos (\beta-\alpha) = 0$ exactly,
and hence $\alpha=\beta- \frac{\pi}{2}$. We adopt the alignment limit 
throughout this paper, which leads to a simplified picture in which the
couplings between Higgs and gauge bosons, normalised to the SM as above, become
simply
\begin{eqnarray}
\hat g_{hAZ}= \hat g_{h H^\pm W}= \hat g_{AVV}= \hat g_{HVV}= 0\ , ~~~ 
\hat g_{HAZ}= \hat g_{H H^\pm W}= \hat g_{A H^\pm W} = \hat g_{hVV} = 1 \, .
\label{Higgscpl}
\end{eqnarray}
In contrast, the Higgs interactions with fermions are model--dependent in a
2HDM, and there are two model options commonly discussed in the
literature~\cite{Branco:2011iw}: Type--II, in which one field generates the
masses of isospin down--type  fermions and the other the masses of up--type
quarks, and Type--I, in which one field generates the masses of all fermions. 
However, there is a simplification in the alignment limit $\alpha=\beta-
\frac{\pi}{2}$, as the $h$ couplings to all fermions are SM--like,  $\hat g_{h
f\bar f}=1$, while the $H$ and $A$ couplings to a given fermion species are
exactly the same, $\hat g_{H f\bar f} =\hat g_{A f\bar f}$.  Thus, one has for
the $\Phi=H,A$ couplings to third-generation fermions: 
\begin{eqnarray}
{\rm Type\!-\!I}~:~  \hat  g_{\Phi t \bar t }= \cot\beta \, , \ \hat  g_{\Phi b \bar b } =  \hat  g_{\Phi \tau\tau } =  \cot\beta  \, , \\
{\rm Type\!-\!II}~:~  \hat  g_{\Phi t \bar t }= \cot\beta \, , \ \hat  g_{\Phi b \bar b } =  \hat  g_{\Phi \tau\tau } =  \tan\beta \, .
\label{Phiff}
\end{eqnarray}
At high $\tan\beta$ values, $\tb \gsim 10$, all couplings are suppressed in
Type--I models, whereas in Type--II models the $\Phi b \bar b$ and $\Phi
\tau\tau$ couplings are instead enhanced. At low $\tb$ values, $\tb \lsim 5$,
the $\Phi t\bar t$ couplings are not suppressed and the $\Phi b\bar b$ and $\Phi
\tau \tau$ couplings  are not strongly enhanced in Type--II models.  Since $m_t
\gg m_b, m_\tau$, the  $\Phi t\bar t$ couplings are then much larger than all
other fermion couplings.  This is a second important simplification: at low
$\tb$, all heavy Higgs couplings to fermions except those to the top quarks can
be ignored and, looking at eq.~(\ref{Phiff}), one sees that there is no
difference  between the two types of 2HDM in this case~\footnote{In fact, this
similarity between models extends to two other possibilities that are also
discussed, namely Type--III and Type--IV models \cite{Branco:2011iw}, which
differ from Type--I and II scenarios only in the couplings of the tau lepton,
which can be ignored at low $\tb$ values, for our purposes.}.    

Note that, contrary to the supersymmetric case to be discussed later where one
generally assumes $1 \leq  \tb \lsim 50$,  values of $\tb$ smaller than unity
are in principle possible in a general 2HDM.  However, for the top quark  Yukawa
coupling $\propto m_t / \tb$ to remain perturbative, one needs to impose the
bound $\tb \gsim 1/3$. 

Another important difference between the MSSM and the general 2HDM is that, 
whereas in the former case one has in most cases $M_H \sim M_A \sim M_{H^\pm}$, 
in the latter case the masses $M_A, M_H$ and $M_{H^\pm}$ are still free
parameters and can, in principle, be widely different.  However, high--precision
data, especially the fact that the $\rho$ parameter must be very close to
unity,  constrain the mass splittings between some of these states. For
instance, for a given $M_A$ value and independently of $\tb$,  one finds after
imposing the constraints from electroweak data that one of the  two masses $M_H$
or $M_{H^\pm}$ must be almost degenerate with $M_A$ (within $\sim 10$\%), while
the other mass can be widely different; see Fig.~11 of
Ref.~\cite{Haller:2018nnx}.

In our numerical analyses,  since our main concern is the study of the $H/A
\rightarrow t\bar t$ process,  we assume that the neutral states $H$ and $A$ are
almost degenerate in mass,  $|M_H / M_A - 1| < 0.2$ (but for completeness, we
have also included results for larger mass splitting up to $M_H-M_A=200$~GeV),
whereas the charged Higgs boson is somewhat heavier: $M_{H^\pm} \gsim M_A, M_H$,
and hence does not affect the phenomenology of $H$ and $A$ states in the present
context. 

To summarize, we propose in this paper a simple 2HDM benchmark in which the
processes $pp \! \rightarrow \! \Phi \! \rightarrow \! t\bar{t}$ can be  studied at the LHC, while retaining the main characteristic model features:
\begin{eqnarray}
{\rm 2HDM}: M_h=125~{\rm GeV} \, , \ \alpha=\beta - \frac{\pi}{2} \, , \  
M_H \approx M_A  \, , \ M_{H^\pm} \geq {\rm max}(M_H,M_A) \, ,
\end{eqnarray}
in which the $h$ state has a mass of 125 GeV and SM--like couplings as favoured
by LHC Higgs data, and the relation between the $A,H,H^\pm$ masses satisfies the
constraints from electroweak precision data, which do not allow Higgs to Higgs
plus gauge boson decays to occur. The alignment limit and  $M_H=M_A (1 \pm
10)\%$  and $M_{H^\pm} \geq {\rm max}(M_H,M_A)$ assumptions  make that all the
non--fermionic decay channels can be ignored. One can  thus concentrate on the
$\Phi \rightarrow t\bar t$ decays with branching ratios close to unity in the
interesting range  $1/3 \leq \tb \leq 5$.

\subsubsection{The (h)MSSM}

A widely studied incarnation of the 2HDM scenario is the MSSM, which is
essentially a 2HDM of Type II in which supersymmetry imposes strong constraints
on the Higgs sector,  so that only two parameters are independent at tree
level,  namely $M_A$ and $\tb$. However, when the radiative corrections in the
Higgs sector are included, in particular the dominant loop contributions from
the top and stop quarks that have strong couplings to the Higgs bosons, many
additional supersymmetric parameters will enter the parameterization. This is,
for instance, the case of the supersymmetry-breaking scale,  chosen to be the
geometric  average of the two stop masses $M_{S}= \sqrt {m_{\tilde t_1}
m_{\tilde t_2} }$,  the stop trilinear couplings $A_{t}$ and the higgsino mass
$\mu$;  the corrections due to other supersymmetric parameters are much smaller.

These radiative corrections are very important in particular in the CP--even
neutral Higgs sector as they can shift the lightest $h$ mass from the
tree--level value $M_h \leq M_Z \cos2\beta \leq M_{Z}$ to the one $M_{h}=125$
GeV that has been measured at the LHC. The neutral CP--even Higgs masses mix via
an angle $\alpha$ that diagonalises the mass eigenstates, leading to $H=
\Phi_1^0 \cos\alpha + \Phi_2^0 \sin\alpha$ and $h= -\Phi_1^0 \sin\alpha +
\Phi_2^0 \cos\alpha$, where $\Phi_{1,2}^0$ denote the neutral CP--even
components of the physical Higgs fields $\Phi_1, \Phi_2$ in the
current-eigenstate basis. The radiative corrections  are captured by a general
$2\times 2$ matrix $\Delta {\cal M}_{ij}^2$ in which only the $\Delta{\cal
M}^{2}_{22}$ entry is relevant in most cases (in particular if the $\mu$
parameter is small). It involves the stop--top sector correction that dominates
by far~\cite{Okada:1990vk,Ellis:1990nz,Haber:1990aw,Chankowski:1991md}:
\begin{eqnarray}
\Delta{\cal M}^{2}_{22} \approx \Delta M_h^2|^{t/\tilde{t}}_{\rm 1loop} \sim 
\frac{3m_t^4}{2\pi^2v^2} \bigg[ \log \frac{M_{S}^2} {m_{t}^2} +\frac{X_{t}^{2}}{M_{S}^{2}}
- \frac{X_{t}^{4} }{12 M_{S}^{4}} \bigg] \;,
\end{eqnarray}
where $M_S$ is the supersymmetry-breaking scale and $X_t=A_t-\mu/\tb$ is the
stop mixing parameter. It has been advocated~\cite{Djouadi:2013uqa,
Bagnaschi:2015hka} that, in this case,  one can simply trade $\Delta {\cal
M}^{2}_{22}$ for the known $M_h$ value using 
\begin{eqnarray}
\Delta {\cal M}^{2}_{22}= \frac{M_{h}^2(M_{A}^2  + M_{Z}^2 -M_{h}^2) -
 M_{A}^2 M_{Z}^2 \cos^2 2\beta } { M_{Z}^2 \cos^{2}{\beta}  +M_{A}^2
 \sin^{2}{\beta} -M_{h}^2} \;. 
\label{m22-hMSSM}
\end{eqnarray}
One can then simply write $M_{H}$ and $\alpha$ in terms of $M_{A},\tan\beta$ and $M_{h}$:
\begin{eqnarray}
M_{H}^2 &= &\frac{(M_{A}^2+M_{Z}^2-M_{h}^2)(M_{Z}^2 \cos^{2}{\beta}
+M_{A}^2 \sin^{2}{\beta}) - M_{A}^2 M_{Z}^2 \cos^{2}{2\beta} } {M_{Z}^2
\cos^{2}{\beta}+M_{A}^2 \sin^{2}{\beta} - M_{h}^2} \, , \\
\alpha &=& -\arctan\left(\frac{ (M_{Z}^2+M_{A}^2) \cos{\beta}
    \sin{\beta}} {M_{Z}^2 \cos^{2}{\beta}+M_{A}^2 \sin^{2}{\beta} - M_{h}^2}\right) \, .
\label{wide} 
\end{eqnarray}
In the case of the $H^\pm$  masses, the radiative corrections are small at 
large $M_A$, and one simply has $M_{H^\pm} \simeq \sqrt { M_A^2 + M_W^2}$ to a good approximation.

This is the hMSSM approach which has been shown to provide a  very good
approximation to the MSSM Higgs sector \cite{Djouadi:2013uqa}.   An important
property of the MSSM  is the decoupling limit that occurs for $M_A \! \gg \!
M_Z$, but it is in practice reached already for $M_A \gsim 350$ GeV for any valueof $\tb$.  In this limit, one  automatically has $\alpha\rightarrow \beta-
\frac{\pi}{2}$  for the CP--even mixing angle, i.e., exactly as in the alignment
limit of the 2HDM discussed above. The 125--GeV $h$ state then has SM--like
Higgs couplings: $\hat g_{hVV}= \hat g_{hff}= 1$, while the couplings of the
heavier $\Phi=H/A$ states to  gauge bosons vanish: $\hat g_{\Phi VV}=0$, and
those to fermions depend only on $\tb$ and are given by eq.~(\ref{Phiff}).  

The other important property of the MSSM Higgs sector decoupling limit is that,
besides decoupling from the massive gauge bosons,  the heavy CP--odd $A$,
CP--even $H$ and the charged $H^\pm$ bosons become  almost degenerate in mass:
$M_H \! \approx \! M_{H^\pm} \! \approx \! M_A$.  This is, in fact, the only
difference between the MSSM close to the decoupling limit and the 2HDM close to
the alignment limit:  in the latter case, the masses $M_A, M_H$ and $M_{H^\pm}$
are free parameters and could be very different. However, for $H/A$, this is not
the case in our 2HDM benchmark scenario, as we have chosen $M_H \approx M_A$,
which makes the situation similar to the MSSM. Note however that in the MSSM, we
will assume $\tb \gsim 1$, contrary to the 2HDM case~\footnote{An advantage of
the hMSSM approach is that it allows one to describe the low $\tan\beta$ region
of the MSSM that has been overlooked because, for SUSY scales of order 1 TeV,
values $\tan\beta <3$ were excluded because they led to an $h$ mass that is
smaller than 125 GeV. The price to pay is that, for such low $\tan\beta$ values,
one has to assume that the supersymmetry-breaking scale $M_S \gg 1$ TeV, and
hence that the model is fine-tuned. Moreover, care has to be taken not to enter
regimes for small values of $\tan\beta$ that cannot be accommodated with the
MSSM as pointed out in Refs.~\cite{Djouadi:2015jea,Bagnaschi:2015hka}.}.

\subsection{Additional matter in the  \texorpdfstring{$gg \rightarrow \Phi$}{gg -> H/A} loop}

\subsubsection{Additional vector-like quark contributions to \texorpdfstring{$gg \rightarrow \Phi$}{gg -> H/A}}

We consider later the possibility that additional vector-like quarks (VLQs) may
contribute to the triangular loop diagram  for $gg \rightarrow \Phi$. In this
case, the first line in eq.~(\ref{Dicus}) for the background is unchanged, but
the second and third lines are modified as follows:
\begin{eqnarray}
 \frac{ {\rm d}\hat \sigma_S }{{\rm d} z} &=&  \frac{3 \alpha_s^2 G_F^2 m_t^2}{8192 \pi^3}  \hat s^2  \sum_\Phi  \frac{ \hat \beta_t^{p_\Phi} | \hat g_{\Phi t\bar t} \sum_{Q} \hat g_{\Phi QQ}  A_{1/2}^\Phi (\tau_Q) |^2} {(s- M_\Phi^2) ^2+ \Gamma_\Phi^2 M_\Phi^2}  \, , \nonumber \\
 \frac{ {\rm d}\hat \sigma_I }{{\rm d} z} &=&  - \frac{\alpha_s^2 G_F m_t^2}{64\sqrt 2  \pi}  \frac {1}{1- \hat \beta_t^2 z^2} {\rm Re} \bigg[ \sum_\Phi \frac{ \hat \beta_t^{p_\Phi} \hat g_{\Phi t\bar t} \sum_{Q} \hat g_{\Phi QQ} A_{1/2}^\Phi (\tau_Q) } {s- M_\Phi^2+i \Gamma_\Phi M_\Phi} \bigg] \, ,
\end{eqnarray}
where the sum over the quarks $Q$ includes the top quark contribution (among the
Standard Model quarks) as well as the additional VLQs.

For simplicity, we will assume in our simulations that all the VLQs have the
same mass as the constraints from precision electroweak data would require this
to be the case if they are in an electroweak doublet. Under this assumption, the
effect of the VLQs can be parameterized by their mass and the number of VLQ
species, $N_Q$.

\subsubsection{Stop squark contributions to \texorpdfstring{$gg \rightarrow \Phi$}{gg -> H/A}}

In a supersymmetric theory the effective $gg \rightarrow \Phi$ coupling may also
receive significant contributions from top quark superpartners 
\cite{Djouadi:2005gj,Gunion:1989we,Djouadi:1998az,Dawson:1996xz,Harlander:2004tp,Muhlleitner:2006wx}.
In this case the second and third lines in eq.~(\ref{Dicus}) are modified as
follows:
\begin{eqnarray}
 \frac{ {\rm d}\hat \sigma_S }{{\rm d} z} &=&  \frac{3 \alpha_s^2 G_F^2 m_t^2}{8192 \pi^3} 
\hat s^2  \sum_\Phi  \frac{ \hat \beta_t^{p_\Phi} \hat g_{\Phi t\bar t}^2 | \sum_{Q} \hat g_{\Phi QQ} 
A_{1/2}^\Phi (\tau_Q)  + {\cal A}_{SUSY}^{\Phi} |^2} {(s- M_\Phi^2) ^2+ \Gamma_\Phi^2 M_\Phi^2}  \, , \nonumber \\
 \frac{ {\rm d}\hat \sigma_I }{{\rm d} z} &=&  - \frac{\alpha_s^2 G_F m_t^2}{64\sqrt 2  \pi} 
\frac {1}{1- \hat \beta_t^2 z^2} {\rm Re} \bigg[ \sum_\Phi \frac{ \hat \beta_t^{p_\Phi} 
\hat g_{\Phi t\bar t} \big( \sum_{Q} \hat g_{\Phi QQ} A_{1/2}^\Phi (\tau_Q) + {\cal A}_{SUSY}^{\Phi} \big)}
{s- M_\Phi^2+i \Gamma_\Phi M_\Phi} \bigg] \, ,
\end{eqnarray}
where~\cite{Gunion:1989we,Djouadi:2005gj}:
\begin{eqnarray}
{\cal A}_{SUSY}^{H} &=& \sum_{i=1}^{2} \frac{\hat g_{H \tilde{t}_{i} \tilde{t}_{i}}}{m_{\tilde{t}_{i}}^2}  
A_{0}^H (\tau_{\tilde{t}_{i}}) \, , \quad 
{\cal A}_{SUSY}^{A} = 0 \, .
\end{eqnarray}
The form factors $A_{0}^{\Phi}$ for the contributions of spin--$0$ particles as
functions of the variable $\tau_{\tilde{t}_{i}} \equiv M_H^2 / 4
m_{\tilde{t}_{i}}^2$, using  the function $f$ defined in
eq.~(\ref{eq:formfactors}), is given in the CP--even $H$ case by 
\begin{equation}
 A_{0}^{H}(\tau) \;  = \;  -\left[  \tau  - f(\tau)\right]  \tau^{-2} \, \, .
\end{equation}
The relevant Feynman rules defining the couplings of the neutral heavy Higgs
bosons $A, H$ of the MSSM to the stops, $g_{\Phi \tilde{t}_{i}\tilde{t}_{i}}$, 
can be written in the form
\begin{equation}
g_{\Phi \tilde{t}_{i}\tilde{t}_{i}} = (-i) \left(2\sqrt{ \sqrt{2} G_{F}} \right) \hat g_{\Phi \tilde{t}_{i}\tilde{t}_{i}} \, .
\end{equation}
The CP--odd Higgs boson has vanishing coupling to identical stop pairs.
In the limit $M_A \gg M_{Z}$, the heavier CP--even Higgs boson couples to same stop pairs (which contribute to the $gg \rightarrow H$ coupling)  through the following simple expressions:
\begin{eqnarray}
\hat g_{H \tilde{t}_1 \tilde{t}_1} &=& \sin 2\beta m_{Z}^2 \left[ \frac{1}{2} \cos^2 \theta_t - \frac{2}{3} s^2_W \cos 2 \theta_t \right] - \frac{m_t^2}{\tan\beta} - \frac{1}{2} \sin 2\theta_t  m_t Y_t  \;, \nonumber \\
\hat g_{H \tilde{t}_2 \tilde{t}_2 } &=&  \sin 2\beta m_{Z}^2 \left[ \frac{1}{2} \sin^2 \theta_t + \frac{2}{3} s_W^2 \cos 2 \theta_t \right] - \frac{m_t^2}{\tan\beta}  + \frac{1}{2} \sin 2\theta_t m_t Y_t  \;. 
\label{gHstst}
\end{eqnarray}
The squark mixing angle, $\theta_t$, is defined by
\begin{eqnarray}
\sin 2\theta_t = \frac{2 m_t X_t}{m_{\tilde{t}_1}^2 - 
m_{\tilde{t}_2}^2} \ \ , \ \ \cos 2 \theta_t = \frac{m_{\tilde{t}_L}^2 - 
m_{\tilde{t}_R}^2 } {m_{\tilde{t}_1}^2 - m_{\tilde{t}_2}^2} \, .
\end{eqnarray}
The stop sector can be parametrized by the three inputs $m_{\tilde{t}_L}$ and
$m_{\tilde{t}_R}$, the soft SUSY breaking stop masses, and $X_t =  A_t -
\mu/\tan\beta$, the stop mixing parameter. The physical stop masses are denoted
by $m_{\tilde{t}_1}$ and $m_{\tilde{t}_2}$. The above couplings also involve the
parameter $Y_t$
\begin{eqnarray}
Y_t &=& \mu + \frac{A_t}{\tan\beta} 
 =\mu+\frac{X_t}{\tan\beta}+\frac{\mu}{\tan\beta^2} \, .
\end{eqnarray}
 
For sufficiently large $Y_t$ values, the coupling of the heavier CP--even Higgs
$H$ to stops are strongly enhanced and can be larger than its coupling to the
top quark. In the limit $M_A \gg M_{Z}$, the couplings eq.~(\ref{gHstst}) of the
heavier CP--odd Higgs boson to the stops simplify to:
\begin{eqnarray}
\hat g_{H \tilde{t}_1 \tilde{t}_1} &=& - \frac{1}{2} \sin 2\theta_t  m_t Y_t \;  , \quad 
\hat g_{H \tilde{t}_2 \tilde{t}_2 } = \frac{1}{2} \sin 2\theta_t m_t Y_t \;.
\label{gAstst}
\end{eqnarray}

In the following, we consider the so called ``light-stop'' MSSM benchmark
scenario which is promoted by the LHC Higgs Cross Section Working group; see
Refs.\cite{MSSM_LHCHXS-lightstop-updated,Carena:2013ytb,Bahl:2018zmf}. The
benefit of this benchmark is to accommodate the constraints on the SM Higgs mass
while still allowing light stops within the TeV regime. In this scenario, one
fixes the SUSY parameters $M_{1}=340$~GeV, $M_{2}=\mu=400$~GeV and the stop
mixing parameter $X_t=1$~TeV. The trilinear couplings are adjusted such that
$A_t = A_b = A_\tau= X_t + \mu / \tan\beta$. The parameter $Y_t$ in this
scenario is then  
\begin{eqnarray}
Y_t^{\rm light~stop}&=&400+\frac{1000}{\tan\beta}+\frac{400}{\tan^ 2\beta} \; {\rm GeV} \; .
\end{eqnarray}
In the scan we performed in the $[M_A,\tan\beta$] plane in our analysis, the lighter stop has a mass around $324$~GeV and the heavier stop a mass around $671$~GeV.

\subsection{Phenomenology at the LHC}

We turn now to the phenomenology of these Higgs sectors and, in particular, that
of the heavier neutral Higgs states, since the phenomenology of the lightest $h$
boson is, by construction, essentially that of a 125--GeV SM Higgs boson, which
is consistent with measurements performed  so far at the LHC. 

\begin{figure}[!h]
 \centering
 \includegraphics[width=0.49\textwidth]{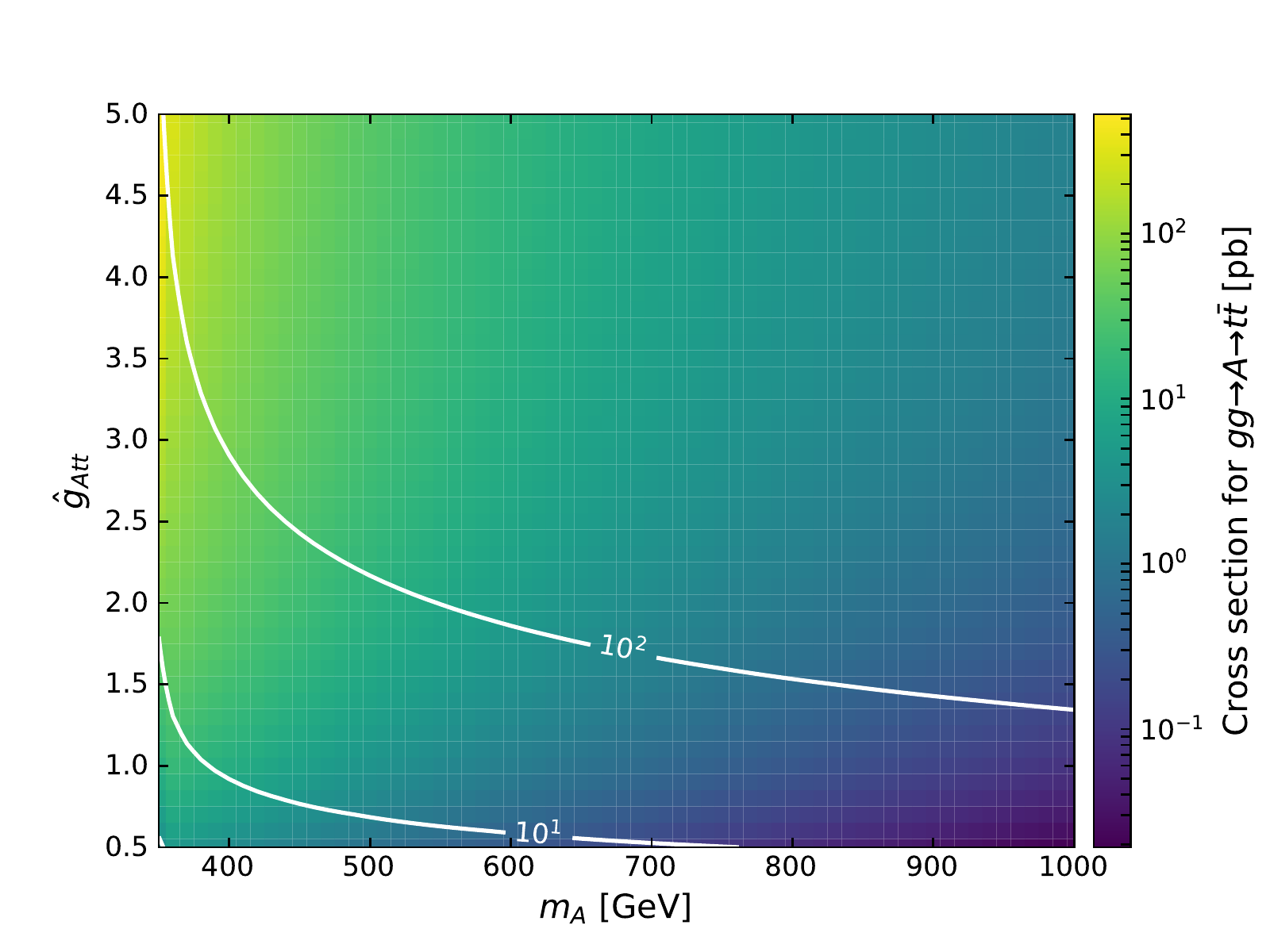}
 \includegraphics[width=0.49\textwidth]{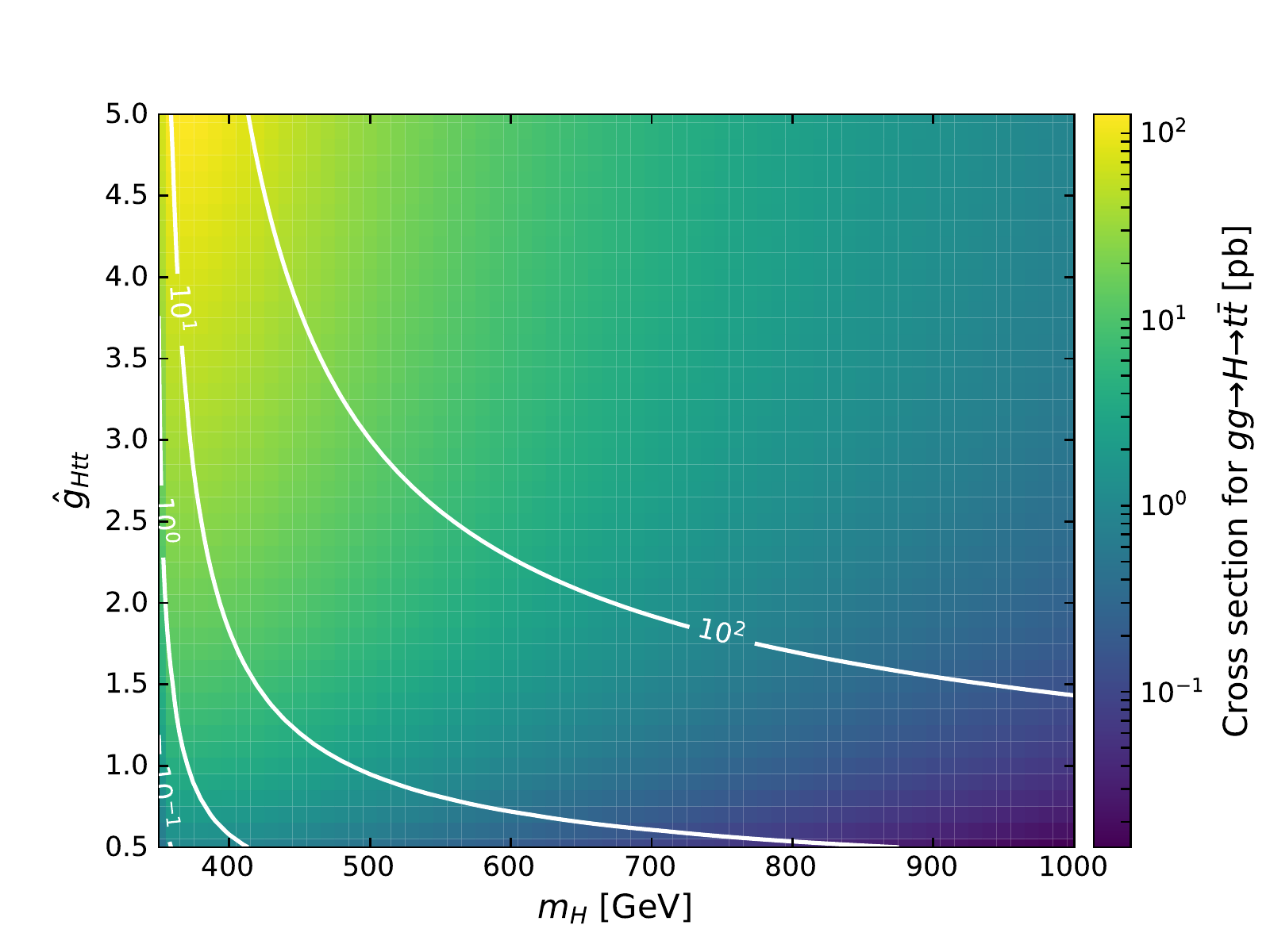} \\
 \includegraphics[width=0.49\textwidth]{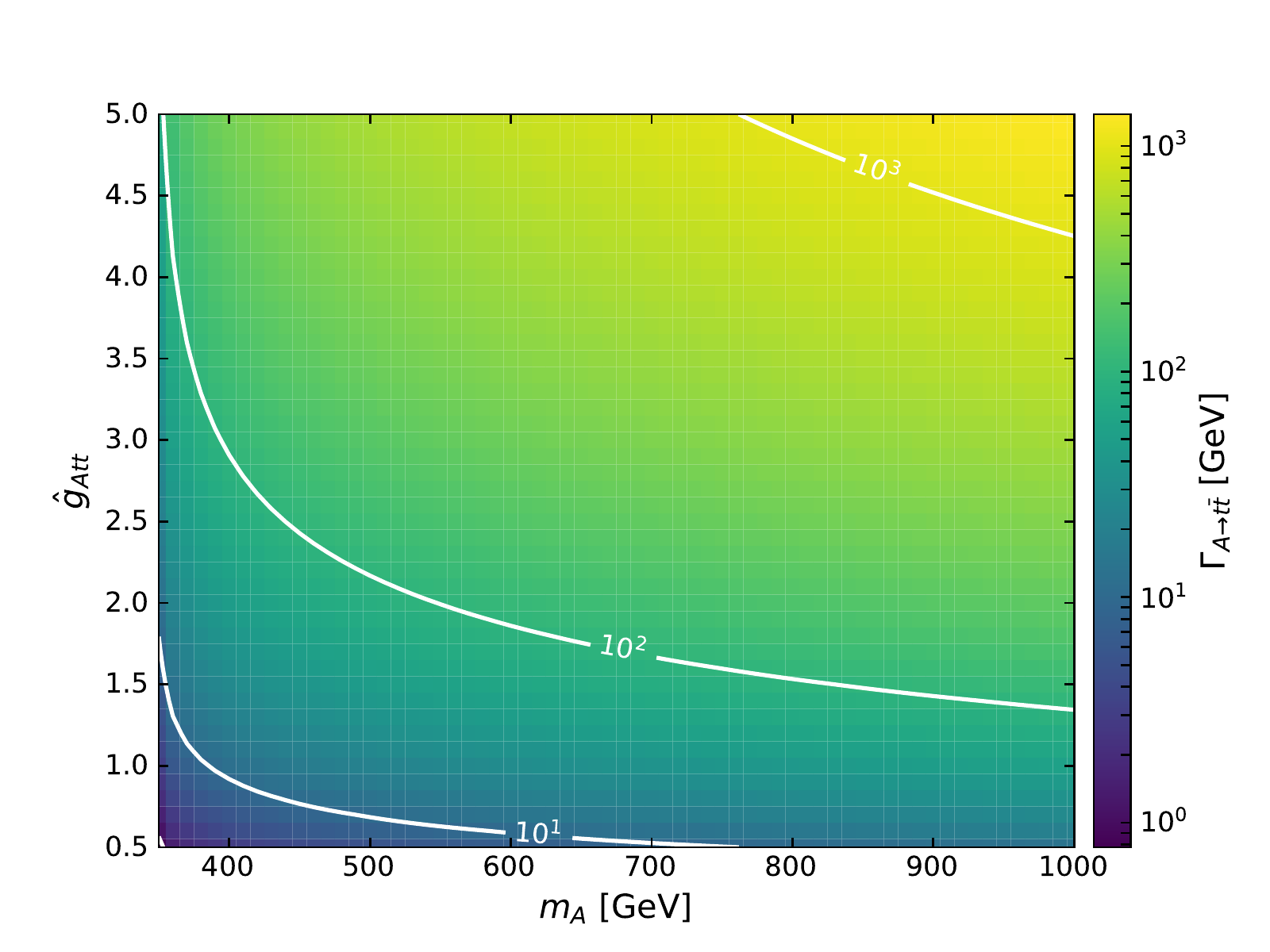}
 \includegraphics[width=0.49\textwidth]{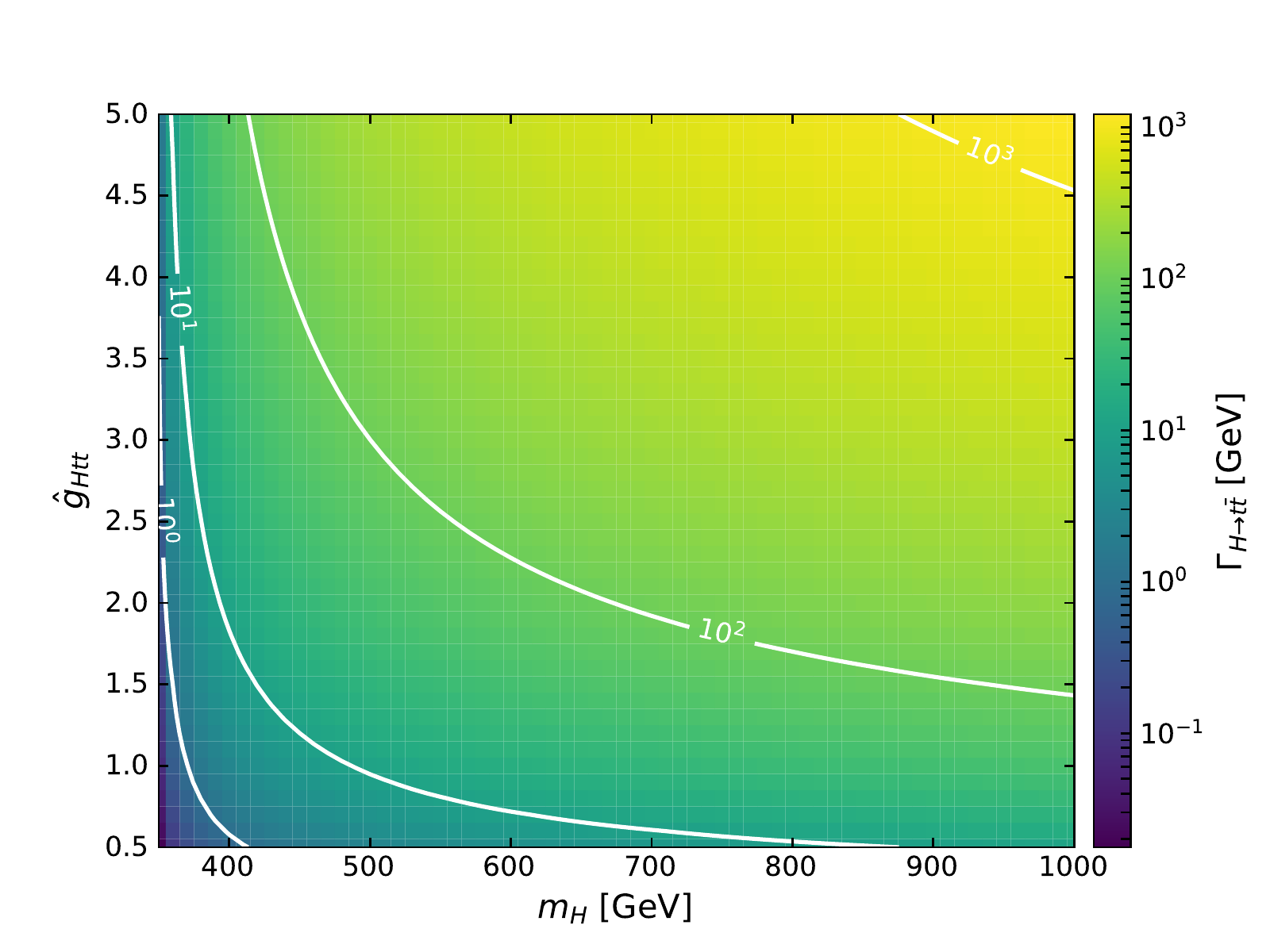}
 \caption{\it The $gg \to \Phi \to \ttbar$ production cross sections at the 13~TeV LHC and the decay widths $\Gamma (\Phi \to \ttbar)$ of a heavy pseudoscalar $A$ (left) and an heavy scalar $H$
(right).}
 \label{onlySP}
\end{figure}

First, the phenomenology of the $H/A$ bosons in the case of the SM with an extra
singlet (pseudo)scalar is simple, when neglecting the $H/A$ couplings to
fermions other than the heavy top quark and ignoring the coupling of the scalar
$H$ boson to massive gauge bosons as well as to light $h$ bosons (the 
pseudoscalar or CP--odd $A$ does not couple to these states at tree--level), as
would be the case in the other models that we consider later, as a consequence
of either decoupling or alignment. The only way to produce such $H/A$ bosons at
the LHC would be through the $gg\to H/A$ mechanism, which proceeds mainly via
top quark triangle diagrams and, to a much lesser extent, in associated 
production with top quark pairs, $pp \to t\bar t H/A$, which has a cross section
that is at least two order of magnitude lower than gluon-fusion. Once produced,
the $H/A$ bosons decay mainly into top quark pairs, as the decays into two
gluons, two photons or a photon and a $Z$ boson (again through  loops of top
quarks, as well as $W^\pm$ loops) will have extremely small rates especially in
the later two cases.  Hence one would have a decay branching ratio BR($\Phi \to
t\bar t) \simeq 1$ and the total decay widths of the two bosons would be almost
identical to the $t\bar t$ partial decay widths. Fig.~\ref{onlySP} displays the
$13$~TeV production cross sections and the $\Gamma (\Phi \to t\bar t)$ decay
widths for both the scalar and pseudoscalar in the case of the SM supplemented
by an isosinglet (pseudoscalar) Higgs boson, as a function of the boson mass and
their reduced coupling to the top quark, $\hat g_{\Phi t \bar t}$.

The phenomenology of the Higgs sector of the MSSM has been discussed at length
in the past, and we refer to Ref.~\cite{Djouadi:2015jea} for a recent account.
In the high--$\tb$ regime, $\tb \gsim 10$, the situation for the $H/A$ states is
simple: they are mainly produced via the $gg$ and $b\bar b$ fusion processes and
decay into either $b\bar b$ $(\approx 90\%)$ or $\tau^+\tau^-$ $(\approx 10\%)$
final states.  The searches for $\tau\tau$ resonances at high invariant masses
performed by ATLAS and CMS set strong constraints on the parameter space and, 
for instance, for $M_A \!  \approx \! M_H \approx \! 750$ GeV,  values of
$\tan\beta \gsim 10$ are excluded~\cite{Aaboud:2017sjh,Sirunyan:2018zut}. At
small and intermediate values of $\tb \lsim 10$ and for $H/A$ masses below the
$t\bar t$ threshold,  one is not yet in the decoupling regime and there is a
plethora of interesting channels  to be considered in the search for the neutral
$\Phi$ bosons,  e.g., $H\rightarrow WW,ZZ,hh$ and $A\rightarrow hZ$
\cite{Djouadi:2015jea}.  Some of these channels have been studied by the LHC
collaborations, and some exclusion limits  have been set in the hMSSM
\cite{Aad:2015pla,CMS:2016qbe,ATLAShMSSM}. However, the low--$\tb$ region $\tb
\lsim 5$ with  $H/A$ masses above the $t\bar t$ threshold has not been probed
experimentally~\footnote{The ATLAS collaboration has also published a search for
$\Phi \to \ttbar$~\cite{Aaboud:2017hnm} that was performed with 20.3 /fb at a
centre-of-mass energy $\sqrt{s} = 8$~TeV. The results were interpreted in the
context of a Type-II two--Higgs doublet model where either $A$ or $H$ is very
heavy, or both are degenerate in mass, $M_A = M_H$. However, this search is not
very sensitive to the models we study.}.

Nevertheless, from the phenomenological point of view,  the situation in the
case $\tb \lsim 5$ and $M_\Phi \gsim 350$ GeV, is also rather simple, and can be
summarized by noting the following points \cite{Djouadi:2015jea}:\vspace*{-2mm} 

\begin{itemize}

\item As the only significant Higgs coupling is the Yukawa coupling to top
quarks, $ \propto m_t/ (v \tb)$,  the only relevant decay mode of the $\Phi=
H/A$ states is into $t \bar t$ final states.\vspace*{-2mm} 

\item The total widths, generated mostly by these decays: $\Gamma_\Phi \approx
\Gamma(\Phi \! \rightarrow \! t\bar t)$,   scale like $M_\Phi/\tan^2 \beta$ and
are $\gsim 1$ GeV.  For $\tb \! \approx \! 1$,  they are ${\cal O}(5)$\% of the
$\Phi$ masses.\vspace*{-2mm} 

\item As the $\Phi$ states do not couple to massive gauge bosons, and only
weakly to all fermions except for tops, the main Higgs production channel is the
gluon-fusion process $gg\rightarrow \Phi$,  in which the top quark loop
generates the dominant contribution.\vspace*{-2mm}  \end{itemize}

\begin{figure}[!h]
\mbox{\hspace*{-6mm} \includegraphics[scale=0.3]{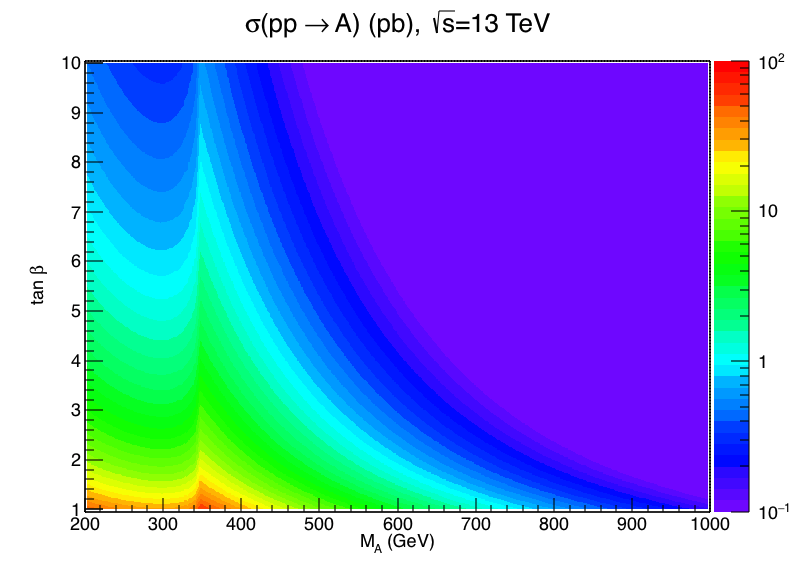}~\includegraphics[scale=0.3]{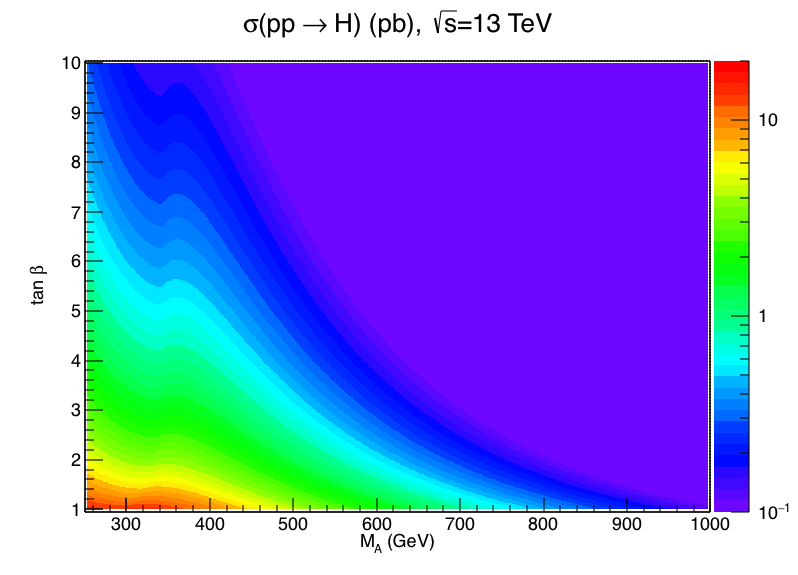} }\\[2mm]
\mbox{\hspace*{-5mm}
\includegraphics[scale=0.3]{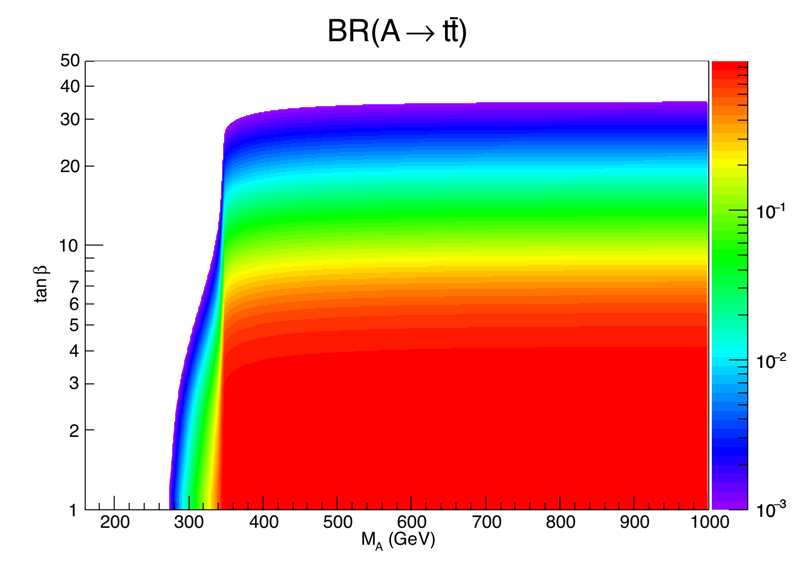}~~\includegraphics[scale=0.3]{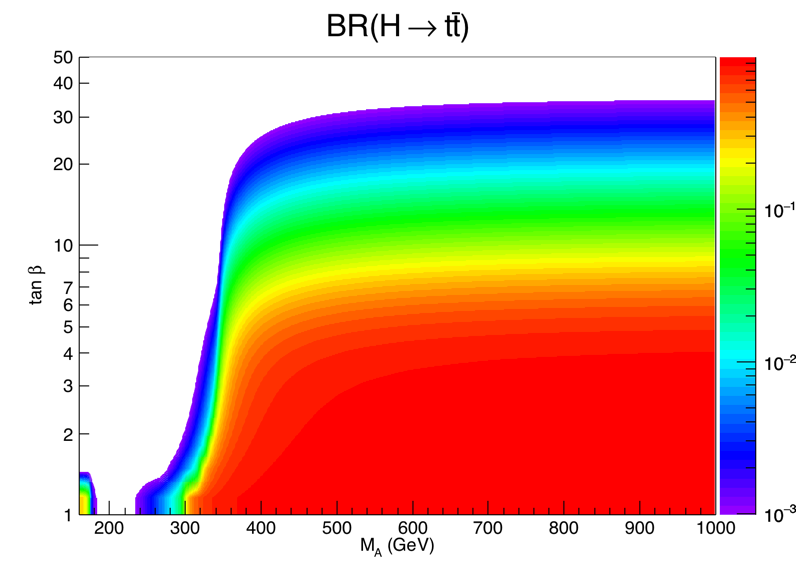} }\\[2mm]
\mbox{\hspace*{-5mm}
\includegraphics[scale=0.3]{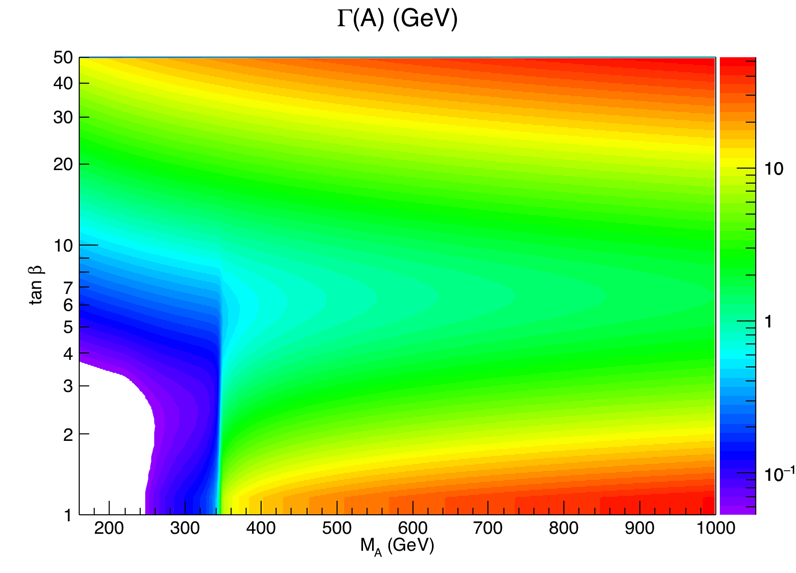}~~\includegraphics[scale=0.3]{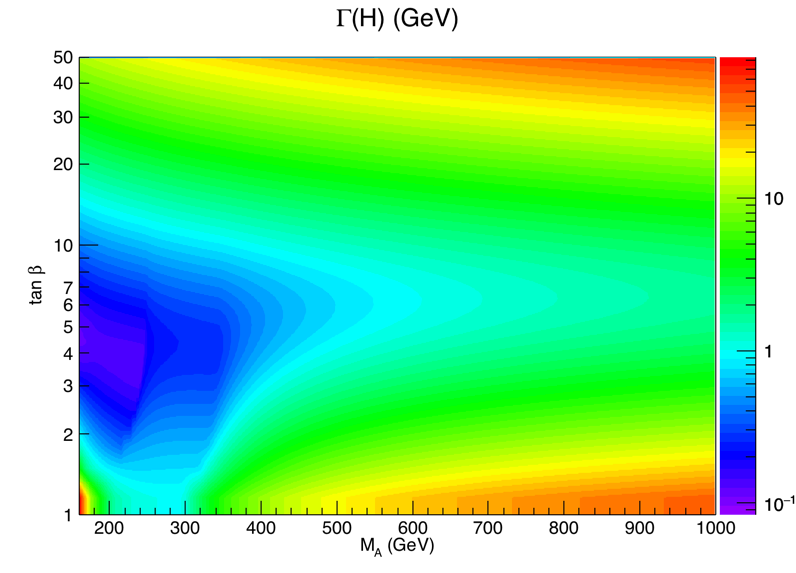} }
\vspace*{-.5cm}
\caption{\it The $gg \rightarrow \Phi$ production cross sections at the 13 TeV
LHC,   the $\Phi \rightarrow t \bar t$ branching  ratios and the total decay
widths $\Gamma_\Phi$ of the  heavier MSSM Higgs bosons $A$ (left) and $H$
(right) in the $[M_{A}, \tan\beta]$ plane, as predicted by the hMSSM.}
\label{rates-MSSM}
\vspace*{-2mm}
\end{figure}

The production cross sections $\sigma(gg\rightarrow \Phi)$ at the 13 TeV LHC, 
the branching ratio BR$(\Phi \rightarrow t \bar t)$ and the total decay widths
$\Gamma_\Phi$  are shown in Fig.~\ref{rates-MSSM} in the $[M_A, \tb]$ parameter
plane for $\Phi=A$ (left) and $\Phi=H$ (right), assuming the mass $M_{H}$ and
and the angle $\alpha$ as given in the hMSSM. One clearly sees that, indeed, for
$\tb \lsim 3$--5 and $M_\Phi \gsim 400$ GeV,  the search channel $gg\rightarrow
\Phi \rightarrow t\bar t$ is worth investigating at the LHC, as the production
rate is significant. Note also that, in the MSSM, since one is not exactly in
the  decoupling limit for small $\tb$ values and $M_A$ not very large,
differences are visible between the $H/A$ plots in Fig.~\ref{rates-MSSM}. 

As discussed previously, the situation could in principle be slightly different
in a 2HDM for two reasons.  First, the alignment limit is not exact and, for
$\cos^2(\beta -\alpha) \approx 0.1$,  decay channels such as  $A\rightarrow hZ$
and $H\rightarrow WW,ZZ,hh$ would be possible. However, for $M_\Phi \gsim 400$
GeV and $\tb \lsim 3$,  the $\Phi \rightarrow t \bar t$ decays are overwhelming.
The branching ratios for these additional channels are very small,  and can be
ignored in a first approximation.

\begin{figure}[!h]
\mbox{\hspace*{-5mm} \includegraphics[scale=0.3]{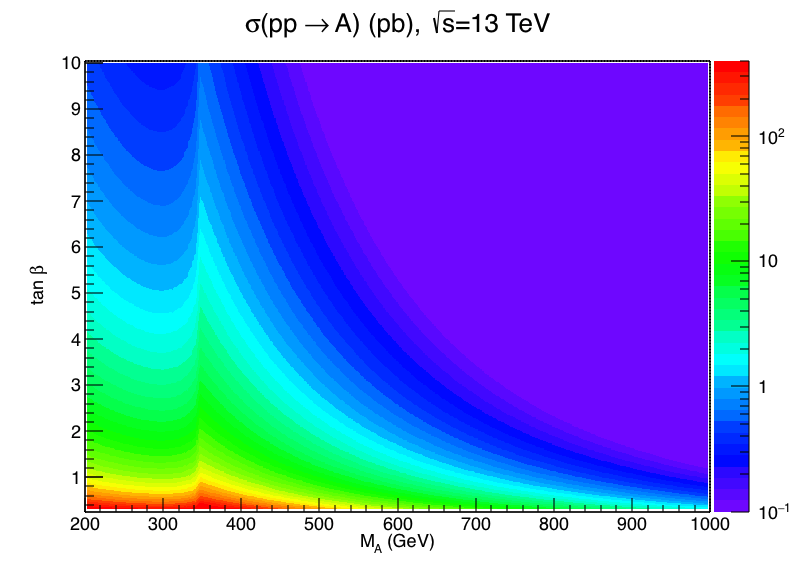}~~\includegraphics[scale=0.3]{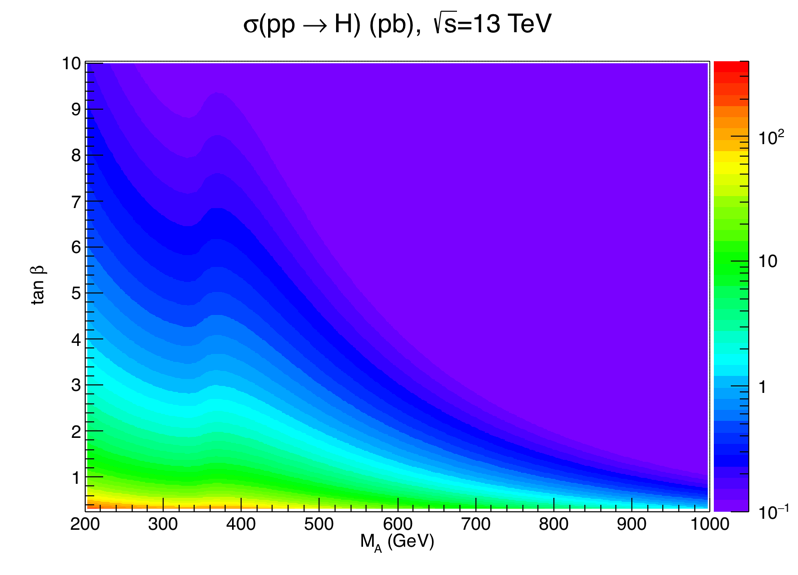} }\\[2mm]
\mbox{\hspace*{-5mm}
\includegraphics[scale=0.3]{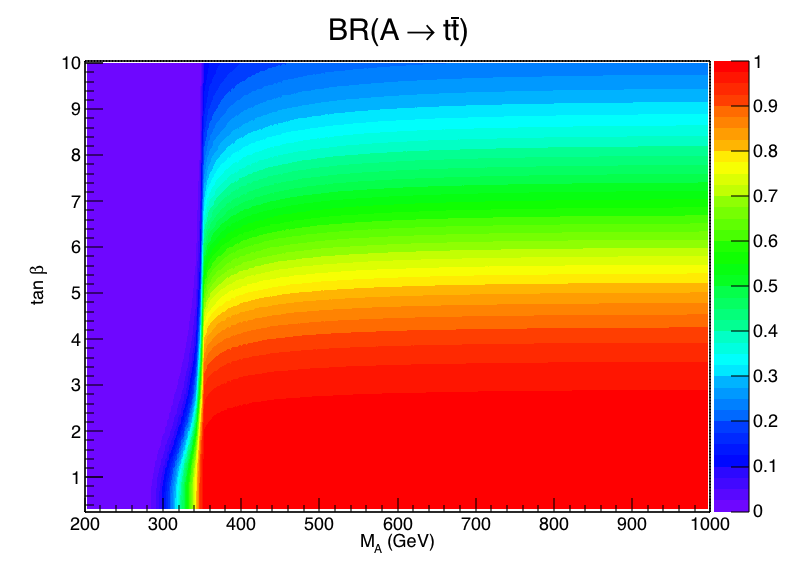}~~\includegraphics[scale=0.3]{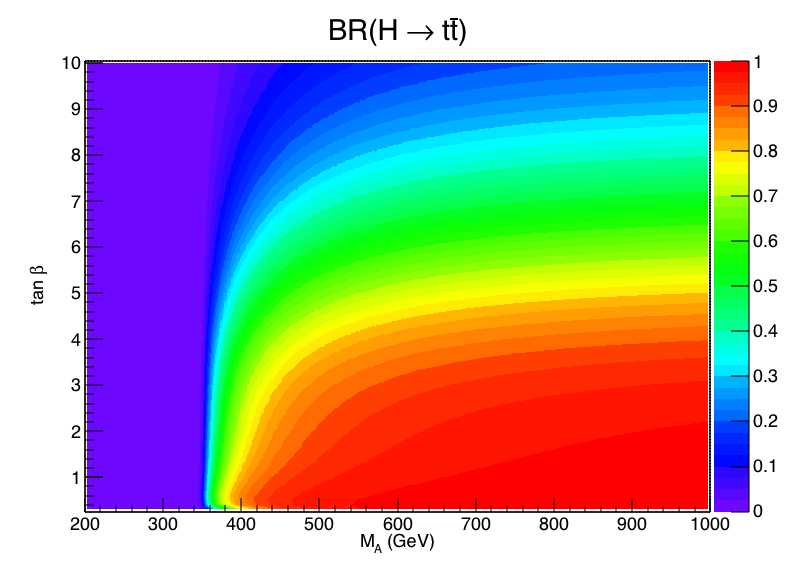} }\\[2mm]
\mbox{\hspace*{-5mm}
\includegraphics[scale=0.3]{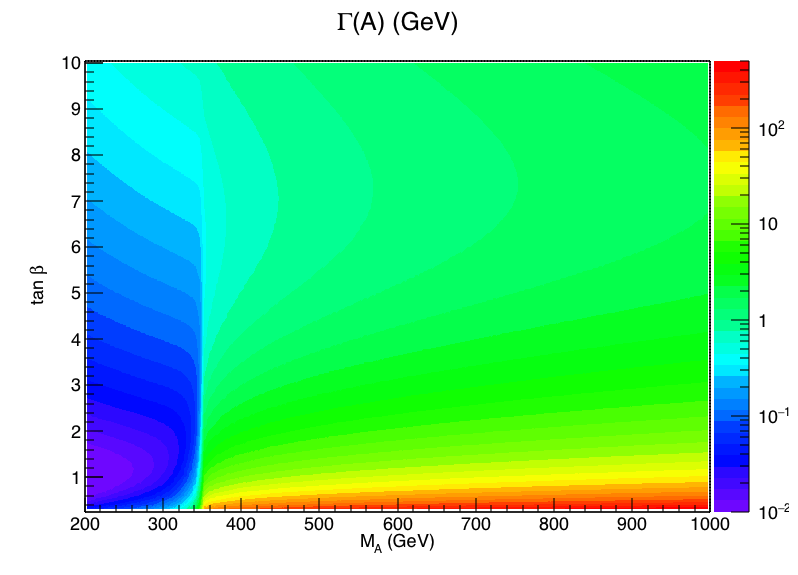}~~\includegraphics[scale=0.3]{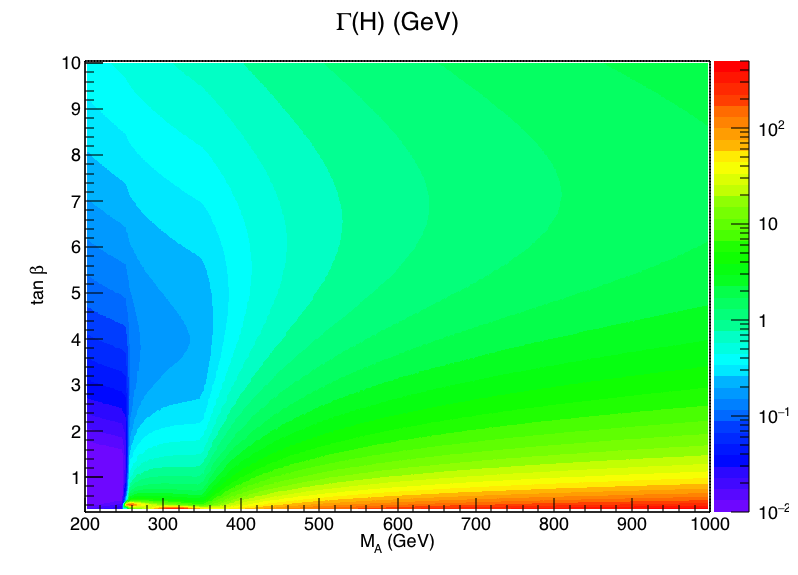} }
\vspace*{-.5cm}
\caption{\it The $gg \rightarrow \Phi$ production cross sections at the 13 TeV
LHC,   the $\Phi \rightarrow t \bar t$ branching  ratios and the total decay
widths $\Gamma_\Phi$ of the 2HDM CP--odd $A$ boson (left) and heavy CP--even $H$ boson (right) in the $[M_{A}, \tan\beta]$ plane, assuming a fixed mass splitting
$M_{H}=M_{A}$.}
\label{rates-2HDM}
\vspace*{-2mm}
\end{figure}

A second difference is that, as also noted previously, one does not have
automatic mass degeneracy  between the heavy $A,H$ and $H^\pm$ states in a 
general 2HDM, so other decay channels with unsuppressed couplings might occur, 
as shown by the second set of couplings in eq.~(\ref{Higgscpl}):  $A\rightarrow
HZ, H^\pm W$ and/or $H\rightarrow AZ, A W^\pm$. When kinematically accessible
these decays can be very important, in particular at relatively high $\tb$
values, and could suppress drastically the branching ratios for $A/H \rightarrow
t\bar t$ decays. In particular, BR($\Phi \rightarrow t\bar t)$ drops drastically
when the channels $H/A \rightarrow  A/H+Z$ are kinematically open.  The $H^\pm$
state is expected to be rather heavy as a results of constraints from heavy
flavor physics \cite{Arbey:2017gmh},  but its impact if $\Phi \rightarrow H^\pm
W^\mp$ were kinematically open  would have been exactly the same. 

Nevertheless, as already mentioned, the constraints from  high--precision
electroweak data require two Higgs masses to be sufficiently close in mass for 
some of these decays to be kinematically closed at the two--body level.  These
constraints, added to those on the light $h$ state from LHC data, motivate our
benchmark scenario, which is close to the  alignment limit that suppresses most
of these decay channels and $|M_H / M_A - 1| \lesssim 0.1$, while $M_{H^\pm}
\geq {\rm max}(M_H,M_A)$, in which case the neutral Higgs to charged Higgs decay
channels are kinematically closed. In this benchmark, one can concentrate on the
interesting $\Phi \rightarrow t\bar t$ decays with branching ratios close to
unity in the interesting range  $1/3 \leq \tb \leq 5$.  

The production cross sections $\sigma(gg\rightarrow \Phi)$ at the 13 TeV LHC,
the branching ratio BR$(\Phi \rightarrow t \bar t)$  and the total decay widths
$\Gamma_\Phi$ in our 2HDM benchmark scenarios with $M_{H}=M_{A}$ are shown in
Fig.~\ref{rates-2HDM} in the $[M_A, \tb]$ parameter plane for $\Phi=A$ (left)
and $\Phi=H$ (right).  We also considered in our analysis various mass
splittings between the 2HDM CP--even and CP--odd heavy Higgs bosons, namely
$M_{H}-M_{A}=10, 50, 100, 200$~GeV.

\section{Simulation of experimental sensitivity}
\label{sec:Simulation}

In this Section we compute the expected sensitivity  and exclusion potential for
each signal hypothesis, based on the distribution of the invariant mass of the
\ttbar~system, \mtt. The SM~\ttbar{} production process is by far the dominant
background in the targeted $\ell + \text{jets}$ final state and the only one
considered in this study. It is described with the help of a Monte--Carlo (MC)
simulation. Distributions of \mtt{} for the BSM contribution are built via a
computationally-efficient approximation  that uses MC simulations of a small
number of signal hypotheses to construct an economical parameterization. The
statistical analysis takes into account the main systematic uncertainties
affecting the \mtt\ distribution. All the software code used in this study is
made available at~\cite{analysis-code}.

We model SM $pp \rightarrow t\bar t$ events with decaying top quarks at leading
order using {\tt MadGraph} {\tt \_aMC@NLO~2.6.0}~\cite{Alwall:2014hca}. Only
final states with exactly one electron or muon are considered. The factorization
and renormalisation scales are set to $\frac12 \mtt $, following the choice in
Ref.~\cite{Hespel:2016qaf}. The parton distribution functions (PDFs) are taken
from the set {\tt PDF4LHC15\_nlo\_30\_pdfas}~\cite{Butterworth:2015oua},  as
provided in {\tt LHAPDF~6.1.6}~\cite{Buckley:2014ana}. We have generated $5
\times 10^6$ events for a nominal mass of the top quark of 173\,GeV,  and the
same number for each of the $\pm 0.5$\,GeV variations in its mass. Showering and
hadronisation for the produced events are performed using {\tt
Pythia~8.230}~\cite{Sjostrand:2007gs}  with the {\tt Monash~2013}
tune~\cite{Skands:2014pea}. Stable particles are clustered into jets using the
anti-$k_\text{T}$ algorithm~\cite{Cacciari:2008gp} with a cone size of 0.4. For
convenience, the last two steps are done with the help of the {\tt
Delphes~3.4.1} framework~\cite{deFavereau:2013fsa},  which, however, is not
employed to simulate the detector response. To account for higher-order
corrections in SM~\ttbar{} production, the sample is normalized to the 
inclusive cross section computed at the $\text{NNLO} + \text{NNLL}$ precision
with the program  \texttt{Top++} 2.0~\cite{Czakon:2011xx}. This corresponds to
applying a flat $K$-factor of 2.0.

We produced MC samples for reference signal hypotheses in the hMSSM with  $M_A =
400, 500, 600, 700$, and 1000~GeV and various values of $\tan\beta$. A custom
{\tt MadGraph} model was used, which is realized as the SM with two  additional
neutral Higgs bosons that are pure CP eigenstates. The effective coupling to
gluons was implemented following Ref.~\cite{Spira:1995rr},  with only top quarks
included in the loop~\footnote{The selection efficiency parameterized as a
function \mtt~ does not change significantly when there are extra loop
contributions from VLQs or stops squarks.}. The masses of the additional Higgs
bosons, their total widths and couplings to top quarks are free parameters of
this model,  and they are fixed according to the model predictions for each
point in the $[M_A, \tan\beta]$ plane. Apart from the differences in the models,
signal events are generated in the same way as for the SM~\ttbar{} events
described above. Samples for the resonant part of the BSM contribution and the
interference are produced independently. Owing to the destructive nature of the
interference in certain regions of the phase space, some events receive negative
weights,  and the total cross section for the interference sample can be
negative. For every value of $M_A$, around $10^6$ events have been generated for
each  CP eigenstate for both the resonant and interference parts.

The event selection made was dictated by the targeted $\ell + \text{jets}$ final
state. The only charged lepton in the final state is required to have $\pt >
30$\,GeV and $|\eta| < 2.4$. Each event must contain at least four jets with
$\pt > 20$\,GeV and $|\eta| < 2.4$. Jets that overlap with the lepton within
$\Delta R < 0.4$ are removed. Two of the jets must be matched to the $b$~quarks,
thus emulating $b$-tagging. No selection on missing~\pt{} is applied, since the
$pp \rightarrow \text{jets}$ background is not considered in this study. The
combined efficiency of lepton identification and $b$-tagging, which is estimated
to be 30\%~\cite{Sirunyan:2018fpa, CMS-DP-2018-017, CMS-DP-2018-030,
Sirunyan:2017ezt}, is included with the event weights. In order to emulate the
effects of detector resolution, Gaussian smearing is applied to the parton-level
invariant mass \mtt{} in the events selected. Two benchmark values for the
\mtt~resolution are considered, namely 10\% and 20\%.

Generating MC samples for each signal hypothesis would not be practical.
Instead, \mtt~distributions for the BSM contribution are constructed starting
from the parton-level cross sections~$\hat \sigma(\hat s)$  given in
Section~\ref{Sec:Models} (with the $K$-factors included as described in
Section~\ref{Sec:Process}), and the cross sections are first convoluted with the
PDFs. The differential cross section in \mtt{} can be computed as
\begin{linenomath}
\begin{equation}
 \label{Eq:PDFConv}
 \frac{\mathrm d\sigma}{\mathrm d\mtt} \; = \; 2 \sqrt{\hat{s}}\, F(\hat s, s) \cdot \hat\sigma(\hat s),
\end{equation}
\end{linenomath}
with $s = (13\,\text{TeV})^2$ being the squared centre-of-mass collision energy
and
\begin{linenomath}
\begin{equation}
 \label{Eq:PDFConvF}
 F(\hat s, s) \; \equiv \; \frac{1}{s} \int_{\hat s / s}^{1} f_g(x) f_g\!\left(\frac{\hat s}{sx}\right) \frac{\mathrm dx}{x},
\end{equation}
\end{linenomath}
where $f_g(x)$ is the gluon PDF and $x$ is the fraction of the proton's momentum
carried by the gluon. The integration over one of the two $x$ variables has
already been carried out,  benefiting from the fact that the parton-level cross
section depends only on $\hat s$. Since the function $F$~eq.~\eqref{Eq:PDFConvF}
does not depend on the parton-level cross section,  it can be precomputed,
permitting a very fast convolution eq.~\eqref{Eq:PDFConv}.

\begin{figure}[!ht]
 \centering
 \includegraphics[width=0.8\textwidth]{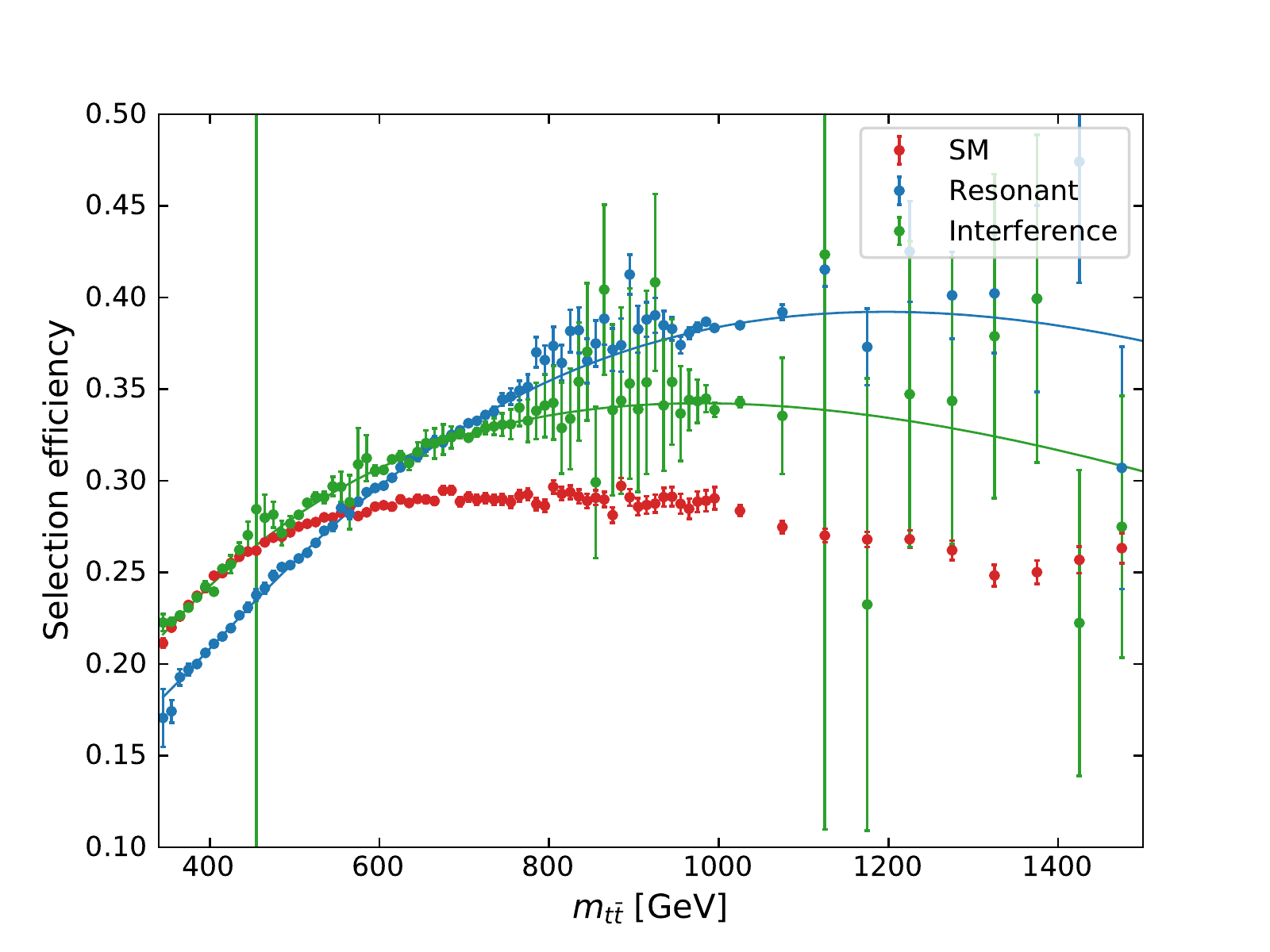}
\caption{\it Efficiency of the event selection as a function of the
 parton-level invariant mass \mtt, as computed with  MC samples for the SM $gg \rightarrow \ttbar$ background, the resonant BSM contribution, and the interference.  Event weights  representing the efficiency of lepton identification and $b$-tagging are not included.}
\label{Fig:SelEff}
\end{figure}

To mimic the event selection, $\mathrm d\sigma / \mathrm d\mtt$ is multiplied by
the selection efficiency. It is computed as a function of the parton-level~\mtt\
from the MC samples for the reference signal hypotheses. The results are shown
in Fig.~\ref{Fig:SelEff}. As can be seen, the efficiencies for the resonant part
and the interference differ from each other  as well as from SM $gg \rightarrow
\ttbar$, which is included for completeness. The signal efficiencies obtained
are fitted with a cubic function of $\ln\mtt$,  which has been found empirically
to provide an accurate description. These functions are also plotted in
Fig.~\ref{Fig:SelEff}. The branching ratio for the targeted final state, which
is $8/27$, is also included,  together with the 30\% efficiency of lepton
identification and $b$-tagging. The resulting combined fitted selection
efficiency is denoted by $\epsilon(\mtt)$ in the following.

\begin{figure}[!ht]
 \centering
 \includegraphics[width=0.8\textwidth]{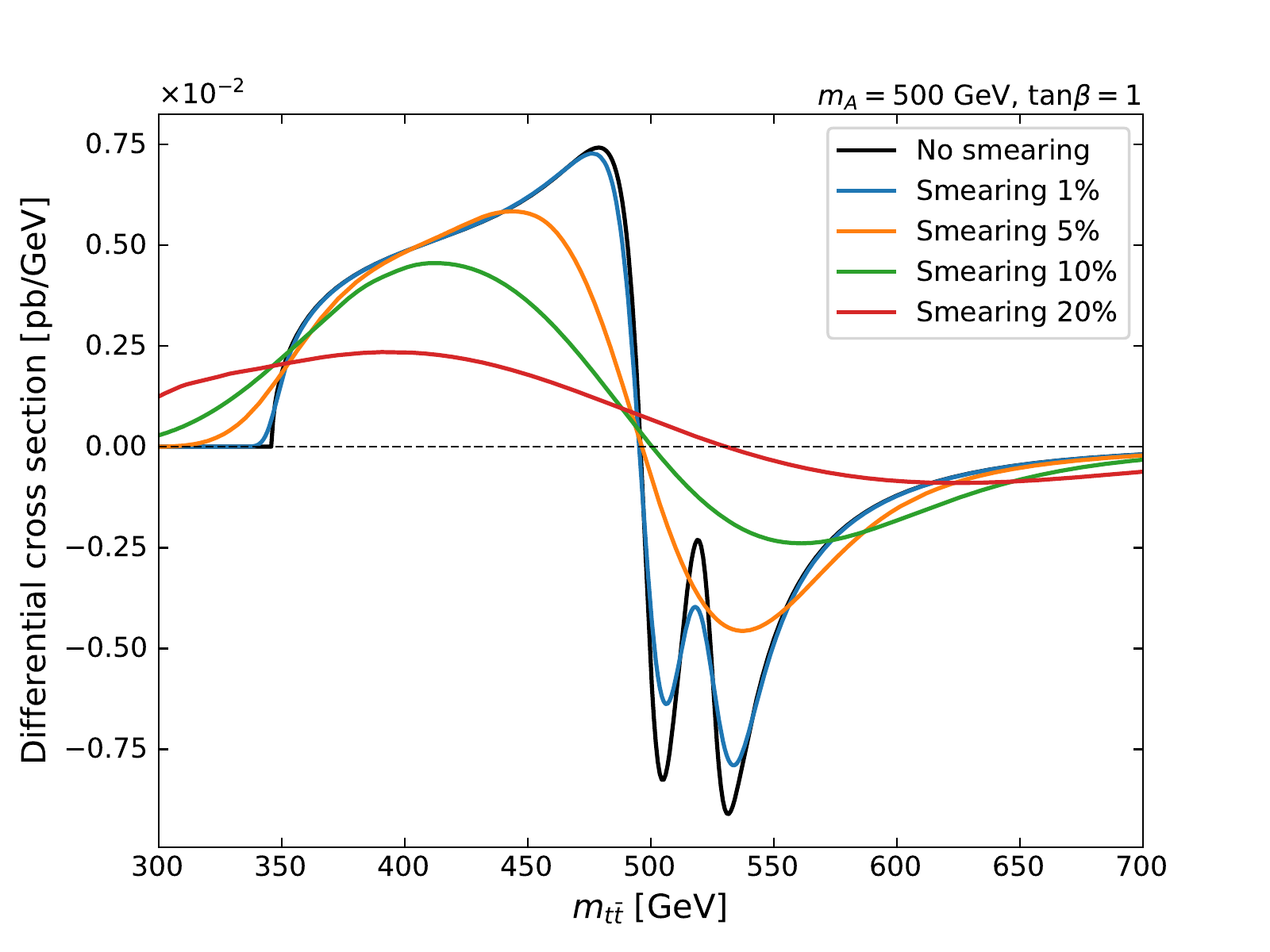}
 \caption{\it The full BSM contribution to the \mtt~spectrum for the hMSSM with $M_A = 500$\,GeV, $\tan\beta = 1$, as obtained for different values of the \mtt~resolution.}
 \label{Fig:SmearedMtt}
\end{figure}

Finally, the parton-level~\mtt{} is smeared using a convolution with a Gaussian kernel:
\begin{linenomath}
\begin{equation}
 \label{Eq:MttSmearing}
 \frac{\mathrm d\tilde\sigma}{\mathrm d\mtt} \; = \; \int \frac{\mathrm d\sigma}{\mathrm d\mtt'} \epsilon(\mtt') \cdot \frac{1}{\sqrt{2\pi\, (r \cdot \mtt')^2}} \exp\left(-\frac{(\mtt - \mtt')^2}{2\, (r \cdot \mtt')}\right) \,\mathrm d\mtt' \, ,
\end{equation}
\end{linenomath}
where $r$ is the specified \mtt~resolution. The integral is computed
numerically, truncating it to the segment $\mtt \cdot (1 \pm 3r)$. The resulting
differential cross section is plotted in Fig.~\ref{Fig:SmearedMtt} for a
representative  hMSSM scenario and various values of the resolution. We see that
the fine details of the mass distribution are lost when the smearing is 10\% or
more, but a peak-and-dip structure is still potentially observable.

The statistical analysis exploits the distribution of the smeared \mtt{} after
the (emulated) event selection. The distribution is described by a histogram
covering the range $350$~GeV $< \mtt < 1200$\,GeV  and contains 24 bins with
relative widths of about 5\%. The distribution for \ttbar{} events due to the SM
is constructed from the MC sample directly, whilst the distribution for the BSM
component is built for each signal hypothesis by integrating  $\mathrm
d\tilde\sigma / \mathrm d\mtt$~eq.~\eqref{Eq:MttSmearing} over each bin of the
histogram. Both distributions are scaled to the desired integrated luminosity.

The \mtt~distributions are affected by various systematic uncertainties that we
estimate as follows. In the signal, the renormalisation scale in the matrix
element is varied by a factor~2 in each direction, simultaneously for both the
resonant part and the interference. A number of uncertainties in the SM~\ttbar{}
background are considered. The overall rate is varied by 10\%, which represents
theoretical uncertainties in the inclusive cross section  and some experimental
uncertainties, such as the uncertainty in the $b$-tagging
efficiency~\cite{Sirunyan:2017ezt}, that affect mostly the overall rate. Values
of \mtt{} are rescaled as $(1 \pm \alpha)\,\mtt$, $\alpha = 0.01$, which serves 
as a conservative proxy for the uncertainty in the jet energy
scale~\cite{Khachatryan:2016kdb}. Renormalisation and factorization scales in
the matrix element are varied independently by factors of 2. Since theoretical
uncertainties in the overall \ttbar~rate have already been covered by the
corresponding uncertainty,  the distributions obtained in each variation are
rescaled so that the inclusive cross section does not change. The variation in
the renormalisation scale is done simultaneously with its counterpart in the BSM
component. The renormalisation scale used in the final-state radiation, which is
controlled by the Pythia parameter \texttt{TimeShower:renormMultFac}, is also
varied by a factor of 2, using MC samples processed with alternative Pythia
configurations. The mass of the top quark is varied by $\pm
0.5$\,GeV~\cite{Khachatryan:2015hba}, using the corresponding dedicated MC
samples. Finally, 30 uncertainties provided in the PDF set as well as the
variation of $\alpha_s$ in PDF are also included. The MC statistical
uncertainties are ignored, effectively assuming that they are much smaller than
statistical uncertainties in the data.

Several systematic variations for the SM~\ttbar{} events (most notably, those
constructed from dedicated MC samples) are affected by statistical fluctuations.
The fluctuations are suppressed by smoothing the relative deviations from the
nominal SM~\ttbar{}  distribution using a version of the {\tt LOWESS}
algorithm~\cite{Cleveland79, Cleveland88}. The value of the relative deviation
in each bin of the distribution is replaced by the result of the weighted
least-squares linear fit, in which nearby bins receive larger weights than those
that are further away from the bin of interest.

The statistical analysis is implemented in the {\tt RooStats}
framework~\cite{Moneta:2010pm}. The model, which includes all the uncertainties
described above, is constructed with the  help of the {\tt HistFactory}
toolkit~\cite{Cranmer:2012sba}. The combined BSM contribution is scaled by an
auxiliary `signal strength' parameter~$\mu$, such that with $\mu = 1$ the model
describes the full \mtt~distribution  with the nominal BSM component, and $\mu =
0$ reproduces the SM. Test statistics defined in Ref.~\cite{Cowan:2010js}, which
are based on the profile likelihood ratio, are used,  and their distributions
are described according to the asymptotic formulae provided in the same
reference. For each signal hypothesis, two kinds of results are obtained.

First, the expected significance is computed on the Asimov data
set~\cite{Cowan:2010js} with $\mu = 1$. In this case the null hypothesis is $\mu
= 0$, the alternative is $\mu \geqslant 0$, and the test statistic~$q_0$
(eq.~(12) in Ref.~\cite{Cowan:2010js}) is used. Second, the exclusion of $\mu =
1$ is tested on the background-only ($\mu = 0$) Asimov data set. Here, the null
hypothesis is $\mu \geqslant 1$, the alternative is $\mu = 0$,  and the test
statistic $\tilde q_\mu$ (eq.~(16) in Ref.~\cite{Cowan:2010js}) is used. The
CL$_s$ criterion~\cite{Read:2002hq, Junk:1999kv} is employed,  and the null
hypothesis is said to be excluded at the 95\% confidence level  if $CL_s = p_0 /
(1 - p_1) < 0.05$, where $p_{0,1}$ are the $p$-values under the null and
alternative hypotheses.

\section{Results}
\label{Sec:Results}

The potential sensitivity to the models discussed in Section~\ref{Sec:Models}
is  evaluated here following the procedure described above. As already
mentioned, two benchmark values are used for the relative \mtt~resolution: 10
and 20\%. For the integrated luminosity~\intlumi{}, three values are considered:
150\,fb$^{-1}$, 450\,fb$^{-1}$,  and 3\,ab$^{-1}$, which correspond to the data
collected at Run~2 and the targets at Run~3 and and HL-LHC (for a single
experiment). In total, this gives six experimental scenarios. It should be
noted, however, that the assumed event selection and experimental uncertainties 
are likely to be not fully representative beyond LHC Run~2.

As already mentioned, we have computed cross sections for the BSM scenarios
using the code {\tt SusHi} (version
$1.6.1$)~\cite{Harlander:2012pb,Harlander:2016hcx} from which we obtain the
signal NNLO $K$--factor, for each point of the grid in our two--dimensional
parameter planes. The branching ratios and total widths have been computed with
{\tt HDECAY} (version $6-52$)
\cite{Djouadi:1997yw,Djouadi:2006bz,Djouadi:2018xqq}) and {\tt FeynHiggs}
(version 2.14.3) \cite{Hahn:2013ria,Bahl:2018qog}.

\subsection{The SM with an extra singlet (pseudo)scalar}

\begin{figure}[!ht]
 \centering
\mbox{ \includegraphics[width=0.5\textwidth]{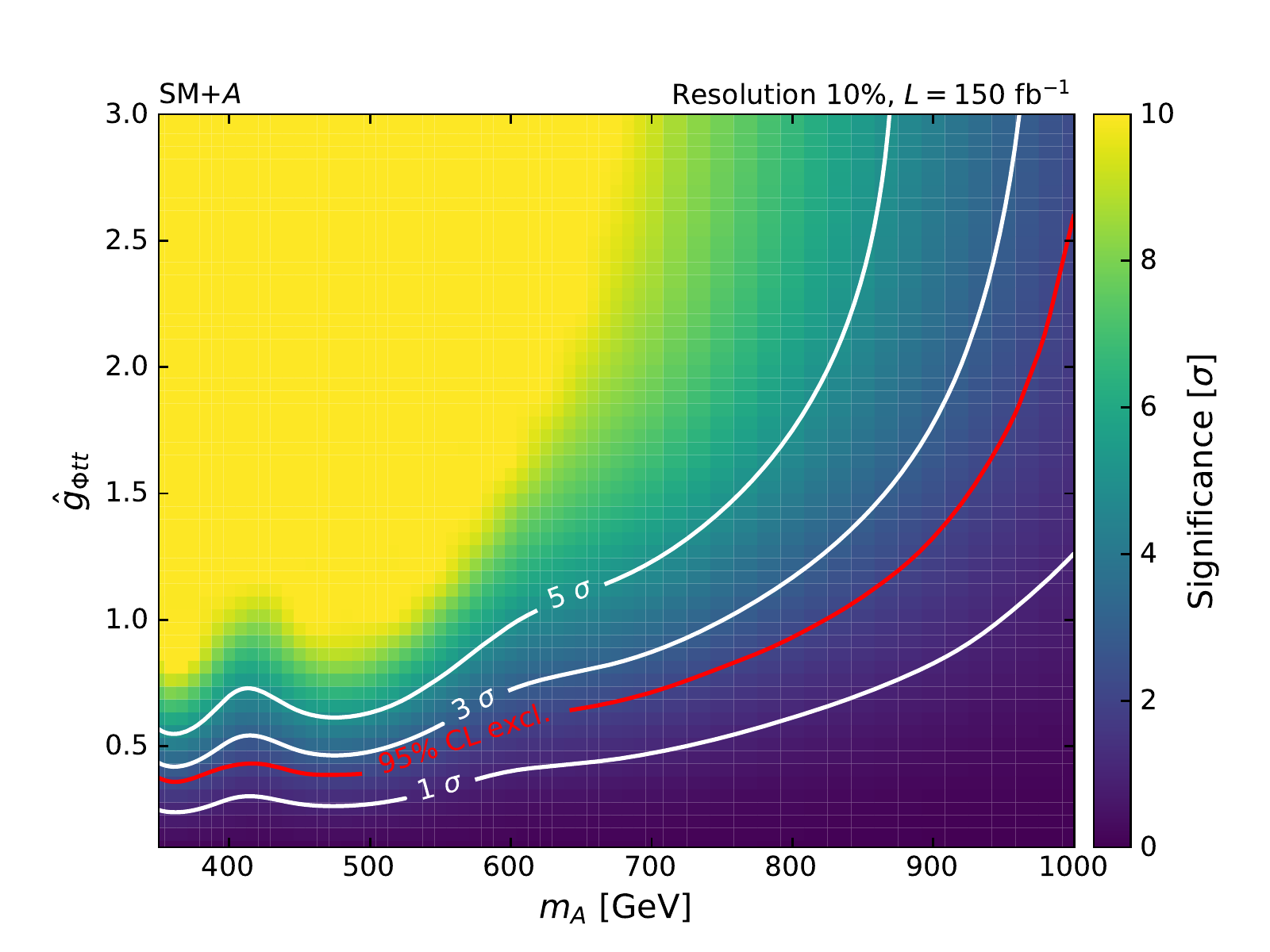}
 \includegraphics[width=0.5\textwidth]{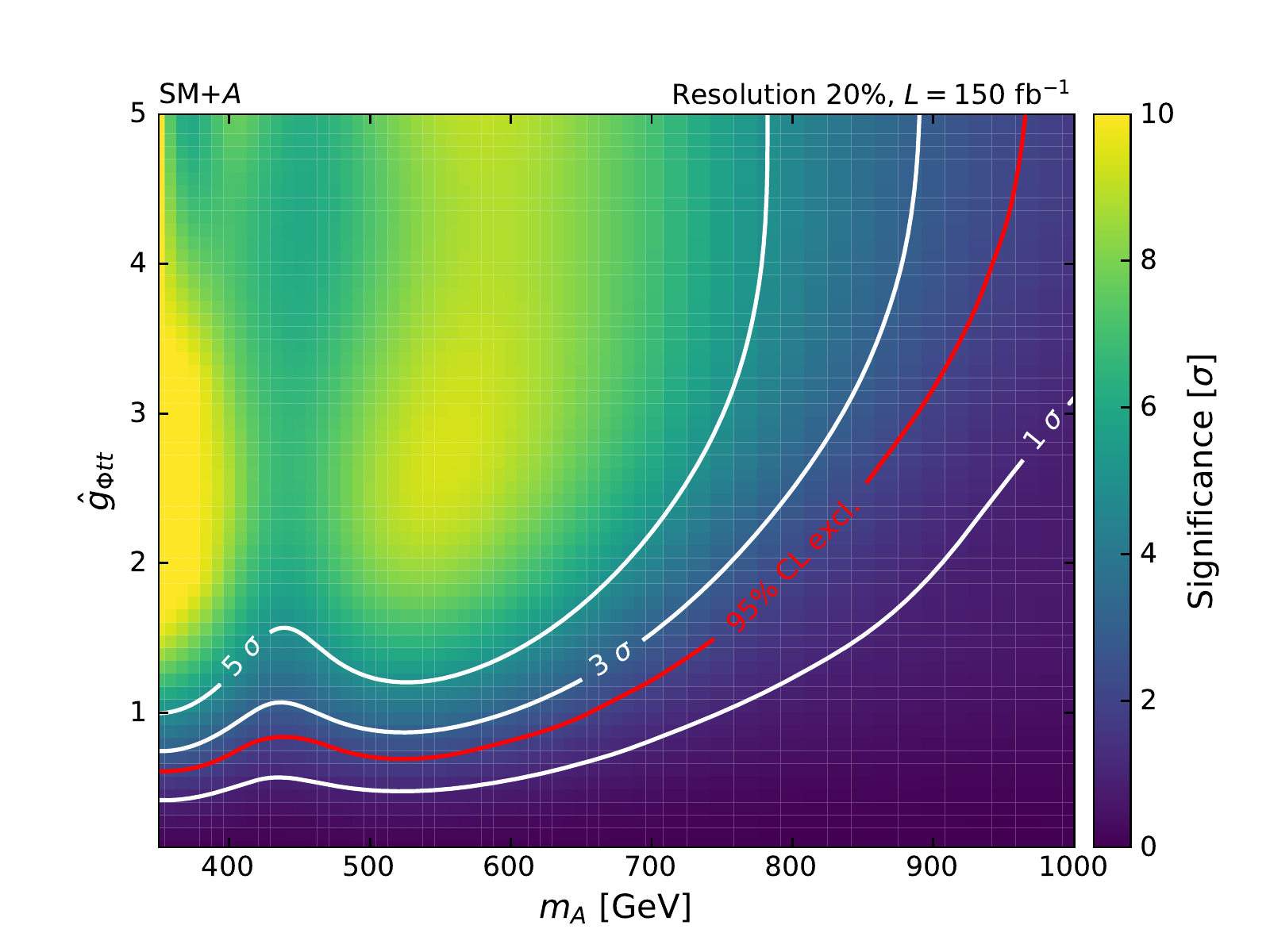} }\\
\mbox{ \includegraphics[width=0.5\textwidth]{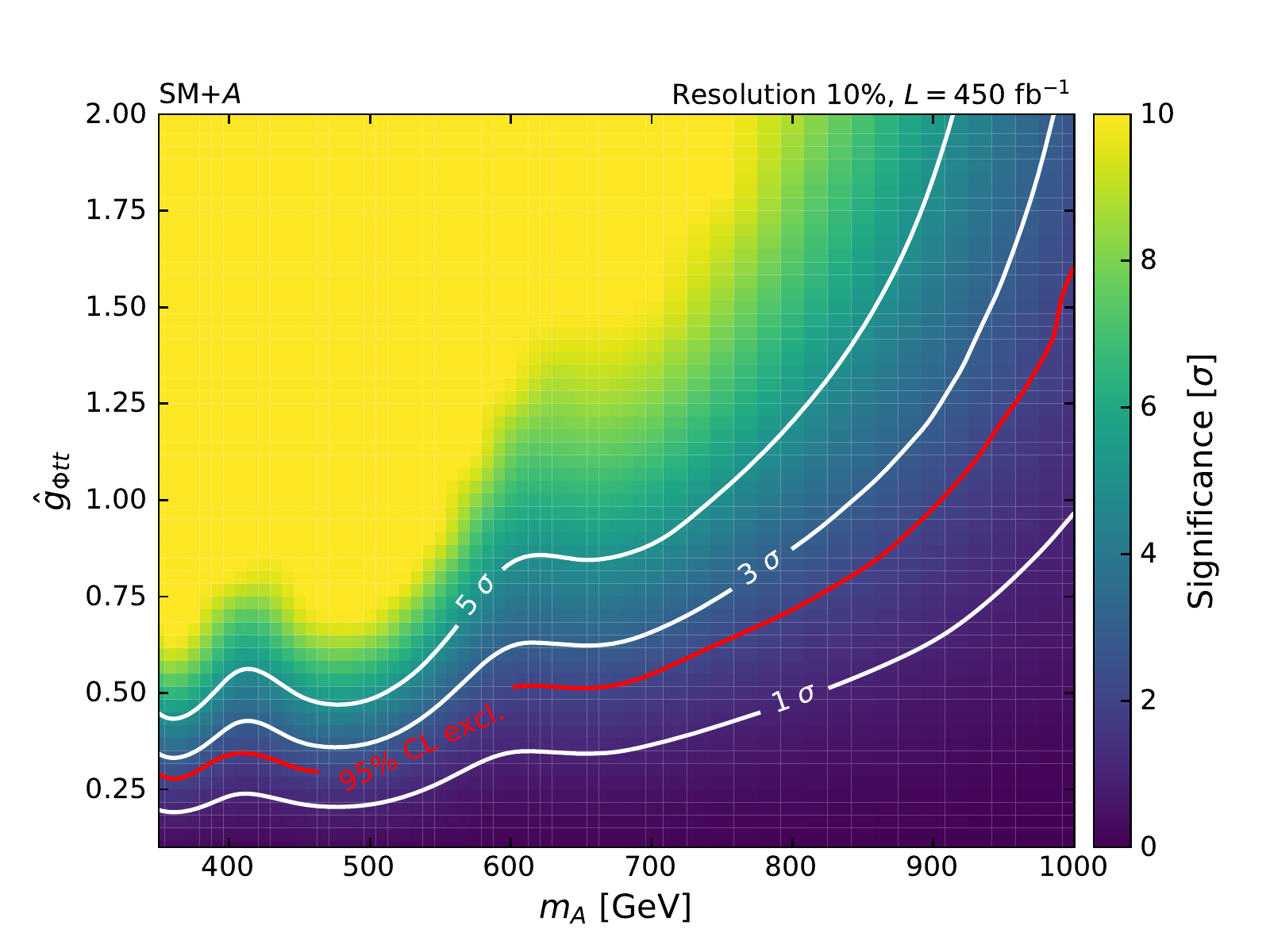}
 \includegraphics[width=0.5\textwidth]{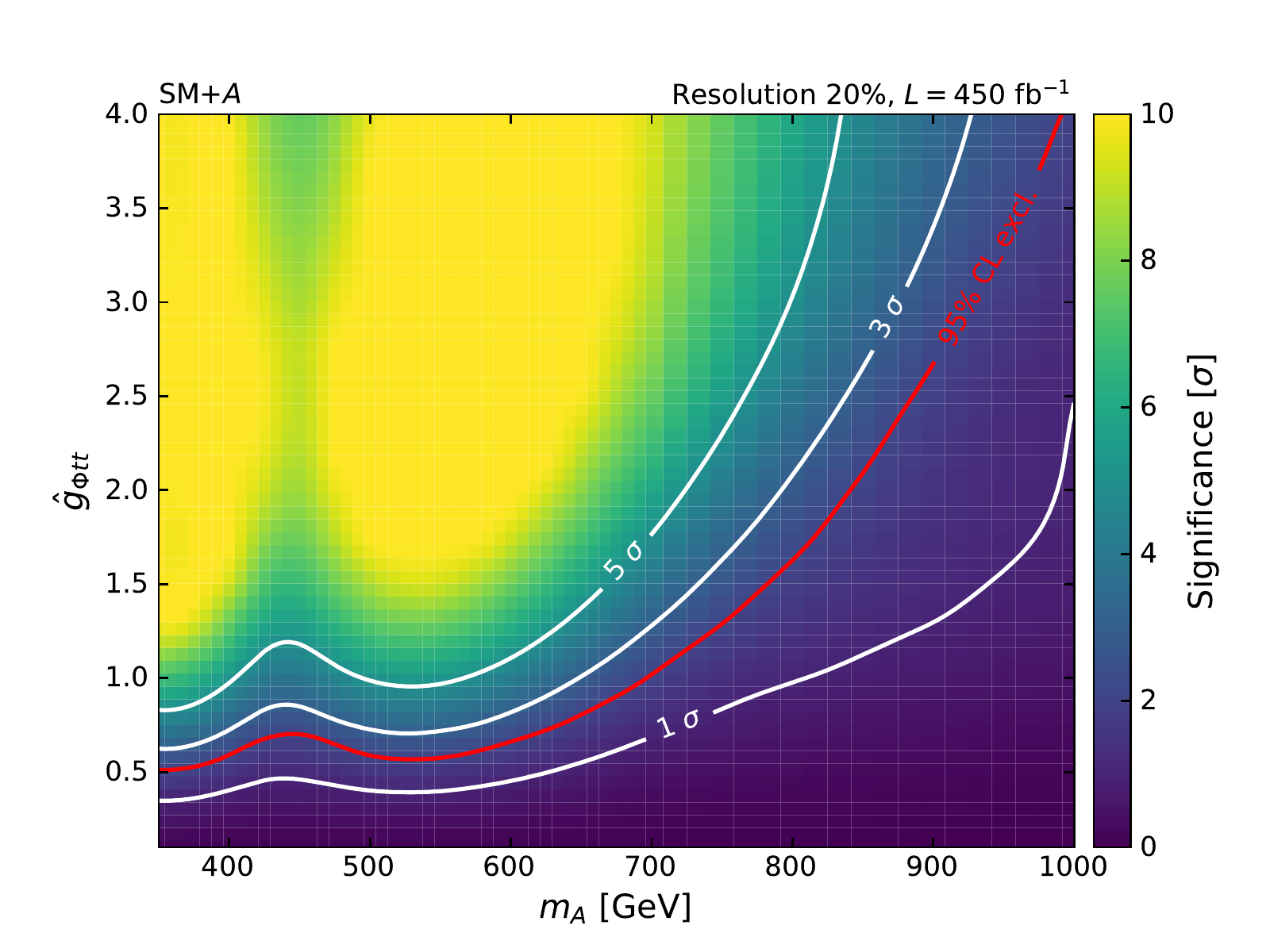} }\\
\mbox{ \includegraphics[width=0.5\textwidth]{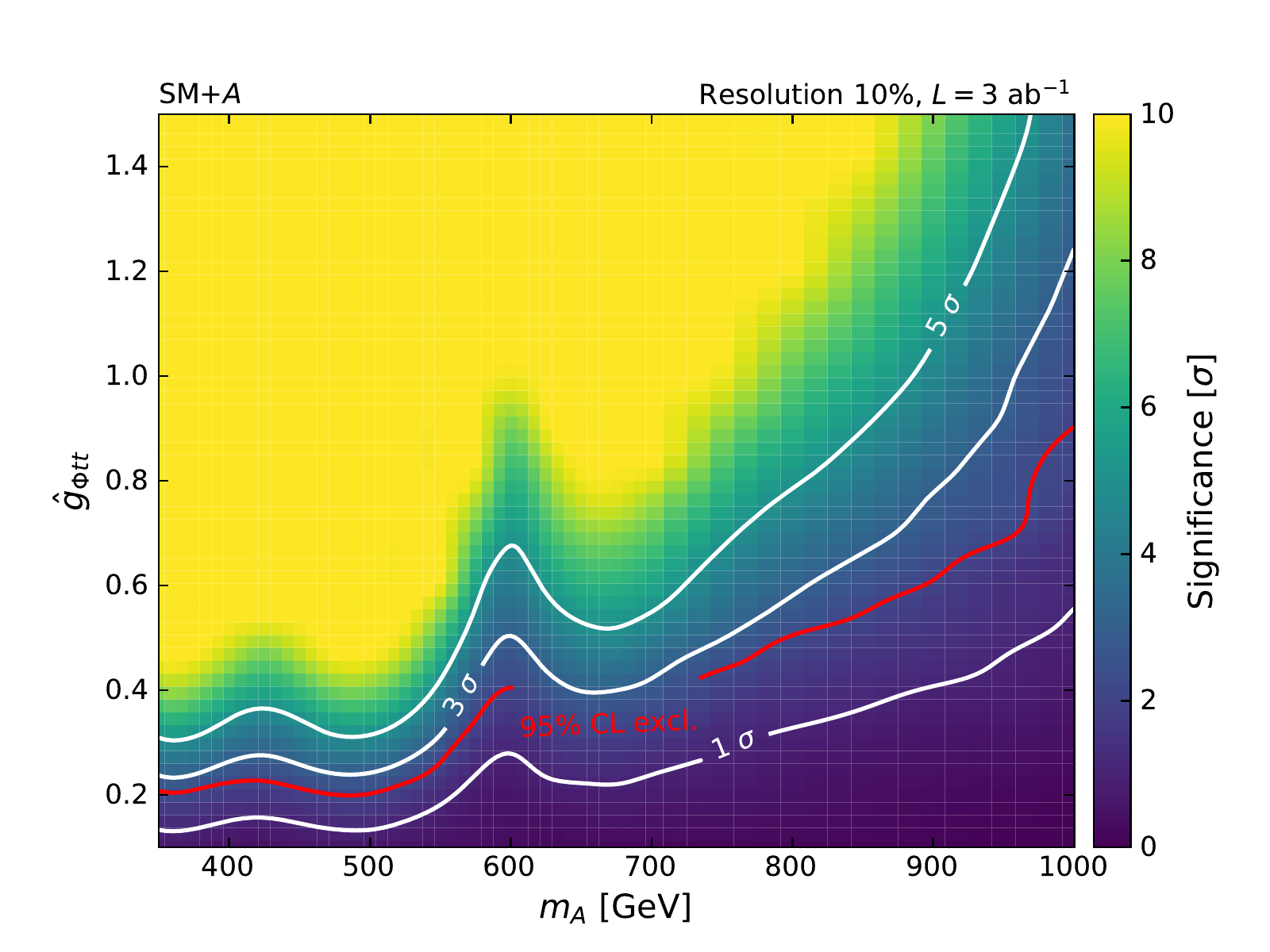}
 \includegraphics[width=0.5\textwidth]{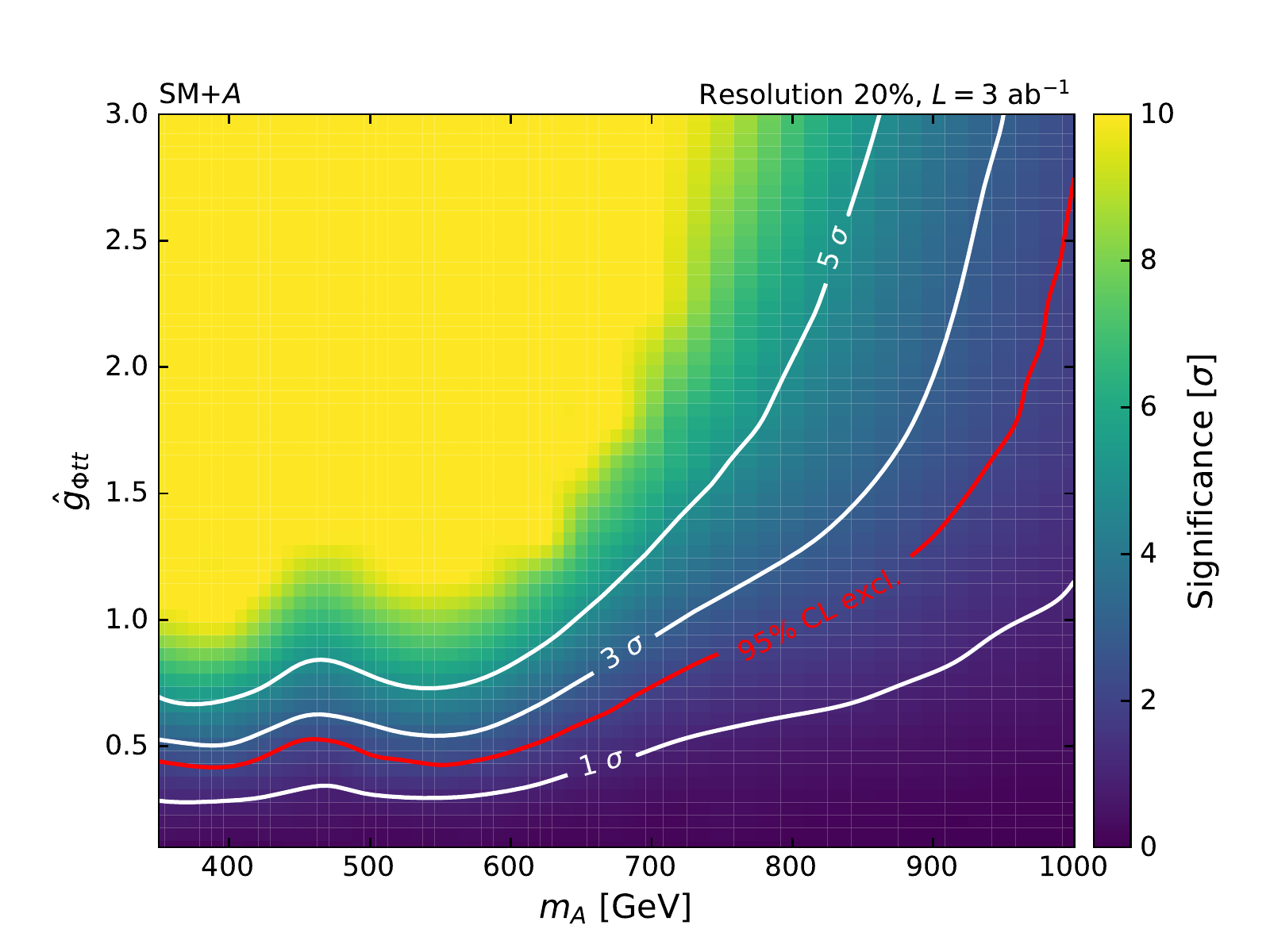} }
 \caption{\it Expected significance and exclusion potential for the $\text{SM} + A$ model in the six experimental scenarios
 considered, as described in the legends and in the text: 10 (20)\% mass resolution in the left (right) panels and 150 (450) (3000)/fb of integrated luminosity in the top (middle) (bottom) row of panels, as indicated in the legend.  Values of significance in excess of $10\,\sigma$ are clipped.}
 \label{Fig:Res_SMA}
\end{figure}

\begin{figure}[!ht]
 \centering
\mbox{ \includegraphics[width=0.5\textwidth]{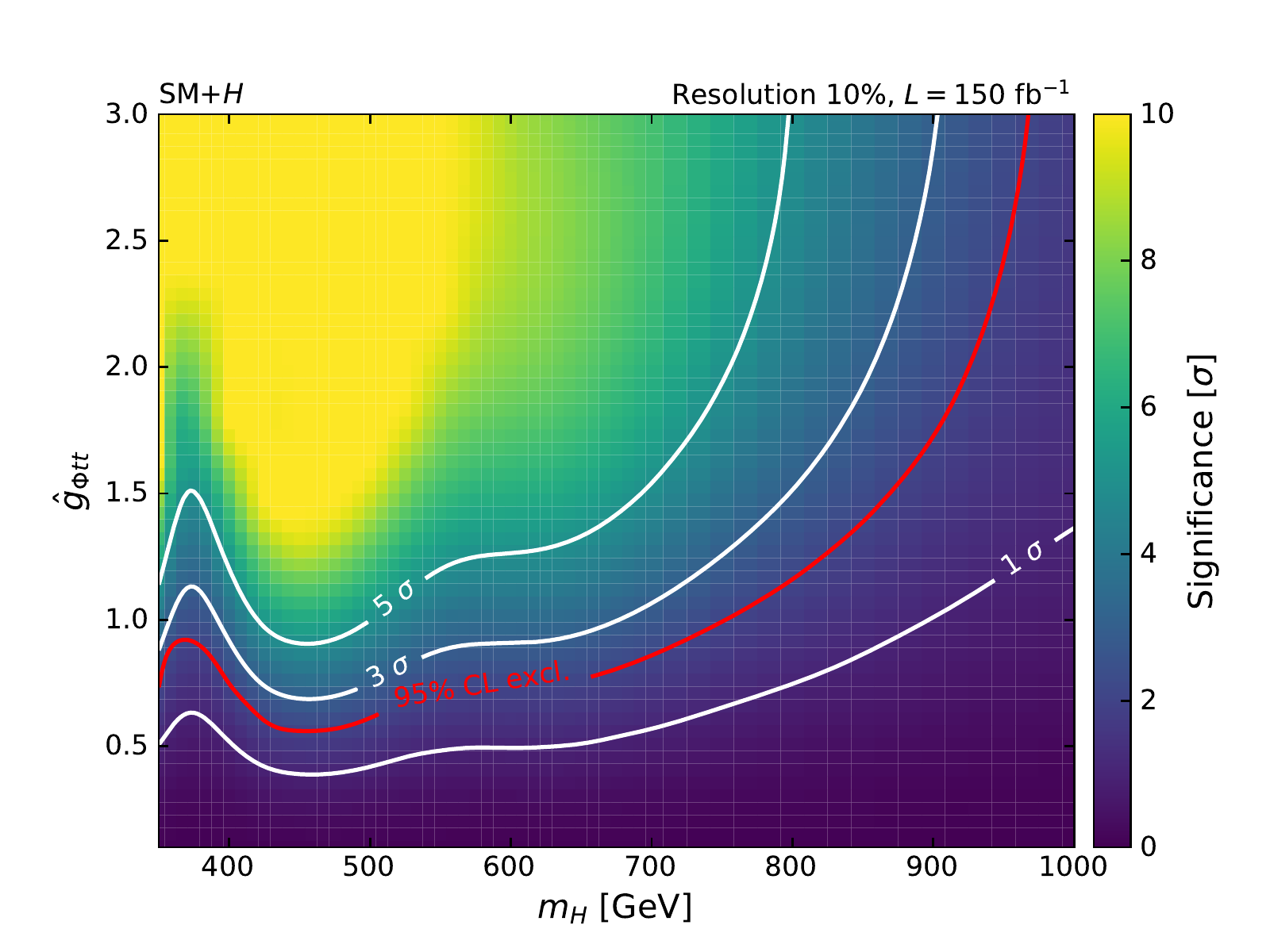}
 \includegraphics[width=0.5\textwidth]{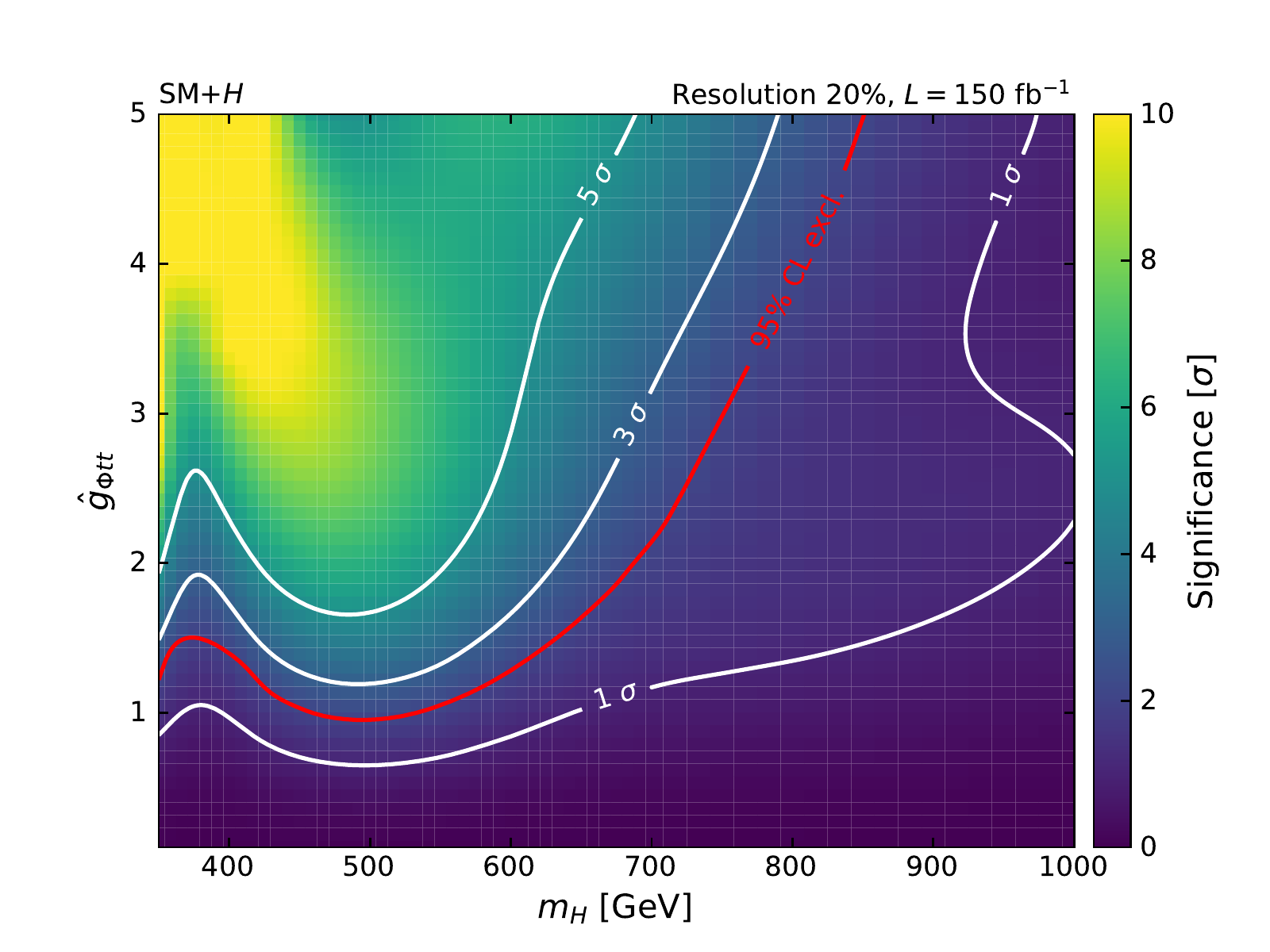} }\\
\mbox{ \includegraphics[width=0.5\textwidth]{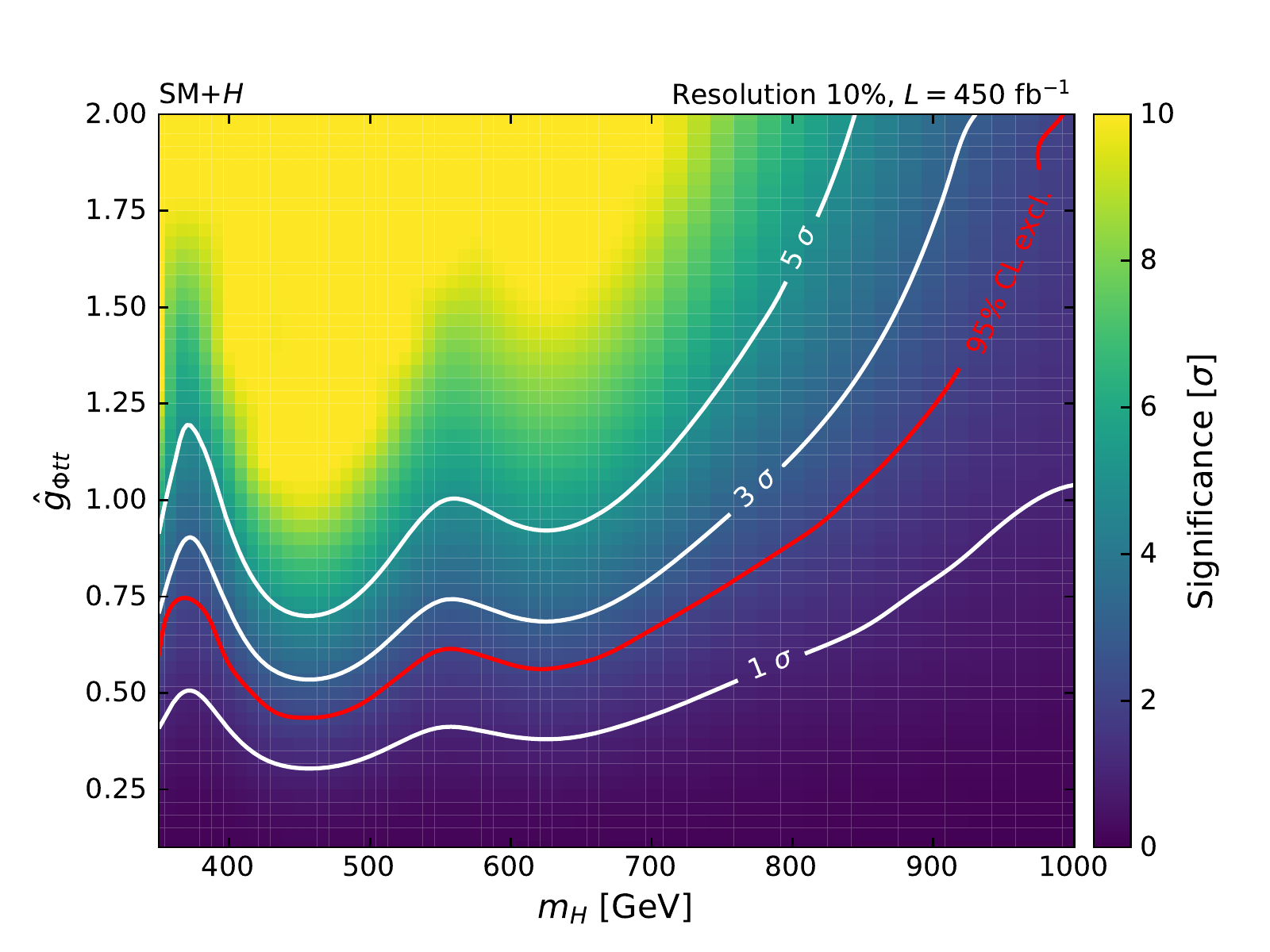}
 \includegraphics[width=0.5\textwidth]{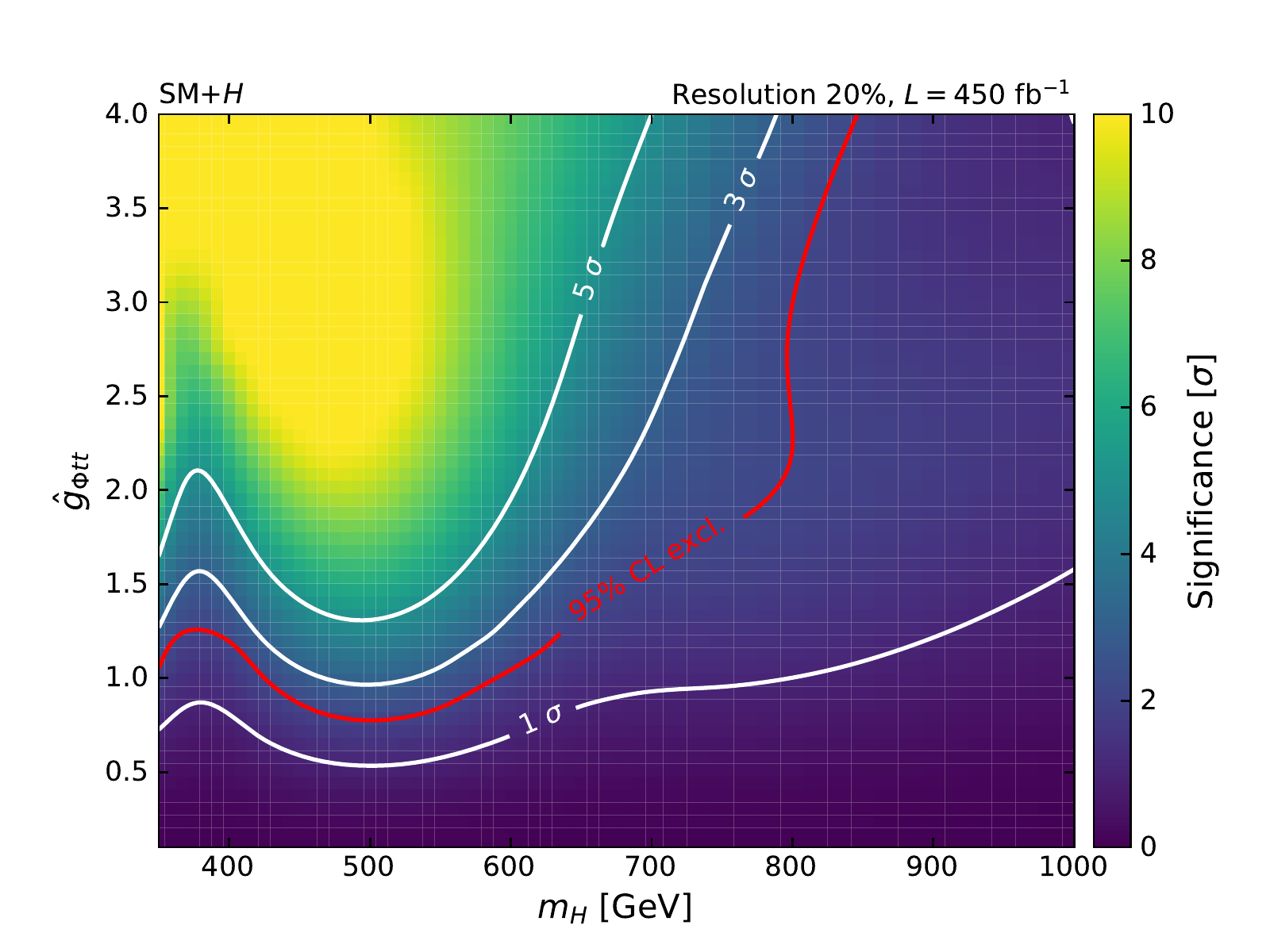} }\\
\mbox{ \includegraphics[width=0.5\textwidth]{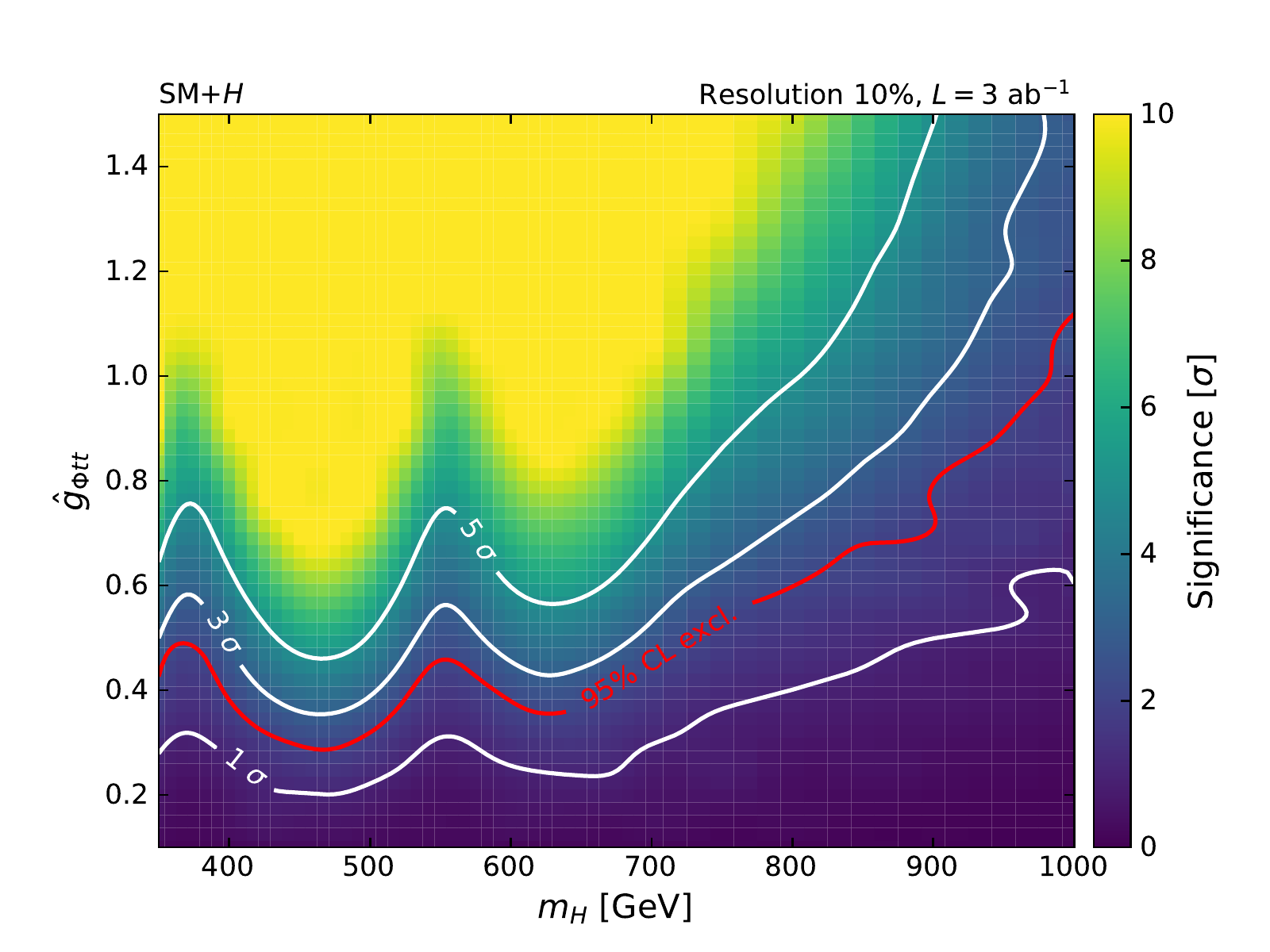}
 \includegraphics[width=0.5\textwidth]{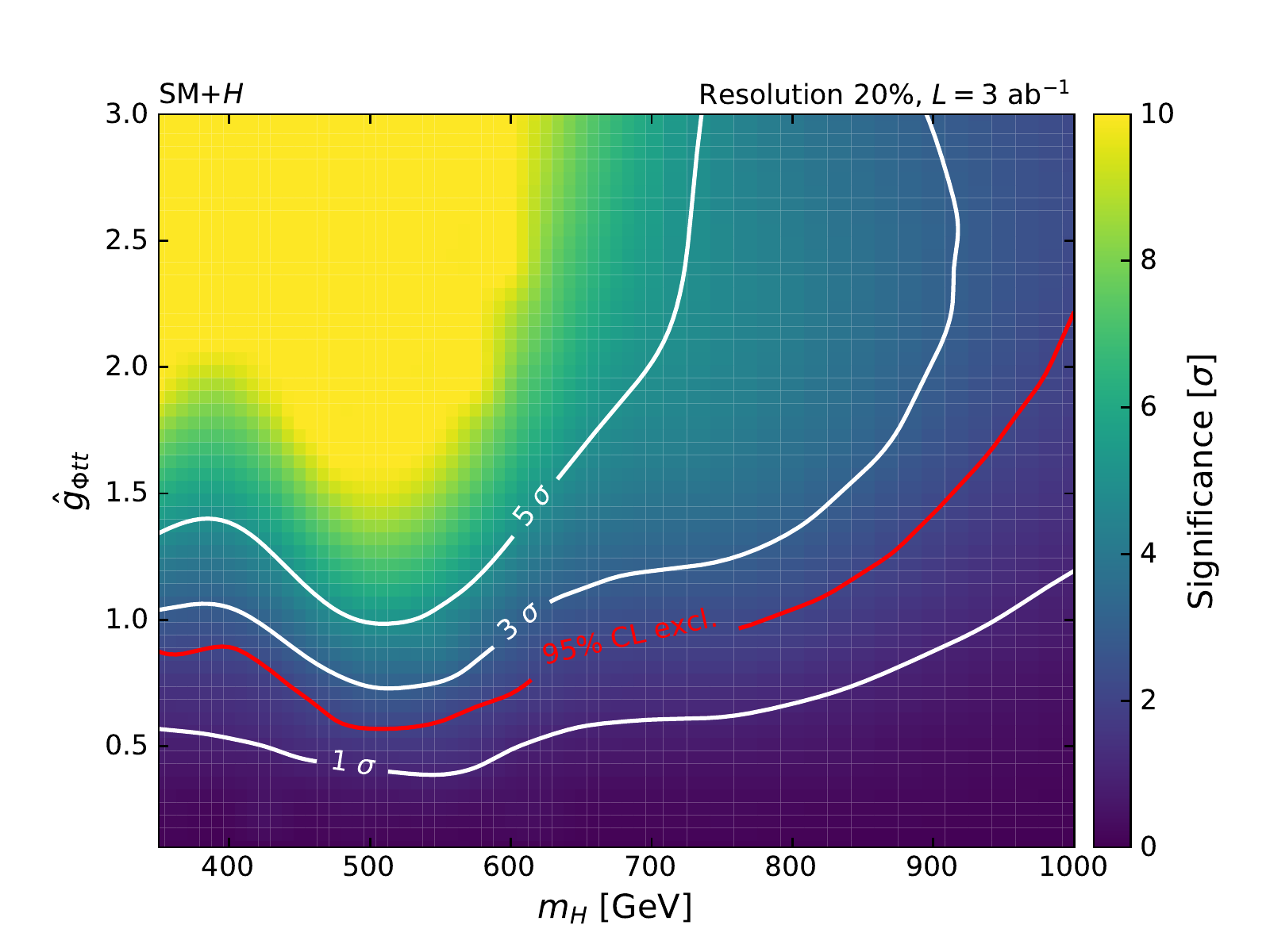}}
 \caption{\it Expected significance and exclusion potential for the $\text{SM} + H$ model in the same six experimental scenarios of Fig.~\ref{Fig:Res_SMH}. Values of significance in excess of $10\,\sigma$ are clipped.}
 \label{Fig:Res_SMH}
\end{figure}

Figures~\ref{Fig:Res_SMA} and \ref{Fig:Res_SMH} show the results for the
$\text{SM} + \Phi$ models,  where $\Phi = A$ or $H$ is a singlet CP--odd or
--even heavy Higgs boson. They are computed, for each of the six experimental
scenarios, as functions of the mass of the  heavy Higgs boson and its reduced
coupling to top quarks. The left plots assume an experimental mass resolution of
10\%, and the right plots assume 20\%. The top row assumes an integrated
luminosity of 150/fb, corresponding to the end of Run~2, the middle row assumes
450/fb, corresponding to the end of Run~3, and the bottom row assumes 3000/fb,
corresponding to the completion of the HL-LHC programme. In each case, the
expected significance is shown with a colour map, and contours corresponding to
values of 1, 3, and 5\,$\sigma$ are plotted in white. The red contour indicates
the boundary of the region in which each point is excluded at the 95\% CL.

Let us consider the implications of the results in Fig.~\ref{Fig:Res_SMA} for a
pseudoscalar $A$ with a $t \bar t$ coupling of the same magnitude as the SM
Higgs boson. In the case with a mass resolution of 10\%, the 5-$\sigma$
discovery sensitivity reaches 600~GeV already with $\intlumi = 150/\text{fb}$,
increasing to 750 (880)~GeV with $\intlumi = 450$ $(3000)/\text{fb}$. With a
mass resolution of 20\% there is no 5-$\sigma$ discovery sensitivity with
150/fb, and the sensitivities with 450 (3000)/fb are reduced to 580 (650)~GeV
with 450 (3000)/fb, with a gap around 450~GeV for 450/fb.  The potential
exclusion ranges from 650~GeV with 20\% mass resolution and 150/fb of integrated
luminosity to over 1~TeV with 10\% mass resolution and 3000/fb of integrated
luminosity. It is encouraging that the search for effects in the $t \bar t$ mass
spectrum has such high sensitivity in the case of a pseudoscalar boson $A$.

In the case of the heavy scalar $H$ shown in Fig.~\ref{Fig:Res_SMH} the
sensitivity is somewhat reduced compared to the case of the pseudoscalar $A$,
because of the lower production rate due first to a smaller value of the form
factor $A_{1/2}^H$ compared to  $A_{1/2}^A$ and then to the suppression by a
$\beta^3$ factor in eqs.~(\ref{Dicus}), as opposed to $\beta$ for the
pseudoscalar. In particular, there is a significant loss of sensitivity at low
mass that implies, for example, that there are holes in the 5-$\sigma$ discovery
potential at low masses $\sim 400$~GeV for 150 and 450/fb of integrated
luminosity, even if a mass resolution of 10\% is assumed. There is also a
degradation in exclusion and discovery sensitivity at large masses. For example,
in the most optimistic case of 10\% mass resolution and $\intlumi = 3/\text{ab}$
the 5-$\sigma$ discovery sensitivity (95\% CL exclusion) corresponds to $M_H
\sim 800$ (980)~GeV. Nevertheless, the reach of the search for effects in the $t
\bar t$ mass spectrum is impressive also in the scalar case.

\subsection{The Type~II 2HDM}

\begin{figure}[!ht]
 \centering
\mbox{ \includegraphics[width=0.5\textwidth]{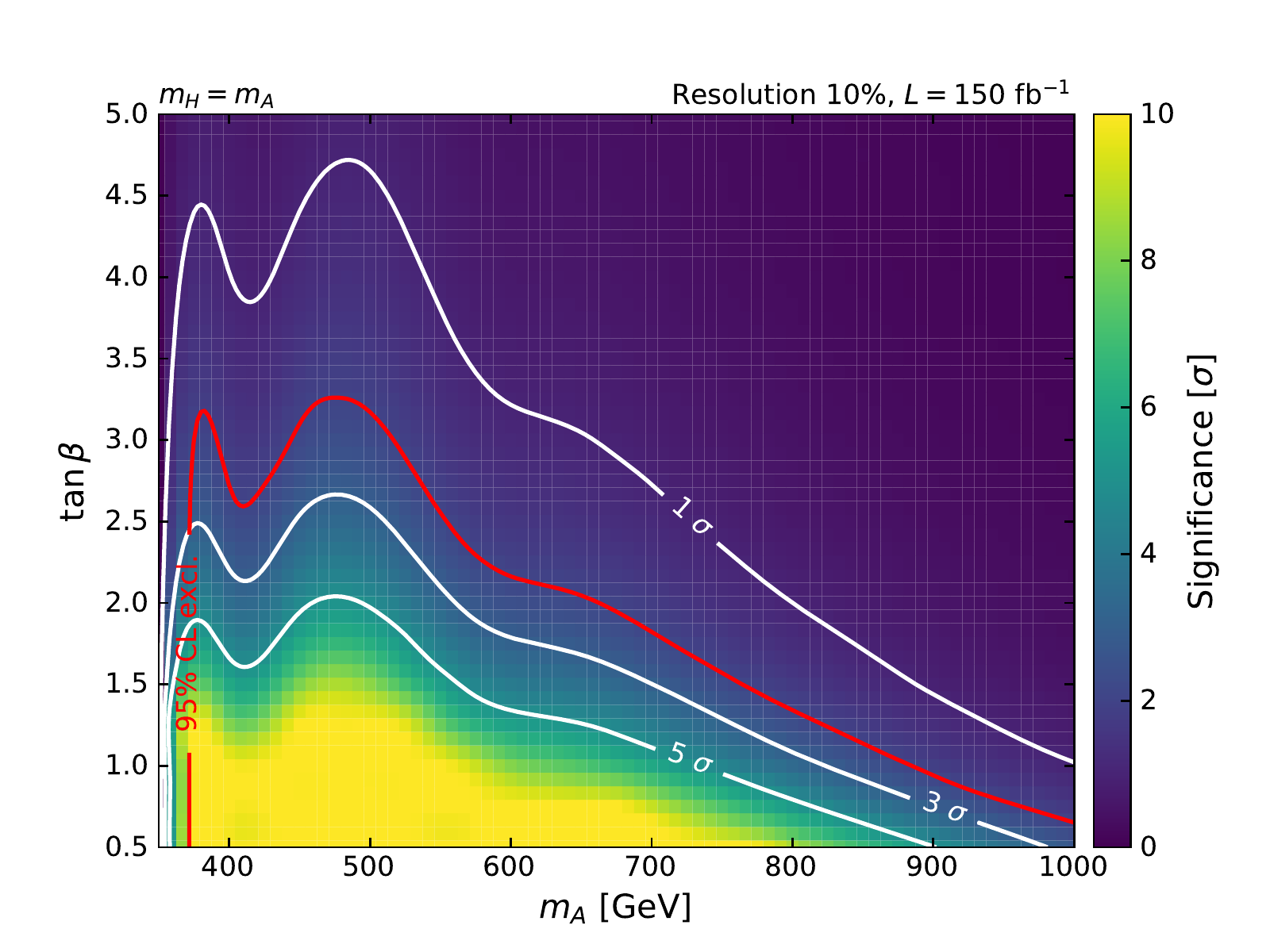}
 \includegraphics[width=0.5\textwidth]{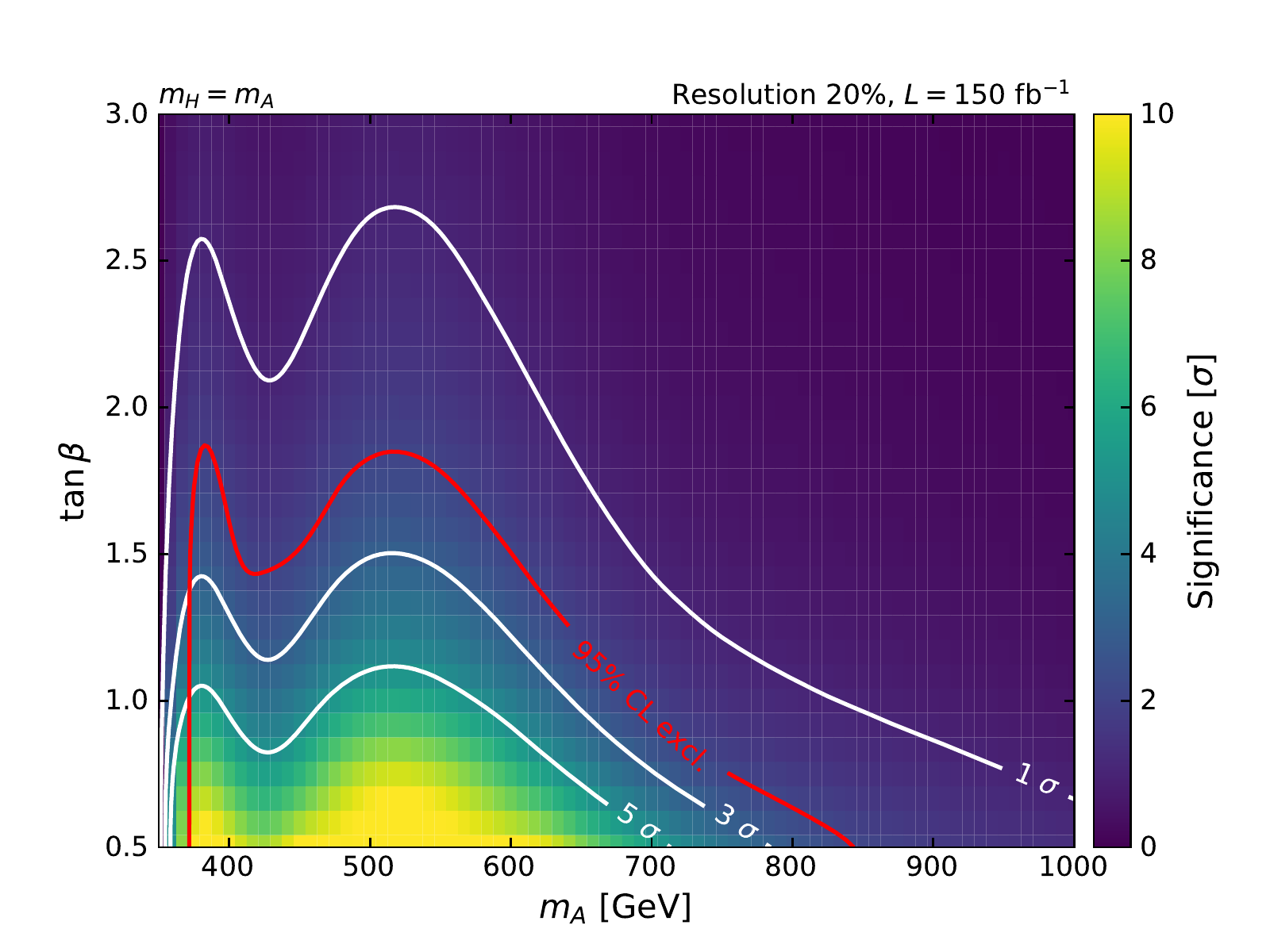} }\\
\mbox{ \includegraphics[width=0.5\textwidth]{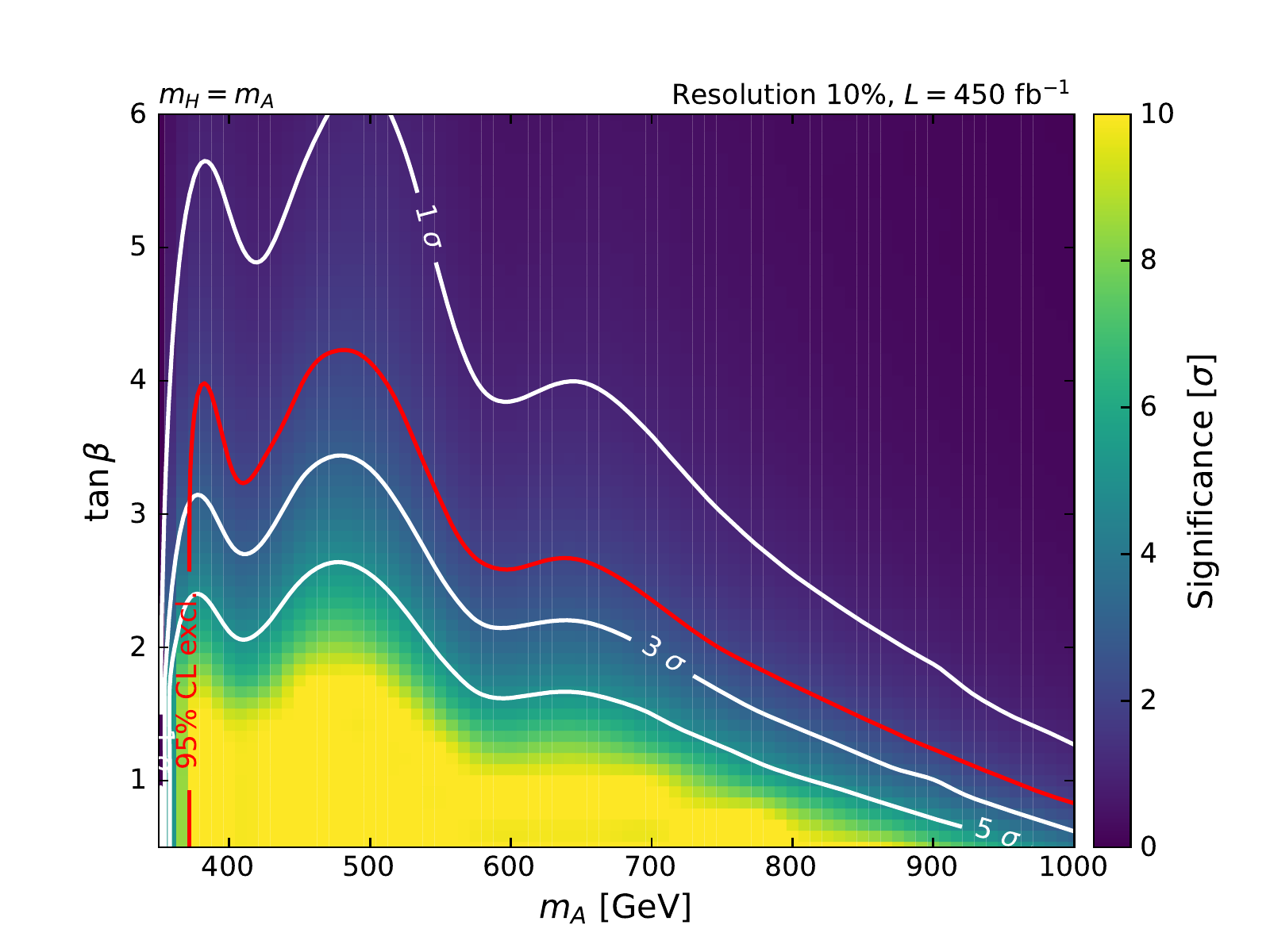}
 \includegraphics[width=0.5\textwidth]{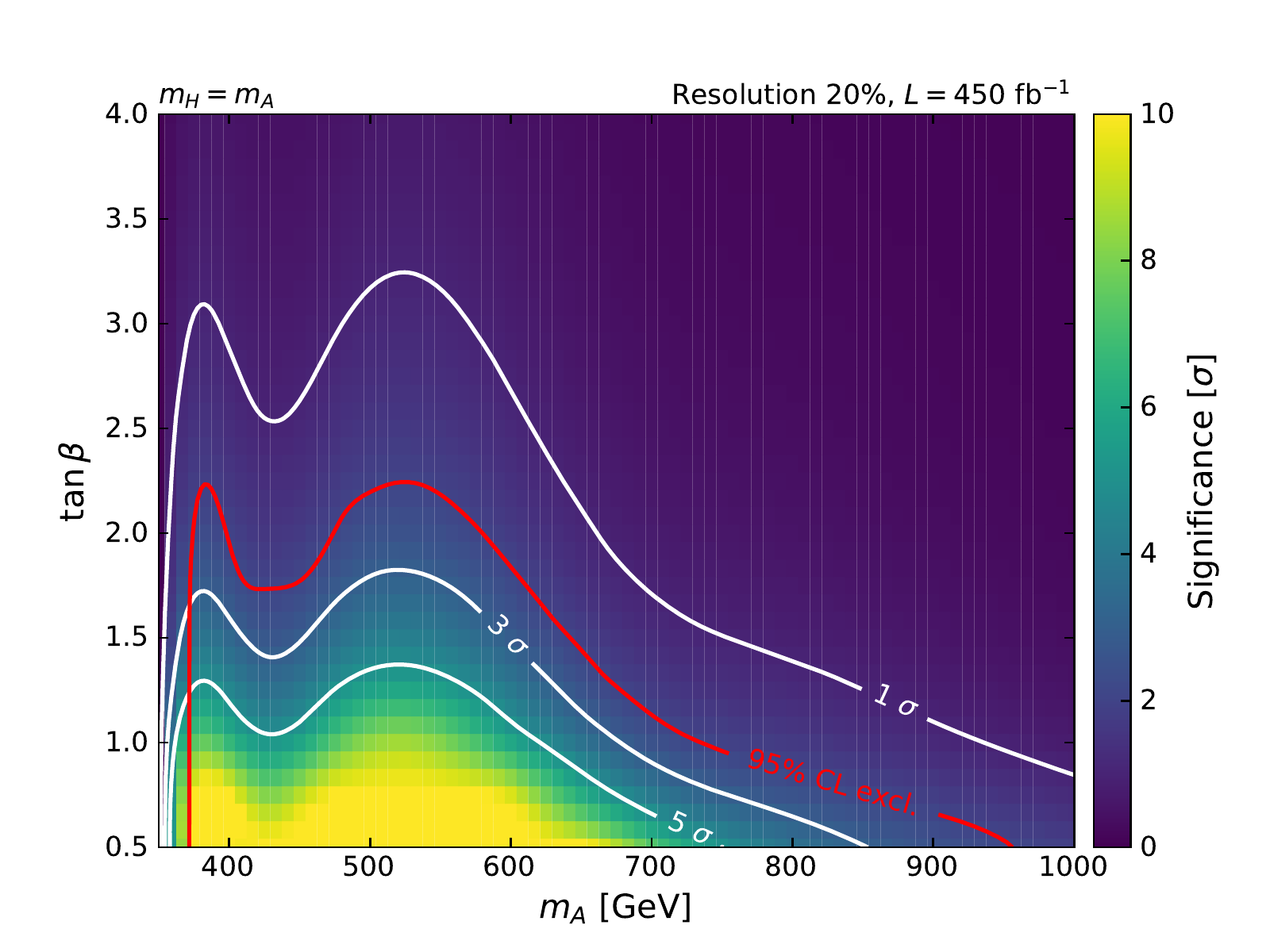} }\\
\mbox{ \includegraphics[width=0.5\textwidth]{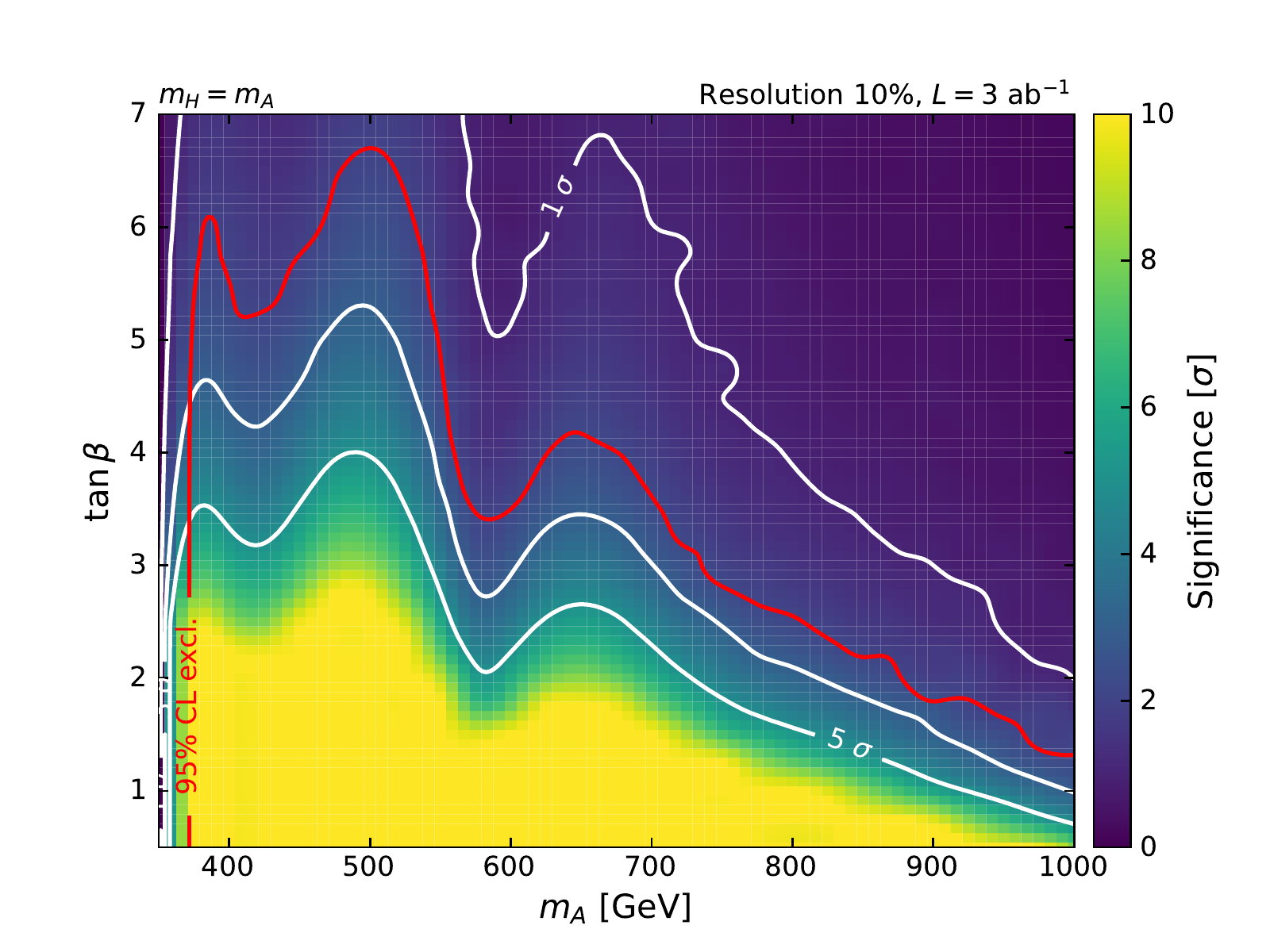}
 \includegraphics[width=0.5\textwidth]{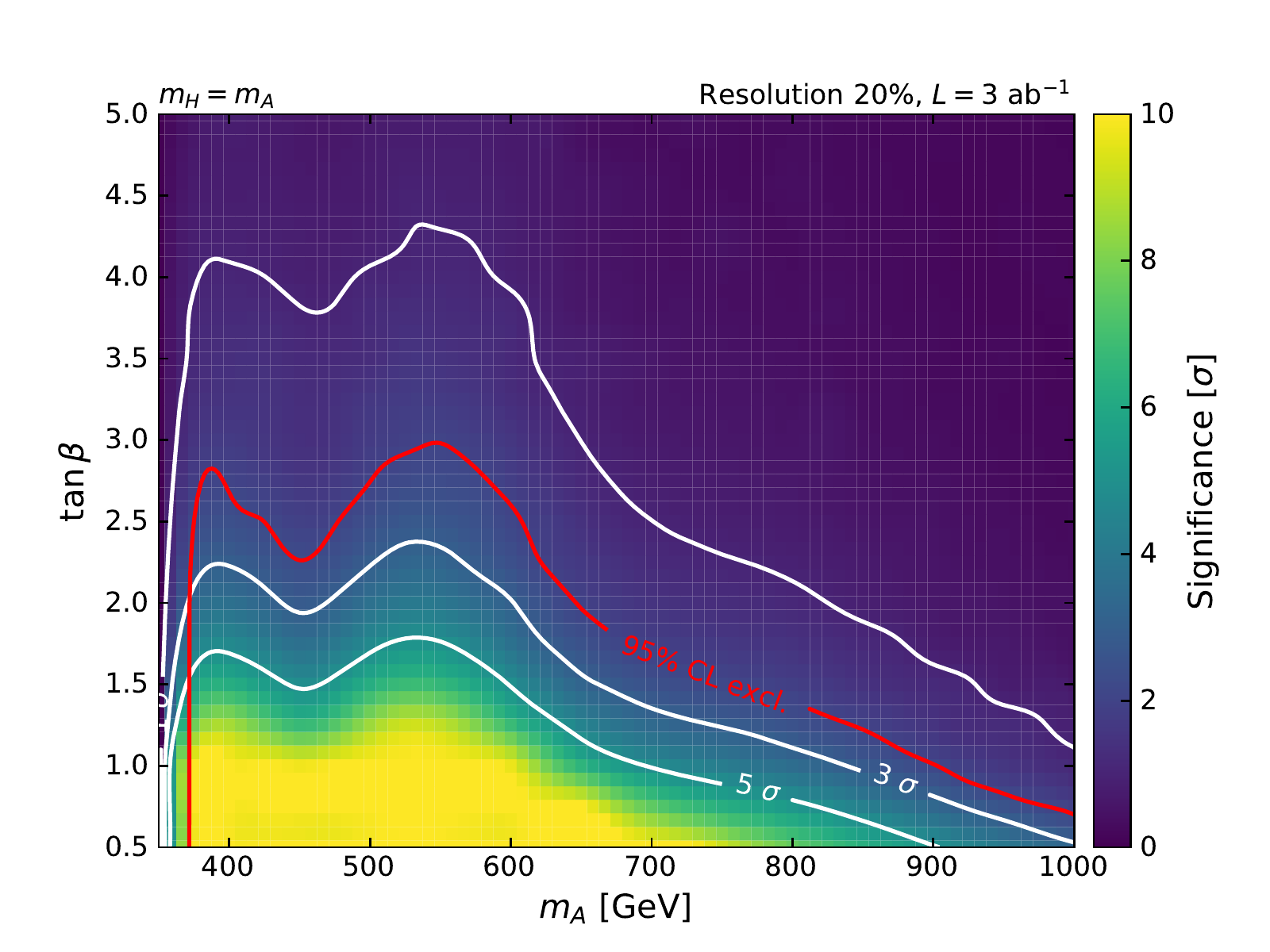}}
  \caption{\it Expected significance and exclusion potential for a Type~II 2HDM
 assuming the mass degeneracy $M_H=M_A$ in the same six experimental scenarios as considered in Fig.~\ref{Fig:Res_SMH}. Values of significance in excess of $10\,\sigma$ are clipped. A resolution of $10\%$ and $20\%$ is assumed for the left- and the right-hand side plots, respectively.}
\label{Fig:Res_2HDM1}
\end{figure}

Fig.~\ref{Fig:Res_2HDM1} presents results for a Type~II 2HDM in the $[M_A, \tb]$
plane, assuming  $M_{H^{\pm}}\ge \text{max}(M_A,M_H)$ and equal scalar and
pseudoscalar masses $M_H=M_A$, with the same mass resolution and integrated
luminosity scenarios as previously. As expected, the experimental sensitivity is
restricted to relatively low values of $\tb$, and we consider the illustrative
case of $\tb = 2$. When the mass resolution is 10\% and $\intlumi =
150/\text{fb}$ points with masses up to 670~GeV can be excluded, but there is no
5-$\sigma$ discovery sensitivity. With 450 (3000)/fb of integrated luminosity,
the exclusion region extends to 750 (900)~GeV, and there is 5-$\sigma$ discovery
sensitivity up to 540 (730)~GeV. On the other hand, if the mass resolution is
20\%, the expected significance on the line $\tb = 2$ does not reach the
discovery level in any of the integrated luminosity scenarios, and no point on
the line is excluded for 150/fb. However, with $\intlumi = 450$ (3000)/fb masses
up to 580 (650)~GeV can be excluded. This example highlights, therefore, the
importance of optimizing the $t \bar t$ mass resolution.

\begin{figure}[!ht]
 \centering
\mbox{ \includegraphics[width=0.5\textwidth]{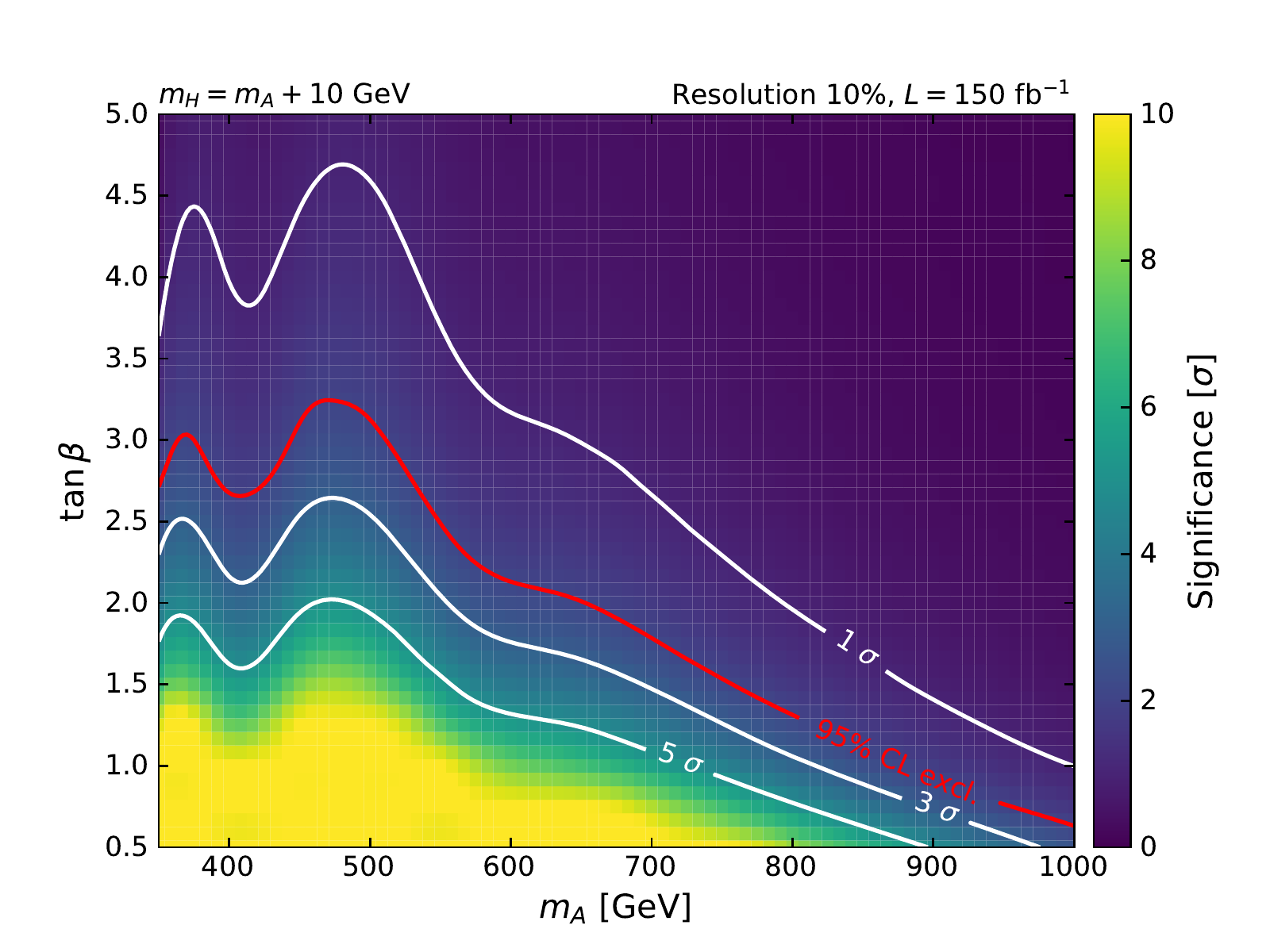}
 \includegraphics[width=0.5\textwidth]{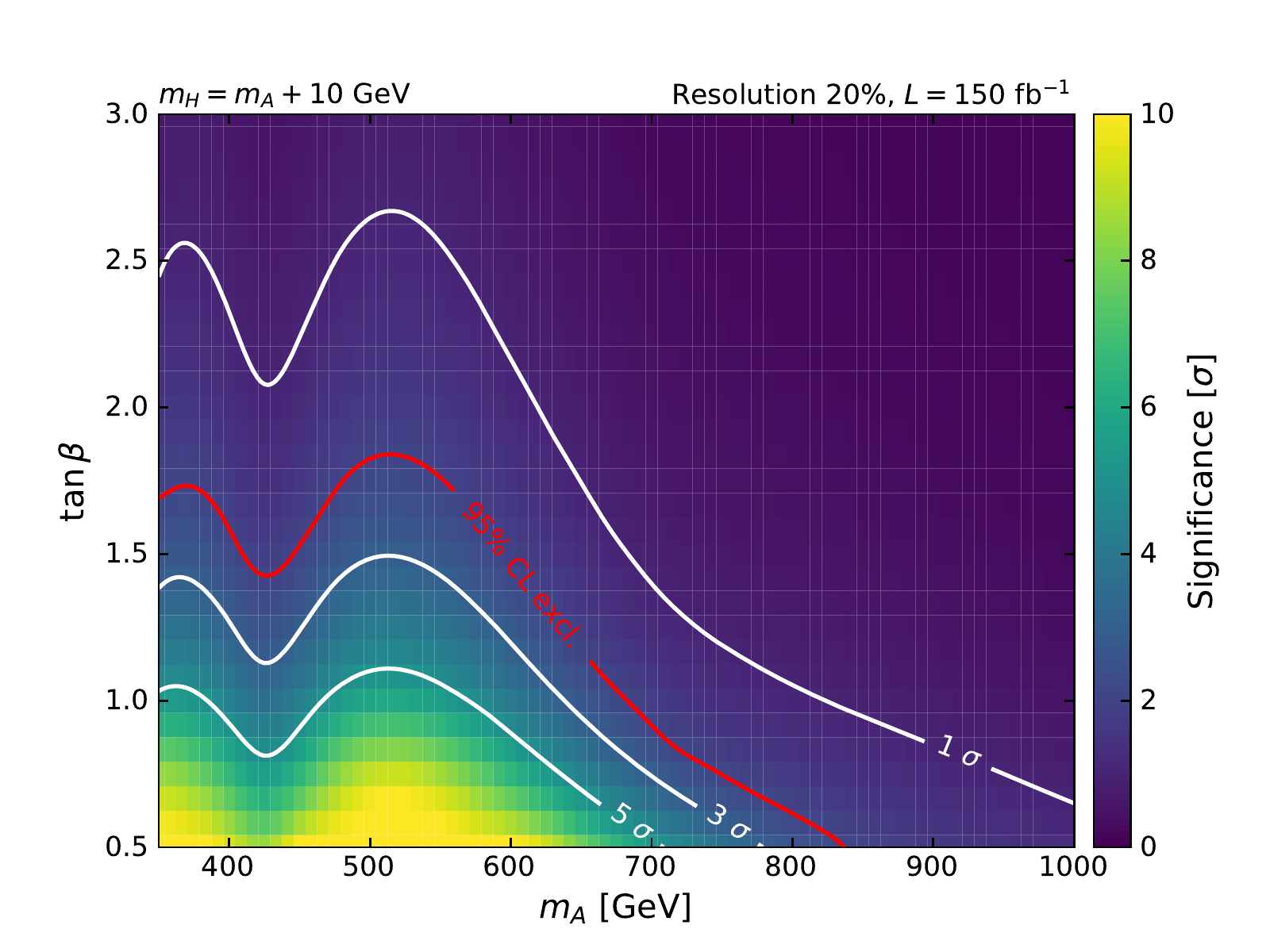} }\\
\mbox{ \includegraphics[width=0.5\textwidth]{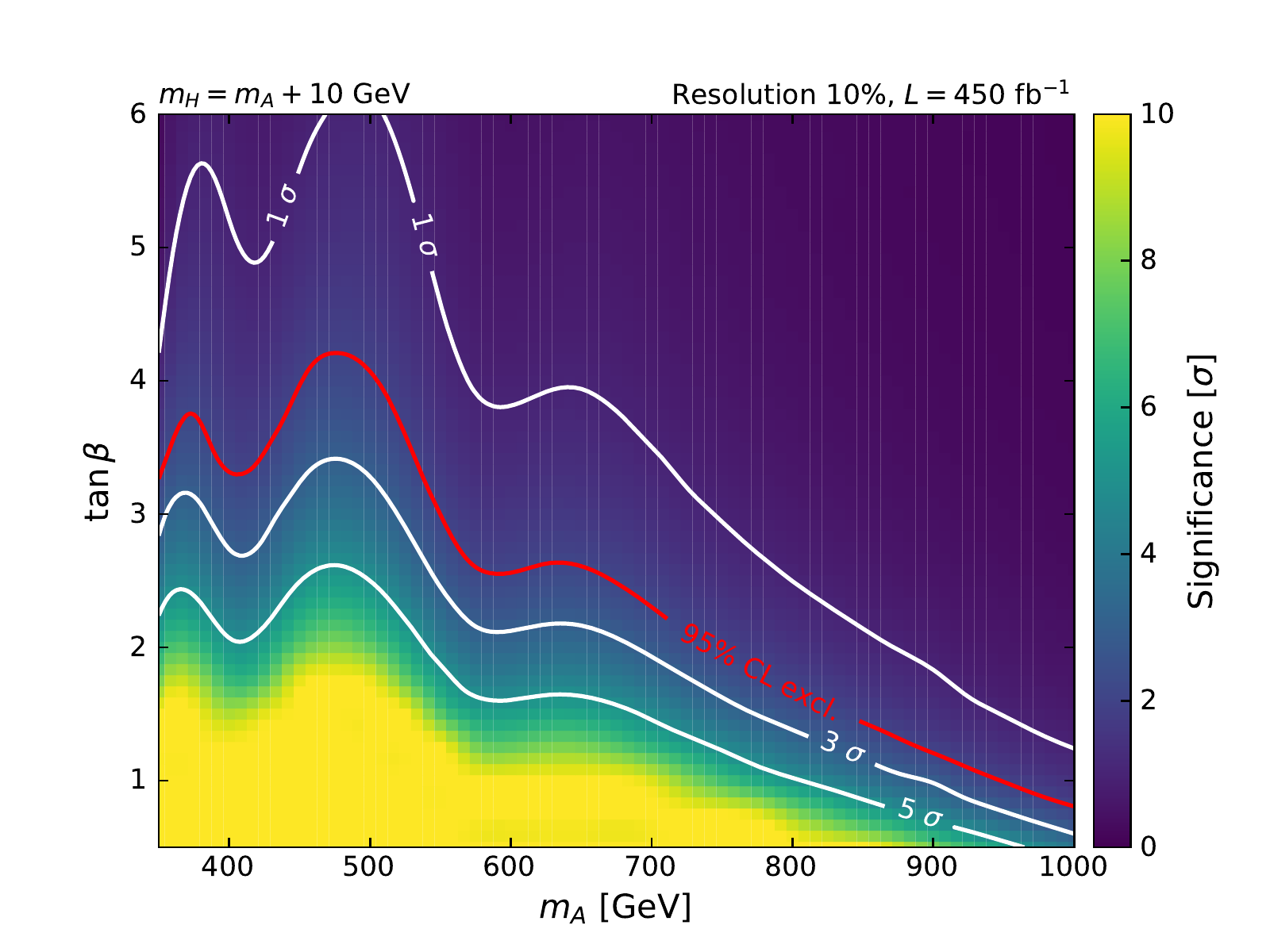}
 \includegraphics[width=0.5\textwidth]{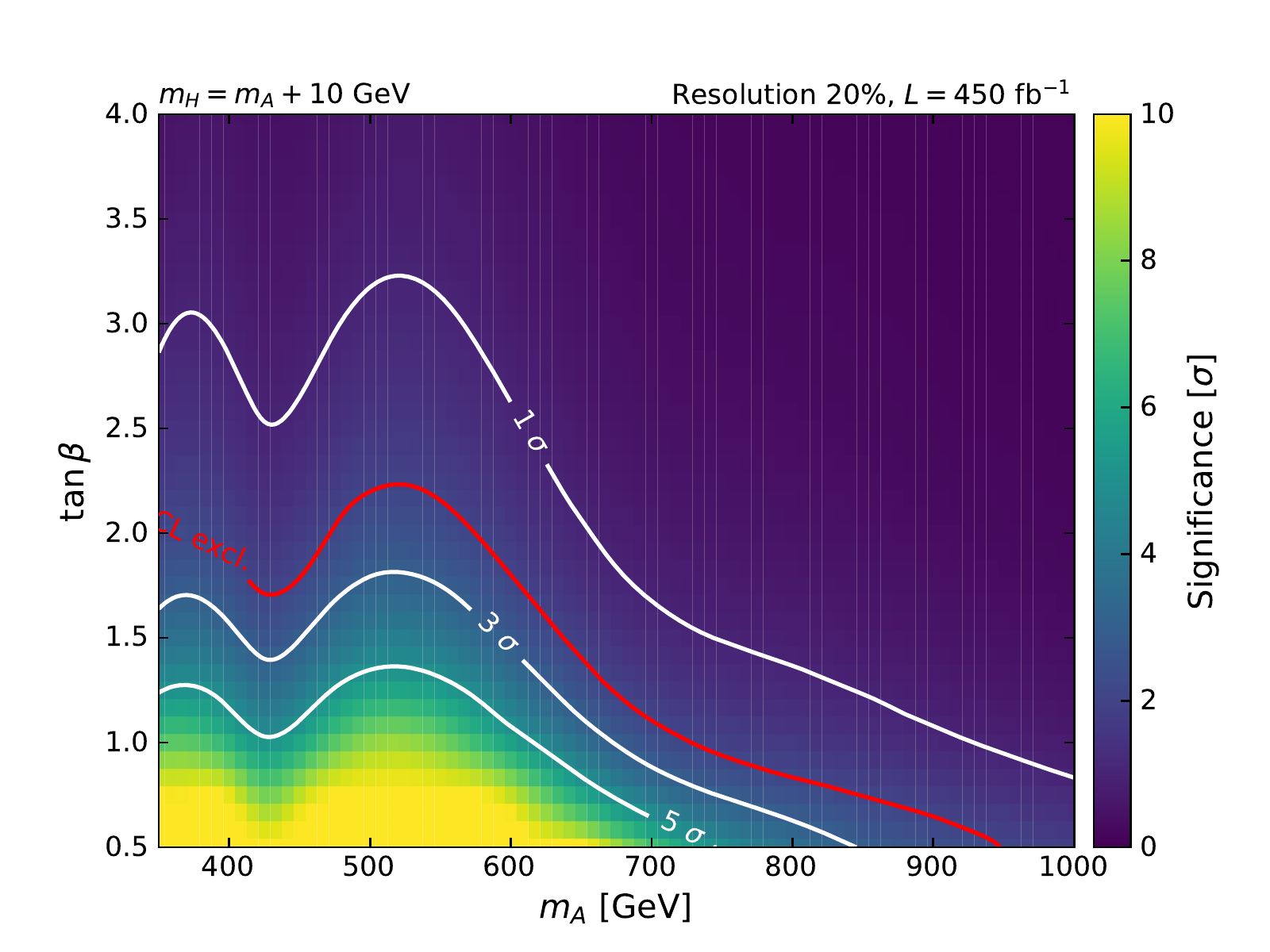} }\\
\mbox{ \includegraphics[width=0.5\textwidth]{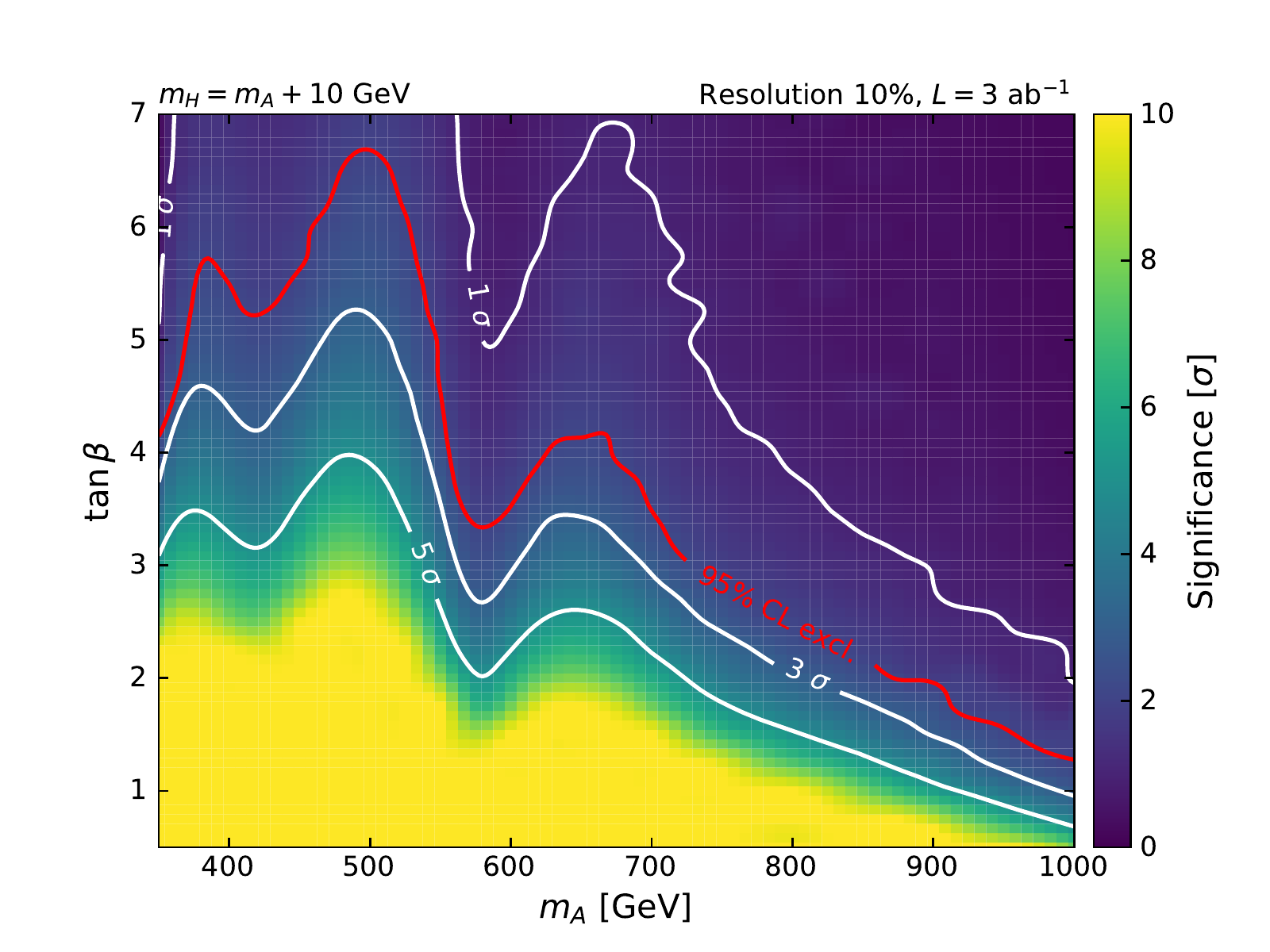}
 \includegraphics[width=0.5\textwidth]{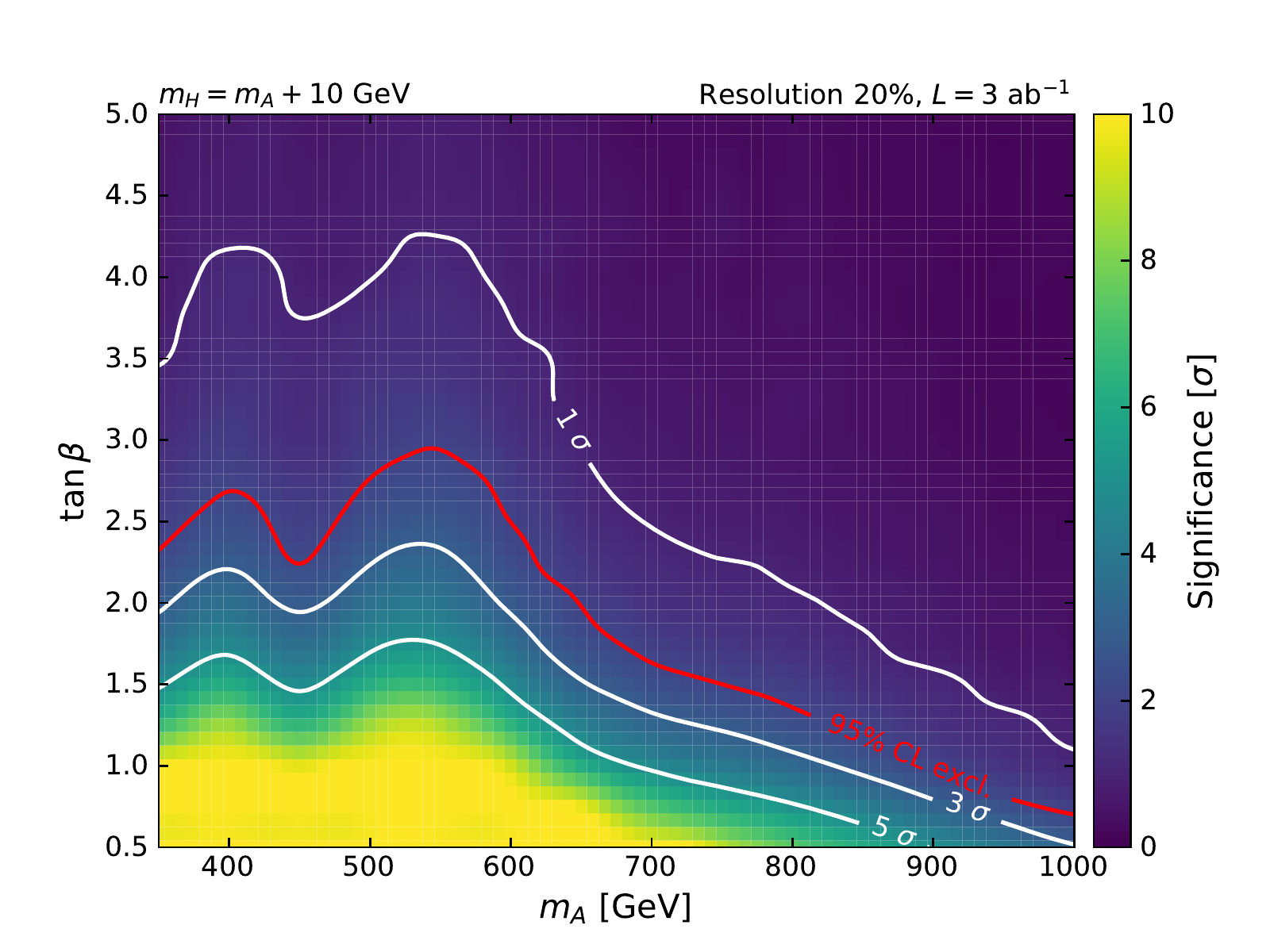}}
  \caption{\it Expected significance and exclusion potential for a Type~II 2HDM
 with a mass splitting $M_H-M_A=10$~GeV in the same six experimental scenarios as considered in Fig.~\ref{Fig:Res_SMH}. Values of significance in excess of $10\,\sigma$ are clipped. A resolution of $10\%$ and $20\%$ is assumed for the left- and the right-hand side plots,  respectively.}
\label{Fig:Res_2HDM2}
\end{figure}

\begin{figure}[!ht]
 \centering
\mbox{ \includegraphics[width=0.5\textwidth]{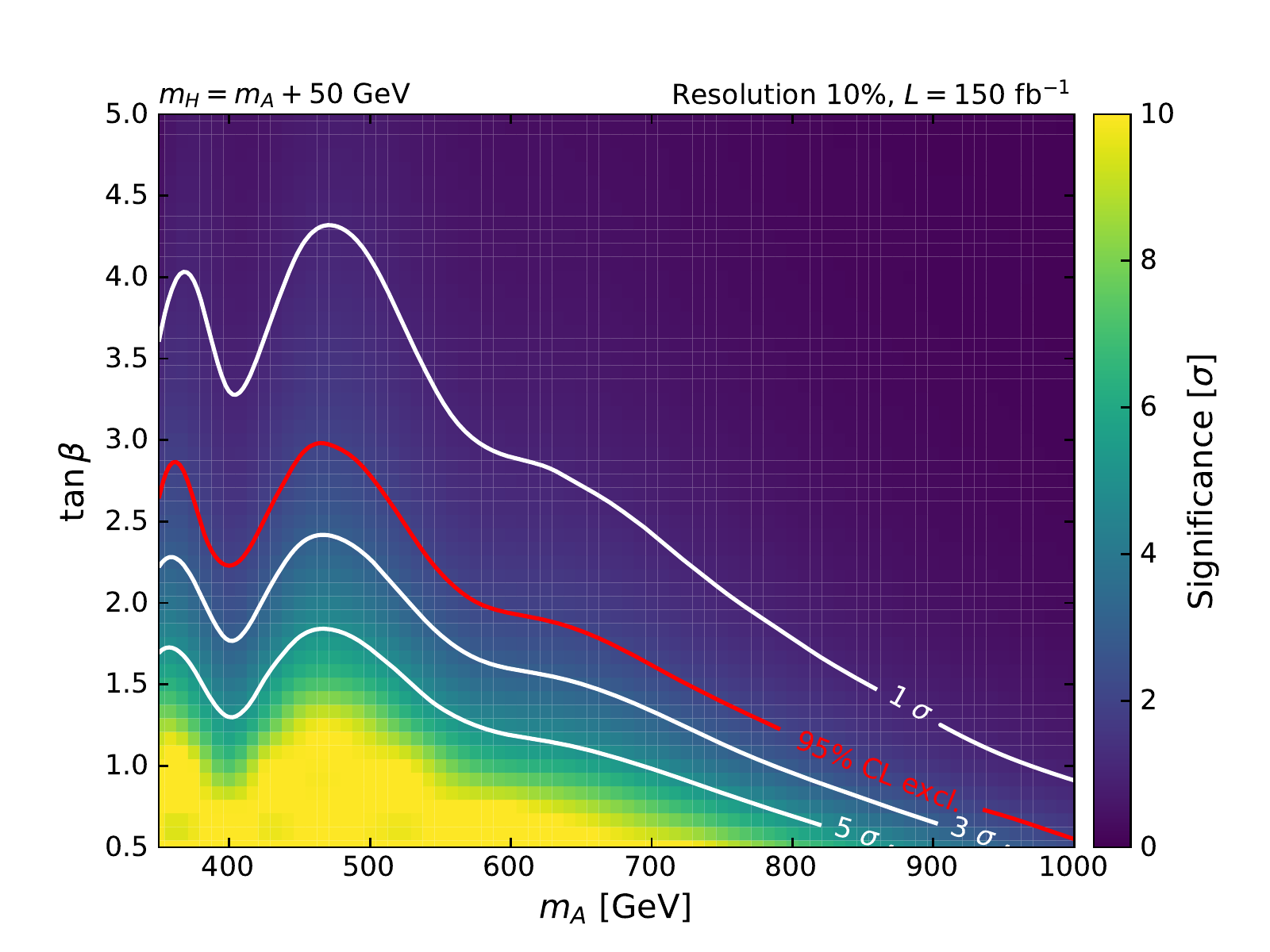}
 \includegraphics[width=0.5\textwidth]{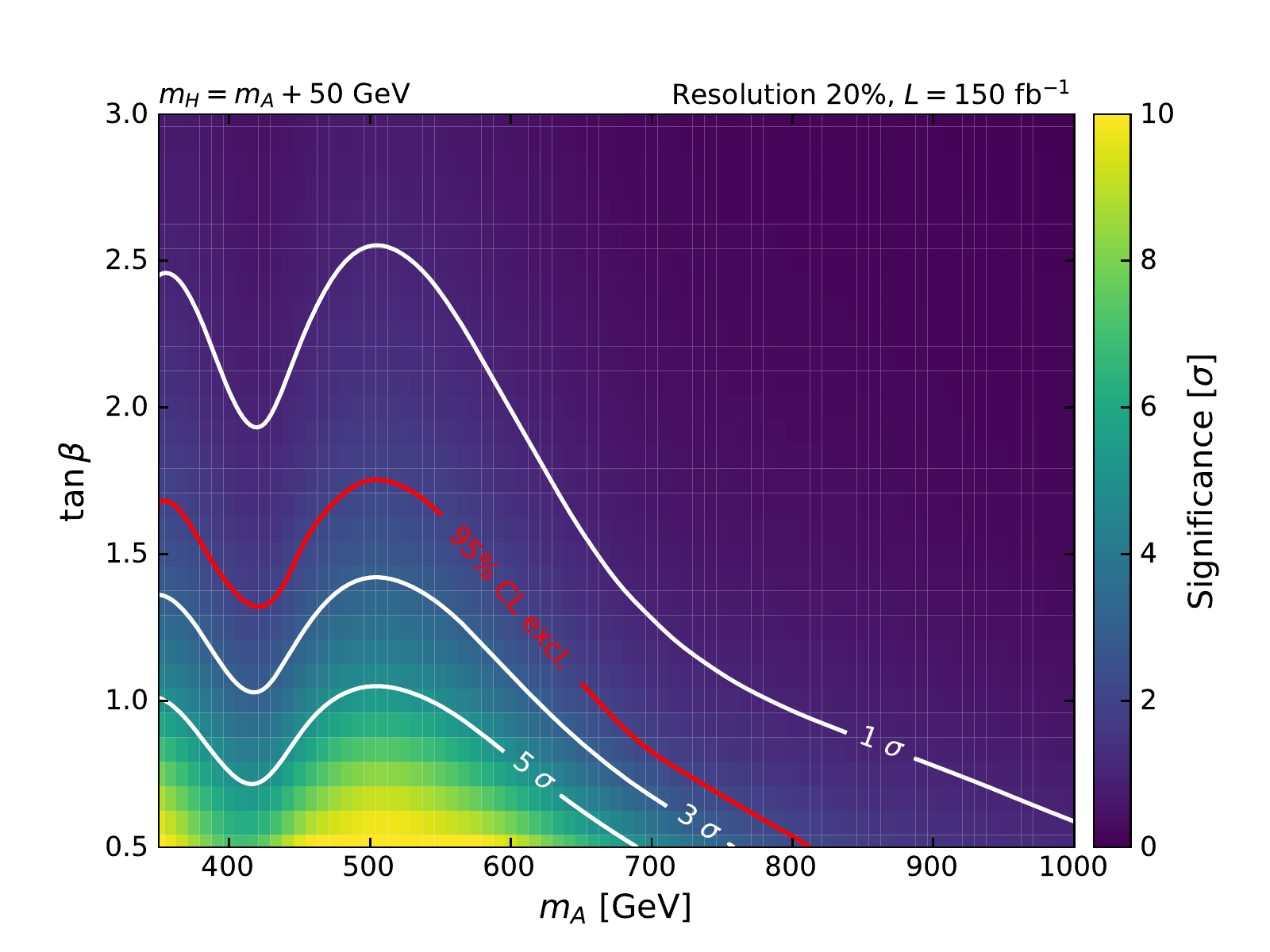} }\\
\mbox{ \includegraphics[width=0.5\textwidth]{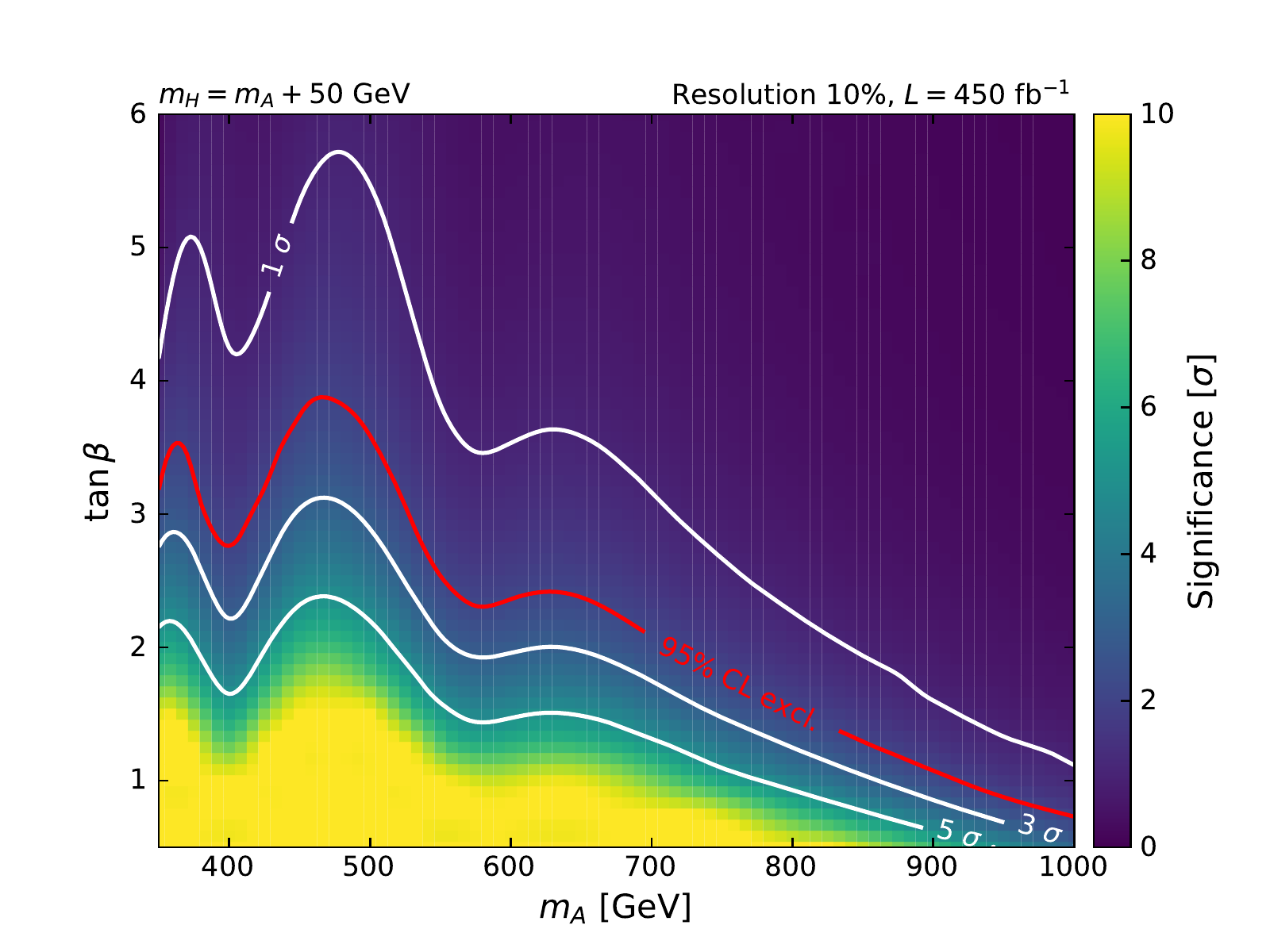}
 \includegraphics[width=0.5\textwidth]{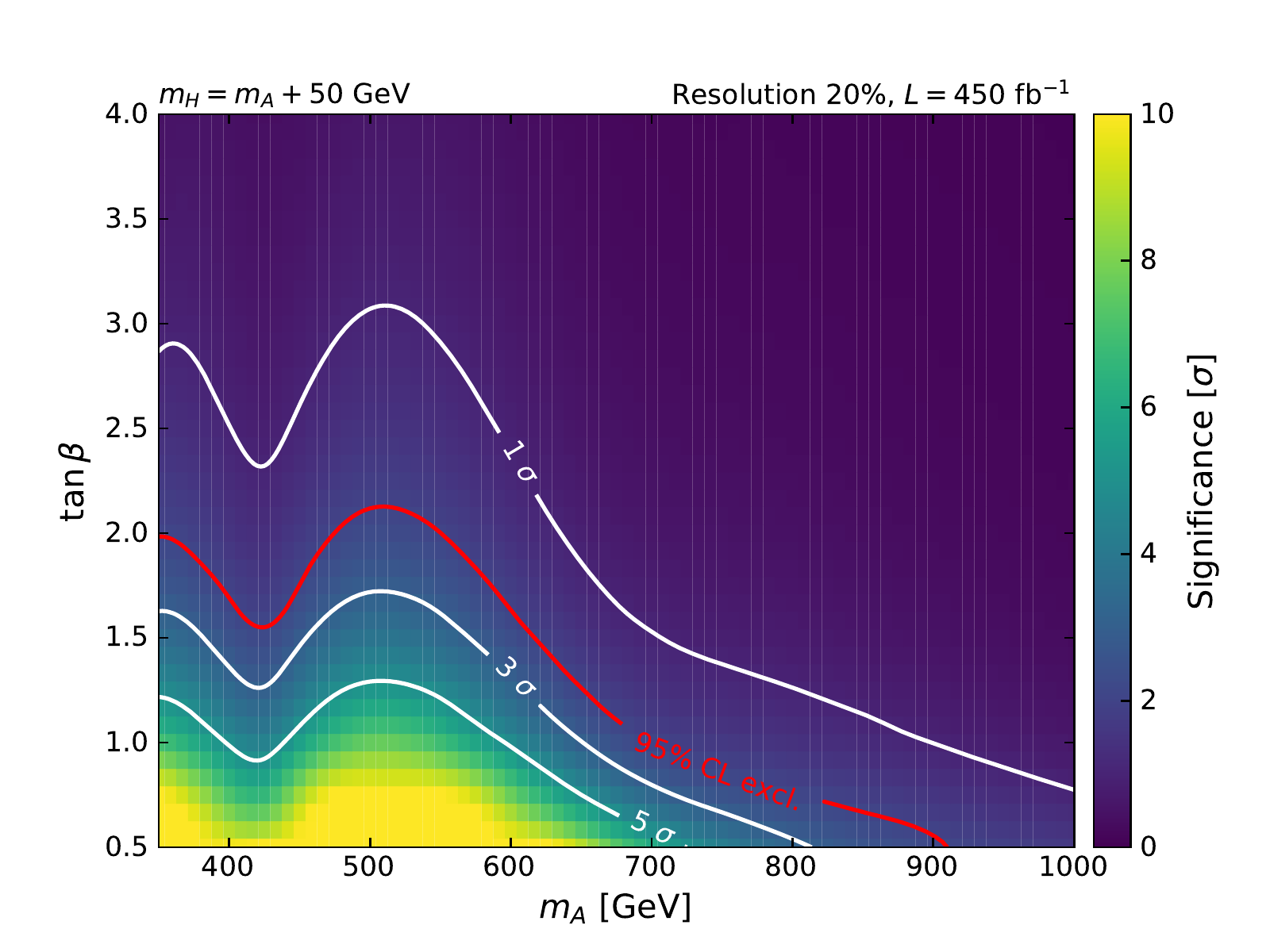} }\\
\mbox{ \includegraphics[width=0.5\textwidth]{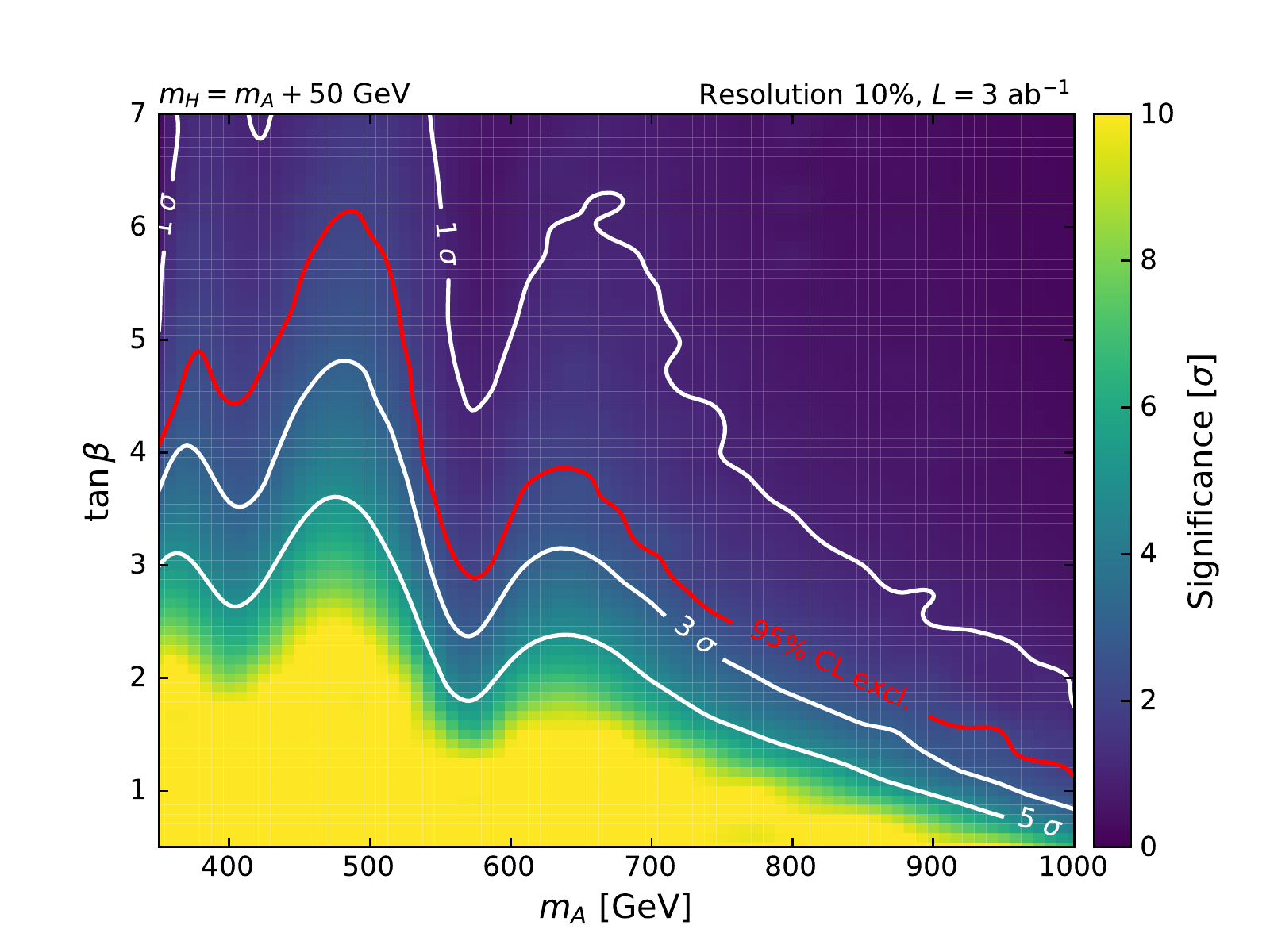}
 \includegraphics[width=0.5\textwidth]{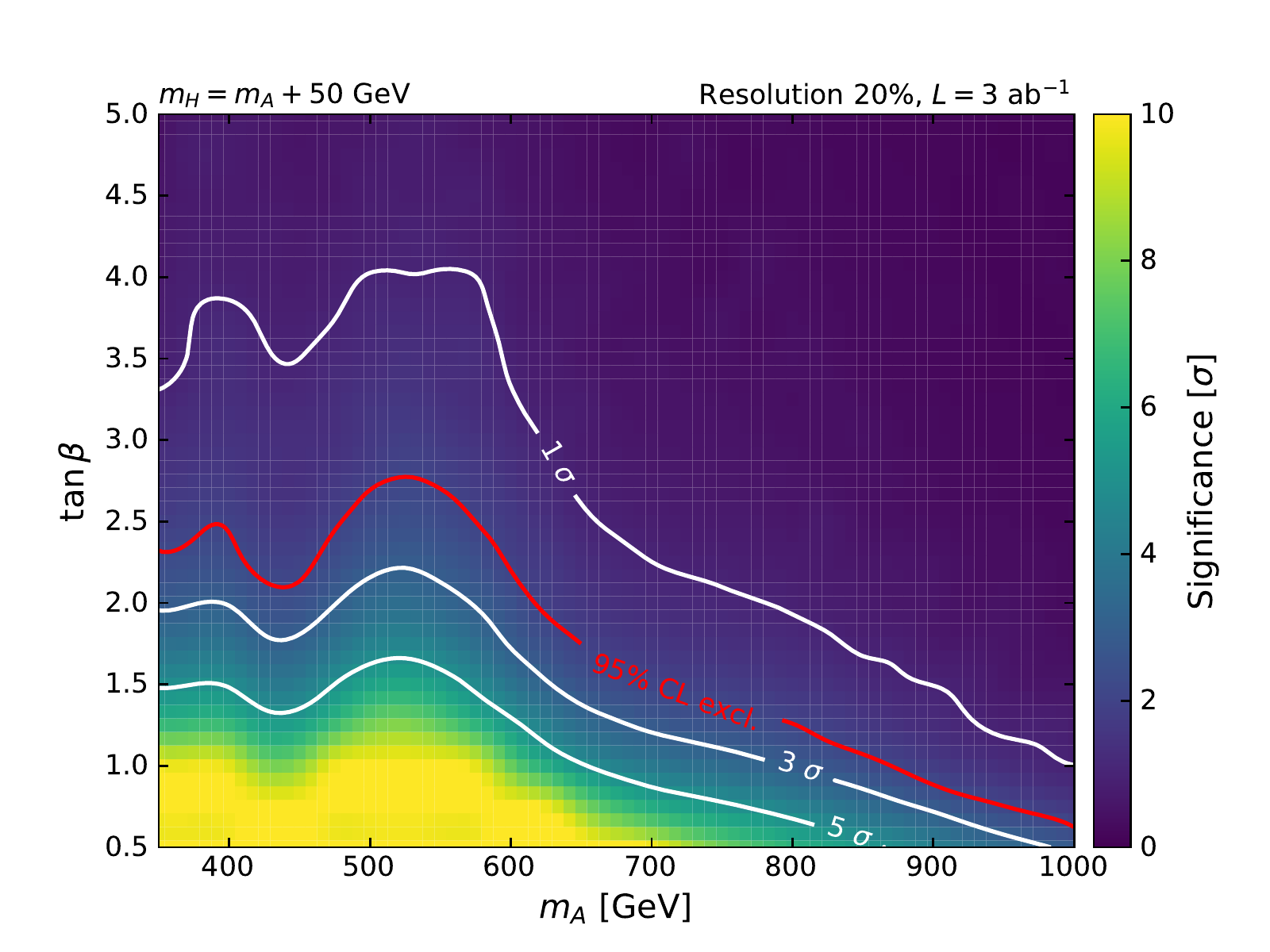}}
  \caption{\it Expected significance and exclusion potential for a Type~II 2HDM
 with a mass splitting $M_H-M_A=50$~GeV in the same six experimental scenarios as considered in Fig.~\ref{Fig:Res_SMH}. Values of significance in excess of $10\,\sigma$ are clipped. A resolution of $10\%$ and $20\%$ is assumed for the left- and the right-hand side plots,  respectively.}
\label{Fig:Res_2HDM3}
\end{figure}

\begin{figure}[!ht]
 \centering
\mbox{ \includegraphics[width=0.5\textwidth]{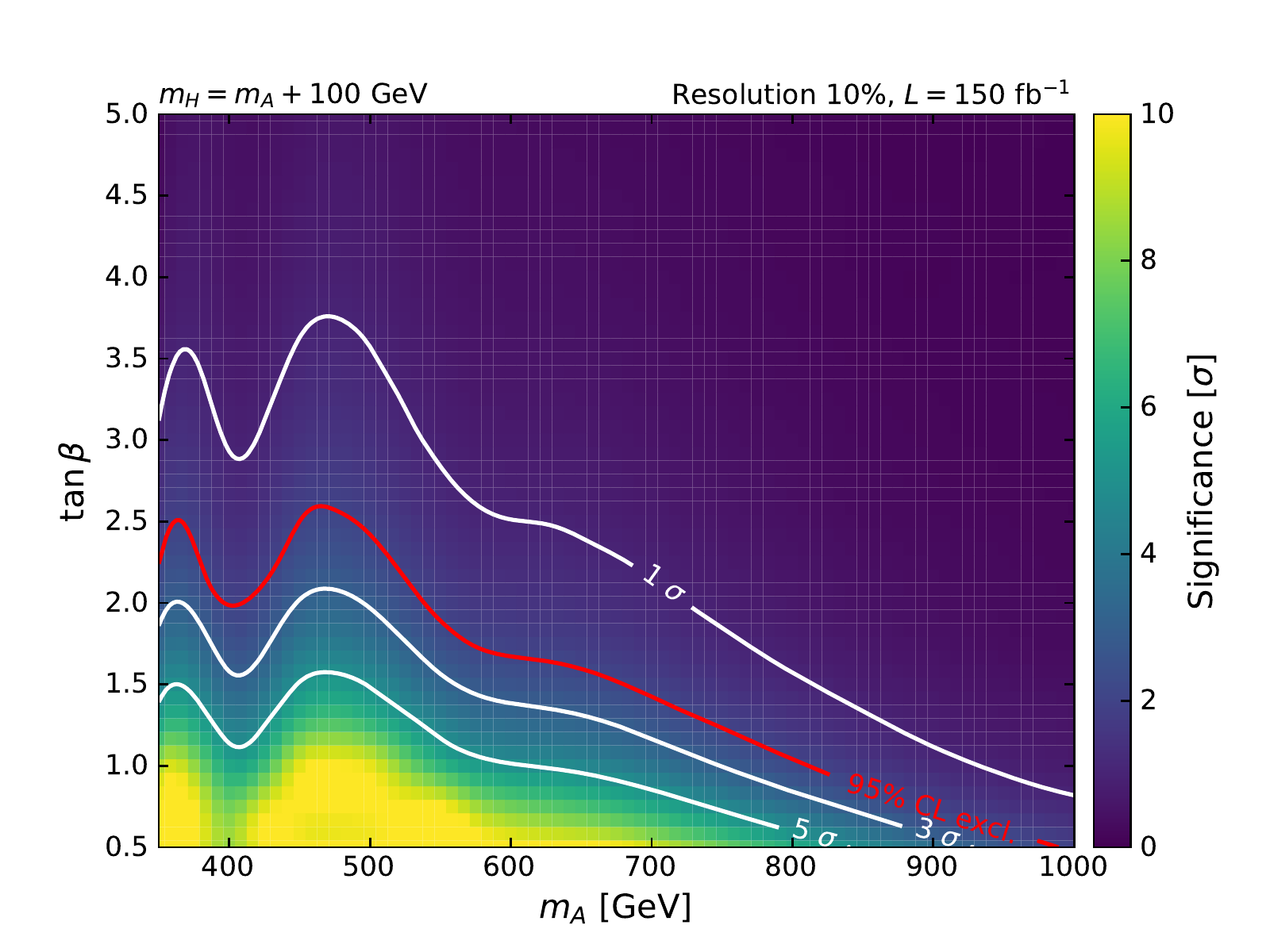}
 \includegraphics[width=0.5\textwidth]{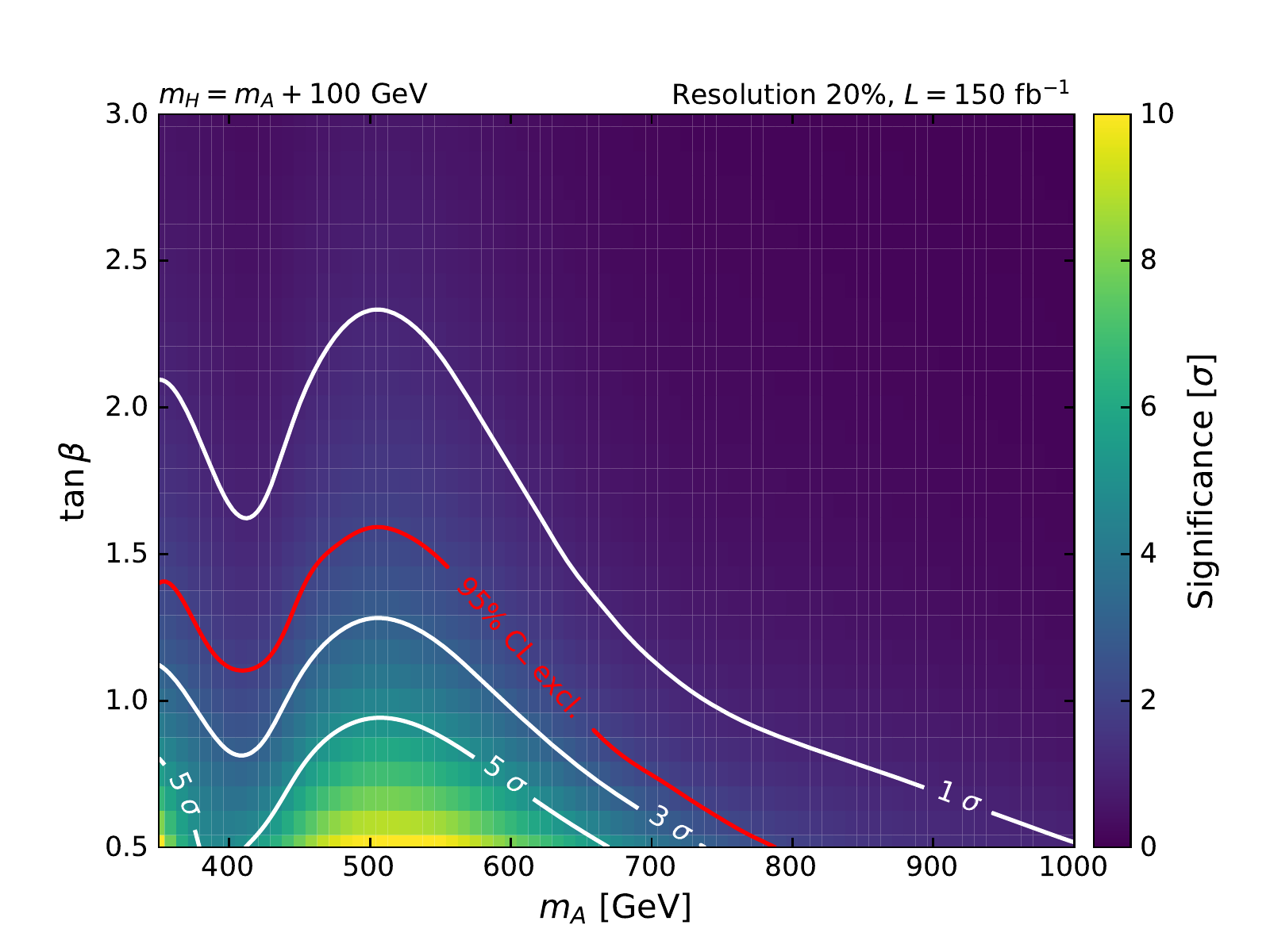} }\\
\mbox{ \includegraphics[width=0.5\textwidth]{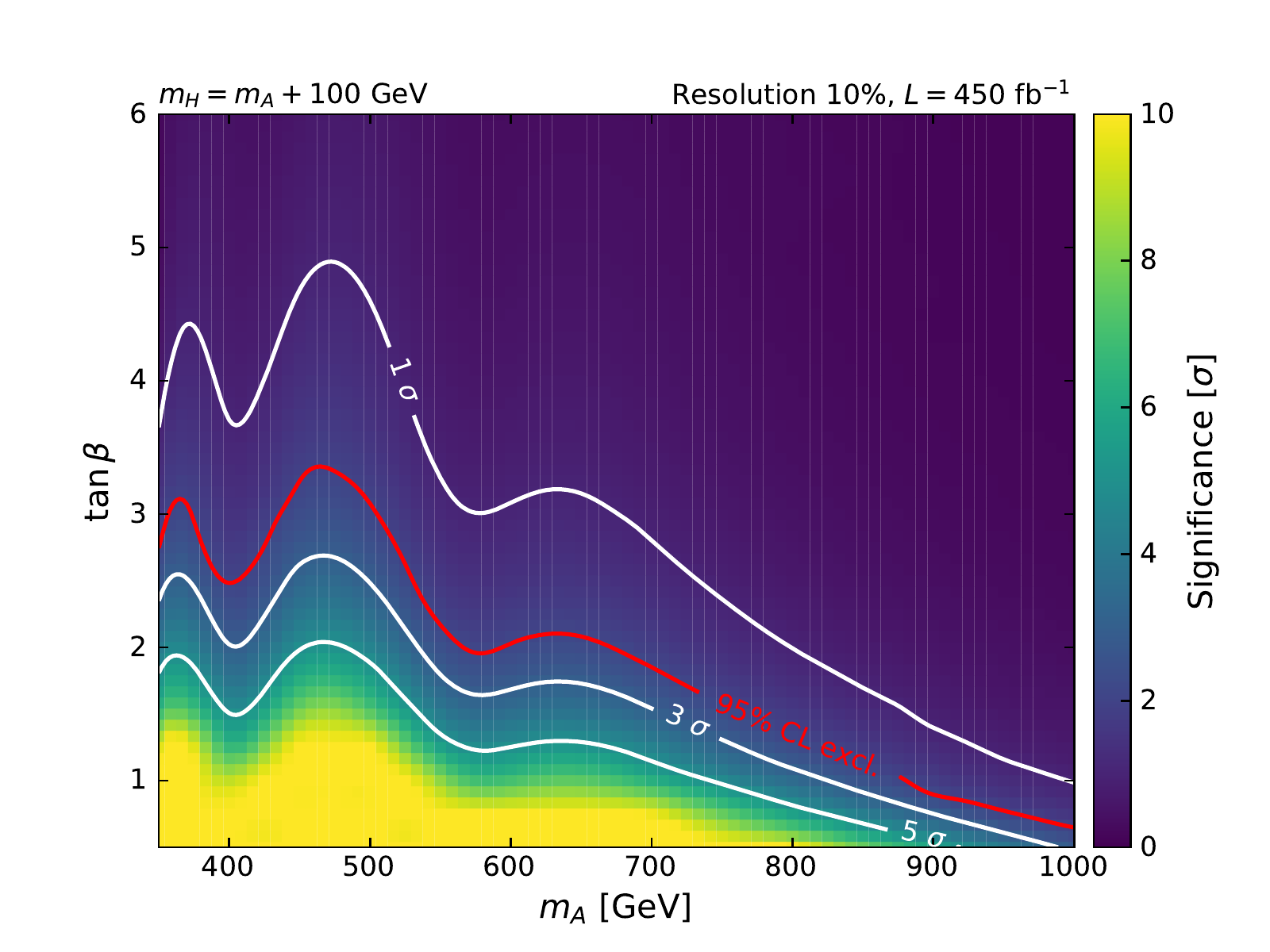}
 \includegraphics[width=0.5\textwidth]{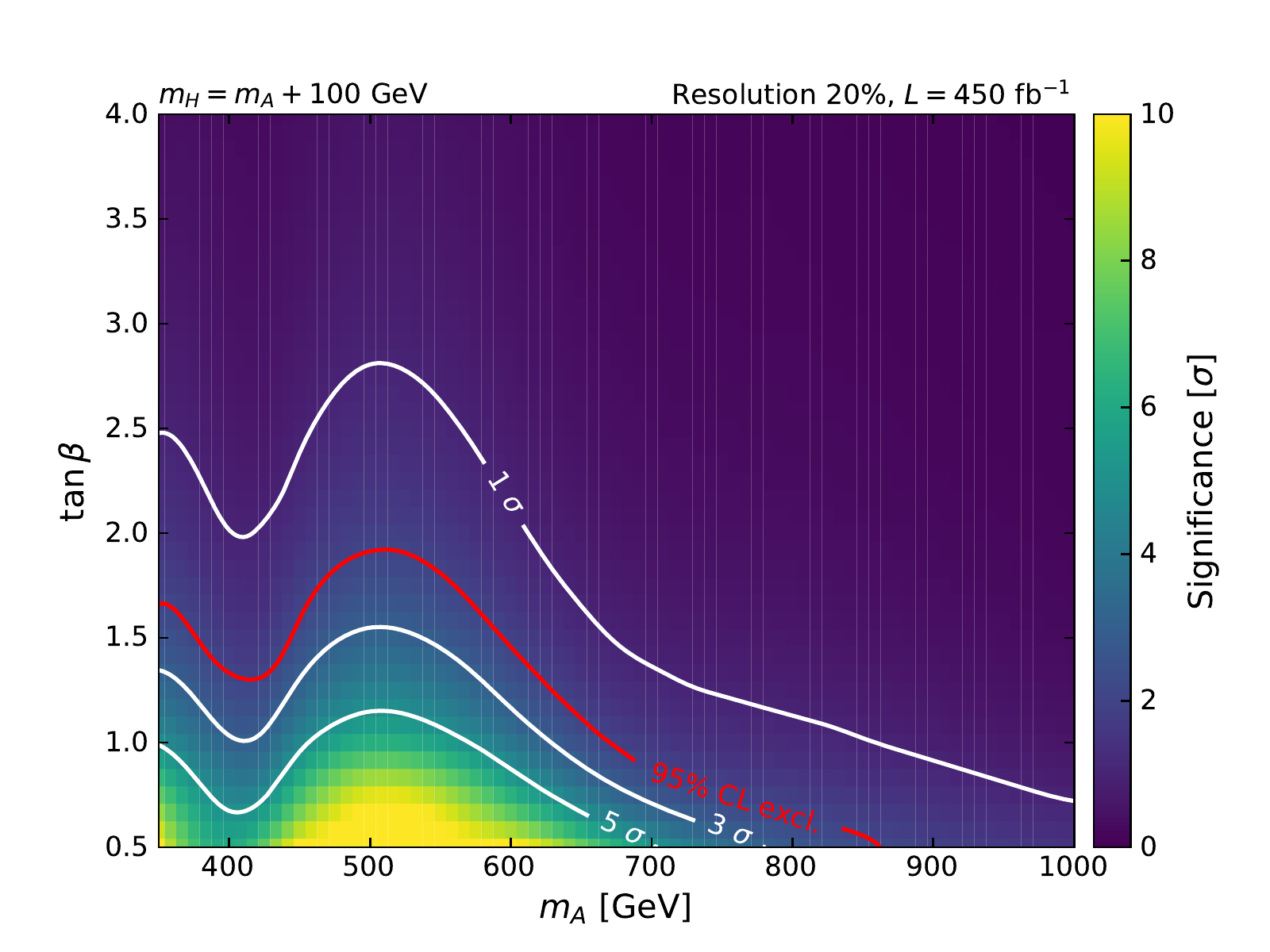} }\\
\mbox{ \includegraphics[width=0.5\textwidth]{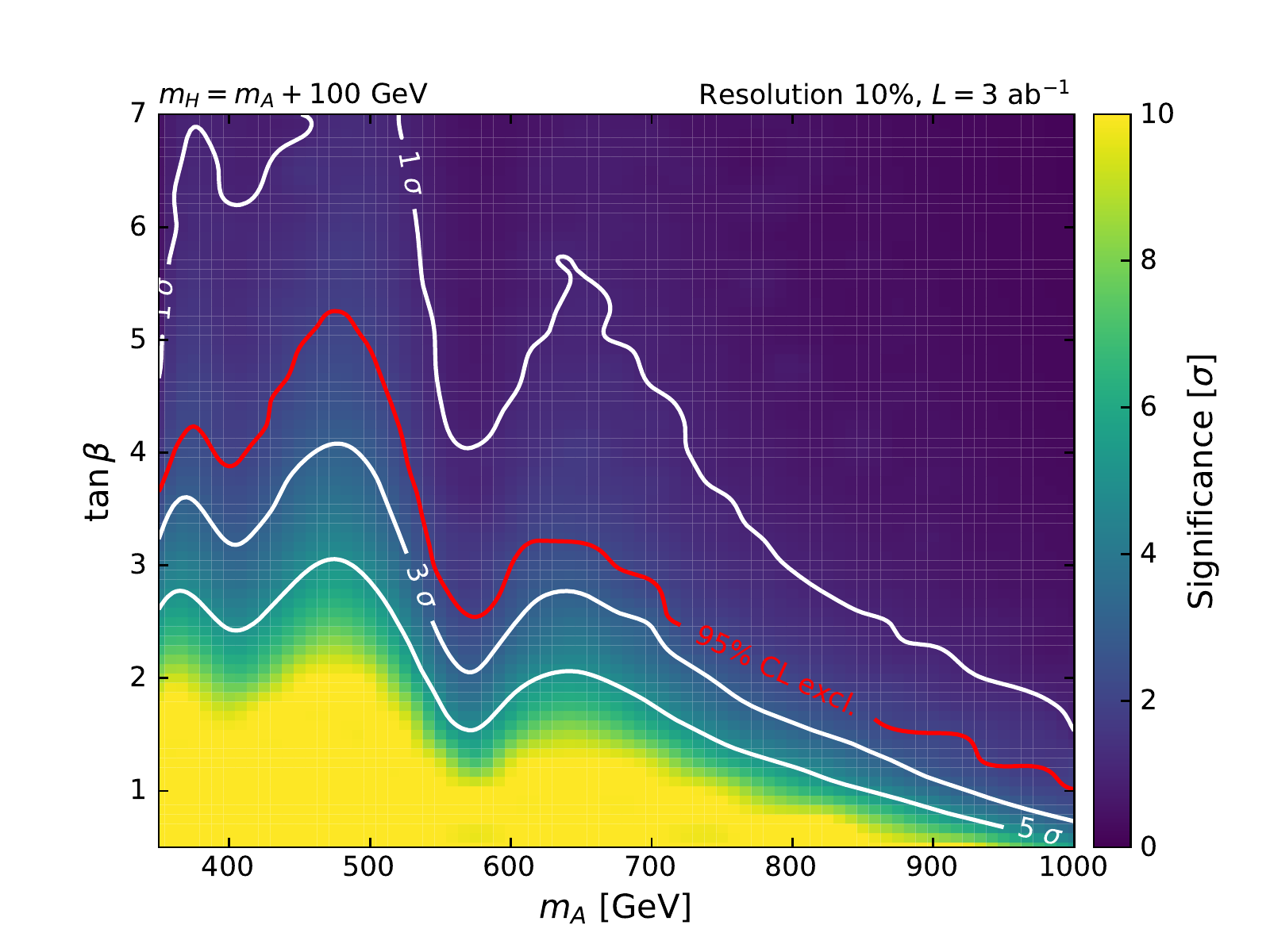}
 \includegraphics[width=0.5\textwidth]{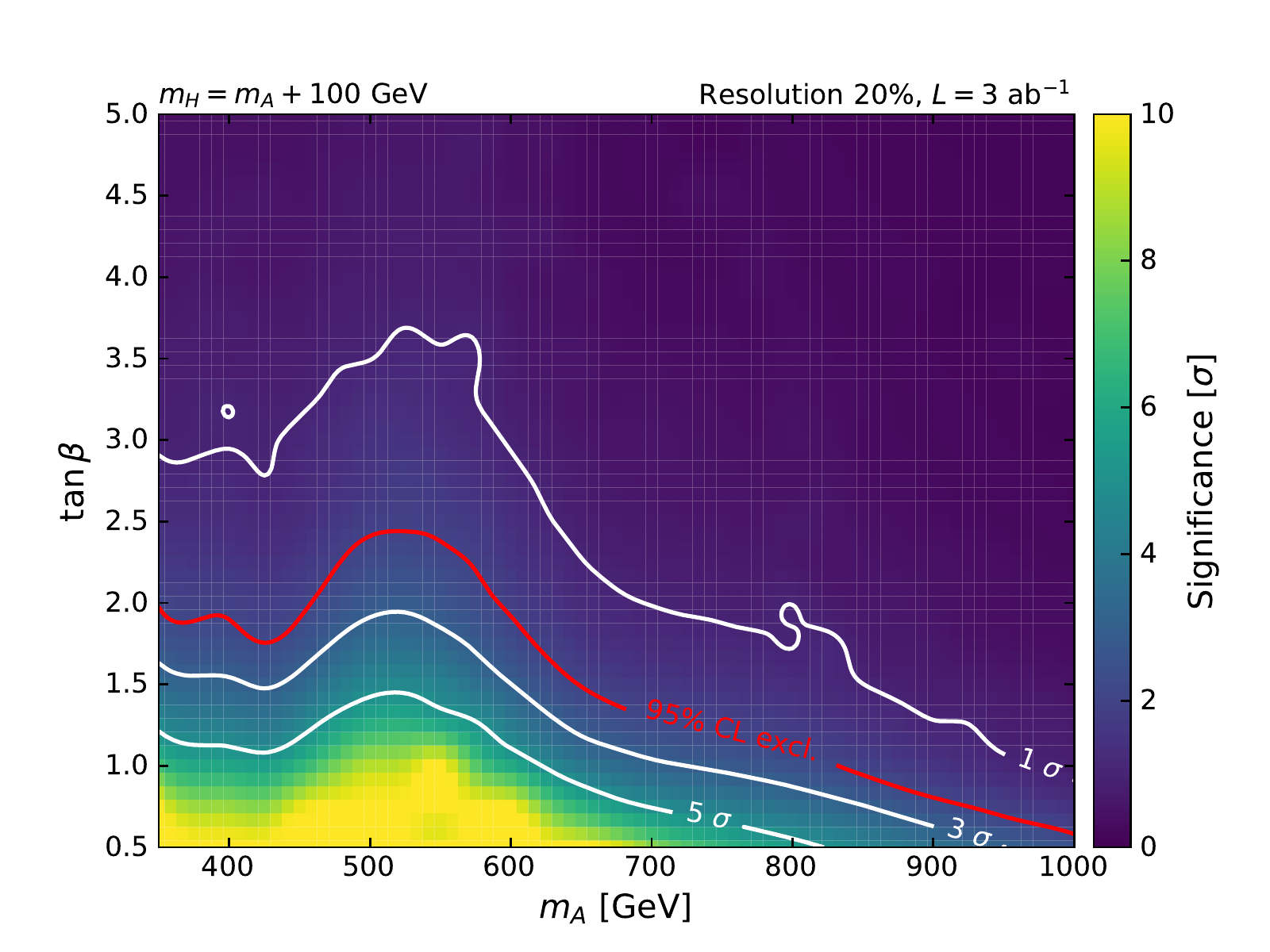}}
  \caption{\it Expected significance and exclusion potential for a Type~II 2HDM
 with a mass splitting $M_H-M_A=100$~GeV in the same six experimental scenarios as  considered in Fig.~\ref{Fig:Res_SMH}. Values of significance in excess of $10\,\sigma$ are clipped. A resolution of $10\%$ and $20\%$ is assumed for the left- and the right-hand side plots, respectively.}
\label{Fig:Res_2HDM4}
\end{figure}

\begin{figure}[!ht]
 \centering
\mbox{ \includegraphics[width=0.5\textwidth]{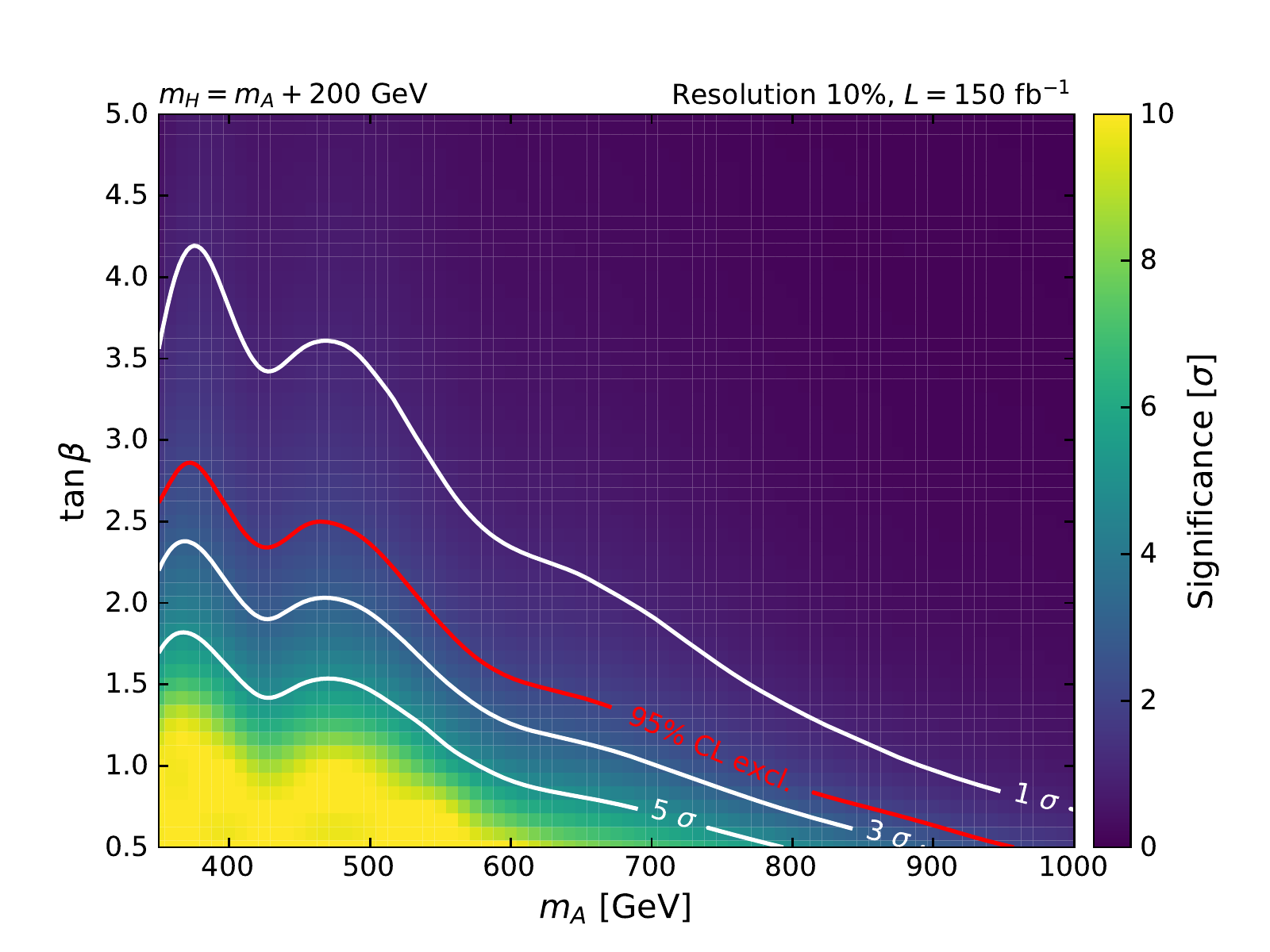}
 \includegraphics[width=0.5\textwidth]{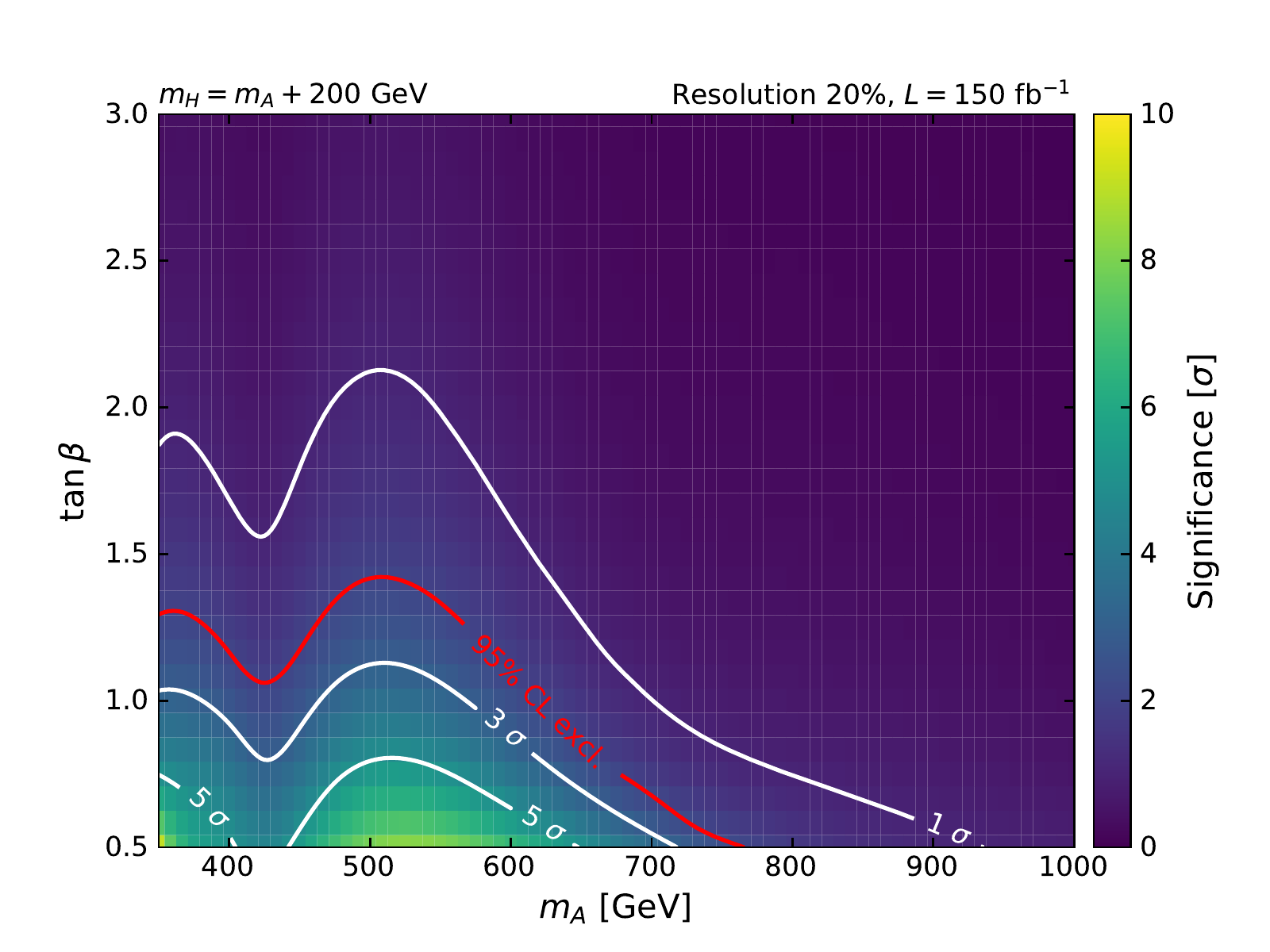} }\\
\mbox{ \includegraphics[width=0.5\textwidth]{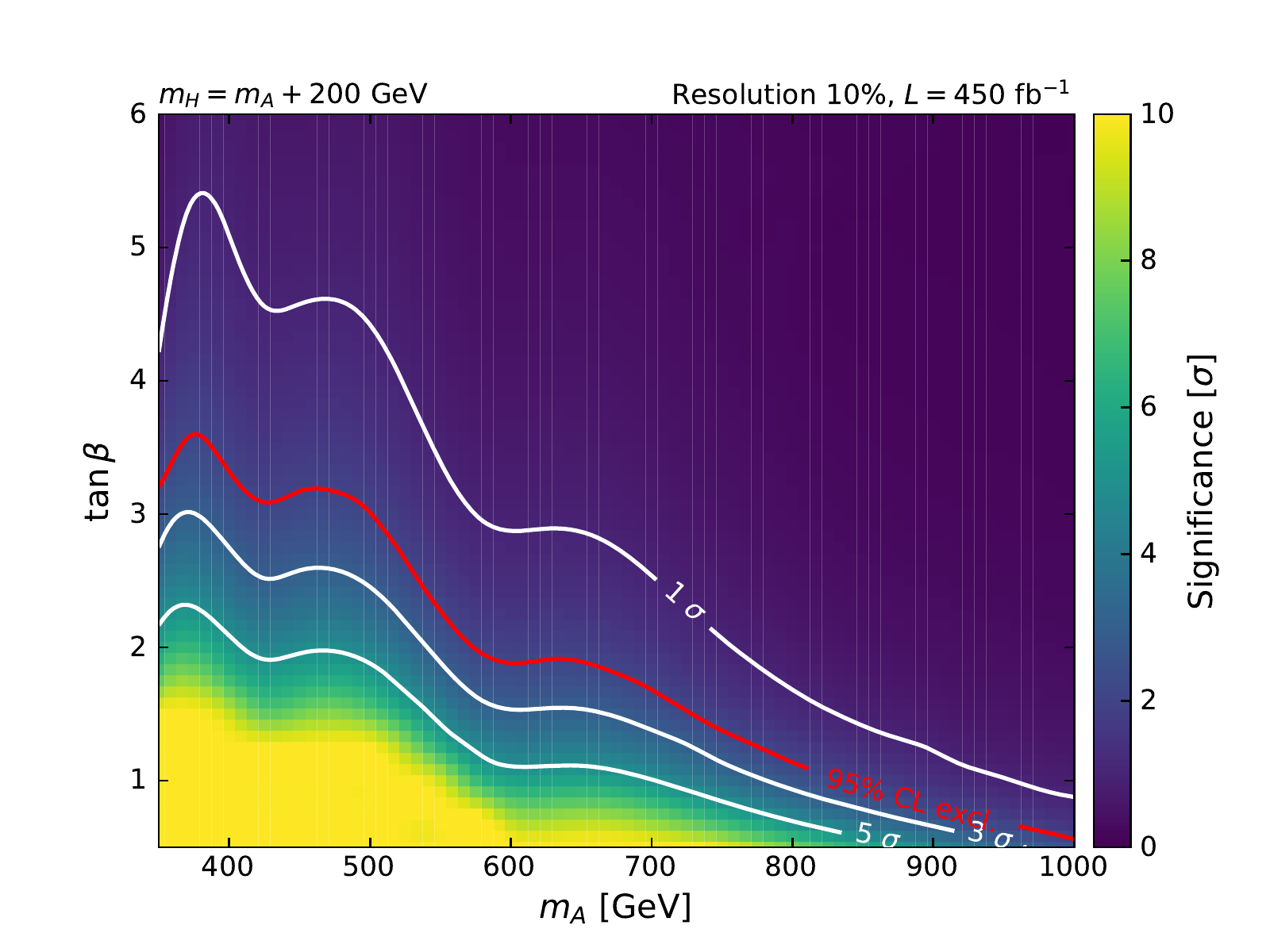}
 \includegraphics[width=0.5\textwidth]{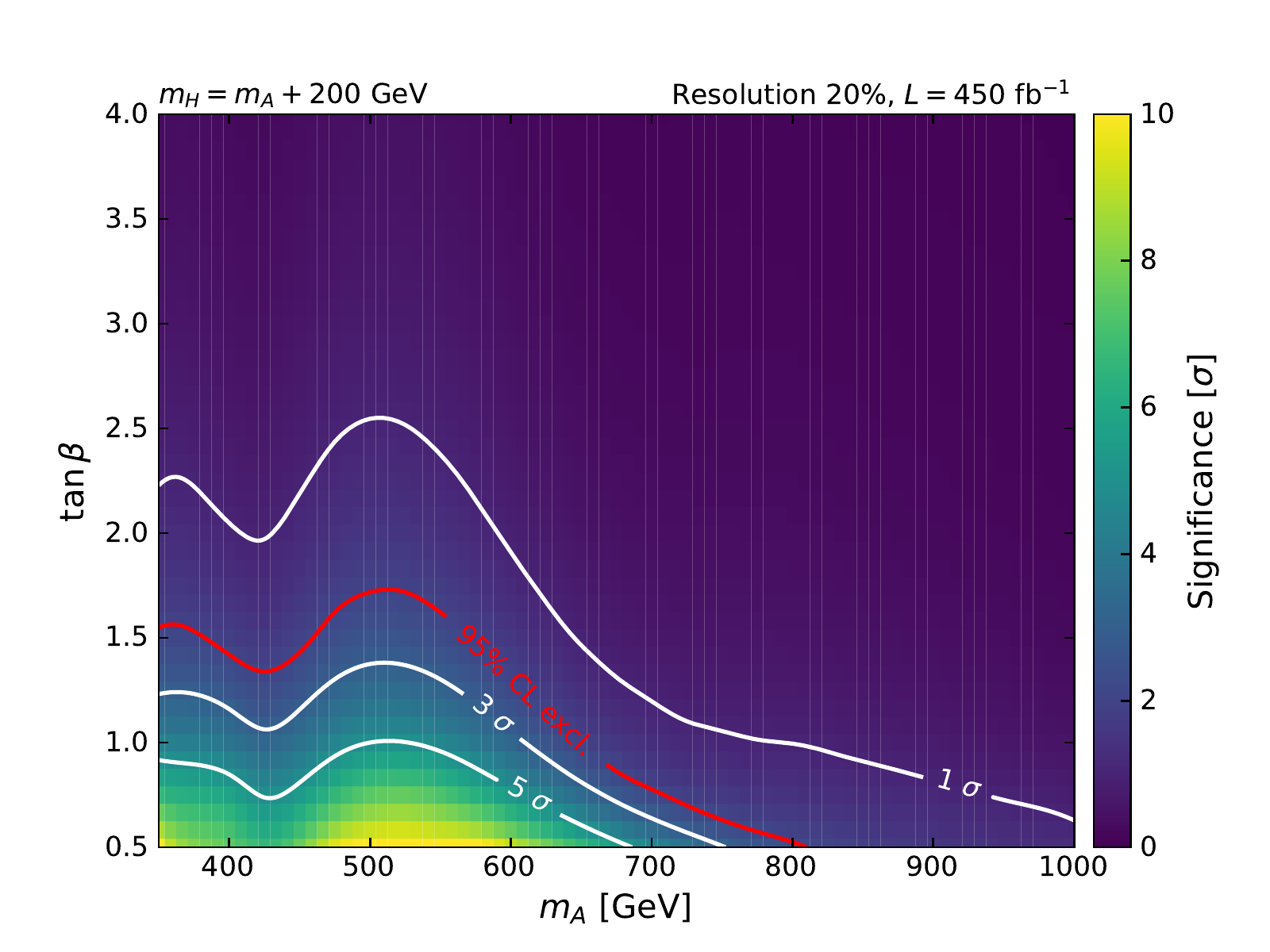} }\\
\mbox{ \includegraphics[width=0.5\textwidth]{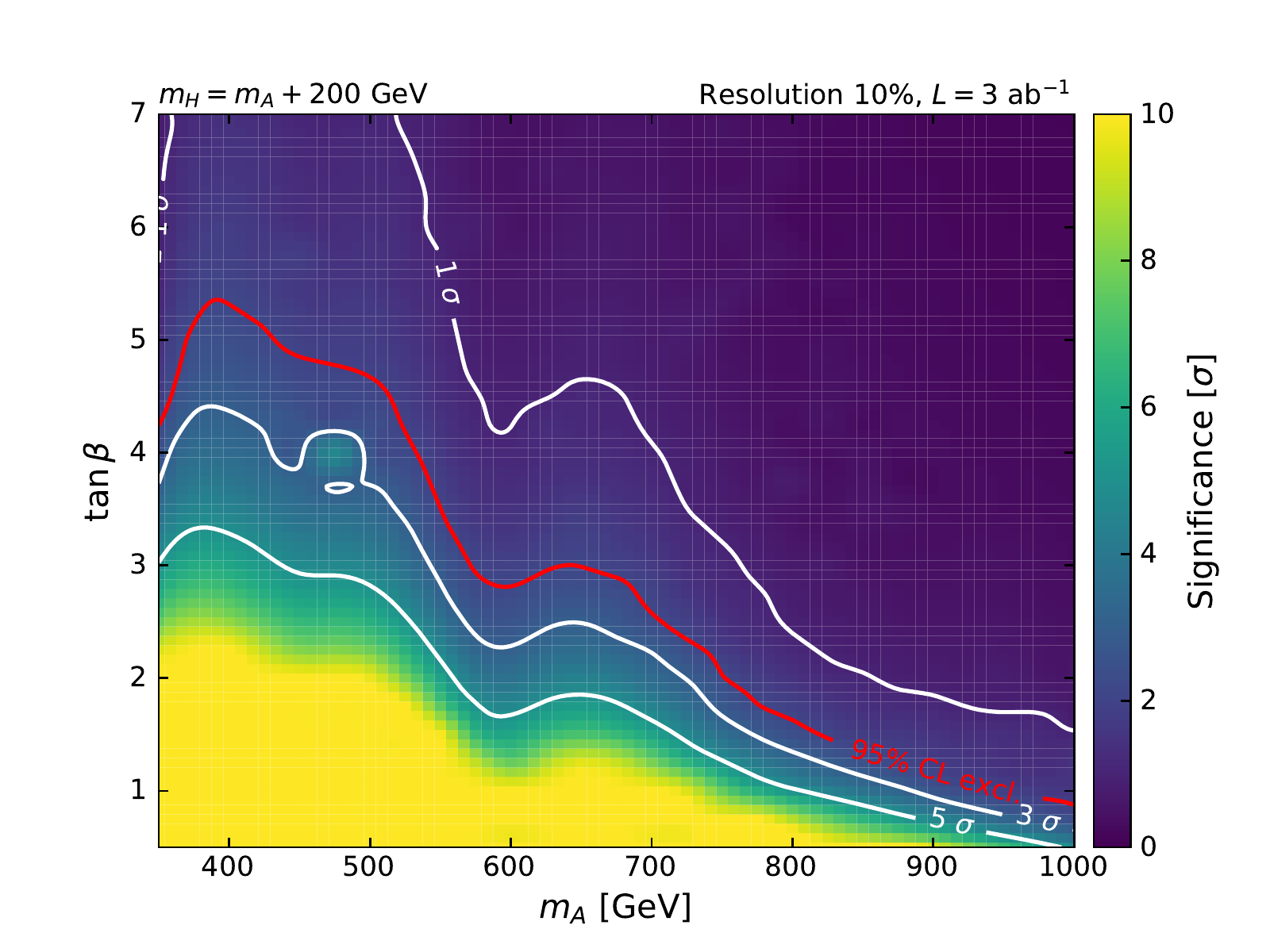}
 \includegraphics[width=0.5\textwidth]{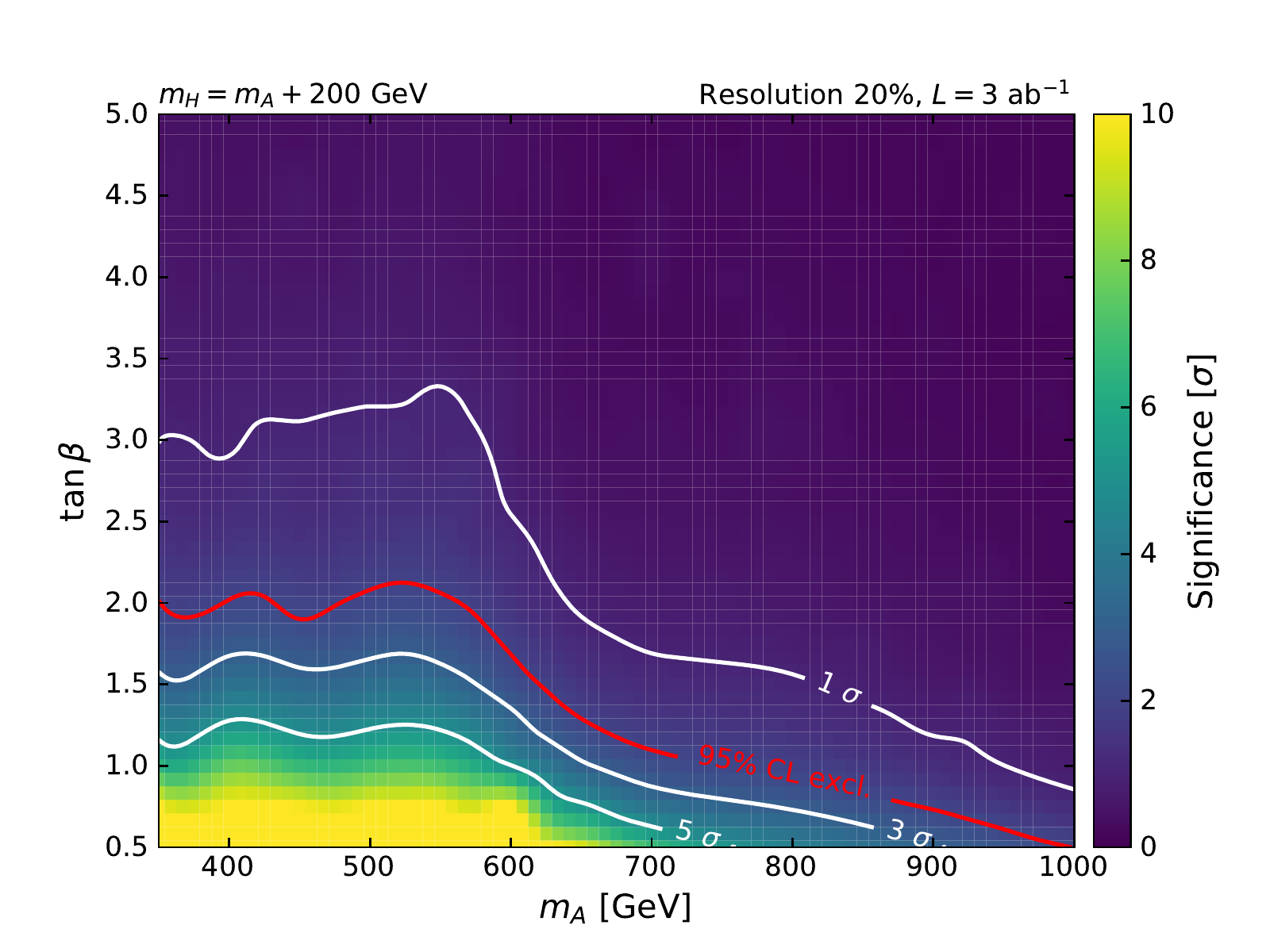}}
  \caption{\it Expected significance and exclusion potential for a Type~II 2HDM
with a mass splitting $M_H-M_A=200$~GeV in the same six experimental scenarios as considered in Fig.~\ref{Fig:Res_SMH}. Values of significance in excess of $10\,\sigma$ are clipped. A resolution of $10\%$ and $20\%$ is assumed for the left- and the right-hand side plots, respectively.}
\label{Fig:Res_2HDM5}
\end{figure}

In Figures~\ref{Fig:Res_2HDM2}, \ref{Fig:Res_2HDM3}, \ref{Fig:Res_2HDM4} and 
\ref{Fig:Res_2HDM5}, we present results for a Type~II 2HDM in the $[M_A, \tb]$
plane, assuming  $M_{H^{\pm}}\ge \text{max}(M_A,M_H)$ and a fixed mass splitting
$M_H-M_A = 10$, $50$, $100$ and $200$~GeV, with the same mass resolution and
integrated luminosity scenarios as previously.

\begin{figure}[!ht]
 \centering
\mbox{ \includegraphics[width=0.45\textwidth]{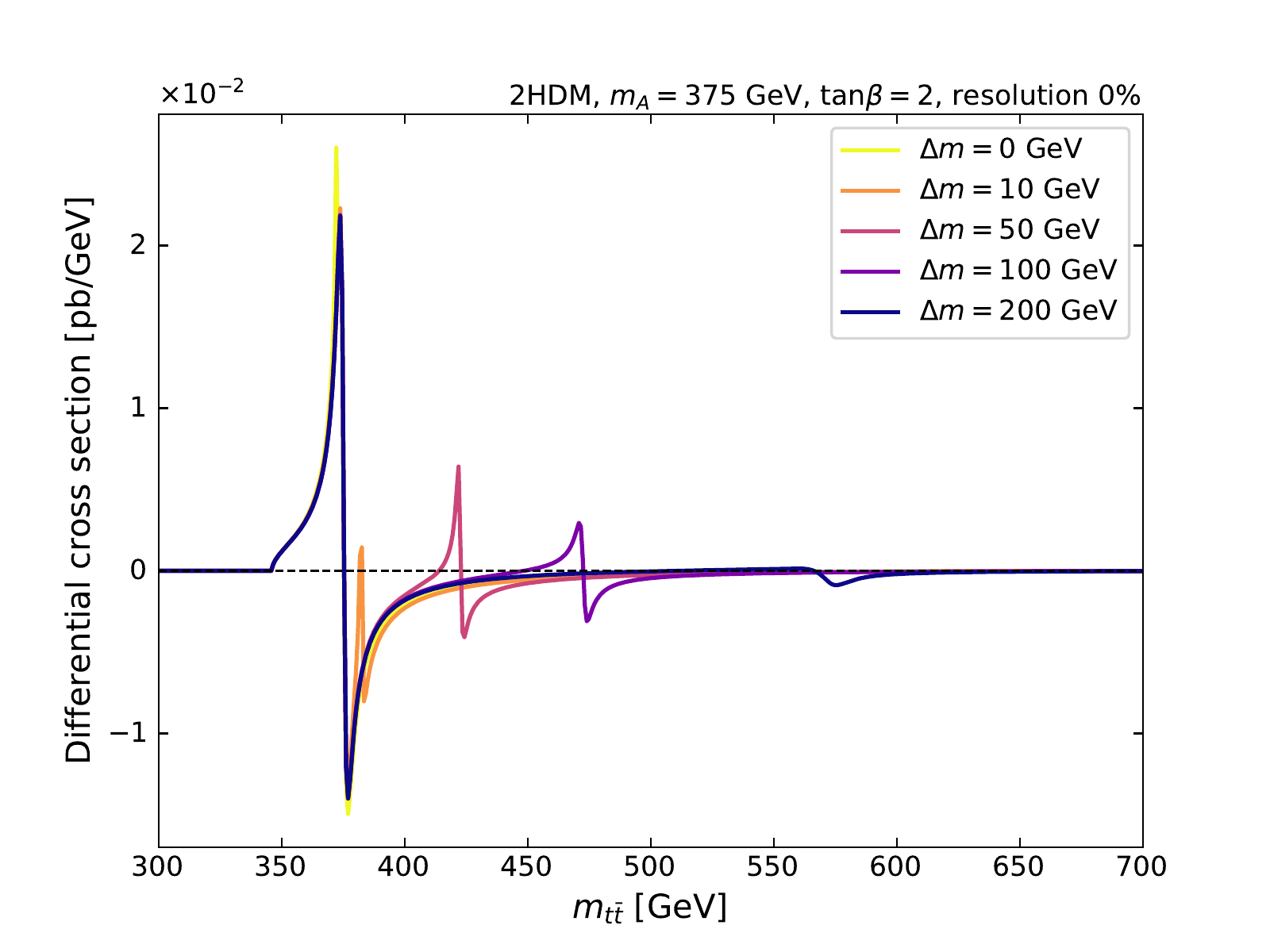}
 \includegraphics[width=0.45\textwidth]{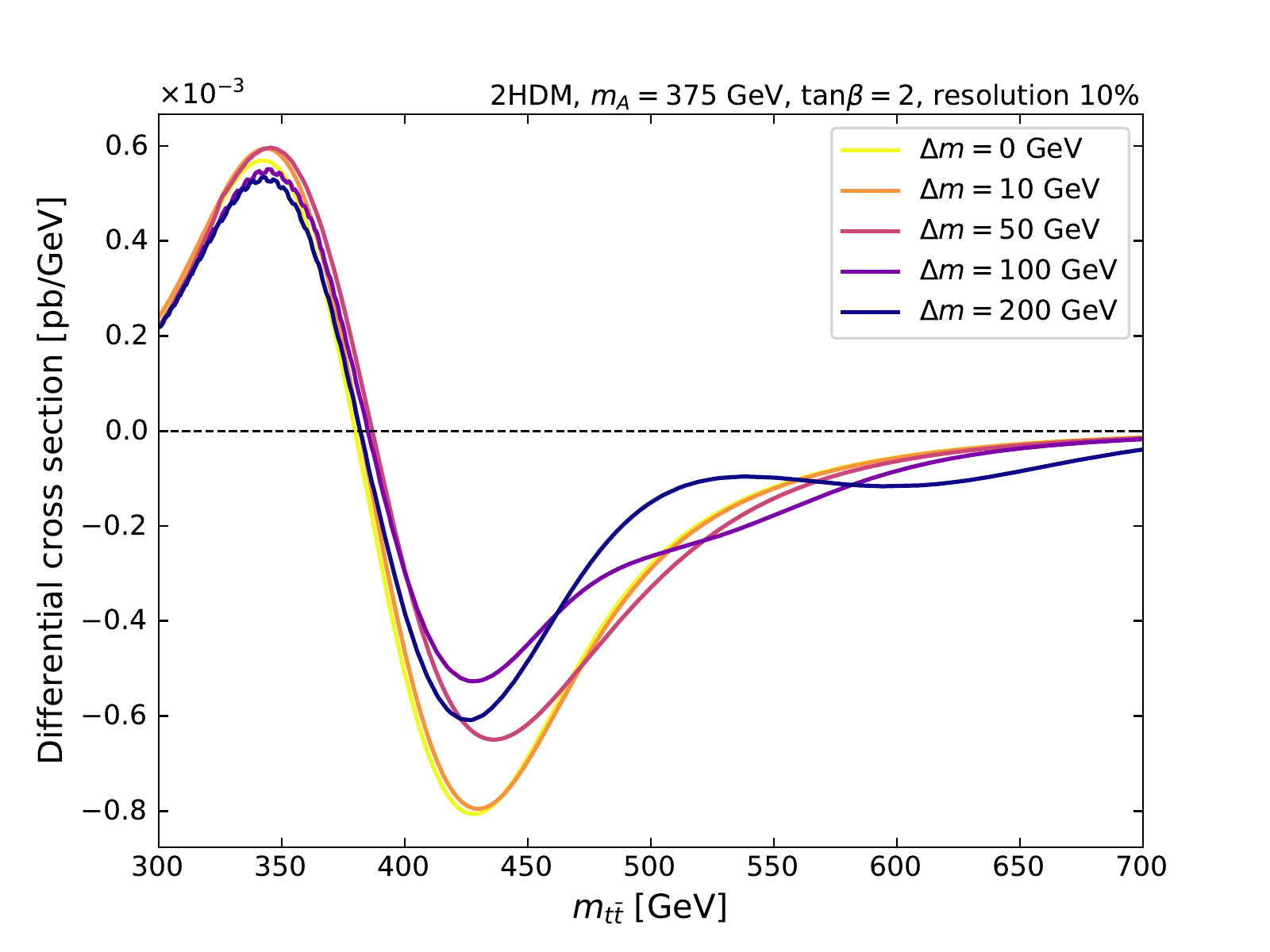} }
  \caption{\it The differential cross section with perfect resolution (left) or $10\%$ smearing (right) for several mass splitting $M_H-M_A=0,10,50,100, 200$~GeV.}
\label{Fig:2HDM_deg}
\end{figure}

An increase in the mass separation between $M_H$ and $M_A$ leads initially to a
degradation in the sensitivity, because the $m_{t\bar{t}}$ spectrum is dominated
by the structure for the CP--odd state, while the peak--dip structure for the
CP--even state lies in the dip for the CP--odd one and partially cancels it out.
This is illustrated by the two plots in Fig.~\ref{Fig:2HDM_deg}, one with a
perfect resolution, the other with $10\%$ smearing. On the other hand, when the
mass separation becomes large enough, the structures from the two states do not
overlap so much, and the sensitivity increases again.

\subsection{The hMSSM}

\begin{figure}[!ht]
 \centering
\mbox{ \includegraphics[width=0.5\textwidth]{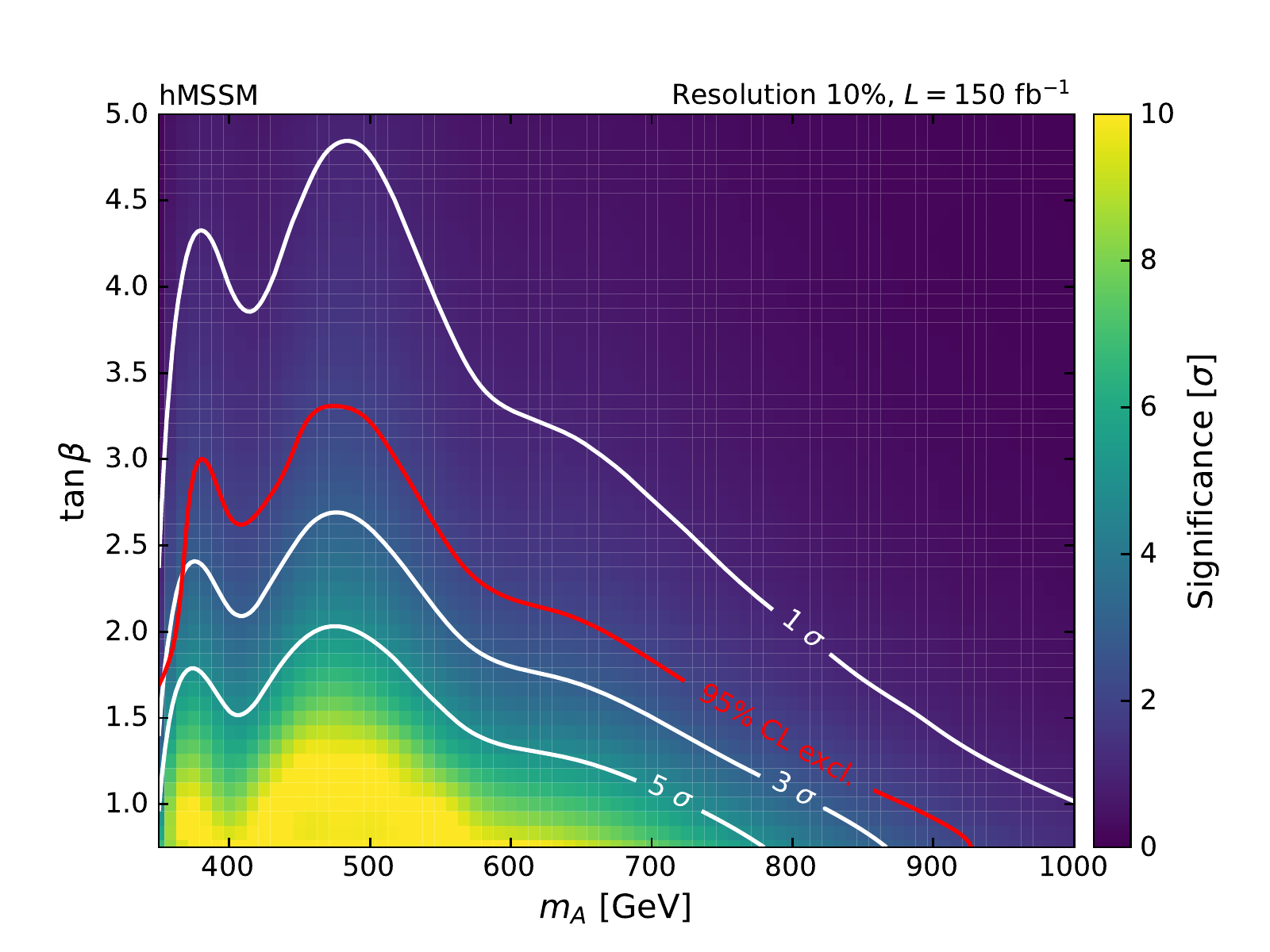}
 \includegraphics[width=0.5\textwidth]{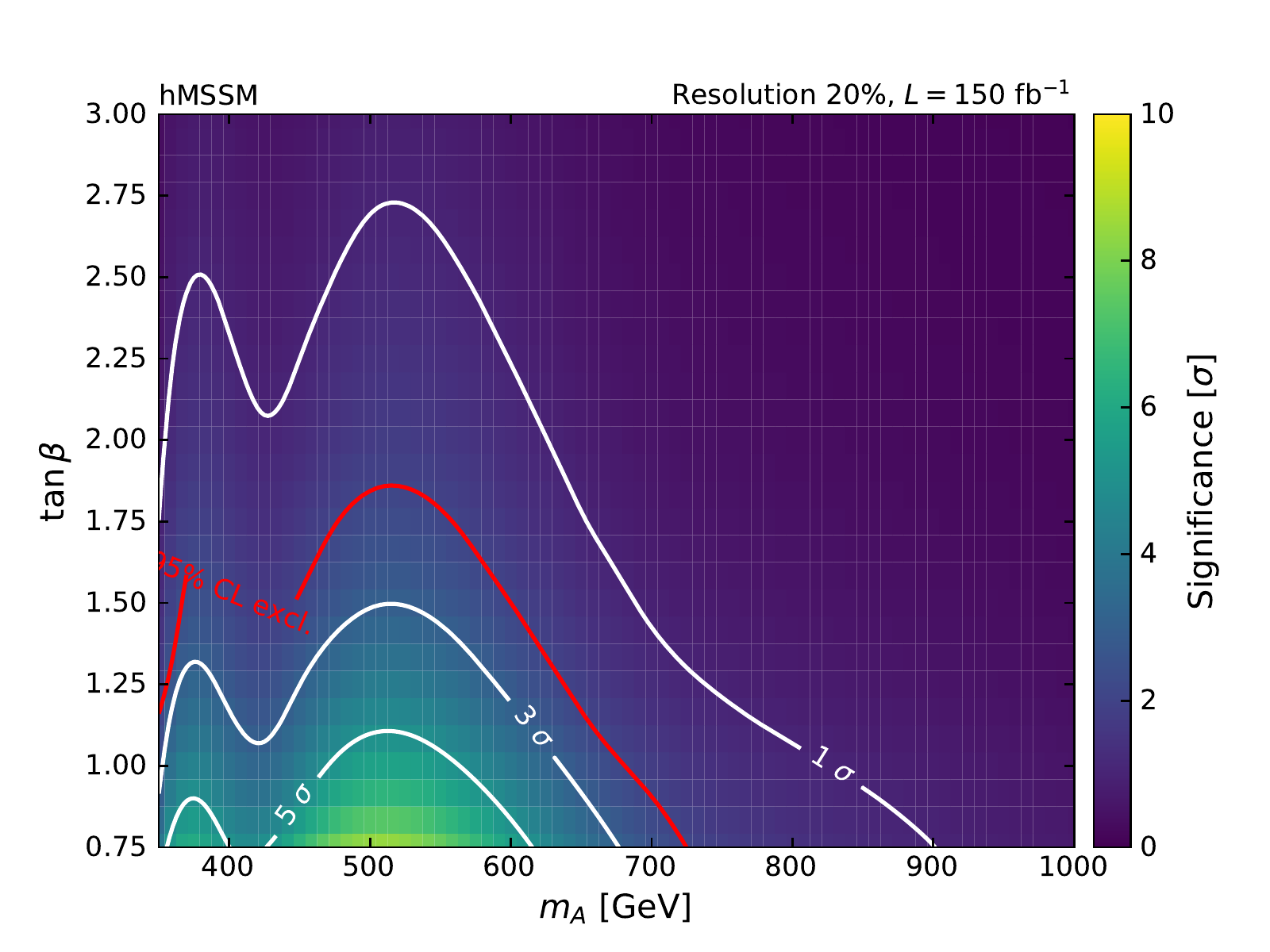} }\\
\mbox{ \includegraphics[width=0.5\textwidth]{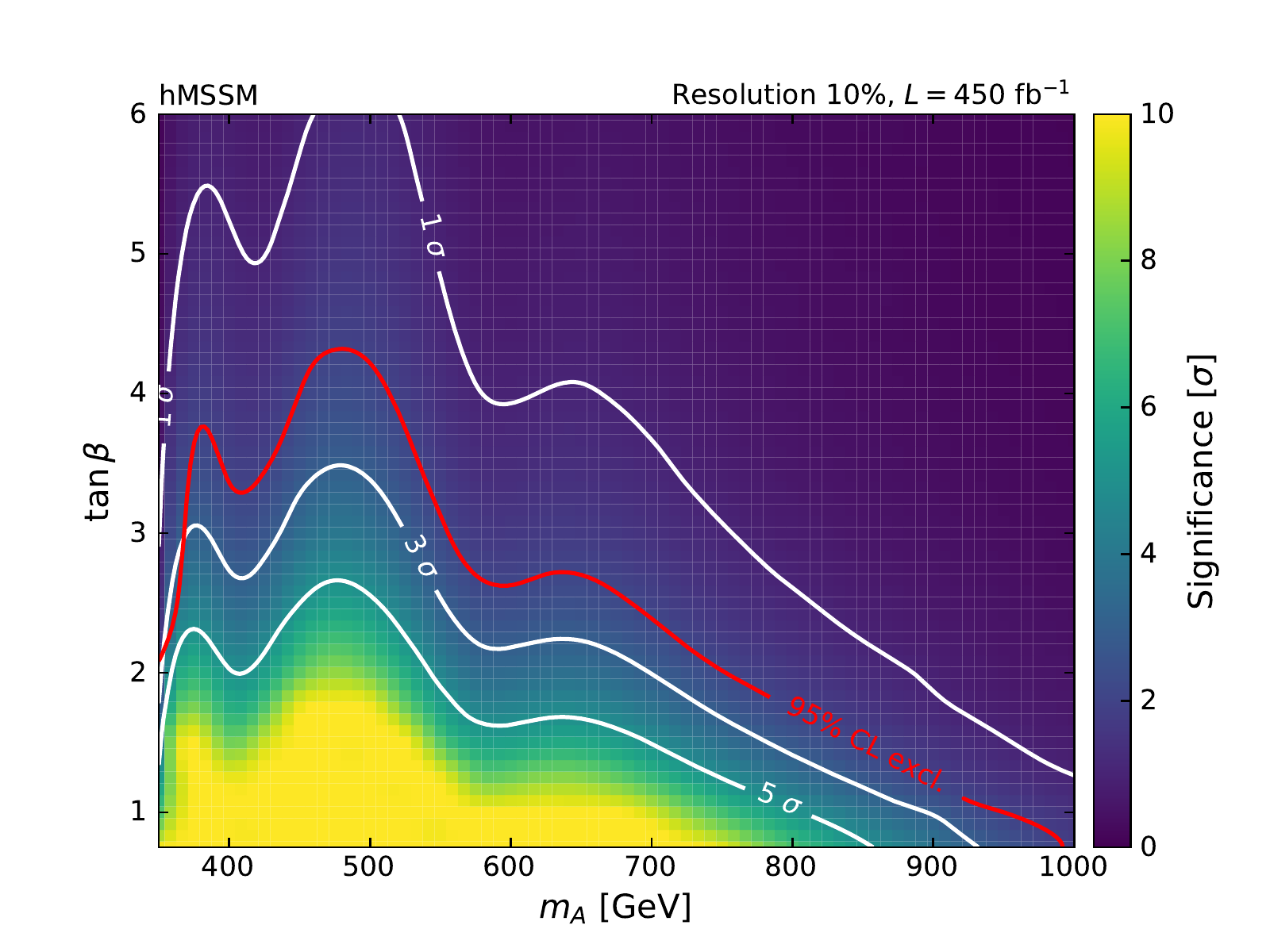}
 \includegraphics[width=0.5\textwidth]{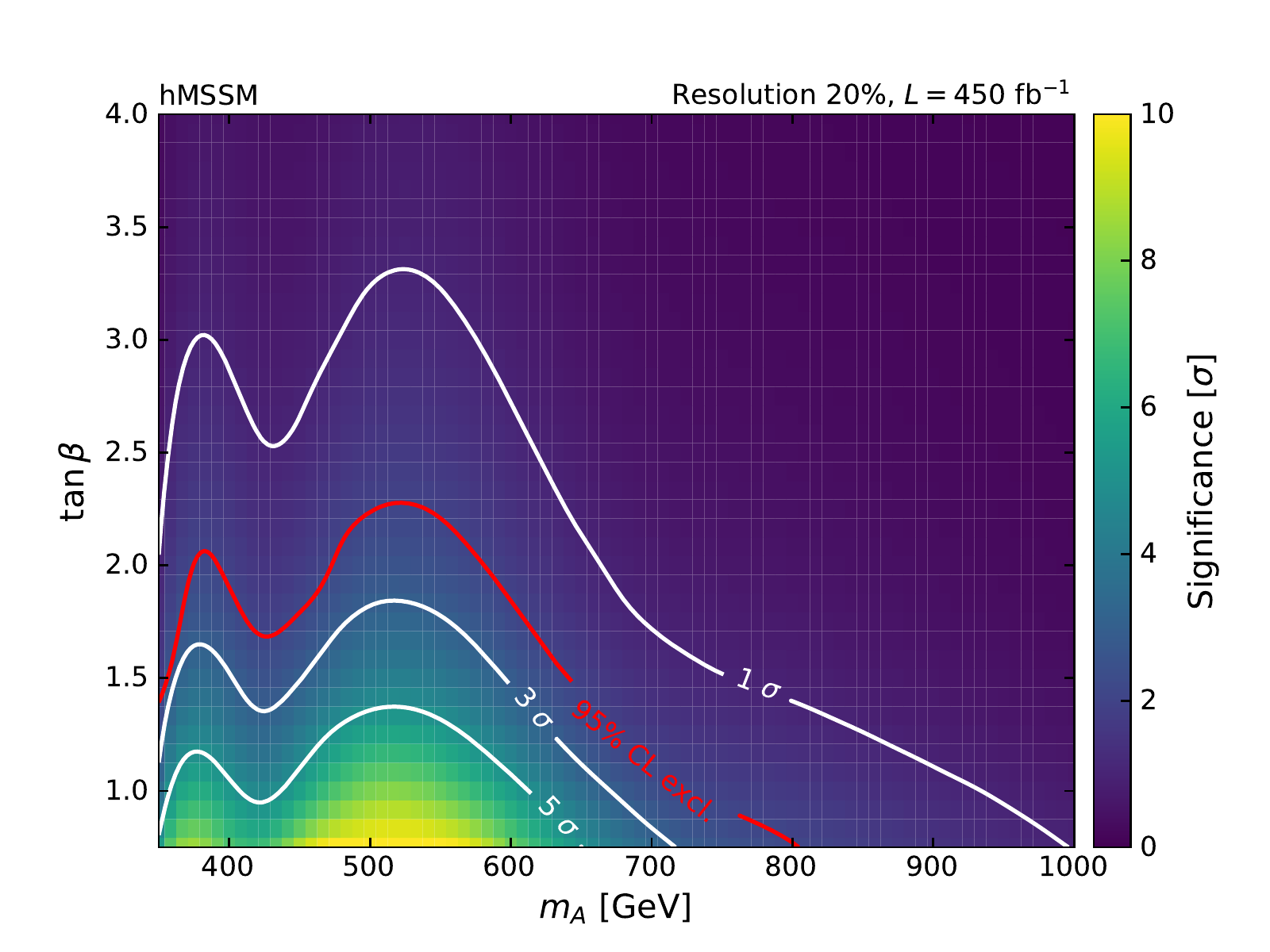} }\\
\mbox{ \includegraphics[width=0.5\textwidth]{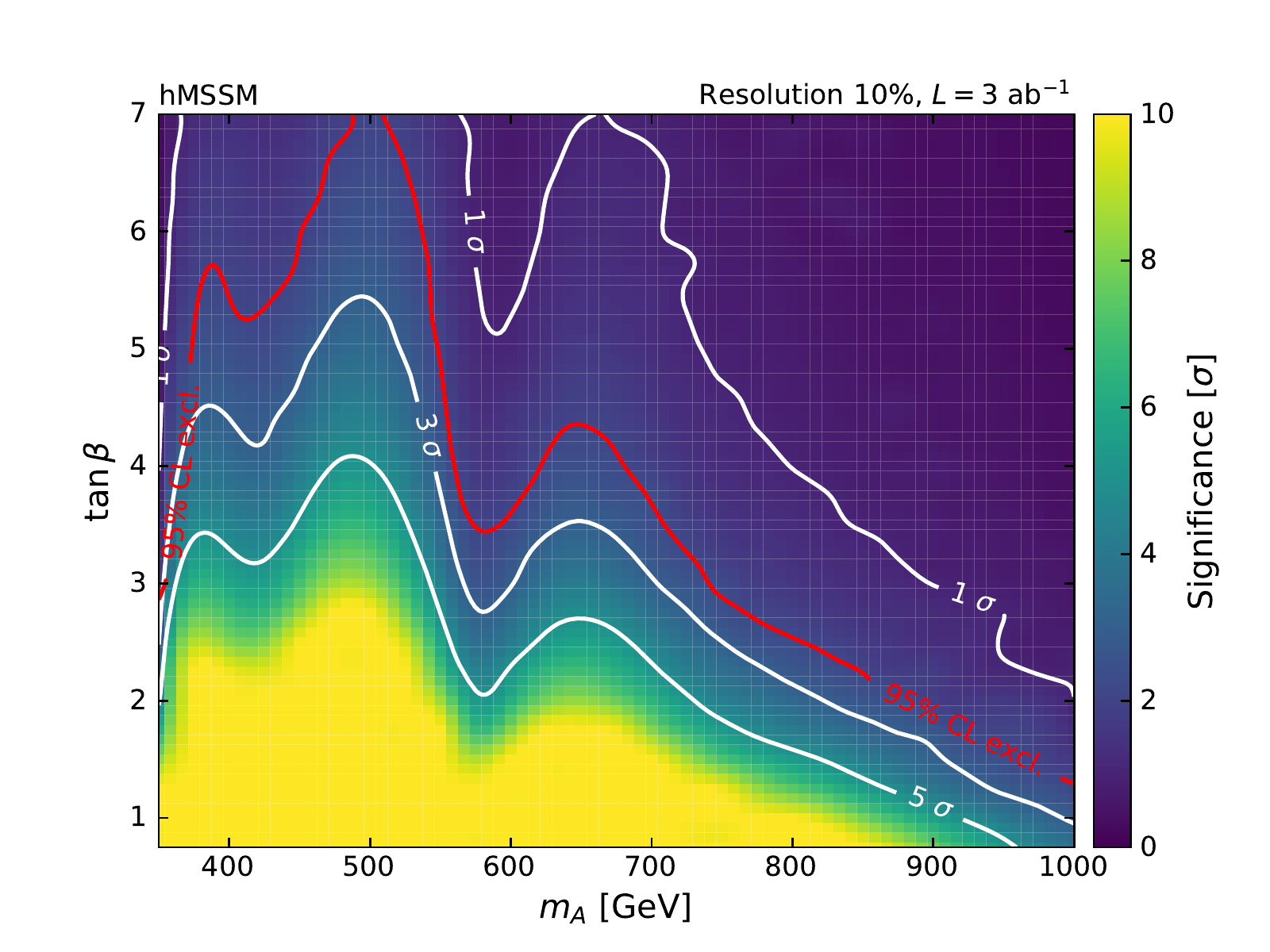}
 \includegraphics[width=0.5\textwidth]{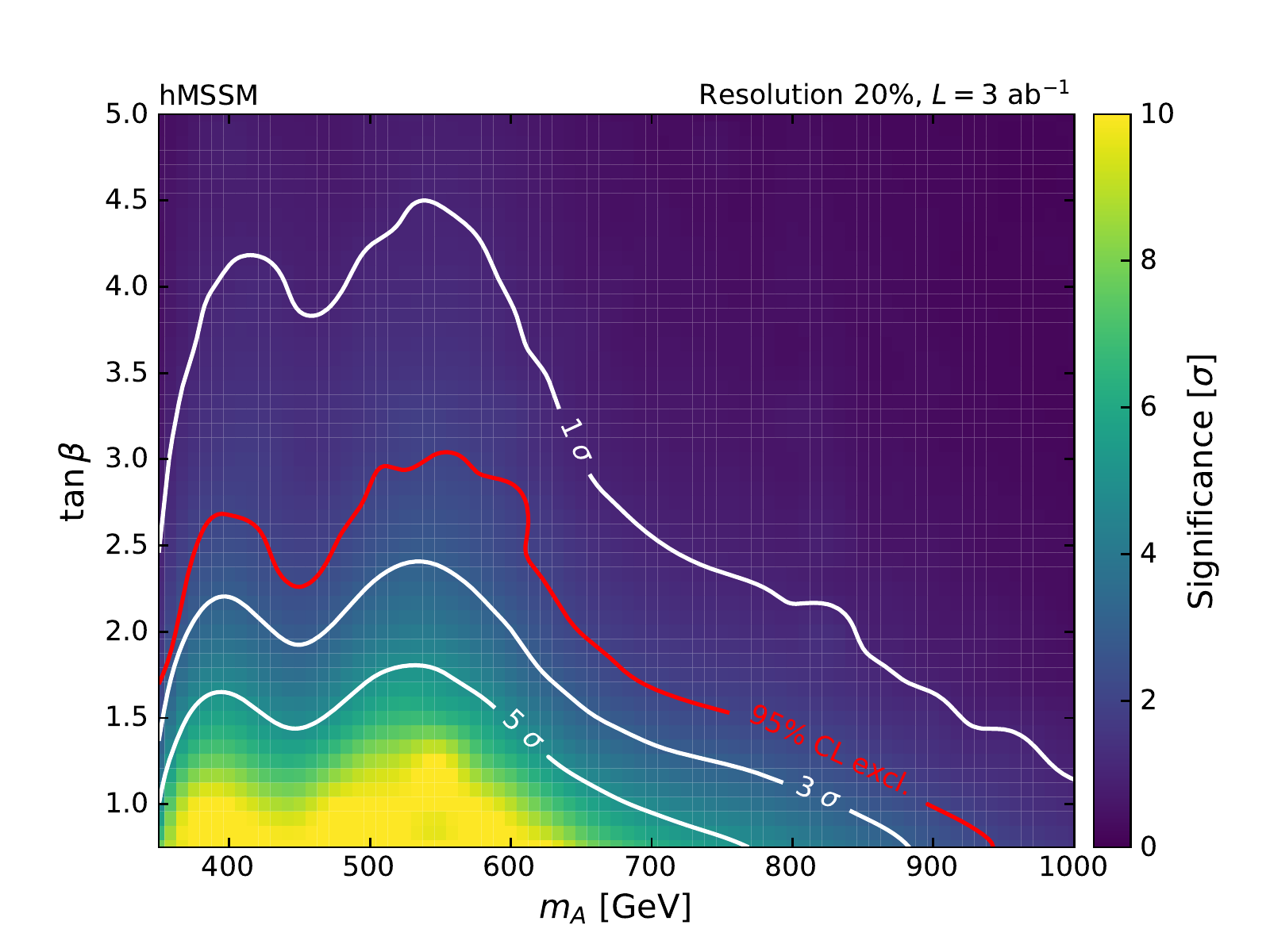} }
 \caption{\it Expected significance and exclusion potential for the hMSSM in the same six experimental scenarios as considered in Fig.~\ref{Fig:Res_SMH}. Values of significance in excess of $10\,\sigma$ are clipped. A resolution of $10\%$ and $20\%$ is assumed for the left- and right plots, respectively.}
 \label{Fig:Res_hMSSM}
\end{figure}

Fig.~\ref{Fig:Res_hMSSM} presents results for the hMSSM in the $[M_A, \tb]$
plane, adopting again the same mass resolution and integrated luminosity
scenarios as previously. The results for the 95\% CL exclusion and 5-$\sigma$
expected significance in these different scenarios are similar to those in the
Type~II 2HDM analyzed previously.

\subsection{Additional vector--like quark contributions to \texorpdfstring{$gg \rightarrow \Phi$}{gg -> H/A}}

\begin{figure}[!ht]
 \centering
 \includegraphics[width=0.7\textwidth]{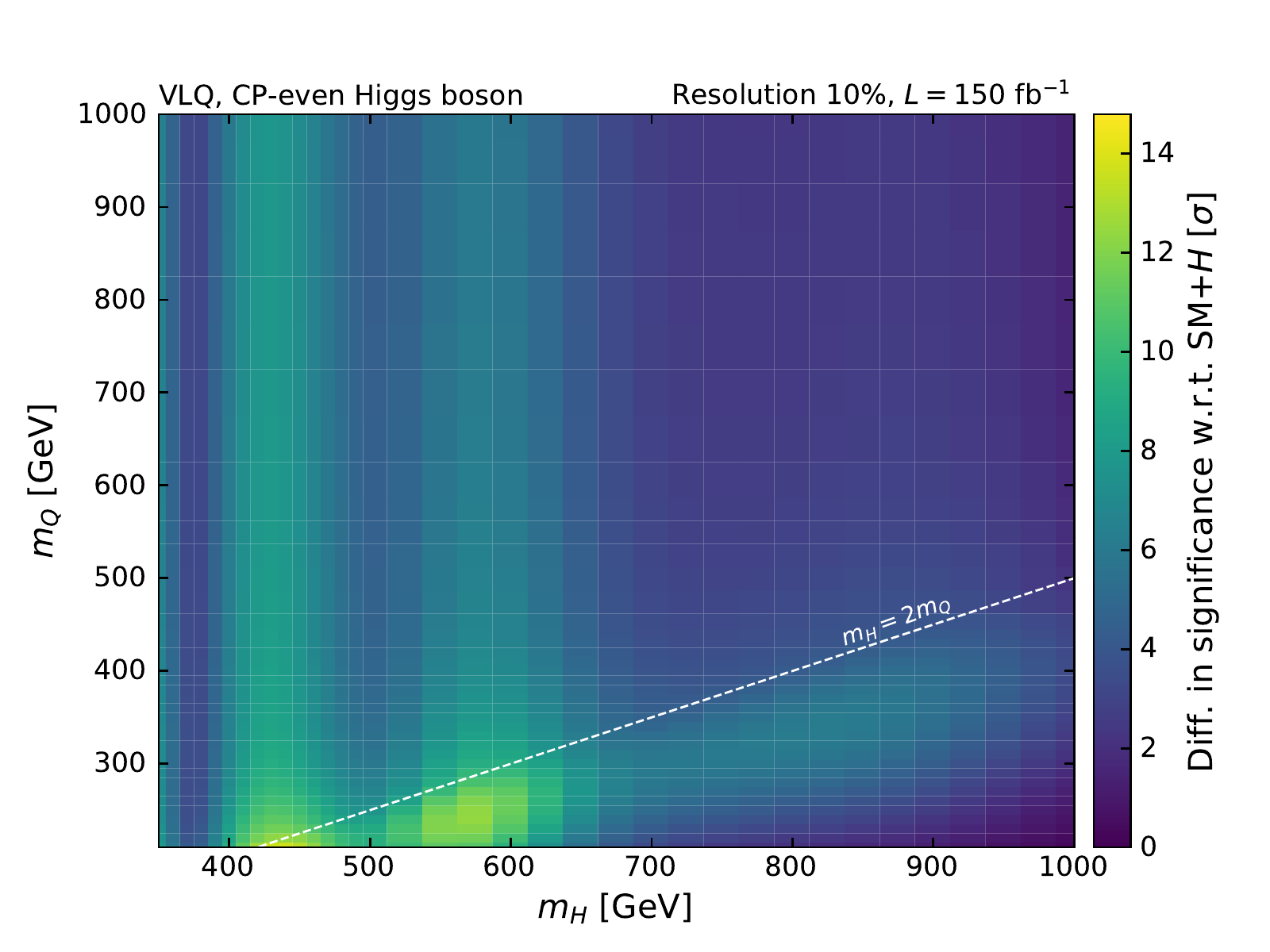}
 \caption{\it The difference in expected significance between the models with and without the contribution of VLQ to the $H t \bar t$ form-factor, computed for a CP--even heavy scalar Higgs boson with $\hat g_{H Q \bar Q} = \hat g_{H t \bar t} = 1$ and $N_Q = 1$.  Below the dashed line, decays $H \rightarrow Q\bar Q$ are kinematically allowed.}
 \label{Fig:Res_VLQ}
\end{figure}

The number of free parameters in the model with additional vector--like quarks
(VLQs) is too large for a comprehensive scan. However, it is interesting to see
in an illustrative example how the contribution from VLQs in the loop changes
the sensitivity compared to the $\text{SM} + \Phi$ model, where only top quarks
are considered in the loop. In this scenario, we do not compute the exact
NNLO $K$--factor taking into account the contribution of the
additional vector--like quarks. Instead, as a first approximation, we set the
signal NNLO $K$--factor to two, the typical NNLO correction to the $gg
\rightarrow \Phi$ production cross section. Fig.~\ref{Fig:Res_VLQ} shows results
for a single representative experimental scenario with a mass resolution of 10\%
and 150/fb of integrated luminosity. We consider the case of a CP--even heavy
Higgs boson $H$ with $\hat g_{\Phi Q \bar Q} = \hat g_{\Phi t \bar t} = 1$ and a
single VLQ species. We find an increase in significance in the VLQ model
compared with the SM+$H$ model over all the $[M_H, M_{VLQ}]$ plane. Also shown
is a dashed line where $M_H = 2 M_{VLQ}$, below which $H \rightarrow Q\bar Q$
decays are kinematically allowed. The region where the decays to VLQ are allowed
is problematic since the total width becomes then larger than $\Gamma (\Phi \to
t\bar t)$ that we assume to be the total width of the new resonance. However in
that region, direct VLQ pair production is likely to provide more distinctive
signatures for larger $M_{\Phi}$.

%\newpage

\subsection{Stop squark contributions to \texorpdfstring{$gg \rightarrow \Phi$}{gg -> H/A}}

Within the MSSM, the predictions for $gg \to H$ are sensitive to the parameters
of the stop sector, and specifically to the mass of the lighter stop, $\tilde
t_1$. Therefore, we have also studied interference in the case of a light stop
in order to highlight the possible effects of stop squark contributions to $gg
\rightarrow H$. We obtain the necessary inputs from the Higgs sector and the
stop sector using the code {\tt FeynHiggs} (version 2.14.3) \cite{Hahn:2013ria,
Bahl:2018qog}, and use them in the ``light-stop'' MSSM benchmark scenario
described in~\cite{MSSM_LHCHXS-lightstop-updated,Carena:2013ytb,Bahl:2018zmf}.
In this scenario the lighter stop has a mass around $324$~GeV and the heavier
stop a mass around $671$~GeV in our scan of  the $[M_A,\tan\beta$] plane. This
value of $m_{\tilde t_1}$ is close to the lower limit from direct LHC searches
in the case of a compressed spectrum with a small difference between the masses
of the stop and the lightest supersymmetric particle. We present results only
for $M_{H} < 2  m_{\tilde t_1}$, since direct stop pair production is likely to
provide more distinctive signatures for larger $M_{H}$. The NNLO $K$--factor has
been computed individually for each point of the grid in the  $[M_A, \tb]$
parameter plane, using the code {\tt SusHi} (version
$1.6.1$)~\cite{Harlander:2012pb,Harlander:2016hcx}.

\begin{figure}[!]
 \centering
\mbox{ \includegraphics[width=0.5\textwidth]{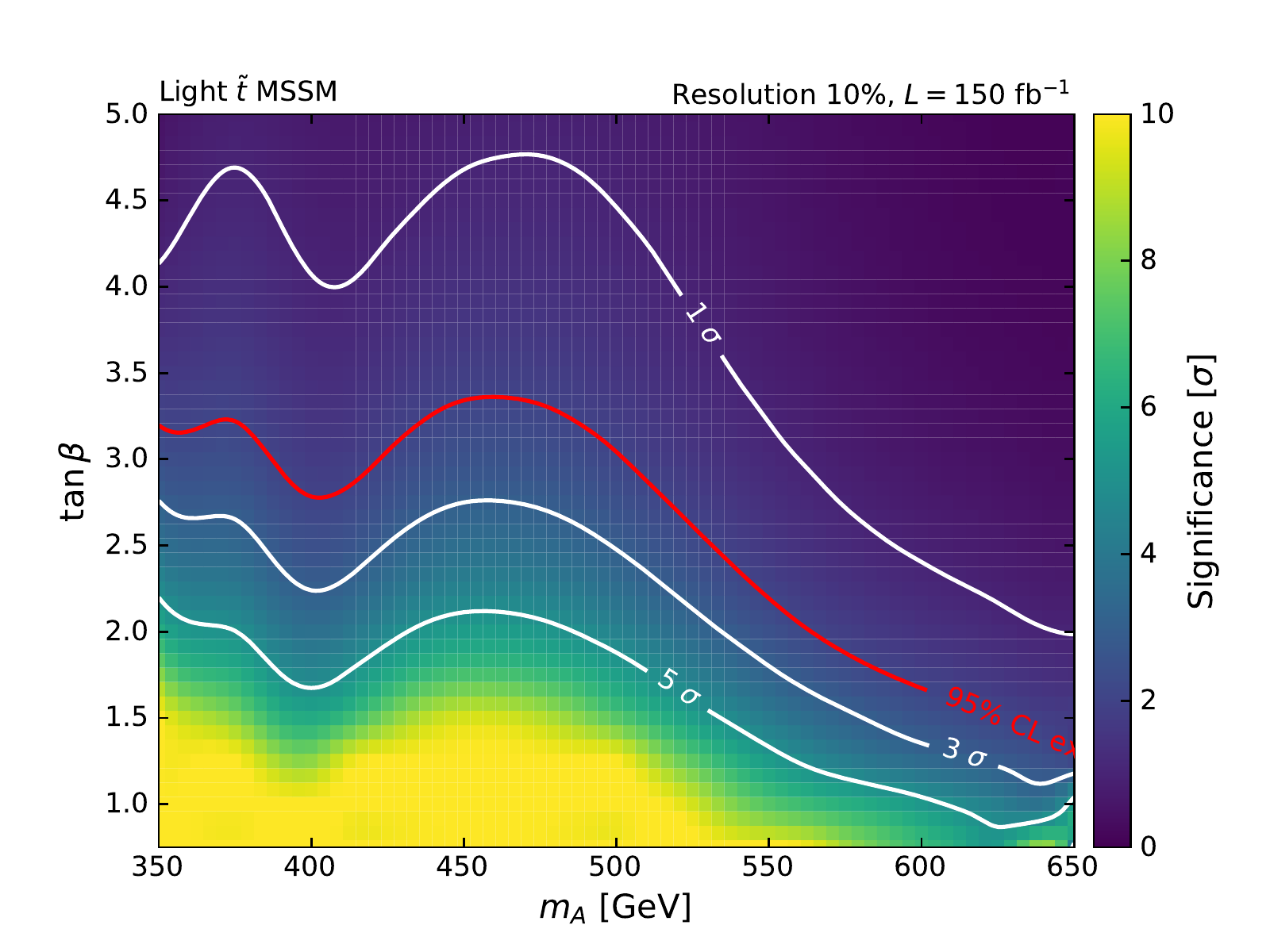}
 \includegraphics[width=0.5\textwidth]{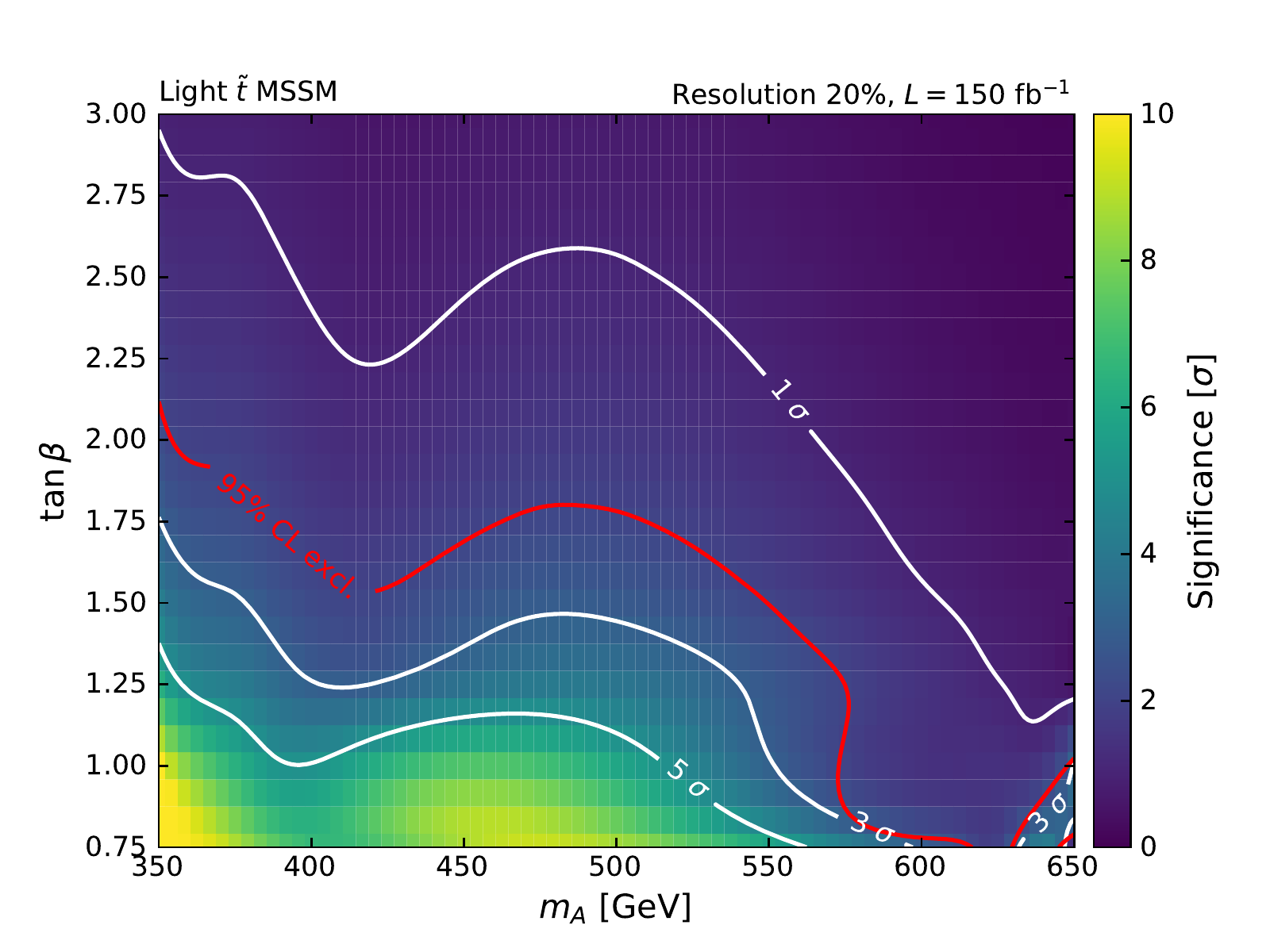} }\\
\mbox{ \includegraphics[width=0.5\textwidth]{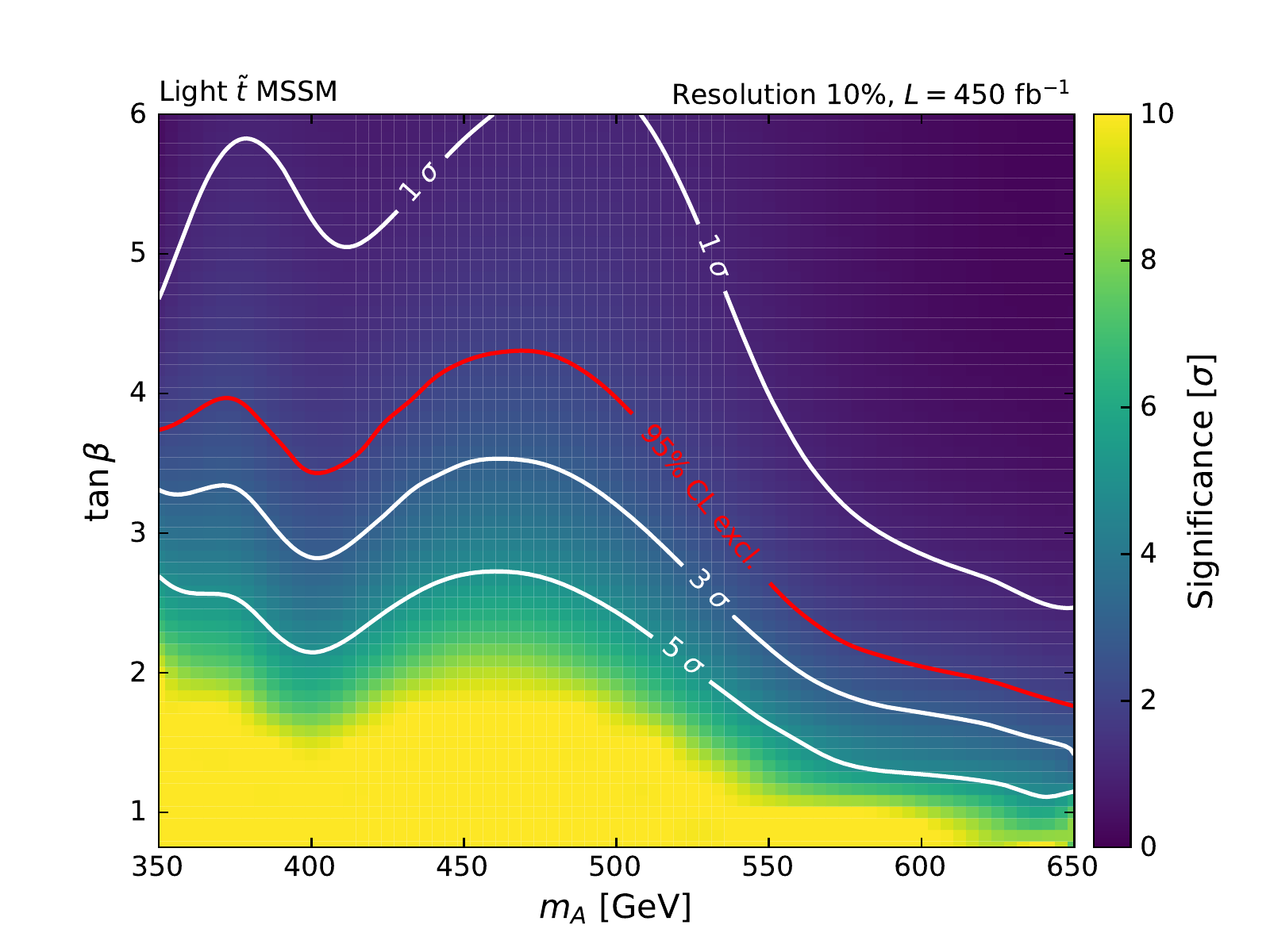}
 \includegraphics[width=0.5\textwidth]{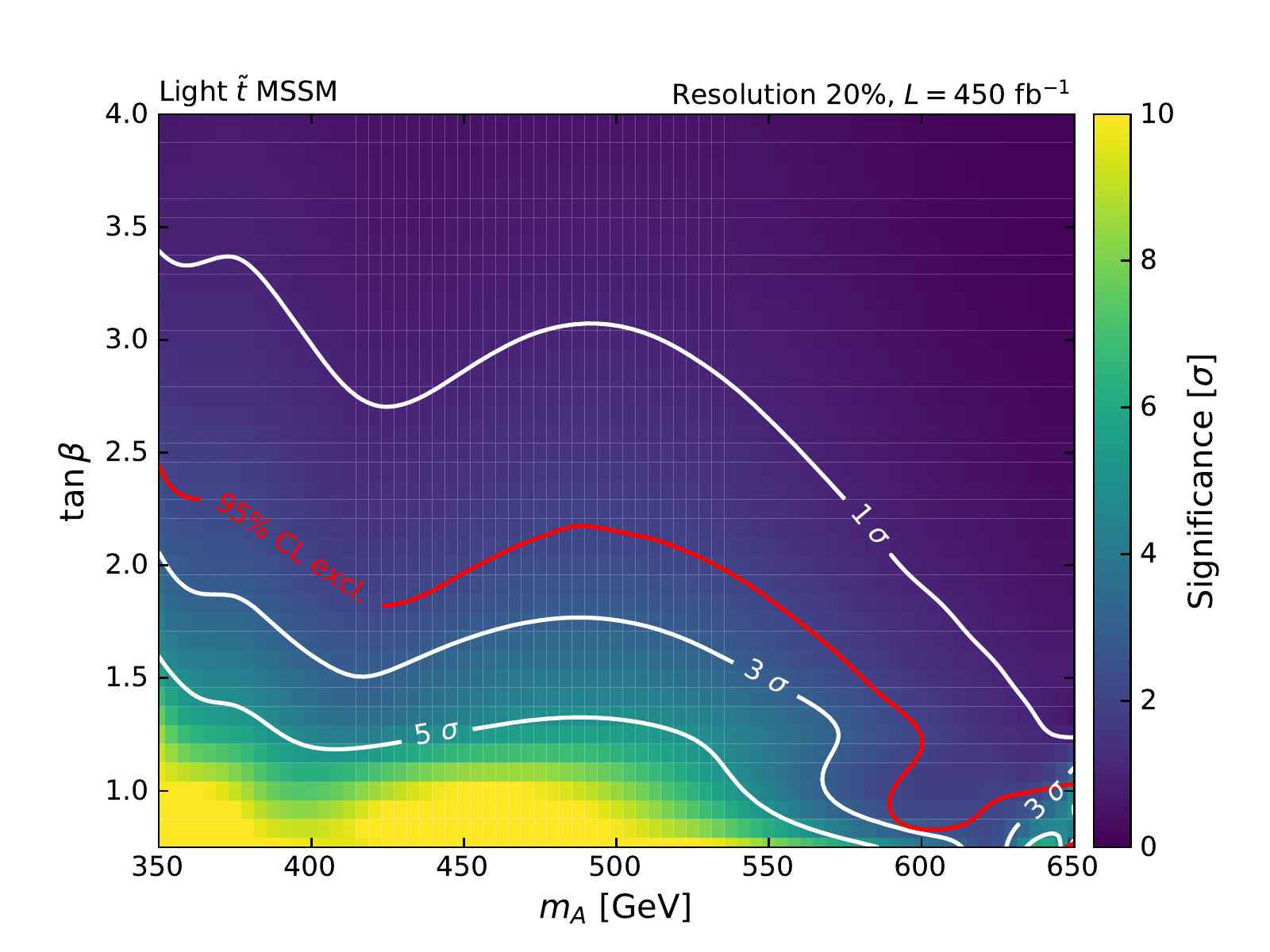} }\\
\mbox{ \includegraphics[width=0.5\textwidth]{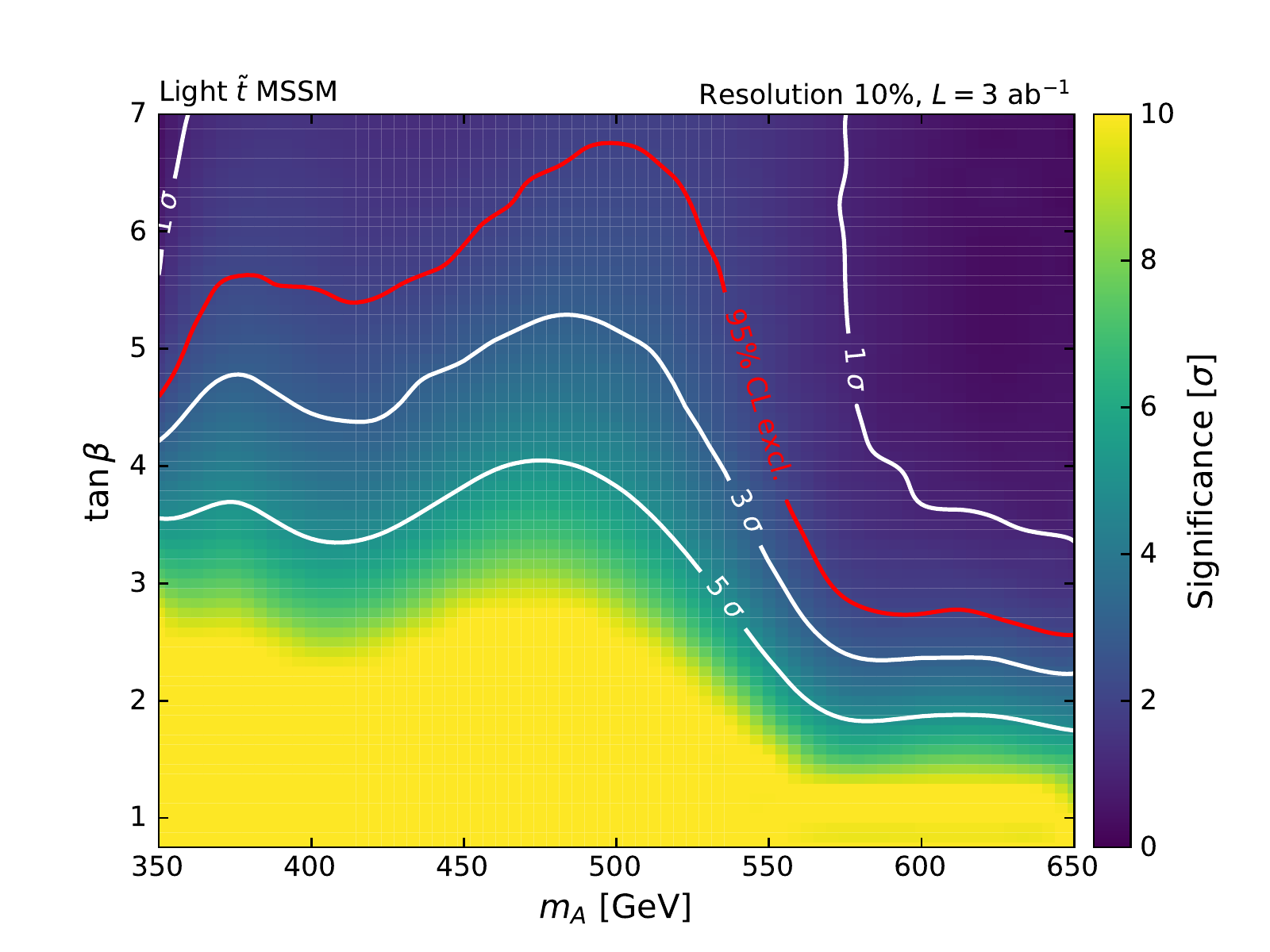}
 \includegraphics[width=0.5\textwidth]{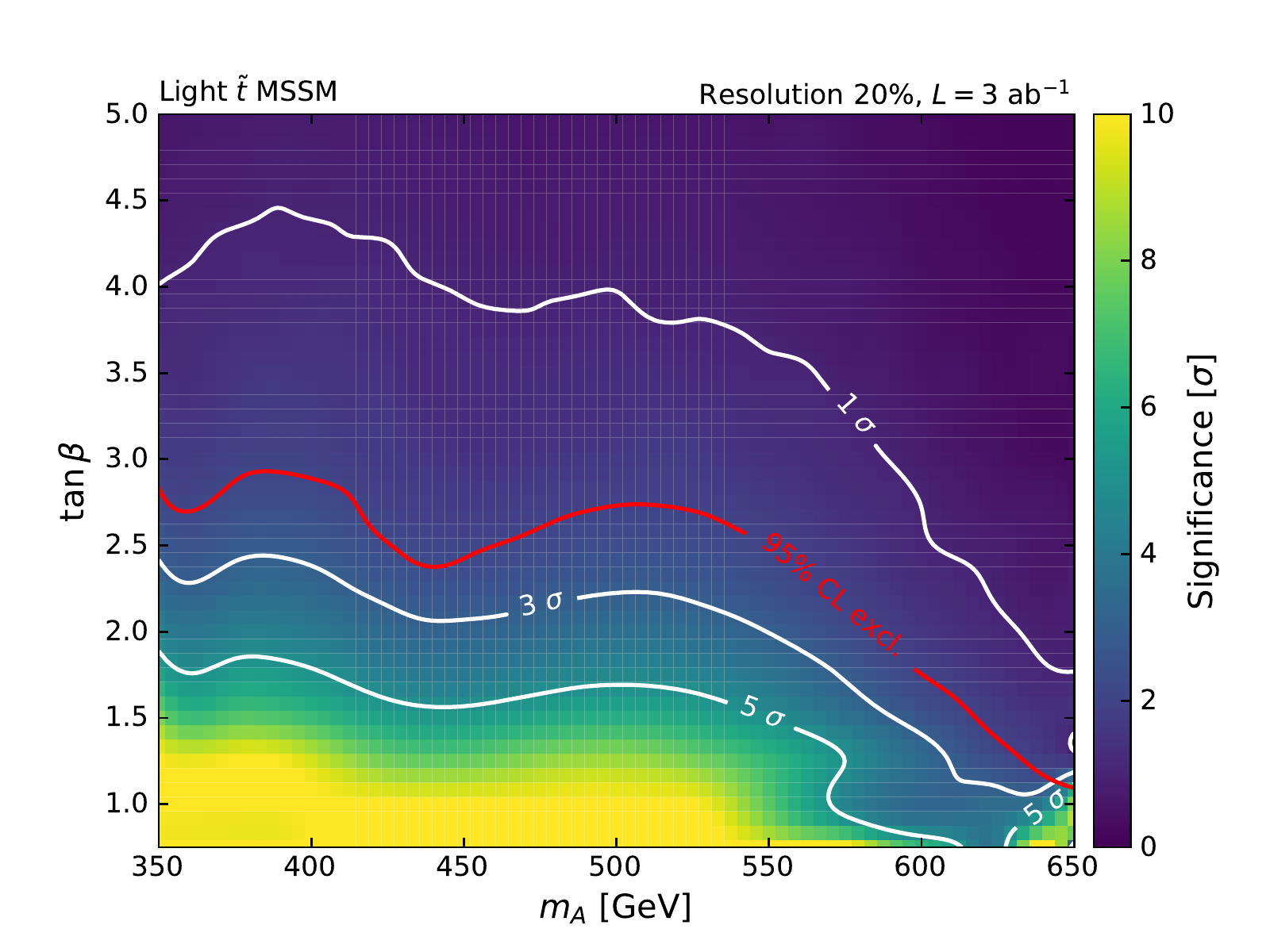} }
 \caption{\it Expected significance and exclusion potential for the  light-stop MSSM benchmark in the same six experimental scenarios as considered in Fig.~\ref{Fig:Res_SMH}. Values of significance in excess of $10\,\sigma$ are clipped. A resolution of $10\%$ and $20\%$ is assumed for the left- and the right-hand side plots, respectively.}
 \label{Fig:Res_MSSM}
\end{figure}

As seen in Fig.~\ref{Fig:Res_MSSM}, we find significant exclusion and discovery
potentials for $M_{H} \lesssim 500$~GeV and smaller values of $\tan \beta$. {As
expected, the experimental sensitivity is restricted to relatively low values of
$\tb$, and we consider the illustrative case of $\tb = 2$. In the case when the
mass resolution is 10\% and the integrated luminosity is 150/fb, the 95\% CL
exclusion extends to a mass of 550~GeV, with a 5-$\sigma$ discovery sensitivity
up to a mass of $\sim480$~GeV. With $\intlumi = 450$ (3000)/fb, the expected
exclusion reaches 600~GeV ($\geq 2 m_{\tilde t_1}$), and there is 5-$\sigma$
discovery sensitivity up to 520 (550)~GeV. On the other hand, if the mass
resolution is 20\% there is no discovery sensitivity in any of the integrated
luminosity scenarios, and no point is excluded for $\intlumi = 150/\text{fb}$.
However, values of $M_H$ up to 530 (580)~GeV can be excluded with 450 (3000)/fb.
This example highlights one more time, therefore, the importance of optimizing
the $t \bar t$ mass resolution.}

\section{Summary}
\label{Sec:Summary}

We have explored in this paper the prospective sensitivity of LHC measurements
of the $t{\bar t}$ invariant-mass spectrum to various BSM scenarios with
additional heavy (pseudo)scalar bosons, via the peak-and-dip features induced by
interference between BSM and SM contributions to the production amplitude that
are illustrated at the parton level in Fig.~\ref{Fig:PartonXSec}. The scenarios
studied included models with one additional singlet (pseudo)scalar boson, a
Type-II 2HDM, the hMSSM, a model with a massive vector--like quark, and models
with a light stop squark. Our analysis is based on a realistic assessment of
detector performance, taking into account the event selection efficiency shown
in Fig.~\ref{Fig:SelEff} and the $t{\bar t}$ invariant-mass resolution, which
smears the parton-level features as illustrated in Fig.~\ref{Fig:SmearedMtt}. 

We have presented results for benchmark scenarios with resolutions of 10 or 20\%
and integrated LHC luminosities ranging from 150/fb (corresponding to the Run 2
data set for a single experiment) to 3000/fb (corresponding to the luminosity
target for HL-LHC). As could be expected, we find that the ranges of model
parameter spaces susceptible to possible exclusion or 5-$\sigma$ discovery at
the LHC are larger for smaller $t{\bar t}$  invariant-mass resolution and the
target HL-LHC luminosity. However, in all the models we find regions of
discovery/exclusion at larger values of the (pseudo)scalar-$t {\bar t}$
coupling, corresponding in 2HDM models such as variants of the MSSM to smaller
values of $\tan \beta$, as illustrated in Fig.~\ref{Fig:Res_SMA} and subsequent
figures.

Our results underline the importance of taking interference effects into account
when interpreting the results of LHC searches for resonance effects in the
$t{\bar t}$ invariant-mass spectrum. The fact that  interferences yield
generically dips as well as bumps may serve as a valuable diagnostic tool for
distinguishing  between interpretations of any features in terms of spin-0 or
spin-1 resonances. Although our analysis has not been at a detector--level
simulation, we think that we have incorporated sufficient features for our
results to provide a realistic estimate of the physics potential of searches for
peak-and-dip structures, and we look forward to the results of experimental
analyses using the full LHC Run 2 dataset and the higher integrated luminosities
to be obtained in the future.

\subsubsection*{Acknowledgements}

This work is supported in part by the Estonian Research Council via Mobilitas
Pluss grants. The work of JE was supported in part by the STFC Grant
ST/P000258/1. The authors are grateful to S\'{e}bastien Brochet and St\'{e}phane
Perri\`es for sharing the \texttt{MadGraph} model employed to simulate the
signal process in the $\text{SM} + \Phi$ model, to Sabine Cr\'ep\'e-Renaudin for
useful discussions, to Stefan Liebler for valuable help and recommendations
regarding the use of {\tt SusHi} and the ``light-stop'' MSSM benchmark scenario
and to S\'{e}bastien Wertz for a valuable suggestion on simulating the
interference.

\bibliographystyle{style}
\bibliography{interference}

\providecommand{\href}[2]{#2}\begingroup\raggedright\begin{thebibliography}{100}%
\makeatletter
\providecommand{\hrefCMSnoop }[0]{\@secondoftwo}%
\makeatother
\providecommand{\doi}{\texttt{doi:}\begingroup \urlstyle{tt}\Url}

\bibitem{Aad:2012tfa}
\hrefCMSnoop {} {{ ATLAS} Collaboration, ``{Observation of a new particle in
  the search for the Standard Model Higgs boson with the ATLAS detector at the
  LHC}'',} \textit{ Phys. Lett.} \textbf{ B716} (2012) 1--29,
  \href{http://dx.doi.org/10.1016/j.physletb.2012.08.020}{\doi{10.1016/j.physletb.2012.08.020}},
\href{http://www.arXiv.org/abs/1207.7214}{\texttt{ arXiv:1207.7214}}.
%%CITATION = ARXIV:1207.7214;%%.

\bibitem{Chatrchyan:2012xdj}
\hrefCMSnoop {} {{ CMS} Collaboration, ``{Observation of a new boson at a mass
  of 125 GeV with the CMS experiment at the LHC}'',} \textit{ Phys. Lett.}
  \textbf{ B716} (2012) 30--61,
  \href{http://dx.doi.org/10.1016/j.physletb.2012.08.021}{\doi{10.1016/j.physletb.2012.08.021}},
\href{http://www.arXiv.org/abs/1207.7235}{\texttt{ arXiv:1207.7235}}.
%%CITATION = ARXIV:1207.7235;%%.

\bibitem{Aaboud:2017sjh}
\hrefCMSnoop {} {{ ATLAS} Collaboration, ``{Search for additional heavy neutral
  Higgs and gauge bosons in the ditau final state produced in 36 fb$^{-1}$ of
  pp collisions at $ \sqrt{s}=13 $ TeV with the ATLAS detector}'',} \textit{
  JHEP} \textbf{ 01} (2018) 055,
  \href{http://dx.doi.org/10.1007/JHEP01(2018)055}{\doi{10.1007/JHEP01(2018)055}},
\href{http://www.arXiv.org/abs/1709.07242}{\texttt{ arXiv:1709.07242}}.
%%CITATION = ARXIV:1709.07242;%%.

\bibitem{Sirunyan:2018exx}
\hrefCMSnoop {} {{ CMS} Collaboration, ``{Search for high-mass resonances in
  dilepton final states in proton-proton collisions at $\sqrt{s}=$ 13 TeV}'',}
  \textit{ JHEP} \textbf{ 06} (2018) 120,
  \href{http://dx.doi.org/10.1007/JHEP06(2018)120}{\doi{10.1007/JHEP06(2018)120}},
\href{http://www.arXiv.org/abs/1803.06292}{\texttt{ arXiv:1803.06292}}.
%%CITATION = ARXIV:1803.06292;%%.

\bibitem{Khachatryan:2017wny}
\hrefCMSnoop {} {{ CMS} Collaboration, ``{Search for leptophobic Z' bosons
  decaying into four-lepton final states in proton--proton collisions at $\sqrt
  s$ =8TeV}'',} \textit{ Phys. Lett.} \textbf{ B773} (2017) 563--584,
  \href{http://dx.doi.org/10.1016/j.physletb.2017.08.069}{\doi{10.1016/j.physletb.2017.08.069}},
\href{http://www.arXiv.org/abs/1701.01345}{\texttt{ arXiv:1701.01345}}.
%%CITATION = ARXIV:1701.01345;%%.

\bibitem{Aaboud:2018jux}
\hrefCMSnoop {} {{ ATLAS} Collaboration, ``{Search for vector-boson resonances
  decaying to a top quark and bottom quark in the lepton plus jets final state
  in $pp$ collisions at $\sqrt{s} = 13$ TeV with the ATLAS detector}'',}
  \textit{ Phys. Lett.} \textbf{ B788} (2019) 347--370,
  \href{http://dx.doi.org/10.1016/j.physletb.2018.11.032}{\doi{10.1016/j.physletb.2018.11.032}},
\href{http://www.arXiv.org/abs/1807.10473}{\texttt{ arXiv:1807.10473}}.
%%CITATION = ARXIV:1807.10473;%%.

\bibitem{Sirunyan:2018lbg}
\hrefCMSnoop {} {{ CMS} Collaboration, ``{Search for a $W'$ boson decaying to a
  $\tau$ lepton and a neutrino in proton-proton collisions at $\sqrt{s} = 13$
  TeV}'',} \textit{ Submitted to: Phys. Lett.} (2018)
\href{http://www.arXiv.org/abs/1807.11421}{\texttt{ arXiv:1807.11421}}.
%%CITATION = ARXIV:1807.11421;%%.

\bibitem{Aaboud:2018mna}
\hrefCMSnoop {} {{ ATLAS} Collaboration, ``{Search for squarks and gluinos in
  final states with hadronically decaying $\tau$-leptons, jets, and missing
  transverse momentum using $pp$ collisions at $\sqrt{s}$ = 13 TeV with the
  ATLAS detector}'',} \textit{ Submitted to: Phys. Rev.} (2018)
\href{http://www.arXiv.org/abs/1808.06358}{\texttt{ arXiv:1808.06358}}.
%%CITATION = ARXIV:1808.06358;%%.

\bibitem{Sirunyan:2018lul}
\hrefCMSnoop {} {{ CMS} Collaboration, ``{Searches for pair production of
  charginos and top squarks in final states with two oppositely charged leptons
  in proton-proton collisions at $\sqrt{s}=$ 13 TeV}'',} \textit{ JHEP}
  \textbf{ 11} (2018) 079,
  \href{http://dx.doi.org/10.1007/JHEP11(2018)079}{\doi{10.1007/JHEP11(2018)079}},
\href{http://www.arXiv.org/abs/1807.07799}{\texttt{ arXiv:1807.07799}}.
%%CITATION = ARXIV:1807.07799;%%.

\bibitem{Caola:2013yja}
\hrefCMSnoop {} {F.~Caola and K.~Melnikov, ``{Constraining the Higgs boson
  width with ZZ production at the LHC}'',} \textit{ Phys. Rev.} \textbf{ D88}
  (2013) 054024,
  \href{http://dx.doi.org/10.1103/PhysRevD.88.054024}{\doi{10.1103/PhysRevD.88.054024}},
\href{http://www.arXiv.org/abs/1307.4935}{\texttt{ arXiv:1307.4935}}.
%%CITATION = ARXIV:1307.4935;%%.

\bibitem{Kauer:2012hd}
\hrefCMSnoop {} {N.~Kauer and G.~Passarino, ``{Inadequacy of zero-width
  approximation for a light Higgs boson signal}'',} \textit{ JHEP} \textbf{ 08}
  (2012) 116,
  \href{http://dx.doi.org/10.1007/JHEP08(2012)116}{\doi{10.1007/JHEP08(2012)116}},
\href{http://www.arXiv.org/abs/1206.4803}{\texttt{ arXiv:1206.4803}}.
%%CITATION = ARXIV:1206.4803;%%.

\bibitem{Dixon:2003yb}
\hrefCMSnoop {} {L.~J. Dixon and M.~S. Siu, ``{Resonance continuum interference
  in the diphoton Higgs signal at the LHC}'',} \textit{ Phys. Rev. Lett.}
  \textbf{ 90} (2003) 252001,
  \href{http://dx.doi.org/10.1103/PhysRevLett.90.252001}{\doi{10.1103/PhysRevLett.90.252001}},
\href{http://www.arXiv.org/abs/hep-ph/0302233}{\texttt{ arXiv:hep-ph/0302233}}.
%%CITATION = HEP-PH/0302233;%%.

\bibitem{Martin:2013ula}
\hrefCMSnoop {} {S.~P. Martin, ``{Interference of Higgs diphoton signal and
  background in production with a jet at the LHC}'',} \textit{ Phys. Rev.}
  \textbf{ D88} (2013), no.~1, 013004,
  \href{http://dx.doi.org/10.1103/PhysRevD.88.013004}{\doi{10.1103/PhysRevD.88.013004}},
\href{http://www.arXiv.org/abs/1303.3342}{\texttt{ arXiv:1303.3342}}.
%%CITATION = ARXIV:1303.3342;%%.

\bibitem{Jung:2015sna}
\hrefCMSnoop {} {S.~Jung, Y.~W. Yoon, and J.~Song, ``{Interference effect on a
  heavy Higgs resonance signal in the $\gamma \gamma$ and $ZZ$ channels}'',}
  \textit{ Phys. Rev.} \textbf{ D93} (2016), no.~5, 055035,
  \href{http://dx.doi.org/10.1103/PhysRevD.93.055035}{\doi{10.1103/PhysRevD.93.055035}},
\href{http://www.arXiv.org/abs/1510.03450}{\texttt{ arXiv:1510.03450}}.
%%CITATION = ARXIV:1510.03450;%%.

\bibitem{Gaemers:1984sj}
\hrefCMSnoop {} {K.~J.~F. Gaemers and F.~Hoogeveen, ``{Higgs Production and
  Decay Into Heavy Flavors With the Gluon Fusion Mechanism}'',} \textit{ Phys.
  Lett.} \textbf{ 146B} (1984) 347--349,
\href{http://dx.doi.org/10.1016/0370-2693(84)91711-8}{\doi{10.1016/0370-2693(84)91711-8}}.
%%CITATION = PHLTA,146B,347;%%.

\bibitem{Dicus:1994bm}
\hrefCMSnoop {} {D.~Dicus, A.~Stange, and S.~Willenbrock, ``{Higgs} decay to
  top quarks at hadron colliders'',} \textit{ Phys. Lett. B} \textbf{ 333}
  (1994) 126--131,
  \href{http://dx.doi.org/10.1016/0370-2693(94)91017-0}{\doi{10.1016/0370-2693(94)91017-0}},
\href{http://www.arXiv.org/abs/hep-ph/9404359}{\texttt{ arXiv:hep-ph/9404359}}.
%%CITATION = HEP-PH/9404359;%%.

\bibitem{Moretti:2012mq}
\hrefCMSnoop {} {S.~Moretti and D.~A. Ross, ``{On the top-antitop invariant
  mass spectrum at the LHC from a Higgs boson signal perspective}'',} \textit{
  Phys. Lett.} \textbf{ B712} (2012) 245--249,
  \href{http://dx.doi.org/10.1016/j.physletb.2012.04.074}{\doi{10.1016/j.physletb.2012.04.074}},
\href{http://www.arXiv.org/abs/1203.3746}{\texttt{ arXiv:1203.3746}}.
%%CITATION = ARXIV:1203.3746;%%.

\bibitem{Frederix:2007gi}
\hrefCMSnoop {} {R.~Frederix and F.~Maltoni, ``{Top pair invariant mass
  distribution: A Window on new physics}'',} \textit{ JHEP} \textbf{ 01} (2009)
  047,
  \href{http://dx.doi.org/10.1088/1126-6708/2009/01/047}{\doi{10.1088/1126-6708/2009/01/047}},
\href{http://www.arXiv.org/abs/0712.2355}{\texttt{ arXiv:0712.2355}}.
%%CITATION = ARXIV:0712.2355;%%.

\bibitem{Hespel:2016qaf}
\hrefCMSnoop {} {B.~Hespel, F.~Maltoni, and E.~Vryonidou, ``Signal background
  interference effects in heavy scalar production and decay to a top-anti-top
  pair'',} \textit{ JHEP} \textbf{ 10} (2016) 016,
  \href{http://dx.doi.org/10.1007/JHEP10(2016)016}{\doi{10.1007/JHEP10(2016)016}},
\href{http://www.arXiv.org/abs/1606.04149}{\texttt{ arXiv:1606.04149}}.
%%CITATION = ARXIV:1606.04149;%%.

\bibitem{Jung:2015etr}
\hrefCMSnoop {} {S.~Jung, J.~Song, and Y.~W. Yoon, ``{How Resonance-Continuum
  Interference Changes 750 GeV Diphoton Excess: Signal Enhancement and Peak
  Shift}'',} \textit{ JHEP} \textbf{ 05} (2016) 009,
  \href{http://dx.doi.org/10.1007/JHEP05(2016)009}{\doi{10.1007/JHEP05(2016)009}},
\href{http://www.arXiv.org/abs/1601.00006}{\texttt{ arXiv:1601.00006}}.
%%CITATION = ARXIV:1601.00006;%%.

\bibitem{Djouadi:2016ack}
\hrefCMSnoop {} {A.~Djouadi, J.~Ellis, and J.~Quevillon, ``{Interference
  effects in the decays of spin-zero resonances into $\gamma \gamma$ and $
  t\overline{t} $}'',} \textit{ JHEP} \textbf{ 07} (2016) 105,
  \href{http://dx.doi.org/10.1007/JHEP07(2016)105}{\doi{10.1007/JHEP07(2016)105}},
\href{http://www.arXiv.org/abs/1605.00542}{\texttt{ arXiv:1605.00542}}.
%%CITATION = ARXIV:1605.00542;%%.

\bibitem{Branco:2011iw}
G.~C. Branco\hrefCMSnoop {} { {et~al.}, ``{Theory and phenomenology of
  two-Higgs-doublet models}'',} \textit{ Phys. Rept.} \textbf{ 516} (2012)
  1--102,
  \href{http://dx.doi.org/10.1016/j.physrep.2012.02.002}{\doi{10.1016/j.physrep.2012.02.002}},
\href{http://www.arXiv.org/abs/1106.0034}{\texttt{ arXiv:1106.0034}}.
%%CITATION = ARXIV:1106.0034;%%.

\bibitem{Djouadi:2005gj}
\hrefCMSnoop {} {A.~Djouadi, ``{The Anatomy of electro-weak symmetry breaking.
  II. The Higgs bosons in the minimal supersymmetric model}'',} \textit{ Phys.
  Rept.} \textbf{ 459} (2008) 1--241,
  \href{http://dx.doi.org/10.1016/j.physrep.2007.10.005}{\doi{10.1016/j.physrep.2007.10.005}},
\href{http://www.arXiv.org/abs/hep-ph/0503173}{\texttt{ arXiv:hep-ph/0503173}}.
%%CITATION = HEP-PH/0503173;%%.

\bibitem{Djouadi:2013uqa}
A.~Djouadi\hrefCMSnoop {} { {et~al.}, ``{The post-Higgs MSSM scenario: Habemus
  MSSM?}'',} \textit{ Eur. Phys. J.} \textbf{ C73} (2013) 2650,
  \href{http://dx.doi.org/10.1140/epjc/s10052-013-2650-0}{\doi{10.1140/epjc/s10052-013-2650-0}},
\href{http://www.arXiv.org/abs/1307.5205}{\texttt{ arXiv:1307.5205}}.
%%CITATION = ARXIV:1307.5205;%%.

\bibitem{Djouadi:2015jea}
A.~Djouadi\hrefCMSnoop {} { {et~al.}, ``{Fully covering the MSSM Higgs sector
  at the LHC}'',} \textit{ JHEP} \textbf{ 06} (2015) 168,
  \href{http://dx.doi.org/10.1007/JHEP06(2015)168}{\doi{10.1007/JHEP06(2015)168}},
\href{http://www.arXiv.org/abs/1502.05653}{\texttt{ arXiv:1502.05653}}.
%%CITATION = ARXIV:1502.05653;%%.

\bibitem{Accomando:2011eu}
E.~Accomando\hrefCMSnoop {} { {et~al.}, ``{Interference effects in heavy
  $W'$-boson searches at the LHC}'',} \textit{ Phys. Rev.} \textbf{ D85} (2012)
  115017,
  \href{http://dx.doi.org/10.1103/PhysRevD.85.115017}{\doi{10.1103/PhysRevD.85.115017}},
\href{http://www.arXiv.org/abs/1110.0713}{\texttt{ arXiv:1110.0713}}.
%%CITATION = ARXIV:1110.0713;%%.

\bibitem{Accomando:2013sfa}
E.~Accomando\hrefCMSnoop {} { {et~al.}, ``{$Z'$ at the LHC: Interference and
  Finite Width Effects in Drell-Yan}'',} \textit{ JHEP} \textbf{ 10} (2013)
  153,
  \href{http://dx.doi.org/10.1007/JHEP10(2013)153}{\doi{10.1007/JHEP10(2013)153}},
\href{http://www.arXiv.org/abs/1304.6700}{\texttt{ arXiv:1304.6700}}.
%%CITATION = ARXIV:1304.6700;%%.

\bibitem{Djouadi:1991sx}
A.~Djouadi\hrefCMSnoop {} { {et~al.}, ``{Signals of new gauge bosons at future
  $e^+ e^-$ colliders}'',} \textit{ Z. Phys.} \textbf{ C56} (1992) 289--300,
\href{http://dx.doi.org/10.1007/BF01555527}{\doi{10.1007/BF01555527}}.
%%CITATION = ZEPYA,C56,289;%%.

\bibitem{Djouadi:2007eg}
\hrefCMSnoop {} {A.~Djouadi, G.~Moreau, and R.~K. Singh, ``{Kaluza-Klein
  excitations of gauge bosons at the LHC}'',} \textit{ Nucl. Phys.} \textbf{
  B797} (2008) 1--26,
  \href{http://dx.doi.org/10.1016/j.nuclphysb.2007.12.024}{\doi{10.1016/j.nuclphysb.2007.12.024}},
\href{http://www.arXiv.org/abs/0706.4191}{\texttt{ arXiv:0706.4191}}.
%%CITATION = ARXIV:0706.4191;%%.

\bibitem{Djouadi:2013vqa}
\hrefCMSnoop {} {A.~Djouadi and J.~Quevillon, ``{The MSSM Higgs sector at a
  high $M_{SUSY}$: reopening the low tan$\beta$ regime and heavy Higgs
  searches}'',} \textit{ JHEP} \textbf{ 10} (2013) 028,
  \href{http://dx.doi.org/10.1007/JHEP10(2013)028}{\doi{10.1007/JHEP10(2013)028}},
\href{http://www.arXiv.org/abs/1304.1787}{\texttt{ arXiv:1304.1787}}.
%%CITATION = ARXIV:1304.1787;%%.

\bibitem{Aad:2015pla}
\hrefCMSnoop {} {{ ATLAS} Collaboration, ``{Constraints on new phenomena via
  Higgs boson couplings and invisible decays with the ATLAS detector}'',}
  \textit{ JHEP} \textbf{ 11} (2015) 206,
  \href{http://dx.doi.org/10.1007/JHEP11(2015)206}{\doi{10.1007/JHEP11(2015)206}},
\href{http://www.arXiv.org/abs/1509.00672}{\texttt{ arXiv:1509.00672}}.
%%CITATION = ARXIV:1509.00672;%%.

\bibitem{CMS:2016qbe}
\href {http://cds.cern.ch/record/2142432} {{ CMS} Collaboration,
``{Summary results of high mass BSM Higgs searches using CMS run-I data}'',}.
%%CITATION = CMS-PAS-HIG-16-007;%%.

\bibitem{deBlas:2016ojx}
J.~de~Blas\hrefCMSnoop {} { {et~al.}, ``{Electroweak precision observables and
  Higgs-boson signal strengths in the Standard Model and beyond: present and
  future}'',} \textit{ JHEP} \textbf{ 12} (2016) 135,
  \href{http://dx.doi.org/10.1007/JHEP12(2016)135}{\doi{10.1007/JHEP12(2016)135}},
\href{http://www.arXiv.org/abs/1608.01509}{\texttt{ arXiv:1608.01509}}.
%%CITATION = ARXIV:1608.01509;%%.

\bibitem{Haller:2018nnx}
J.~Haller\hrefCMSnoop {} { {et~al.}, ``{Update of the global electroweak fit
  and constraints on two-Higgs-doublet models}'',} \textit{ Eur. Phys. J.}
  \textbf{ C78} (2018), no.~8, 675,
  \href{http://dx.doi.org/10.1140/epjc/s10052-018-6131-3}{\doi{10.1140/epjc/s10052-018-6131-3}},
\href{http://www.arXiv.org/abs/1803.01853}{\texttt{ arXiv:1803.01853}}.
%%CITATION = ARXIV:1803.01853;%%.

\bibitem{Djouadi:1995gv}
\hrefCMSnoop {} {A.~Djouadi, J.~Kalinowski, and P.~M. Zerwas, ``{Two and
  three-body decay modes of SUSY Higgs particles}'',} \textit{ Z. Phys.}
  \textbf{ C70} (1996) 435--448,
  \href{http://dx.doi.org/10.1007/s002880050121}{\doi{10.1007/s002880050121}},
\href{http://www.arXiv.org/abs/hep-ph/9511342}{\texttt{ arXiv:hep-ph/9511342}}.
%%CITATION = HEP-PH/9511342;%%.

\bibitem{Bernreuther:1997gs}
\hrefCMSnoop {} {W.~Bernreuther, M.~Flesch, and P.~Haberl, ``{Signatures of
  Higgs bosons in the top quark decay channel at hadron colliders}'',} \textit{
  Phys. Rev.} \textbf{ D58} (1998) 114031,
  \href{http://dx.doi.org/10.1103/PhysRevD.58.114031}{\doi{10.1103/PhysRevD.58.114031}},
\href{http://www.arXiv.org/abs/hep-ph/9709284}{\texttt{ arXiv:hep-ph/9709284}}.
%%CITATION = HEP-PH/9709284;%%.

\bibitem{Barger:2006hm}
\hrefCMSnoop {} {V.~Barger, T.~Han, and D.~G.~E. Walker, ``{Top Quark Pairs at
  High Invariant Mass: A Model-Independent Discriminator of New Physics at the
  LHC}'',} \textit{ Phys. Rev. Lett.} \textbf{ 100} (2008) 031801,
  \href{http://dx.doi.org/10.1103/PhysRevLett.100.031801}{\doi{10.1103/PhysRevLett.100.031801}},
\href{http://www.arXiv.org/abs/hep-ph/0612016}{\texttt{ arXiv:hep-ph/0612016}}.
%%CITATION = HEP-PH/0612016;%%.

\bibitem{Barcelo:2010bm}
\hrefCMSnoop {} {R.~Barcelo and M.~Masip, ``{Extra Higgs bosons in $t\bar{t}$
  production at the LHC}'',} \textit{ Phys. Rev.} \textbf{ D81} (2010) 075019,
  \href{http://dx.doi.org/10.1103/PhysRevD.81.075019}{\doi{10.1103/PhysRevD.81.075019}},
\href{http://www.arXiv.org/abs/1001.5456}{\texttt{ arXiv:1001.5456}}.
%%CITATION = ARXIV:1001.5456;%%.

\bibitem{Figy:2011yu}
\hrefCMSnoop {} {T.~Figy and R.~Zwicky, ``{The other Higgses, at resonance, in
  the Lee-Wick extension of the Standard Model}'',} \textit{ JHEP} \textbf{ 10}
  (2011) 145,
  \href{http://dx.doi.org/10.1007/JHEP10(2011)145}{\doi{10.1007/JHEP10(2011)145}},
\href{http://www.arXiv.org/abs/1108.3765}{\texttt{ arXiv:1108.3765}}.
%%CITATION = ARXIV:1108.3765;%%.

\bibitem{Craig:2015jba}
N.~Craig\hrefCMSnoop {} { {et~al.}, ``{The Hunt for the Rest of the Higgs
  Bosons}'',} \textit{ JHEP} \textbf{ 06} (2015) 137,
  \href{http://dx.doi.org/10.1007/JHEP06(2015)137}{\doi{10.1007/JHEP06(2015)137}},
\href{http://www.arXiv.org/abs/1504.04630}{\texttt{ arXiv:1504.04630}}.
%%CITATION = ARXIV:1504.04630;%%.

\bibitem{Carena:2016npr}
\hrefCMSnoop {} {M.~Carena and Z.~Liu, ``{Challenges and opportunities for
  heavy scalar searches in the $ t\overline{t} $ channel at the LHC}'',}
  \textit{ JHEP} \textbf{ 11} (2016) 159,
  \href{http://dx.doi.org/10.1007/JHEP11(2016)159}{\doi{10.1007/JHEP11(2016)159}},
\href{http://www.arXiv.org/abs/1608.07282}{\texttt{ arXiv:1608.07282}}.
%%CITATION = ARXIV:1608.07282;%%.

\bibitem{Djouadi:2005gi}
\hrefCMSnoop {} {A.~Djouadi, ``{The Anatomy of electro-weak symmetry breaking.
  I: The Higgs boson in the standard model}'',} \textit{ Phys. Rept.} \textbf{
  457} (2008) 1--216,
  \href{http://dx.doi.org/10.1016/j.physrep.2007.10.004}{\doi{10.1016/j.physrep.2007.10.004}},
\href{http://www.arXiv.org/abs/hep-ph/0503172}{\texttt{ arXiv:hep-ph/0503172}}.
%%CITATION = HEP-PH/0503172;%%.

\bibitem{Djouadi:1997yw}
\hrefCMSnoop {} {A.~Djouadi, J.~Kalinowski, and M.~Spira, ``{HDECAY: A Program
  for Higgs boson decays in the standard model and its supersymmetric
  extension}'',} \textit{ Comput. Phys. Commun.} \textbf{ 108} (1998) 56--74,
  \href{http://dx.doi.org/10.1016/S0010-4655(97)00123-9}{\doi{10.1016/S0010-4655(97)00123-9}},
\href{http://www.arXiv.org/abs/hep-ph/9704448}{\texttt{ arXiv:hep-ph/9704448}}.
%%CITATION = HEP-PH/9704448;%%.

\bibitem{Baglio:2010ae}
\hrefCMSnoop {} {J.~Baglio and A.~Djouadi, ``{Higgs production at the lHC}'',}
  \textit{ JHEP} \textbf{ 03} (2011) 055,
  \href{http://dx.doi.org/10.1007/JHEP03(2011)055}{\doi{10.1007/JHEP03(2011)055}},
\href{http://www.arXiv.org/abs/1012.0530}{\texttt{ arXiv:1012.0530}}.
%%CITATION = ARXIV:1012.0530;%%.

\bibitem{Spira:1995rr}
\hrefCMSnoop {} {M.~Spira, A.~Djouadi, D.~Graudenz, and P.~M. Zerwas, ``{Higgs
  boson production at the LHC}'',} \textit{ Nucl. Phys.} \textbf{ B453} (1995)
  17--82,
  \href{http://dx.doi.org/10.1016/0550-3213(95)00379-7}{\doi{10.1016/0550-3213(95)00379-7}},
\href{http://www.arXiv.org/abs/hep-ph/9504378}{\texttt{ arXiv:hep-ph/9504378}}.
%%CITATION = HEP-PH/9504378;%%.

\bibitem{Djouadi:1991tka}
\hrefCMSnoop {} {A.~Djouadi, M.~Spira, and P.~M. Zerwas, ``{Production of Higgs
  bosons in proton colliders: QCD corrections}'',} \textit{ Phys. Lett.}
  \textbf{ B264} (1991) 440--446,
\href{http://dx.doi.org/10.1016/0370-2693(91)90375-Z}{\doi{10.1016/0370-2693(91)90375-Z}}.
%%CITATION = PHLTA,B264,440;%%.

\bibitem{Dawson:1990zj}
\hrefCMSnoop {} {S.~Dawson, ``{Radiative corrections to Higgs boson
  production}'',} \textit{ Nucl. Phys.} \textbf{ B359} (1991) 283--300,
\href{http://dx.doi.org/10.1016/0550-3213(91)90061-2}{\doi{10.1016/0550-3213(91)90061-2}}.
%%CITATION = NUPHA,B359,283;%%.

\bibitem{Spira:1993bb}
\hrefCMSnoop {} {M.~Spira, A.~Djouadi, D.~Graudenz, and P.~M. Zerwas, ``{SUSY
  Higgs production at proton colliders}'',} \textit{ Phys. Lett.} \textbf{
  B318} (1993) 347--353,
\href{http://dx.doi.org/10.1016/0370-2693(93)90138-8}{\doi{10.1016/0370-2693(93)90138-8}}.
%%CITATION = PHLTA,B318,347;%%.

\bibitem{Harlander:2002wh}
\hrefCMSnoop {} {R.~V. Harlander and W.~B. Kilgore, ``{Next-to-next-to-leading
  order Higgs production at hadron colliders}'',} \textit{ Phys. Rev. Lett.}
  \textbf{ 88} (2002) 201801,
  \href{http://dx.doi.org/10.1103/PhysRevLett.88.201801}{\doi{10.1103/PhysRevLett.88.201801}},
\href{http://www.arXiv.org/abs/hep-ph/0201206}{\texttt{ arXiv:hep-ph/0201206}}.
%%CITATION = HEP-PH/0201206;%%.

\bibitem{Anastasiou:2002yz}
\hrefCMSnoop {} {C.~Anastasiou and K.~Melnikov, ``{Higgs boson production at
  hadron colliders in NNLO QCD}'',} \textit{ Nucl. Phys.} \textbf{ B646} (2002)
  220--256,
  \href{http://dx.doi.org/10.1016/S0550-3213(02)00837-4}{\doi{10.1016/S0550-3213(02)00837-4}},
\href{http://www.arXiv.org/abs/hep-ph/0207004}{\texttt{ arXiv:hep-ph/0207004}}.
%%CITATION = HEP-PH/0207004;%%.

\bibitem{Ravindran:2003um}
\hrefCMSnoop {} {V.~Ravindran, J.~Smith, and W.~L. van Neerven, ``{NNLO
  corrections to the total cross-section for Higgs boson production in hadron
  hadron collisions}'',} \textit{ Nucl. Phys.} \textbf{ B665} (2003) 325--366,
  \href{http://dx.doi.org/10.1016/S0550-3213(03)00457-7}{\doi{10.1016/S0550-3213(03)00457-7}},
\href{http://www.arXiv.org/abs/hep-ph/0302135}{\texttt{ arXiv:hep-ph/0302135}}.
%%CITATION = HEP-PH/0302135;%%.

\bibitem{Harlander:2002vv}
\hrefCMSnoop {} {R.~V. Harlander and W.~B. Kilgore, ``{Production of a
  pseudoscalar Higgs boson at hadron colliders at next-to-next-to leading
  order}'',} \textit{ JHEP} \textbf{ 10} (2002) 017,
  \href{http://dx.doi.org/10.1088/1126-6708/2002/10/017}{\doi{10.1088/1126-6708/2002/10/017}},
\href{http://www.arXiv.org/abs/hep-ph/0208096}{\texttt{ arXiv:hep-ph/0208096}}.
%%CITATION = HEP-PH/0208096;%%.

\bibitem{Anastasiou:2015ema}
C.~Anastasiou\hrefCMSnoop {} { {et~al.}, ``{Higgs Boson Gluon-Fusion Production
  in QCD at Three Loops}'',} \textit{ Phys. Rev. Lett.} \textbf{ 114} (2015)
  212001,
  \href{http://dx.doi.org/10.1103/PhysRevLett.114.212001}{\doi{10.1103/PhysRevLett.114.212001}},
\href{http://www.arXiv.org/abs/1503.06056}{\texttt{ arXiv:1503.06056}}.
%%CITATION = ARXIV:1503.06056;%%.

\bibitem{Harlander:2012pb}
\hrefCMSnoop {} {R.~V. Harlander, S.~Liebler, and H.~Mantler, ``{SusHi: A
  program for the calculation of Higgs production in gluon fusion and
  bottom-quark annihilation in the Standard Model and the MSSM}'',} \textit{
  Comput. Phys. Commun.} \textbf{ 184} (2013) 1605--1617,
  \href{http://dx.doi.org/10.1016/j.cpc.2013.02.006}{\doi{10.1016/j.cpc.2013.02.006}},
\href{http://www.arXiv.org/abs/1212.3249}{\texttt{ arXiv:1212.3249}}.
%%CITATION = ARXIV:1212.3249;%%.

\bibitem{Harlander:2016hcx}
\hrefCMSnoop {} {R.~V. Harlander, S.~Liebler, and H.~Mantler, ``{SusHi Bento:
  Beyond NNLO and the heavy-top limit}'',} \textit{ Comput. Phys. Commun.}
  \textbf{ 212} (2017) 239--257,
  \href{http://dx.doi.org/10.1016/j.cpc.2016.10.015}{\doi{10.1016/j.cpc.2016.10.015}},
\href{http://www.arXiv.org/abs/1605.03190}{\texttt{ arXiv:1605.03190}}.
%%CITATION = ARXIV:1605.03190;%%.

\bibitem{Bernreuther:2018ynm}
\hrefCMSnoop {} {W.~Bernreuther, L.~Chen, and Z.-G. Si, ``{Differential decay
  rates of CP-even and CP-odd Higgs bosons to top and bottom quarks at NNLO
  QCD}'',} \textit{ JHEP} \textbf{ 07} (2018) 159,
  \href{http://dx.doi.org/10.1007/JHEP07(2018)159}{\doi{10.1007/JHEP07(2018)159}},
\href{http://www.arXiv.org/abs/1805.06658}{\texttt{ arXiv:1805.06658}}.
%%CITATION = ARXIV:1805.06658;%%.

\bibitem{Nason:1987xz}
\hrefCMSnoop {} {P.~Nason, S.~Dawson, and R.~K. Ellis, ``{The Total
  Cross-Section for the Production of Heavy Quarks in Hadronic Collisions}'',}
  \textit{ Nucl. Phys.} \textbf{ B303} (1988) 607--633,
\href{http://dx.doi.org/10.1016/0550-3213(88)90422-1}{\doi{10.1016/0550-3213(88)90422-1}}.
%%CITATION = NUPHA,B303,607;%%.

\bibitem{Beenakker:1988bq}
\hrefCMSnoop {} {W.~Beenakker, H.~Kuijf, W.~L. van Neerven, and J.~Smith,
  ``{QCD Corrections to Heavy Quark Production in p anti-p Collisions}'',}
  \textit{ Phys. Rev.} \textbf{ D40} (1989) 54--82,
\href{http://dx.doi.org/10.1103/PhysRevD.40.54}{\doi{10.1103/PhysRevD.40.54}}.
%%CITATION = PHRVA,D40,54;%%.

\bibitem{Czakon:2013goa}
\hrefCMSnoop {} {M.~Czakon, P.~Fiedler, and A.~Mitov, ``{Total Top-Quark
  Pair-Production Cross Section at Hadron Colliders Through
  $O(\frac{4}{S})$}'',} \textit{ Phys. Rev. Lett.} \textbf{ 110} (2013) 252004,
  \href{http://dx.doi.org/10.1103/PhysRevLett.110.252004}{\doi{10.1103/PhysRevLett.110.252004}},
\href{http://www.arXiv.org/abs/1303.6254}{\texttt{ arXiv:1303.6254}}.
%%CITATION = ARXIV:1303.6254;%%.

\bibitem{BuarqueFranzosi:2017jrj}
\hrefCMSnoop {} {D.~Buarque~Franzosi, E.~Vryonidou, and C.~Zhang, ``{Scalar
  production and decay to top quarks including interference effects at NLO in
  QCD in an EFT approach}'',} \textit{ JHEP} \textbf{ 10} (2017) 096,
  \href{http://dx.doi.org/10.1007/JHEP10(2017)096}{\doi{10.1007/JHEP10(2017)096}},
\href{http://www.arXiv.org/abs/1707.06760}{\texttt{ arXiv:1707.06760}}.
%%CITATION = ARXIV:1707.06760;%%.

\bibitem{Khachatryan:2016vau}
\hrefCMSnoop {} {{ ATLAS, CMS} Collaboration, ``{Measurements of the Higgs
  boson production and decay rates and constraints on its couplings from a
  combined ATLAS and CMS analysis of the LHC pp collision data at $ \sqrt{s}=7
  $ and 8 TeV}'',} \textit{ JHEP} \textbf{ 08} (2016) 045,
  \href{http://dx.doi.org/10.1007/JHEP08(2016)045}{\doi{10.1007/JHEP08(2016)045}},
\href{http://www.arXiv.org/abs/1606.02266}{\texttt{ arXiv:1606.02266}}.
%%CITATION = ARXIV:1606.02266;%%.

\bibitem{Sirunyan:2018koj}
\hrefCMSnoop {} {{ CMS} Collaboration, ``{Combined measurements of Higgs boson
  couplings in proton-proton collisions at $\sqrt{s}=$ 13 TeV}'',} \textit{
  Submitted to: Eur. Phys. J.} (2018)
\href{http://www.arXiv.org/abs/1809.10733}{\texttt{ arXiv:1809.10733}}.
%%CITATION = ARXIV:1809.10733;%%.

\bibitem{Bernon:2015qea}
J.~Bernon\hrefCMSnoop {} { {et~al.}, ``{Scrutinizing the alignment limit in
  two-Higgs-doublet models: mh=125 GeV}'',} \textit{ Phys. Rev.} \textbf{ D92}
  (2015), no.~7, 075004,
  \href{http://dx.doi.org/10.1103/PhysRevD.92.075004}{\doi{10.1103/PhysRevD.92.075004}},
\href{http://www.arXiv.org/abs/1507.00933}{\texttt{ arXiv:1507.00933}}.
%%CITATION = ARXIV:1507.00933;%%.

\bibitem{Okada:1990vk}
\hrefCMSnoop {} {Y.~Okada, M.~Yamaguchi, and T.~Yanagida, ``{Upper bound of the
  lightest Higgs boson mass in the minimal supersymmetric standard model}'',}
  \textit{ Prog. Theor. Phys.} \textbf{ 85} (1991) 1--6,
\href{http://dx.doi.org/10.1143/ptp/85.1.1}{\doi{10.1143/ptp/85.1.1}}.
%%CITATION = PTPKA,85,1;%%.

\bibitem{Ellis:1990nz}
\hrefCMSnoop {} {J.~R. Ellis, G.~Ridolfi, and F.~Zwirner, ``{Radiative
  corrections to the masses of supersymmetric Higgs bosons}'',} \textit{ Phys.
  Lett.} \textbf{ B257} (1991) 83--91,
\href{http://dx.doi.org/10.1016/0370-2693(91)90863-L}{\doi{10.1016/0370-2693(91)90863-L}}.
%%CITATION = PHLTA,B257,83;%%.

\bibitem{Haber:1990aw}
\hrefCMSnoop {} {H.~E. Haber and R.~Hempfling, ``{Can the mass of the lightest
  Higgs boson of the minimal supersymmetric model be larger than m(Z)?}'',}
  \textit{ Phys. Rev. Lett.} \textbf{ 66} (1991) 1815--1818,
\href{http://dx.doi.org/10.1103/PhysRevLett.66.1815}{\doi{10.1103/PhysRevLett.66.1815}}.
%%CITATION = PRLTA,66,1815;%%.

\bibitem{Chankowski:1991md}
\hrefCMSnoop {} {P.~H. Chankowski, S.~Pokorski, and J.~Rosiek, ``{Charged and
  neutral supersymmetric Higgs boson masses: Complete one loop analysis}'',}
  \textit{ Phys. Lett.} \textbf{ B274} (1992) 191--198,
\href{http://dx.doi.org/10.1016/0370-2693(92)90522-6}{\doi{10.1016/0370-2693(92)90522-6}}.
%%CITATION = PHLTA,B274,191;%%.

\bibitem{Bagnaschi:2015hka}
\hrefCMSnoop {} {E.~Bagnaschi {et~al.},
``{Benchmark scenarios for low $\tan \beta$ in the MSSM}'',}.
%%CITATION = LHCHXSWG-2015-002;%%.

\bibitem{Gunion:1989we}
\hrefCMSnoop {} {J.~F. Gunion, H.~E. Haber, G.~L. Kane, and S.~Dawson, ``{The
  Higgs Hunter's Guide}'',} \textit{ Front. Phys.} \textbf{ 80} (2000)
1--404.
%%CITATION = FRPHA,80,1;%%.

\bibitem{Djouadi:1998az}
\hrefCMSnoop {} {A.~Djouadi, ``{Squark effects on Higgs boson production and
  decay at the LHC}'',} \textit{ Phys. Lett.} \textbf{ B435} (1998) 101--108,
  \href{http://dx.doi.org/10.1016/S0370-2693(98)00784-9}{\doi{10.1016/S0370-2693(98)00784-9}},
\href{http://www.arXiv.org/abs/hep-ph/9806315}{\texttt{ arXiv:hep-ph/9806315}}.
%%CITATION = HEP-PH/9806315;%%.

\bibitem{Dawson:1996xz}
\hrefCMSnoop {} {S.~Dawson, A.~Djouadi, and M.~Spira, ``{QCD corrections to
  SUSY Higgs production: The Role of squark loops}'',} \textit{ Phys. Rev.
  Lett.} \textbf{ 77} (1996) 16--19,
  \href{http://dx.doi.org/10.1103/PhysRevLett.77.16}{\doi{10.1103/PhysRevLett.77.16}},
\href{http://www.arXiv.org/abs/hep-ph/9603423}{\texttt{ arXiv:hep-ph/9603423}}.
%%CITATION = HEP-PH/9603423;%%.

\bibitem{Harlander:2004tp}
\hrefCMSnoop {} {R.~V. Harlander and M.~Steinhauser, ``{Supersymmetric Higgs
  production in gluon fusion at next-to-leading order}'',} \textit{ JHEP}
  \textbf{ 09} (2004) 066,
  \href{http://dx.doi.org/10.1088/1126-6708/2004/09/066}{\doi{10.1088/1126-6708/2004/09/066}},
\href{http://www.arXiv.org/abs/hep-ph/0409010}{\texttt{ arXiv:hep-ph/0409010}}.
%%CITATION = HEP-PH/0409010;%%.

\bibitem{Muhlleitner:2006wx}
\hrefCMSnoop {} {M.~Muhlleitner and M.~Spira, ``{Higgs Boson Production via
  Gluon Fusion: Squark Loops at NLO QCD}'',} \textit{ Nucl. Phys.} \textbf{
  B790} (2008) 1--27,
  \href{http://dx.doi.org/10.1016/j.nuclphysb.2007.08.011}{\doi{10.1016/j.nuclphysb.2007.08.011}},
\href{http://www.arXiv.org/abs/hep-ph/0612254}{\texttt{ arXiv:hep-ph/0612254}}.
%%CITATION = HEP-PH/0612254;%%.

\bibitem{MSSM_LHCHXS-lightstop-updated}
\hrefCMSnoop {} {}\url{https://sushi.hepforge.org/examples.html}.

\bibitem{Carena:2013ytb}
M.~Carena\hrefCMSnoop {} { {et~al.}, ``{MSSM Higgs Boson Searches at the LHC:
  Benchmark Scenarios after the Discovery of a Higgs-like Particle}'',}
  \textit{ Eur. Phys. J.} \textbf{ C73} (2013), no.~9, 2552,
  \href{http://dx.doi.org/10.1140/epjc/s10052-013-2552-1}{\doi{10.1140/epjc/s10052-013-2552-1}},
\href{http://www.arXiv.org/abs/1302.7033}{\texttt{ arXiv:1302.7033}}.
%%CITATION = ARXIV:1302.7033;%%.

\bibitem{Bahl:2018zmf}
H.~Bahl\hrefCMSnoop {} { {et~al.}, ``{MSSM Higgs Boson Searches at the LHC:
  Benchmark Scenarios for Run 2 and Beyond}'',}
\href{http://www.arXiv.org/abs/1808.07542}{\texttt{ arXiv:1808.07542}}.
%%CITATION = ARXIV:1808.07542;%%.

\bibitem{Sirunyan:2018zut}
\hrefCMSnoop {} {{ CMS} Collaboration, ``{Search for additional neutral MSSM
  Higgs bosons in the $\tau\tau$ final state in proton-proton collisions at
  $\sqrt{s}=$ 13 TeV}'',} \textit{ JHEP} \textbf{ 09} (2018) 007,
  \href{http://dx.doi.org/10.1007/JHEP09(2018)007}{\doi{10.1007/JHEP09(2018)007}},
\href{http://www.arXiv.org/abs/1803.06553}{\texttt{ arXiv:1803.06553}}.
%%CITATION = ARXIV:1803.06553;%%.

\bibitem{ATLAShMSSM}
\hrefCMSnoop {}
  {}\url{https://atlas.web.cern.ch/Atlas/GROUPS/PHYSICS/CombinedSummaryPlots/HIGGS/ATLAS_HIGGS5100_BSM_hMSSM_tanb_vs_mA_Summary/ATLAS_HIGGS5100_BSM_hMSSM_tanb_vs_mA_Summary.pdf}.
  \href{http://dx.doi.org/10.5281/zenodo.2485983}{\doi{10.5281/zenodo.2485983}}.

\bibitem{Aaboud:2017hnm}
\hrefCMSnoop {} {{ ATLAS} Collaboration, ``Search for heavy {H}iggs bosons
  {$A/H$} decaying to a top quark pair in $pp$ collisions at $\sqrt{s} =
  8$\,{TeV} with the {ATLAS} detector'',} \textit{ Phys. Rev. Lett.} \textbf{
  119} (2017), no.~19, 191803,
  \href{http://dx.doi.org/10.1103/PhysRevLett.119.191803}{\doi{10.1103/PhysRevLett.119.191803}},
\href{http://www.arXiv.org/abs/1707.06025}{\texttt{ arXiv:1707.06025}}.
%%CITATION = ARXIV:1707.06025;%%.

\bibitem{Arbey:2017gmh}
\hrefCMSnoop {} {A.~Arbey, F.~Mahmoudi, O.~Stal, and T.~Stefaniak, ``{Status of
  the Charged Higgs Boson in Two Higgs Doublet Models}'',} \textit{ Eur. Phys.
  J.} \textbf{ C78} (2018), no.~3, 182,
  \href{http://dx.doi.org/10.1140/epjc/s10052-018-5651-1}{\doi{10.1140/epjc/s10052-018-5651-1}},
\href{http://www.arXiv.org/abs/1706.07414}{\texttt{ arXiv:1706.07414}}.
%%CITATION = ARXIV:1706.07414;%%.

\bibitem{analysis-code}
\hrefCMSnoop {} {}\url{https://github.com/andrey-popov/pheno-htt}.
  \href{http://dx.doi.org/10.5281/zenodo.2485983}{\doi{10.5281/zenodo.2485983}}.

\bibitem{Alwall:2014hca}
J.~Alwall\hrefCMSnoop {} { {et~al.}, ``The automated computation of tree-level
  and next-to-leading order differential cross sections, and their matching to
  parton shower simulations'',} \textit{ JHEP} \textbf{ 07} (2014) 079,
  \href{http://dx.doi.org/10.1007/JHEP07(2014)079}{\doi{10.1007/JHEP07(2014)079}},
\href{http://www.arXiv.org/abs/1405.0301}{\texttt{ arXiv:1405.0301}}.
%%CITATION = ARXIV:1405.0301;%%.

\bibitem{Butterworth:2015oua}
\hrefCMSnoop {} {J.~Butterworth {et~al.}, ``{PDF4LHC} recommendations for {LHC}
  {R}un {II}'',} \textit{ J. Phys. G} \textbf{ 43} (2016) 023001,
  \href{http://dx.doi.org/10.1088/0954-3899/43/2/023001}{\doi{10.1088/0954-3899/43/2/023001}},
\href{http://www.arXiv.org/abs/1510.03865}{\texttt{ arXiv:1510.03865}}.
%%CITATION = ARXIV:1510.03865;%%.

\bibitem{Buckley:2014ana}
A.~Buckley\hrefCMSnoop {} { {et~al.}, ``{LHAPDF}6: parton density access in the
  {LHC} precision era'',} \textit{ Eur. Phys. J. C} \textbf{ 75} (2015) 132,
  \href{http://dx.doi.org/10.1140/epjc/s10052-015-3318-8}{\doi{10.1140/epjc/s10052-015-3318-8}},
\href{http://www.arXiv.org/abs/1412.7420}{\texttt{ arXiv:1412.7420}}.
%%CITATION = ARXIV:1412.7420;%%.

\bibitem{Sjostrand:2007gs}
\hrefCMSnoop {} {T.~Sj{\"o}strand, S.~Mrenna, and P.~Skands, ``A brief
  introduction to {PYTHIA} 8.1'',} \textit{ Comp. Phys. Comm.} \textbf{ 178}
  (2008) 852,
  \href{http://dx.doi.org/10.1016/j.cpc.2008.01.036}{\doi{10.1016/j.cpc.2008.01.036}},
\href{http://www.arXiv.org/abs/0710.3820}{\texttt{ arXiv:0710.3820}}.
%%CITATION = HEP-PH/0603175;%%.

\bibitem{Skands:2014pea}
\hrefCMSnoop {} {P.~Skands, S.~Carrazza, and J.~Rojo, ``Tuning {PYTHIA} 8.1:
  the {M}onash 2013 tune'',} \textit{ Eur. Phys. J.} \textbf{ C74} (2014),
  no.~8, 3024,
  \href{http://dx.doi.org/10.1140/epjc/s10052-014-3024-y}{\doi{10.1140/epjc/s10052-014-3024-y}},
\href{http://www.arXiv.org/abs/1404.5630}{\texttt{ arXiv:1404.5630}}.
%%CITATION = ARXIV:1404.5630;%%.

\bibitem{Cacciari:2008gp}
\hrefCMSnoop {} {M.~Cacciari, G.~P. Salam, and G.~Soyez, ``The anti-$k_t$ jet
  clustering algorithm'',} \textit{ JHEP} \textbf{ 04} (2008) 063,
  \href{http://dx.doi.org/10.1088/1126-6708/2008/04/063}{\doi{10.1088/1126-6708/2008/04/063}},
\href{http://www.arXiv.org/abs/0802.1189}{\texttt{ arXiv:0802.1189}}.
%%CITATION = ARXIV:0802.1189.

\bibitem{deFavereau:2013fsa}
\hrefCMSnoop {} {{ DELPHES 3} Collaboration, ``{DELPHES}~3: {A} modular
  framework for fast simulation of a generic collider experiment'',} \textit{
  JHEP} \textbf{ 02} (2014) 057,
  \href{http://dx.doi.org/10.1007/JHEP02(2014)057}{\doi{10.1007/JHEP02(2014)057}},
\href{http://www.arXiv.org/abs/1307.6346}{\texttt{ arXiv:1307.6346}}.
%%CITATION = ARXIV:1307.6346;%%.

\bibitem{Czakon:2011xx}
\hrefCMSnoop {} {M.~Czakon and A.~Mitov, ``{T}op++: {A} program for the
  calculation of the top-pair cross-section at hadron colliders'',} \textit{
  Comput. Phys. Commun.} \textbf{ 185} (2014) 2930,
  \href{http://dx.doi.org/10.1016/j.cpc.2014.06.021}{\doi{10.1016/j.cpc.2014.06.021}},
\href{http://www.arXiv.org/abs/1112.5675}{\texttt{ arXiv:1112.5675}}.
%%CITATION = ARXIV:1112.5675;%%.

\bibitem{Sirunyan:2018fpa}
\hrefCMSnoop {} {{ CMS} Collaboration, ``Performance of the {CMS} muon detector
  and muon reconstruction with proton-proton collisions at $\sqrt{s} =
  13$\,TeV'',} \textit{ JINST} \textbf{ 13} (2018), no.~06, P06015,
  \href{http://dx.doi.org/10.1088/1748-0221/13/06/P06015}{\doi{10.1088/1748-0221/13/06/P06015}},
\href{http://www.arXiv.org/abs/1804.04528}{\texttt{ arXiv:1804.04528}}.
%%CITATION = ARXIV:1804.04528;%%.

\bibitem{CMS-DP-2018-017}
\hrefCMSnoop {} {{ {CMS}} Collaboration, ``Electron and photon performance in
  {CMS} with the full 2017 data sample and additional 2016 highlights for the
  {CALOR} 2018 conference'',} technical report, 2018.

\bibitem{CMS-DP-2018-030}
\hrefCMSnoop {} {{ {CMS}} Collaboration, ``Electron trigger performance in
  {CMS} with the full 2017 data sample'',} technical report, 2018.

\bibitem{Sirunyan:2017ezt}
\hrefCMSnoop {} {{ CMS} Collaboration, ``Identification of heavy-flavour jets
  with the {CMS} detector in $pp$~collisions at 13\,{TeV}'',} \textit{ JINST}
  \textbf{ 13} (2018), no.~05, P05011,
  \href{http://dx.doi.org/10.1088/1748-0221/13/05/P05011}{\doi{10.1088/1748-0221/13/05/P05011}},
\href{http://www.arXiv.org/abs/1712.07158}{\texttt{ arXiv:1712.07158}}.
%%CITATION = ARXIV:1712.07158;%%.

\bibitem{Khachatryan:2016kdb}
\hrefCMSnoop {} {{ CMS} Collaboration, ``Jet energy scale and resolution in the
  {CMS} experiment in pp~collisions at 8~{TeV}'',} \textit{ JINST} \textbf{ 12}
  (2017) P02014,
  \href{http://dx.doi.org/10.1088/1748-0221/12/02/P02014}{\doi{10.1088/1748-0221/12/02/P02014}},
\href{http://www.arXiv.org/abs/1607.03663}{\texttt{ arXiv:1607.03663}}.
%%CITATION = ARXIV:1607.03663;%%.

\bibitem{Khachatryan:2015hba}
\hrefCMSnoop {} {{ CMS} Collaboration, ``Measurement of the top quark mass
  using proton-proton data at $\sqrt{s} = 7$ and 8\,{TeV}'',} \textit{ Phys.
  Rev. D} \textbf{ 93} (2016), no.~7, 072004,
  \href{http://dx.doi.org/10.1103/PhysRevD.93.072004}{\doi{10.1103/PhysRevD.93.072004}},
\href{http://www.arXiv.org/abs/1509.04044}{\texttt{ arXiv:1509.04044}}.
%%CITATION = ARXIV:1509.04044;%%.

\bibitem{Cleveland79}
\hrefCMSnoop {} {W.~S. Cleveland, ``Robust locally weighted regression and
  smoothing scatterplots'',} \textit{ J. Am. Stat. Assoc.} \textbf{ 74} (1979)
  829, \href{http://dx.doi.org/10.2307/2286407}{\doi{10.2307/2286407}}.

\bibitem{Cleveland88}
\hrefCMSnoop {} {W.~S. Cleveland and S.~J. Devlin, ``Locally-weighted
  regression: {A}n approach to regression analysis by local fitting'',}
  \textit{ J. Am. Stat. Assoc.} \textbf{ 83} (1988) 596,
  \href{http://dx.doi.org/10.2307/2289282}{\doi{10.2307/2289282}}.

\bibitem{Moneta:2010pm}
L.~Moneta\hrefCMSnoop {} { {et~al.}, ``The {R}oo{S}tats project'',} in \textit{
  13$^\text{th}$ International Workshop on Advanced Computing and Analysis
  Techniques in Physics Research (ACAT2010)}, volume ACAT2010, p.~057.
\newblock 2010.
\newblock
\href{http://www.arXiv.org/abs/1009.1003}{\texttt{ arXiv:1009.1003}}.
\newblock
%%CITATION = ARXIV:1009.1003;%%.

\bibitem{Cranmer:2012sba}
\hrefCMSnoop {} {{ {ROOT}} Collaboration, ``{H}ist{F}actory: {A} tool for
  creating statistical models for use with {R}oo{F}it and {R}oo{S}tats'',}
  technical report, 2012.

\bibitem{Cowan:2010js}
\hrefCMSnoop {} {G.~Cowan, K.~Cranmer, E.~Gross, and O.~Vitells, ``Asymptotic
  formulae for likelihood-based tests of new physics'',} \textit{ Eur. Phys. J.
  C} \textbf{ 71} (2011) 1554,
  \href{http://dx.doi.org/10.1140/epjc/s10052-011-1554-0}{\doi{10.1140/epjc/s10052-011-1554-0}},
\href{http://www.arXiv.org/abs/1007.1727}{\texttt{ arXiv:1007.1727}}.
%%CITATION = ARXIV:1007.1727;%%.

\bibitem{Read:2002hq}
\hrefCMSnoop {} {A.~L. Read, ``Presentation of search results: {The} {$CL_s$}
  technique'',} \textit{ J. Phys. G} \textbf{ 28} (2002) 2693--2704,
\href{http://dx.doi.org/10.1088/0954-3899/28/10/313}{\doi{10.1088/0954-3899/28/10/313}}.
%%CITATION = JPAGA,G28,2693;%%.

\bibitem{Junk:1999kv}
\hrefCMSnoop {} {T.~Junk, ``Confidence level computation for combining searches
  with small statistics'',} \textit{ Nucl. Instrum. Meth. A} \textbf{ 434}
  (1999), no.~CARLETON-OPAL-PHYS-99-01, CERN-EP-99-041, 435--443,
  \href{http://dx.doi.org/10.1016/S0168-9002(99)00498-2}{\doi{10.1016/S0168-9002(99)00498-2}},
\href{http://www.arXiv.org/abs/hep-ex/9902006}{\texttt{ arXiv:hep-ex/9902006}}.
%%CITATION = HEP-EX/9902006;%%.

\bibitem{Djouadi:2006bz}
\hrefCMSnoop {} {A.~Djouadi, M.~M. Muhlleitner, and M.~Spira, ``{Decays of
  supersymmetric particles: The Program SUSY-HIT
  (SUspect-SdecaY-Hdecay-InTerface)}'',} \textit{ Acta Phys. Polon.} \textbf{
  B38} (2007) 635--644,
\href{http://www.arXiv.org/abs/hep-ph/0609292}{\texttt{ arXiv:hep-ph/0609292}}.
%%CITATION = HEP-PH/0609292;%%.

\bibitem{Djouadi:2018xqq}
\hrefCMSnoop {} {A.~Djouadi, J.~Kalinowski, M.~Muehlleitner, and M.~Spira,
  ``{HDECAY: Twenty$_{++}$ Years After}'',}
\href{http://www.arXiv.org/abs/1801.09506}{\texttt{ arXiv:1801.09506}}.
%%CITATION = ARXIV:1801.09506;%%.

\bibitem{Hahn:2013ria}
T.~Hahn\hrefCMSnoop {} { {et~al.}, ``{High-Precision Predictions for the Light
  CP -Even Higgs Boson Mass of the Minimal Supersymmetric Standard Model}'',}
  \textit{ Phys. Rev. Lett.} \textbf{ 112} (2014), no.~14, 141801,
  \href{http://dx.doi.org/10.1103/PhysRevLett.112.141801}{\doi{10.1103/PhysRevLett.112.141801}},
\href{http://www.arXiv.org/abs/1312.4937}{\texttt{ arXiv:1312.4937}}.
%%CITATION = ARXIV:1312.4937;%%.

\bibitem{Bahl:2018qog}
H.~Bahl\hrefCMSnoop {} { {et~al.}, ``{Precision calculations in the MSSM
  Higgs-boson sector with FeynHiggs 2.14}'',}
\href{http://www.arXiv.org/abs/1811.09073}{\texttt{ arXiv:1811.09073}}.
%%CITATION = ARXIV:1811.09073;%%.

\end{thebibliography}\endgroup

\end{document}